\documentclass[10pt,leqno]{amsart}
\usepackage{indentfirst,csquotes}

\usepackage{amssymb,amsthm,amsmath}
\usepackage{xcolor,paralist,hyperref,fancyhdr,etoolbox}
\usepackage{graphicx}
\usepackage{natbib}
\usepackage{threeparttable}
\usepackage{algorithm}
\usepackage{algorithmic}
\usepackage{booktabs}
\usepackage{multirow}

\newtheorem{theorem}{Theorem}[section]

\newtheorem{lemma}[theorem]{Lemma}

\newtheorem{corollary}[theorem]{Corollary}

\newtheorem{remark}[theorem]{Remark}

\newcommand{\diag}{\mathrm{diag}}

\newcommand{\Var}{\mathrm{Var}}
\newcommand{\E}{\mathrm{E}}
\DeclareMathOperator*{\argmin}{arg\,min}

\def\bs{\boldsymbol}

\usepackage{tabularx}
\usepackage{bm}

\theoremstyle{definition}
\newtheorem{assumpA}{Assumption}

\theoremstyle{plain}

\newcommand{\tri}{\triangle}
\newcommand{\Bz}{\mathbf{B}(\bs z)}
\newcommand{\Bzz}{\mathbf{B}(\bs z,\bs z')}
\newcommand{\IO}{\mathcal{I}_O}
\newcommand{\Ssp}{\mathbb{S}} 
\newcommand{\Stp}{\mathbb{S}^{\otimes 2}} 
\newcommand{\lam}{\lambda} 

\hypersetup{ colorlinks=true, linkcolor=black, filecolor=black, urlcolor=black, hypertexnames=false }

\begin{document}
\title[SSFR for Surrogate-Augmented Neuroimaging]{Synthetic Surrogate Functional Regression (SSFR): A Robust Framework for Estimating Spatially Varying Covariate Effects in Surrogate-Augmented Neuroimaging} 
\author[G. Wang]{Guannan Wang$^{1}$}
\author[Y. Long]{Yang Long$^{2}$}
\author[S. Yu]{Shan Yu$^{3}$}
\author[Z. Gu]{Zhiling Gu$^{4}$}
\author[L. Wang]{Lily Wang$^{5}$}

\thanks{$^{1}$Department of Mathematics, William \& Mary. Email: \href{mailto:gwang01@wm.edu}{gwang01@wm.edu}}
\thanks{$^{2}$Department of Biostatistics, Yale University. Email: \href{mailto:yang.long@yale.edu}{yang.long@yale.edu}}
\thanks{$^{3}$Department of Statistics, University of Virginia. Email: \href{mailto:sy5jx@virginia.edu}{sy5jx@virginia.edu}}
\thanks{$^{4}$Department of Statistics, University of South Carolina. Email: \href{mailto:zhiling@sc.edu}{zhiling@sc.edu}}
\thanks{$^{5}$Department of Statistics, George Mason University. Email: \href{mailto:lwang41@gmu.edu}{lwang41@gmu.edu}}

\date{\today}

\begin{abstract}
Understanding how covariates, such as demographic, clinical, and genetic variables, relate to high-dimensional spatial imaging outcomes is a fundamental problem in brain imaging studies, particularly when the target modality is observed for only a subset of participants. Although modern generative models can synthesize missing imaging modalities from auxiliary data, incorporating synthetic images into regression analysis requires careful statistical treatment to ensure valid inference. In this paper, we propose a two-stage Synthetic Surrogate Functional Regression (SSFR) framework, a statistical safeguard for incorporating synthetic neuroimages to learn spatially varying covariate effects in the presence of partially observed imaging modalities. By calibrating surrogate information against observed target images, SSFR improves estimation efficiency while remaining robust to discrepancies between the surrogate and target regression functions. We establish the consistency and convergence rate of the proposed estimator and show that first-stage surrogate-model estimation error does not alter its asymptotic rate. We also introduce a voxelwise surrogate explanatory measure $R^2(\bs{z})$ to quantify the fraction of residual target-image variation explained by a candidate surrogate. Together with the target-image observation rate, this measure provides a principled criterion for surrogate screening and provides an interpretable indication of potential efficiency gain. We further develop a weighted bootstrap procedure for constructing simultaneous confidence corridors (SCCs) for the coefficient functions. As surrogate informativeness increases, the SCC widths approach those obtained from complete observations, improving the localization of significant spatially varying covariate effects that cannot be identified using only the observed data. Extensive simulations demonstrate the numerical accuracy and robustness of the proposed methods, including under surrogate-model misspecification and when surrogates are generated by modern image-to-image neural networks. Applied to spatially normalized PET images from ADNI, SSFR yields interpretable coefficient maps and uniformly narrower SCCs than an analysis based only on observed PET, enlarging the region over which age-related metabolic decline is simultaneously significant.
\end{abstract}

\maketitle

\section{Introduction}
\label{SEC:introduction}

Modern neuroimaging studies routinely collect high-dimensional imaging outcomes together with demographic, clinical, and genetic covariates, creating new opportunities to investigate spatially varying associations between subject characteristics and brain imaging phenotypes \citep{zhu2023statistical, huang2017fgwas}. A persistent challenge is modality-level missingness: expensive imaging modalities such as positron emission tomography (PET) are often available for only a subset of participants, whereas structural MRI and clinical measurements are collected much more broadly. For example, in the Alzheimer's Disease Neuroimaging Initiative (ADNI) \citep{Jack2008}, PET is substantially less available than MRI, reducing the effective sample size for PET-based regression analyses.

Recent advances in statistical imputation and deep generative models have made it increasingly feasible to synthesize missing imaging modalities from auxiliary data \citep{he2011functional, ciarleglio2022elucidating, lu2025new, gopinath_synthetic_2024, sohn_learning_2015}. However, considerably less attention has been devoted to valid downstream regression and inference using these generated images. Treating synthetic images as if they were the true missing outcomes ignores potential systematic discrepancies between the generated and target modalities and can produce invalid statistical inference. Consequently, there remains a need for regression methods that can exploit synthetic images while preserving valid inference for the original imaging modality.

To address this problem, we propose a novel \emph{Synthetic Surrogate Functional Regression} (SSFR) framework for image-on-scalar regression with partially observed imaging outcomes under the missing-at-random (MAR) assumption. Throughout, \emph{synthetic} is used in its statistical sense: data generated from a model rather than directly observed \citep{Rubin1993}, so a synthetic surrogate may come from classical imputation, regression-based prediction, or modern generative models alike. Rather than treating such images as imputed observations, SSFR regards them as \emph{surrogate images}, that is, statistically informative proxies for the target imaging modality that provide auxiliary information but are not assumed to be identical to the missing images. The term ``surrogate'' emphasizes that these images enter as auxiliary outcomes rather than direct replacements for the missing responses, allowing SSFR to borrow information from them while preserving inference for the original imaging modality.

A key feature of SSFR is its two-stage estimation procedure. The first stage estimates surrogate coefficient functions using all available surrogate images, while the second stage calibrates the target regression model using the observed target images together with the surrogate-derived predictors. This calibration step corrects potential surrogate-model misspecification while efficiently transferring information from the surrogate data.

To accommodate the irregular geometry of neuroimaging domains, both stages of SSFR are estimated using trivariate penalized splines over triangulations (TPST) \citep{Li:etal:24}, which naturally respect complex anatomical boundaries. Scientific studies also require uncertainty quantification to identify spatial regions exhibiting significant covariate effects. We further develop a weighted bootstrap procedure for constructing simultaneous confidence corridors (SCCs) for the estimated coefficient functions and establish the consistency and convergence rate of the proposed estimator.

The contributions of this paper are fourfold. First, we propose SSFR, a surrogate-assisted image-on-scalar regression framework that remains robust to surrogate-model misspecification while efficiently incorporating partially observed imaging outcomes. Second, we introduce a voxelwise explanatory measure $R^2(\bs{z})$ to quantify surrogate informativeness and guide surrogate selection. Together with the observed-image fraction, this measure characterizes the potential variance reduction of SSFR relative to ISR. Third, we establish consistency and convergence rates for the SSFR estimator, showing that first-stage surrogate estimation does not affect the asymptotic convergence rate. Finally, we develop a weighted bootstrap procedure for SCCs that propagates uncertainty from both estimation stages. Simulation studies and the analysis of ADNI PET data demonstrate substantial efficiency gains over existing approaches while maintaining robustness under surrogate misspecification.

The remainder of this paper is organized as follows. Section \ref{SEC:data} describes the ADNI dataset, image preprocessing procedures, and surrogate image generation. Section \ref{SEC:SSFR} introduces the SSFR framework and the two-stage estimation method based on TPST, together with the voxelwise surrogate explanatory measure $R^2(\bs{z})$. Section \ref{SEC:theory} establishes the consistency and convergence rate of the SSFR estimator. Section \ref{SSEC:bootstrap} develops the weighted bootstrap procedure for SCCs. Section \ref{SEC:simulation} reports simulation results, and Section \ref{SEC:application} applies the proposed methods to the PET data from the ADNI study. Section \ref{SEC:conclusion} concludes the paper. Additional simulations, sensitivity analyses, and complete proofs are provided in the Supplementary Material.

\vspace{-6pt}
\section{ADNI Data, Preprocessing, and Synthetic PET Generation}
\label{SEC:data}

We analyze fluorodeoxyglucose (FDG) PET data from the Alzheimer's Disease Neuroimaging Initiative (ADNI) to study spatially varying associations between subject-level covariates and regional brain metabolism. Although FDG-PET provides a direct and biologically interpretable measure of neuronal activity, it is available for only a small subset of participants, resulting in substantial modality-level missingness. T1-weighted MRI, by contrast, is available for a much larger proportion of the cohort, and we therefore propose a cross-modal procedure that uses subject-specific MRI to generate synthetic PET images for individuals without PET scans. All imaging data are preprocessed to a common anatomical template, normalized for intensity, and standardized for downstream functional regression. This section provides an overview of the analytic cohort (Section \ref{SSEC:dataset}), the preprocessing pipeline (Section \ref{SSEC:preprocess}), and the synthetic PET generation procedure (Section \ref{SSEC:synthgen}).

\vspace{-6pt}
\subsection{Dataset Construction}
\label{SSEC:dataset}

To construct the analysis cohort, we identified ADNI participants with temporally matched FDG-PET and T1-weighted MRI acquired within 60 days of each other, yielding 663 MRI--PET visit pairs from 262 unique participants. All paired visits were used to train the MRI-to-PET synthesis network. For the downstream regression analysis, we retained the earliest paired visit from each participant and linked demographic, cognitive, and genetic covariates measured within the same 60-day window, resulting in 260 participants with complete PET, MRI, and covariate data. An additional 325 participants had baseline MRI and complete covariates but no matched PET scan; synthetic PET images were generated for these individuals and used as surrogate outcomes in the proposed SSFR framework.

\vspace{-6pt}
\subsection{Image Preprocessing}
\label{SSEC:preprocess}

\begin{figure}
\centering
\includegraphics[width=0.7\linewidth]{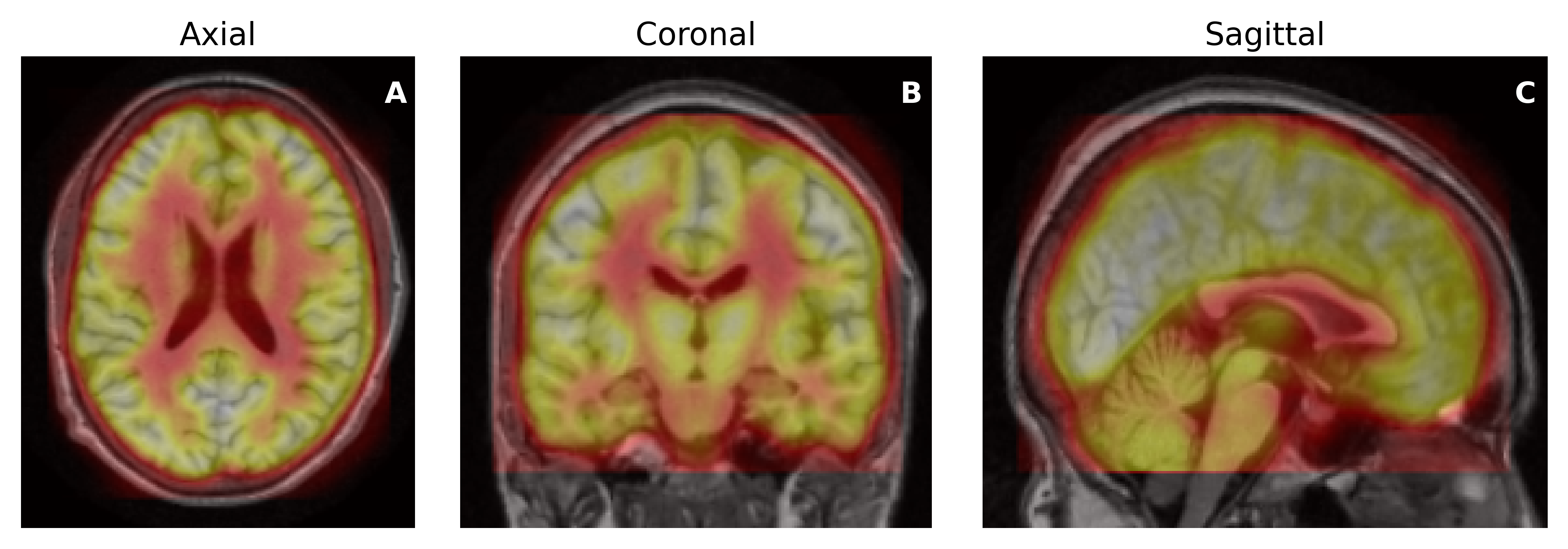} \vspace{-.2in}
\caption{PET and MRI images aligned to MNI space for a representative subject. Axial (a), coronal (b), and sagittal (c) slices of the PET image (warm colors) are overlaid on the corresponding grayscale T1-weighted MRI slices, demonstrating good anatomical correspondence.}
\label{FIG:align}
\end{figure}

All MRI, observed PET, and synthetic PET images are aligned to a common anatomical space following procedures standard in ADNI and multimodal deep-learning studies \citep{Jack2008, Wen2020}. Starting from the ADNI-provided NIfTI images, each MRI is reoriented and registered to the Montreal Neurological Institute (MNI) 152 template using a 12-parameter affine transformation. The corresponding PET scan is then resampled onto the same 2 mm isotropic grid using trilinear interpolation. MRI-only participants are processed using the same pipeline to ensure spatial compatibility. Figure \ref{FIG:align} illustrates the registration quality for a representative subject.

For intensity normalization, MRI volumes are first masked by thresholding nonzero voxels in MNI space, clipped to the 0.5th--99.5th percentiles, and then standardized with within-brain $z$-scoring, following established MRI standardization methods \citep{Nyul1999, Wen2020}. PET scans are normalized using the same brain mask: voxel intensities are clipped to the 1st--99th percentiles within the mask and divided by the 99th percentile, yielding normalized values primarily in $[0,1]$ \citep{Narazani2022, Zhao2023}. Subject-level train/validation/test splits (80\%/10\%/10\%) are defined on the set of 260 participants with at least one MRI--PET pair, ensuring that all visits from a given participant belong to the same split and avoiding information leakage \citep{Wen2020, Narazani2022}.

\vspace{-6pt}
\subsection{Synthetic Image Generation}
\label{SSEC:synthgen}

We train a compact 3D U-Net \citep{Ronneberger2015, Cicek2016} to learn a mapping from T1-weighted MRI to the temporally matched normalized PET image. This architecture, widely used in volumetric medical image segmentation and cross-modal synthesis \citep{Sikka2018, Chen2024, Lee2024}, consists of four downsampling and four upsampling blocks connected by skip pathways. To accommodate the four downsampling stages while preserving computational efficiency for a modest training sample size, each image is padded symmetrically so that all three spatial dimensions are divisible by $2^4 = 16$. Padding is removed after prediction to return volumes to the original MNI grid. Although the downstream statistical analysis focuses on baseline measurements, the MRI-to-PET synthesis network is trained using all available visit-level MRI--PET pairs. This strategy increases the effective training sample size and improves generalization of the learned cross-modal mapping. Visit time is not used as an input to the network, and visits are treated as independent training instances. To prevent information leakage across correlated longitudinal scans, data splitting is performed at the subject level so that all visits from a given participant are confined to a single split.

\begin{figure}
\centering
\includegraphics[width=0.7\linewidth]{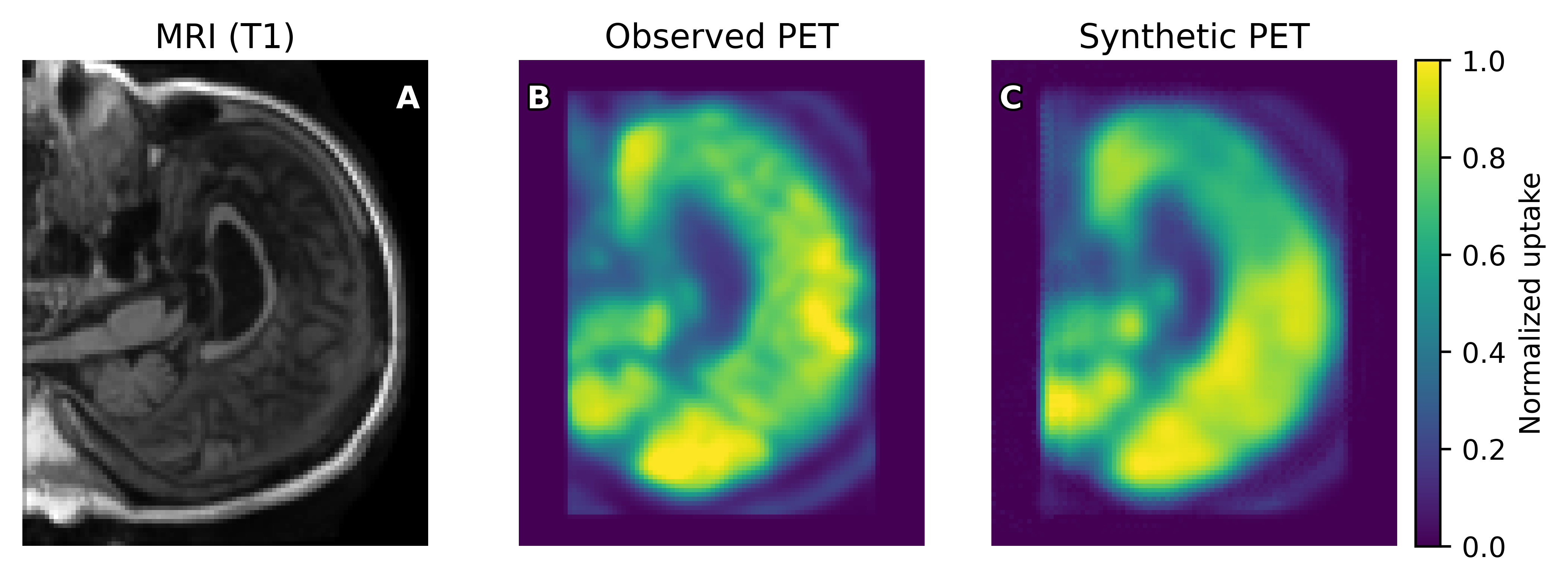}  \vspace{-.2in}
\caption{Observed MRI, observed PET, and synthetic PET for a representative subject. Sagittal T1-weighted MRI slice (a), observed PET slice (b), and synthetic PET slice predicted from the MRI (c). PET images are shown in normalized units with a common color scale.}
\label{FIG:pred}
\end{figure}

The network is trained to predict normalized PET using a voxel-wise mean absolute error (L1) loss, a standard choice in PET synthesis \citep{Sikka2018, Lee2024}. Optimization is performed using the Adam optimizer with a learning rate of $10^{-4}$ and a Reduce-on-Plateau schedule, with the final model selected based on the minimum validation loss. After training, the fitted model is applied to the baseline MRI of every participant to generate synthetic PET images, including both the participants whose scans were used for training and evaluation and the additional MRI-only participants. Once generated, these synthetic PET images are treated as realized surrogate inputs to the SSFR procedure. In the subsequent bootstrap analysis, these realized surrogate images are held fixed, while the two SSFR estimation stages are refitted across bootstrap replicates to quantify the sampling variability of the SSFR estimator. Figure \ref{FIG:pred} compares observed and synthetic PET for a representative subject.

\vspace{-6pt}
\section{Synthetic Surrogate Functional Regression (SSFR) and Its Estimation Method}
\label{SEC:SSFR}
\setlength\abovedisplayskip{4pt plus 3pt}
\setlength\belowdisplayskip{4pt plus 3pt}

This section introduces the SSFR framework for image-on-scalar regression with partially observed imaging outcomes. Rather than treating synthetic images as imputations of the missing responses, SSFR incorporates them as surrogate outcomes within a two-stage TPST-based estimation procedure, enabling efficient estimation while remaining robust to surrogate-model misspecification.

\vspace{-6pt}
\subsection{Joint Functional Regression Model}
\label{SSEC:model}

Let $\Omega \subset \mathbb{R}^3$ denote a bounded spatial domain representing the brain, and let $\bs{z}=(z_1,z_2,z_3)\in\Omega$ represent a spatial location. Consider a dataset of $n$ subjects with complete covariate information, and let $\mathcal{I}_O$ be the index set of subjects for whom the target imaging phenotype is observed, with $n_{O}= |\mathcal{I}_O|$ denoting its cardinality. For each subject $i\in\mathcal{I}_O$, let $Y_i(\bs{z})$ denote the measurement of the target imaging phenotype, and let $\mathbf{X}_i=(X_{i0},\ldots,X_{ip})^\top$ be the associated covariate vector, where $X_{i0} \equiv 1$. We first define the primary \textit{Image-on-Scalar Regression (ISR)} model as:
\begin{equation}
Y_i(\bs{z}) = \mathbf{X}_i^\top \bs{\alpha}_{0}(\bs{z}) + \varepsilon_{O,i}(\bs{z}), ~i\in\mathcal{I}_O, ~\bs{z}\in\Omega,
\label{model:isr}
\end{equation}
where $\bs{\alpha}_{0}(\bs{z}) = (\alpha_{0,0}(\bs{z}), \ldots, \alpha_{0,p}(\bs{z}))^\top$ is a vector of spatially varying coefficient functions, and $\varepsilon_{O,i}(\bs{z})$ is a mean-zero Gaussian process with covariance function $\Sigma_{OO}(\bs{z},\bs{z}')$.

In many imaging studies, the response $Y_i(\bs{z})$ is missing for a large subset of subjects, although covariates and auxiliary imaging modalities are available. To incorporate information from all $n$ subjects, we construct a synthetic surrogate image $\widetilde{Y}_i(\bs{z})$ for every individual using cross-modal prediction or generative AI; see Section \ref{SEC:data} for model details. These surrogate images are then described by the following naive surrogate ISR (SISR) model:
\begin{equation}
\widetilde{Y}_i(\bs{z}) = \mathbf{X}_i^\top \bs{\beta}_{0}(\bs{z}) + \varepsilon_{S,i}(\bs{z}), ~i=1,\ldots,n, ~\bs{z}\in\Omega,
\label{model:surrogate}
\end{equation}
where $\bs{\beta}_{0}(\bs{z})=(\beta_{0,0}(\bs{z}),\ldots,\beta_{0,p}(\bs{z}))^\top$ represents the spatially varying coefficients based on surrogate data and $\varepsilon_{S,i}(\bs{z})$ is the associated error process. The surrogate coefficient functions $\bs{\beta}_{0}(\bs{z})$ may differ from $\bs{\alpha}_{0}(\bs{z})$ due to approximation error or model misspecification in the surrogate generation process. The SSFR framework accounts for this discrepancy by combining information from both observed and surrogate images.

Combining \eqref{model:isr} and \eqref{model:surrogate}, we define the joint functional regression model:
\begin{equation}
\begin{bmatrix}
Y_i(\bs{z}) \\[4pt]
\widetilde{Y}_i(\bs{z})
\end{bmatrix}
=
\begin{bmatrix}
\mathbf{X}_i^\top \bs{\alpha}_0(\bs{z}) \\[4pt]
\mathbf{X}_i^\top \bs{\beta}_0(\bs{z})
\end{bmatrix}
+
\begin{bmatrix}
\varepsilon_{O,i}(\bs{z}) \\[4pt]
\varepsilon_{S,i}(\bs{z})
\end{bmatrix},
\label{EQ:model_joint}
\end{equation}
where $Y_i(\bs{z})$ is observed only for $i\in\mathcal{I}_O$, while $\widetilde{Y}_i(\bs{z})$ is available for all subjects. To account for shared structure between the true and surrogate images, we model the subject-level deviations as a bivariate Gaussian process:
\[
\begin{bmatrix}
\varepsilon_{O,i}(\bs{z}) \\[4pt]
\varepsilon_{S,i}(\bs{z}')
\end{bmatrix}
\sim
\mathcal{N}\!\left(
\mathbf{0},
\begin{bmatrix}
\Sigma_{OO}(\bs{z}, \bs{z}) & \Sigma_{OS}(\bs{z}, \bs{z}') \\[4pt]
\Sigma_{SO}(\bs{z}', \bs{z}) & \Sigma_{SS}(\bs{z}', \bs{z}')
\end{bmatrix}
\right).
\]
To borrow strength from the surrogate data, we compute the conditional expectation of $\varepsilon_{O,i}(\bs{z})$ given $\{\varepsilon_{S,i}(\bs{z}')\}_{\bs{z}'\in\Omega}$, yielding the co-kriging decomposition:
$\varepsilon_{O,i}(\bs{z}) = {\rm E}[\varepsilon_{O,i}(\bs{z}) \mid \{\varepsilon_{S,i}(\bs{z}')\}_{\bs{z}'\in\Omega}] + \zeta_i(\bs{z})$, where the conditional mean is the co-kriging predictor: ${\rm E}[\varepsilon_{O,i}(\bs{z}) \mid \cdot] = \int_{\Omega} U(\bs{z},\bs{z}') \varepsilon_{S,i}(\bs{z}')d\bs{z}'$. Here, $U(\bs{z},\cdot) = \Sigma_{OS}(\bs{z},\cdot)\,\Sigma_{SS}^{-1}$ is the co-kriging weight kernel, which links location $\bs{z}$ to every other location $\bs{z}'$, and the residual term $\zeta_i(\bs{z})$ is a mean-zero process capturing prediction uncertainty. Substituting $\varepsilon_{S,i}(\bs{z}')$ from \eqref{model:surrogate}, we obtain ${\rm E}[\varepsilon_{O,i}(\bs{z}) \mid \cdot] = \int_{\Omega} U(\bs{z},\bs{z}') \{\widetilde{Y}_i(\bs{z}') - \mathbf{X}_i^\top \bs{\beta}_0(\bs{z}')\} d\bs{z}'$. Substituting this back into \eqref{model:isr} gives the SSFR-adjusted model for subjects with observed imaging phenotype:
\[
Y_i(\bs{z}) = \mathbf{X}_i^\top \left\{ \bs{\alpha}_0(\bs{z}) - \int_{\Omega} U(\bs{z},\bs{z}') \bs{\beta}_0(\bs{z}') d\bs{z}' \right\} + \int_{\Omega} U(\bs{z},\bs{z}') \widetilde{Y}_i(\bs{z}') d\bs{z}' + \zeta_i(\bs{z}).
\]

Next, define the adjusted coefficient functions $\bs{\nu}_0(\bs{z})=\bs{\alpha}_0(\bs{z}) - \int_{\Omega} U(\bs{z},\bs{z}') \bs{\beta}_0(\bs{z}') d\bs{z}'$, and let $C(\bs{z}, \widetilde{Y}_i) = \int_{\Omega} U(\bs{z},\bs{z}') \widetilde{Y}_i(\bs{z}') d\bs{z}'$ denote the spatially weighted average of the surrogate image. Thus, we obtain the following surrogate-adjusted functional regression model:
\begin{equation}
Y_i(\bs{z}) = \mathbf{X}_i^\top \bs{\nu}_{0}(\bs{z}) + C(\bs{z}, \widetilde{Y}_i) + \zeta_i(\bs{z}),
\label{model:SSFR}
\end{equation}
where $\bs{\nu}_{0}(\bs{z})$ represents the covariate coefficient functions after adjustment for the surrogate contribution,
while $C(\bs{z}, \widetilde{Y}_i)$ provides a smoothed contribution from the surrogate image, and $\zeta_i(\bs{z})$ represents the residual spatial variation not captured by either source.

\vspace{-6pt}
\subsection{The Two-stage Estimation Method}
\label{SSEC:estimation}

Fitting the ISR model in \eqref{model:isr} directly using only the observed imaging data can be unstable, particularly when the number of observed scans is limited. On the other hand, relying solely on the naive SISR model in \eqref{model:surrogate}, which uses surrogate imaging phenotypes, may introduce bias if the surrogate generation model is misspecified. To address the limitations of both observed-data-only and imputation-based approaches, we propose a two-stage least squares (2SLS) estimation procedure. In the first stage, we fit the SISR model \eqref{model:surrogate} using the full sample of synthetic surrogate images to estimate the spatially varying coefficient functions $\bs{\beta}_0(\bs{z})$. In the second stage, we jointly estimate $\bs{\nu}_{0}(\bs{z})$ and $U(\bs z,\bs z')$ from the SSFR-adjusted model in \eqref{model:SSFR}, using the observed outcomes $Y_i(\bs{z})$, covariates $\mathbf{X}_i$, and the corresponding surrogate images $\widetilde{Y}_i(\bs{z})$. We then combine these estimates with the Stage-1 estimate of $\bs{\beta}_0(\bs{z})$ to recover $\bs\alpha_0$.

To estimate both models \eqref{model:surrogate} and \eqref{model:SSFR}, we employ trivariate penalized spline smoothing over a triangulated domain (TPST), which respects the irregular boundary of the brain. Let $\triangle = \{T_1, \ldots, T_K\}$ be a \textit{triangulation} of $\Omega$, that is, a partition $\Omega = \cup_{k=1}^K T_k$ into tetrahedra,  any pair of which meet at most at a shared vertex, edge, or face. We write $|\triangle|=\max_{T\in\triangle}\mathrm{diam}(T)$ for the mesh size of $\triangle$, that is, the length of its longest edge. Denote by $\mathcal{C}^r(\Omega)$ the set of $r$-times continuously differentiable functions over $\Omega$, where $r$ is a nonnegative integer. Given triangulation $\triangle$, the \textit{spline space of degree $d$ and smoothness $r$ over $\triangle$} is $\mathcal{S}_d^r(\triangle) = \{g \in \mathcal{C}^r(\Omega): g|_T \in \mathcal{P}_d(\bs{z}), \ T \in \triangle\}$, where $\mathcal{P}_d$ is the space of polynomials of total degree at most $d$. Let $\{B_m, m \in \mathcal{M}\}$ denote degree-$d$ trivariate Bernstein basis polynomials for $\mathcal{S}_d^r(\triangle)$. Any $g \in \mathcal{S}_d^r(\triangle)$ can be written as $g(\bs{z}) = \sum_{m \in \mathcal{M}} B_m(\bs{z}) \gamma_m = \mathbf{B}(\bs{z})^\top \bs{\gamma}$, where $\bs{\gamma}$ is the vector of spline coefficients. Continuity constraints between adjacent tetrahedra are enforced via linear constraints $\mathbf{H}\bs{\gamma}=\mathbf{0}$, with details determined by $d$, $r$, and $\triangle$. Owing to the imposed $\mathcal{C}^r(\Omega)$ continuity constraints, the fitted spline is single-valued at locations on shared tetrahedral faces, edges, or vertices, which therefore require no special treatment, while exterior brain-mask boundary points are handled as ordinary one-sided spline evaluations without extrapolation beyond the mask. Full technical details are provided in \cite{Li:etal:24}. Define $\mathcal{G}^{(p+1)}(\triangle) = \left\{\bs{g}=(g_0, \ldots, g_p)^{\top}, g_{\ell} \in \mathcal{S}_d^r(\triangle), \ell = 0, \ldots, p\right\}$.

\vspace{-3pt}
\subsubsection*{Estimation of SISR Model \eqref{model:surrogate}}
We estimate $\beta_{0,\ell}(\cdot)$, $\ell=0,\ldots,p$, in \eqref{model:surrogate} using a penalized spline approach that balances flexibility and smoothness. Specifically, let $D_{z_q}^2 g(\bs{z})$ be the second-order derivative in direction $z_q$, $q = 1, 2, 3$, and define the roughness penalty $\mathcal{E}(g)= \int_{\Omega} \left\{(D_{z_1}^2 g)^2 + (D_{z_2}^2g)^2 + (D_{z_{3}}^2 g)^2\right\}dz_1 dz_2 dz_3$. We consider the following penalized least squares problem:
\begin{equation*}
\label{EQ:objective}
\min_{(\beta_0, \ldots, \beta_p)^{\top} \in \mathcal{G}^{(p+1)}(\triangle)} \sum_{i=1}^n \sum_{j=1}^N \left\{\widetilde{Y}_i(\bs{z}_j) - \sum_{\ell=0}^p X_{i\ell}\beta_{\ell}(\bs{z}_j)\right\}^2 + \sum_{\ell=0}^p\lambda_{\ell}\mathcal{E}(\beta_{\ell}),
\end{equation*}
where $N$ denotes the number of spatial locations (voxels) in the imaging domain $\Omega$, and $\lambda_{\ell}$ is the roughness penalty parameter for the $\ell$th coefficient function.

To approximate each $\beta_{0,\ell}(\cdot)$, we use a common triangulation $\triangle$ and represent each function in the Bernstein-B\'ezier basis as: $\beta_{\ell}(\bs{z}_j) \approx \mathbf{B}_{\ell}^{\top}(\bs{z}_j)\bs{\gamma}_{\ell}$, where $\mathbf{B}_{\ell} = \{B_{\ell m}\}_{m \in \mathcal{M}_{\ell}}$ are the basis functions for $\mathcal{S}_{d_{\ell}}^r(\triangle)$, and $\bs{\gamma}_{\ell}$ is the spline coefficient vector. The smoothness constraints $\mathbf{H}_{\ell}\bs{\gamma}_{\ell} = \mathbf{0}$, $\ell = 0, \ldots, p$, can be removed by applying QR decomposition: $\mathbf{H}_{\ell}^{\top} = \mathbf{Q}_{\ell}\mathbf{R}_{\ell} = (\mathbf{Q}_{1\ell}~\mathbf{Q}_{2\ell}) \binom{\mathbf{R}_{1\ell}}{\mathbf{R}_{2\ell}}$, where $\mathbf{Q}_{\ell}$ is an orthogonal matrix and $\mathbf{R}_{\ell}$ is an upper triangular matrix. The reduced basis $\mathbf{Q}_{2\ell}$ spans the null space of $\mathbf{H}_{\ell}$. Re-parameterizing with $\bs{\gamma}_{\ell} = \mathbf{Q}_{2\ell}\bs{\theta}_{\ell}$ ensures $\mathbf{H}_{\ell}\bs{\gamma}_{\ell} = \mathbf{0}$. Substituting this representation into the penalized regression problem yields the following unconstrained minimization:
\begin{equation}
\sum_{i=1}^n \sum_{j=1}^N \left\{\widetilde{Y}_i(\bs{z}_j) - \sum_{\ell=0}^p X_{i\ell}\mathbf{B}_{\ell}^{\top}(\bs{z}_j) \mathbf{Q}_{2\ell}\bs{\theta}_{\ell}\right\}^2 + \sum_{\ell=0}^p \lambda_{\ell}\bs{\theta}_{\ell}^{\top}\mathbf{D}_{\ell}\bs{\theta}_{\ell},
\label{EQ:PLS}
\end{equation}
where $\mathbf{D}_{\ell} = \mathbf{Q}_{2\ell}^{\top}\mathbf{P}_{\ell}\mathbf{Q}_{2\ell}$ and $\mathbf{P}_{\ell}$ is the block diagonal penalty matrix encoding the roughness operator $\mathcal{E}(\cdot)$. Let $\widehat{\bs{\theta}} = (\widehat{\bs{\theta}}_0^{\top}, \ldots, \widehat{\bs{\theta}}_p^{\top})^{\top}$ be the minimizer of \eqref{EQ:PLS}. Then, the estimated coefficient functions are: $\widehat{\beta}_{\ell}(\bs{z}) = \mathbf{B}_{\ell}(\bs{z})^{\top}\widehat{\bs{\gamma}}_{\ell}$, where $\widehat{\bs{\gamma}}_{\ell} = \mathbf{Q}_{2\ell}\widehat{\bs{\theta}}_{\ell}$.

\vspace{-3pt}
\subsubsection*{Estimation of SSFR-Adjusted Model \eqref{model:SSFR}}

To estimate model \eqref{model:SSFR}, we employ TPST together with tensor-product spline smoothing. Specifically, we approximate $U(\bs{z},\bs{z}')$ by the tensor-product basis functions, ${\mathbf{B}(\bs{z},\bs{z}')= \mathbf{B}(\bs{z})\otimes\mathbf{B}(\bs{z}')=\{B_{m}(\bs{z})B_{m^{\prime}}(\bs{z}'), m,m^{\prime}\in \mathcal{M}\}^{\top}}$. Denote $C_{i,m,m'}(\bs{z})=B_{m}(\bs{z})\int_{\Omega}B_{m^{\prime}}(\bs{z}')\widetilde{Y}_i(\bs{z}')d\bs{z}'$, $m, m'\in \mathcal{M}$, $i=1,\ldots,n$, and $\mathbf{C}_{ij}=\mathbf{C}_{i}(\bs{z}_j)= \left\{C_{i,m,m'}(\bs{z}_j),m,m'\in \mathcal{M}\right\}^{\top}$ at voxel $\bs{z}_j$ for subject $i$. We approximate the vector of the coefficient functions $\bs{\nu}(\bs{z})$ in \eqref{model:SSFR} by trivariate splines as $\bs{\nu}(\bs{z})  \approx {\mathbf{I}_{p+1} \otimes} \mathbf{B}(\bs{z})^{\top} {\bs{a}}$, where $\bs{a}=(\bs{a}_{0}^{\top},\ldots,\bs{a}_{p}^{\top})^{\top}$ is the concatenated vector of spline coefficients. We approximate the kernel $U(\bs{z},\bs{z}')$ by $U(\bs{z},\bs{z}')\approx \mathbf{B}(\bs{z},\bs{z}')^{\top} {\bs{b}}$, where $\bs{b}$ is the coefficient vector.

Under the assumption that the imaging responses are MAR, the estimators of $\bs{a}$ and $\bs{b}$ can be obtained by minimizing the following weighted penalized least squares criterion:
\begin{equation}
L(\bs{a},\bs{b}) = \frac{1}{\sum_{i\in\mathcal{I}_O} w_i}\sum_{i\in\mathcal{I}_O} w_i\sum_{j=1}^N \left(Y_{ij} -\widetilde{\mathbf{X}}^{\top}_{ij}\bs{a} -\mathbf{C}_{ij}^{\top}\bs{b}\right)^{2} + \sum_{\ell=0}^p \lambda_{\ell}\bs{a}_{\ell}^{\top}\mathbf{P}_{\ell}\bs{a}_{\ell} + \tau_U\bs b^\top \bs b,
\label{EQ:LeastSquare}
\end{equation}
subject to smoothness constraints $\mathbf{H}_{\ell}\bs{a}_{\ell}=\mathbf{0}$, $\ell=0,\ldots,p$, and $(\mathbf{H} \otimes \mathbf{H}) \bs{b} = \mathbf{0}$, where $w_i$ is the weight for individual $i\in \mathcal{I}_O$, for example, the inverse of the inclusion probability, $\widetilde{\mathbf{X}}^{\top}_{ij} = \mathbf{X}_i^\top \otimes \mathbf{B}(\bs{z}_j)^{\top}$, and $\mathbf{D}_{\ell}$ is the penalty matrix associated with the smoothness of the $\ell$th coefficient function. 
Here $\tau_U>0$ controls ridge shrinkage of the tensor coefficient vector $\bs b$; the matrices $\mathbf P_\ell$ control roughness of the coefficient functions.
Minimizing (\ref{EQ:LeastSquare}), we obtain the estimator of $(\bs{a}^{\top},\bs{b}^{\top})^{\top}$. The spline estimators of $\bs{\nu}_{0}(\bs{z})$ and $U(\bs{z},\bs{z}')$ are given by: $\widehat{\bs{\nu}} (\bs{z}) = \mathbf{I}_{p+1} \otimes \mathbf{B}(\bs{z})^{\top} \widehat{\bs{a}}, ~\mbox{and}~\widehat{U}(\bs{z},\bs{z}') = \mathbf{B}(\bs{z},\bs{z}')^{\top} \widehat{\bs{b}}$. Finally, the 2SLS estimator of $\bs{\alpha}_{0}(\bs{z})$ is:
\begin{equation}
\widehat{\bs{\alpha}}^{\text{2SLS}}(\bs{z})=\widehat{\bs{\nu}}(\bs{z}) + \int_{\Omega} \widehat{U}(\bs{z},\bs{z}') \widehat{\bs{\beta}}(\bs{z}') d\bs{z}',
\label{EQ:alpha2sls}
\end{equation}
which combines the SSFR-adjusted estimate with a smoothed bias correction derived from the surrogate images. For brevity, we write $\widehat{\bs{\alpha}}(\bs{z})\equiv\widehat{\bs{\alpha}}^{\text{2SLS}}(\bs{z})$ in what follows.

Algorithm \ref{alg:SSFR} summarizes the complete two-stage estimation procedure.

\begin{algorithm}[!ht]
\caption{Two-Stage Synthetic Surrogate Functional Regression (SSFR) Estimation}
\label{alg:SSFR}
\begin{algorithmic}[1]
\STATE \textbf{Input:} Triangulation $\triangle$ of $\Omega$, weights $w_i$, covariates $\mathbf{X}_i$, synthetic surrogate images $\widetilde{Y}_i(\bs{z}_j)$, and observed imaging phenotype $Y_i(\bs{z}_j)$, $i=1,\ldots,n$, $j=1,\ldots,N$.
\STATE \textbf{Stage 1:} Fit the SISR model \eqref{model:surrogate} to $\{\widetilde{Y}_i(\bs{z}_j)\}_{i=1}^n$ using TPST as in \eqref{EQ:PLS}, obtaining $\widehat{\bs\beta}(\bs{z})$.
\STATE \textbf{Stage 2:} Fit the SSFR-adjusted model \eqref{model:SSFR} to $\{Y_i(\bs{z}_j)\}_{i\in\mathcal{I}_O}$ and $\{\widetilde{Y}_i(\bs{z}_j)\}_{i=1}^n$ via the weighted penalized least squares criterion \eqref{EQ:LeastSquare}, obtaining $\widehat{\bs\nu}(\bs{z})$ and $\widehat{U}(\bs{z},\bs{z}')$.
\STATE \textbf{Output:} The SSFR estimator $\widehat{\bs{\alpha}}(\bs{z})=\widehat{\bs{\nu}}(\bs{z}) + \int_{\Omega} \widehat{U}(\bs{z},\bs{z}') \widehat{\bs{\beta}}(\bs{z}') d\bs{z}'$.
\end{algorithmic}
\end{algorithm}

\vspace{-6pt}
\subsection{Measuring Surrogate Informativeness}
\label{SSEC:surrogate_measure}

The statistical efficiency gain from SSFR depends on the amount of residual information the surrogate contains about the target outcome after adjusting for the observed covariates. We therefore assess the informativeness of a candidate surrogate before incorporating it into SSFR. This assessment quantifies the potential efficiency gain and provides a principled criterion for selecting among candidate surrogate images generated from different imaging modalities or machine-learning algorithms.

Because the coefficient functions are estimated in the spline space $\mathcal{S}_d^r(\triangle)$, we assess surrogate informativeness after projecting images onto this space. Voxel-level variation orthogonal to $\mathcal{S}_d^r(\triangle)$ is removed by the TPST fits and does not affect the variance of the resulting estimators. Let $M = \dim\{\mathcal{S}_d^r(\triangle)\}$, let $\widetilde{\mathbf{B}}$ be the $N \times M$ matrix whose $j$th row is $\mathbf{B}(\bs{z}_j)^\top\mathbf{Q}_{2\ell}$, which is common to all $\ell$ under a shared triangulation. Define $\Pi = \widetilde{\mathbf{B}}(\widetilde{\mathbf{B}}^\top\widetilde{\mathbf{B}})^{-1}\widetilde{\mathbf{B}}^\top$, the projection onto $\mathcal{S}_d^r(\triangle)$, and for an image $g$, write $g^{\Pi}=\Pi g$.

Let $\widehat{\bs{t}}_i=(\widehat{t}_{i1},\ldots,\widehat{t}_{iL})^\top$ denote the leading $L$ functional principal component scores of the projected surrogate residuals $\widetilde{R}_i^{\Pi}$, $i=1,\ldots,n$, where $L$ is the rank of the surrogate representation used in the second stage ($L=M$ for the tensor-product kernel in \eqref{EQ:LeastSquare}). At each spatial location $\bs{z}$, let $\widehat{\varepsilon}_{O,i}(\bs{z})$ and $\widehat{\zeta}_i(\bs{z})$, $i\in\mathcal{I}_O$, denote the residuals from the weighted least-squares fits of $Y_i^{\Pi}(\bs{z})$ on $\mathbf{X}_i$ and on $(\mathbf{X}_i^\top,\widehat{\bs{t}}_i^\top)^\top$, respectively, using the weights $w_i$ in \eqref{EQ:LeastSquare}. Accounting for the additional $L$ degrees of freedom introduced by the surrogate scores, we define the local coefficient of partial determination:
\begin{equation}
\widehat{R}^{2}(\bs{z}) = 1 - \frac{ \left\{\sum_{i\in\mathcal{I}_O} w_i\widehat{\zeta}_i(\bs{z})^2\right\}/(n_O-p-1-L) }{ \left\{\sum_{i\in\mathcal{I}_O} w_i\widehat{\varepsilon}_{O,i}(\bs{z})^2\right\}/(n_O-p-1)},
\label{EQ:R2z}
\end{equation}
which estimates $R^2(\bs{z}) = 1 - { \Var\{\zeta_i^{\Pi}(\bs{z})\}}/{\Var\{\varepsilon_{O,i}^{\Pi}(\bs{z})\} }$,
the local proportion of residual target variation explained by the surrogate.

Aggregating across all $N$ voxels yields the variance-weighted image-level partial $R^{2}$:
\begin{equation}
\overline{R^2} = 1 - \frac{ \left\{\sum_{j=1}^{N}\sum_{i\in\mathcal{I}_O} w_i\widehat{\zeta}_i(\bs{z}_j)^2\right\}/(n_O-p-1-L) }{ \left\{\sum_{j=1}^{N}\sum_{i\in\mathcal{I}_O} w_i\widehat{\varepsilon}_{O,i}(\bs{z}_j)^2\right\}/(n_O-p-1) },
\label{EQ:R2bar}
\end{equation}
which summarizes this explanatory power over the entire imaging domain, assigning greater weight to spatial locations with larger residual variation. Larger values indicate that the surrogate explains a greater fraction of the residual target variation and is therefore expected to yield larger efficiency gains in SSFR.

The efficiency gain of SSFR arises from two sources: (i) residual variance reduction, since the conditional target residual covariance satisfies $\Sigma_{\zeta} = \Sigma_{OO} - \Sigma_{OS}\Sigma_{SS}^{-1}\Sigma_{SO} \preceq \Sigma_{OO}$, and (ii) effective sample size amplification, since the surrogate coefficient component is estimated from all $n$ subjects while the discrepancy adjustment uses only the $n_O$ subjects with observed target images. Let $\pi=n_O/n$ denote the observed-image fraction. The corresponding approximate variance and mean integrated squared error (MISE) ratios are $1-(1-\pi)\widehat{R}^{2}(\bs{z})$ and $1-(1-\pi)\overline{R^2}$, respectively. These expressions show how surrogate explanatory power and the proportion of observed target images jointly determine the potential efficiency gain. When all target images are observed, $\pi=1$, these ratios equal one by construction, and $\overline{R^2}$ should be interpreted as a descriptive measure of surrogate explanatory power rather than as a predicted missing-data efficiency gain.

To provide an interpretable benchmark for the potential efficiency gain, we use the following heuristic variance approximation. Suppose that the missingness mechanism is MCAR, the second-stage surrogate representation has fixed rank $L$, the smoothing-bias terms are negligible, and the relevant least-squares degrees of freedom satisfy $n_O-p-L-2>0$. We therefore use the following approximate variance relationship:
\begin{equation}
\frac{ \Var\{\widehat{\alpha}_{\ell}(\bs{z})\} }{ \Var\{\widehat{\alpha}_{\ell,{\rm ISR}}(\bs{z})\} } \approx \{1-R^2(\bs{z})\}c_L+\pi R^2(\bs{z}), \qquad c_L=\frac{n_O-p-2}{n_O-p-L-2},
\label{EQ:VR}
\end{equation}
Here $R^2(z)$ measures the fraction of residual target variation explained by the surrogate, $\pi=n_O/n$, and $c_L$ represents the finite-sample degrees-of-freedom cost of estimating the $L$ surrogate-score coefficients. Equation \eqref{EQ:VR} can be used as a qualitative and approximate guide to the expected efficiency pattern. Under MAR and inverse-probability weighting, the same expression should be regarded as heuristic rather than exact. Related discussion of surrogate-induced residual covariance reduction is provided in Supplement Section C.7.


\vspace{-6pt}
\section{Theoretical Properties}
\label{SEC:theory}

This section establishes the consistency and convergence rate of the SSFR estimator $\widehat{\bs\alpha}(\bs z)$ defined in \eqref{EQ:alpha2sls}. A central question is whether estimation error incurred in the first-stage SISR fit $\widehat{\bs\beta}$ propagates into, and potentially inflates the convergence rate of, the second-stage estimator $\widehat{\bs\alpha}$. We show that under a mild nondegeneracy condition on the surrogate residual process, it does not: $\widehat{\bs\alpha}$ attains the same rate as an oracle estimator that has access to the true $\bs\beta_0$. Complete proofs are given in Supplement Section C. Here we summarize the conditions informally, state the results, and discuss their interpretation.

The formal assumptions are collected in Supplement Section C.1; they impose standard regularity on the triangulation, coefficient functions, and error processes, together with two conditions specific to SSFR: the surrogate deviation must be exogenous and nondegenerate on the spline space (Assumption A5), which identifies the second-stage kernel $U$, and missingness must be MAR with bounded inverse-probability weights (Assumption A6). Assumptions A1--A4 and A7 drive the first-stage estimator; Assumptions A5, A6 and A8 additionally govern the second stage under MAR.

\vspace{-6pt}
\subsection{Consistency and Convergence Rate}

We first record the rate of the first-stage TPST estimator.  \vspace{-3pt}
\begin{theorem}[First-stage consistency]
\label{thm:stage1}
Under Assumptions A1--A4 and A7, the first-stage estimator $\widehat{\bs\beta}$ from \eqref{EQ:PLS} satisfies $\|\widehat\beta_\ell-\beta_{0,\ell}\|_{L_2(\Omega)}^2 = O_p\left(n^{-1}+|\triangle|^{2(d+1)}\right)$, for each $\ell=0,\ldots,p$.
\end{theorem}
Theorem \ref{thm:stage1} follows from the first-stage assumptions listed above and is the standard nonparametric rate for penalized trivariate spline regression over a triangulation \citep{Li:etal:24}: the first term reflects sampling variability across the $n$ subjects, and the second is the approximation bias governed by the spline degree $d$ and the mesh size $|\triangle|$. When $|\triangle|=O(n^{-1/(2(d+1))})$, the two terms balance and the rate simplifies to $O_p(n^{-1})$ for the squared $L_2$ error, i.e., $O_p(n^{-1/2})$ for the $L_2$ error itself. Our main result characterizes how this first-stage error propagates into the second-stage SSFR estimator.  \vspace{-3pt}

\begin{theorem}[Consistency and rate of the SSFR estimator]
\label{thm:stage2}
Under Assumptions A1--A8, the SSFR estimator $\widehat{\bs\alpha}$ from \eqref{EQ:alpha2sls} satisfies, for each $\ell=0,\ldots,p$, $\|\widehat\alpha_\ell-\alpha_{0,\ell}\|_{L_2(\Omega)}^2 = O_p\!\left(n_O^{-1}+|\triangle|^{2(d+1)}\right)$. In particular, the rate is of the same order as that of an oracle estimator constructed with the true surrogate coefficient functions $\bs\beta_0$ known in advance; the first-stage estimation error contributes only to the leading constant, not to the rate.
\end{theorem} \vspace{-3pt}

The proof (Supplement Section C.6) decomposes $\widehat{\bs\alpha}-\bs\alpha_0$ into a second-stage sampling error of order $n_O^{-1/2}+|\triangle|^{d+1}$ and a first-stage propagation term; Assumption A5 ensures the latter is of the same order, so the two balance rather than compound.

\begin{remark}[Role of the surrogate explanatory measure]
\label{rem:surrogate-efficiency}
While Theorem \ref{thm:stage2} shows that the \emph{convergence rate} of $\widehat{\bs\alpha}$ does not depend on surrogate informativeness, the \emph{leading constant} in that rate may depend on. Heuristically, in the co-kriging decomposition of Section \ref{SSEC:model}, the variance reduction achieved by conditioning on the surrogate residual process is governed by the strength of the residual association between the target and surrogate images, which is quantified by the voxelwise partial $R^2$, $R^2(\bs{z})$ in \eqref{EQ:R2z}: a surrogate with $R^2(\bs{z})$ near zero contributes little beyond the observed-data-only estimator, whereas larger values indicate greater potential for variance reduction. The simulation results in Section \ref{SEC:simulation}, in which estimation error decreases monotonically as the surrogate--target association increases, are consistent with this heuristic. The approximate variance relationship in \eqref{EQ:VR} provides an interpretable summary of this connection under the simplifying conditions stated in Section 3.3.
\end{remark} 

\begin{corollary}[Robustness to surrogate misspecification]
\label{cor:misspec}
Theorem~\ref{thm:stage2} holds regardless of whether $\bs\beta_0=\bs\alpha_0$; that is, $\widehat{\bs\alpha}$ remains consistent for $\bs\alpha_0$ at the stated rate even when the surrogate generation model is misspecified, provided Assumption A5 holds for the (possibly biased) surrogate residual process. This distinguishes SSFR from the naive SISR-imputation estimator, which is consistent for $\bs\alpha_0$ only if $\bs\beta_0=\bs\alpha_0$.
\end{corollary}  \vspace{-6pt}

Corollary \ref{cor:misspec} is the theoretical counterpart of the misspecified-surrogate simulation results in Section \ref{SEC:simulation}: because the second stage recalibrates the surrogate-derived predictor against the observed outcomes rather than substituting $\widetilde Y_i$ directly for $Y_i$, a biased surrogate degrades efficiency (through reduced residual explanatory power) but does not introduce bias into $\widehat{\bs\alpha}$.  

\vspace{-6pt}
\section{Weighted Bootstrap SCCs}
\label{SSEC:bootstrap}

Beyond point estimation, statistical inference for the spatially varying coefficient functions $\alpha_{0,\ell}(\bs{z})$, $\ell=0,\ldots,p$, is essential for interpreting SSFR results in practice. In many neuroimaging applications, the scientific objective is not only to estimate covariate effects, but also to identify spatial regions that exhibit significant associations with clinical variables, disease status, or demographic factors. To facilitate such analyses, we construct SCCs that provide simultaneous coverage over the entire imaging domain $\Omega$. Our inferential target is the SSFR estimator conditional on the realized surrogate images; thus, the bootstrap quantifies the sampling variability induced by the SSFR estimation procedure. Unlike pointwise confidence intervals, SCCs allow researchers to make global statements regarding the spatial extent and shape of significant regions while properly accounting for multiplicity across voxels.

Constructing SCCs for SSFR presents a specific challenge because the estimator is obtained through a two-stage procedure involving surrogate-model estimation in Stage 1 and inverse-probability-weighted regression in Stage 2. As a result, deriving an explicit closed-form asymptotic covariance function for $\widehat{\bs\alpha}$ is analytically cumbersome, particularly because Theorem \ref{thm:stage2} shows that the two stages contribute comparably to the estimator's sampling variability. We therefore develop a weighted bootstrap procedure that jointly propagates the sampling variability from both SSFR stages while treating the realized surrogate images as fixed inputs, without requiring an explicit sandwich-type variance estimator or a separate delta-method correction for the first stage.

\vspace{-6pt}
\subsection{Weighted Bootstrap Procedure}
\label{SSSEC:bootstrap-procedure}

Let $B$ denote the number of bootstrap replicates. We generate subject-level bootstrap multipliers $u_i^{(b)} \overset{\mathrm{iid}}{\sim} \mathrm{Exp}(1)$, for $i=1,\ldots,n$ and $b=1,\ldots,B$. For each bootstrap replicate $b$, we refit both stages of the SSFR procedure with these multipliers applied multiplicatively to the per-subject loss contributions: the Stage-1 objective in \eqref{EQ:PLS} for $\bs\gamma_\ell$ is reweighted by $u_i^{(b)}$ for each subject $i=1,\ldots,n$, and the Stage-2 objective in \eqref{EQ:LeastSquare} for $(\bs a,\bs b)$ is reweighted by $w_iu_i^{(b)}$ in place of $w_i$ for every subject $i\in\mathcal I_O$. Denote the resulting replicate estimators by $\widehat{\bs\beta}^{(b)}(\bs z)$, $\widehat{\bs\nu}^{(b)}(\bs z)$, and $\widehat U^{(b)}(\bs z,\bs z')$, obtained with the TPST procedure of Section \ref{SSEC:estimation}. The bootstrap replicate of the SSFR estimator is then
\begin{equation}
\widehat{\bs\alpha}^{(b)}(\bs z) = \widehat{\bs\nu}^{(b)}(\bs z) + \int_\Omega \widehat U^{(b)}(\bs z,\bs z')\,\widehat{\bs\beta}^{(b)}(\bs z')\,d\bs z', \qquad b=1,\ldots,B.
\label{EQ:alphaboot}
\end{equation}

A key design feature is that the same multiplier $u_i^{(b)}$ is applied to subject $i$ in both the Stage-1 and Stage-2 refits, with the Stage-2 weight given by the product $w_iu_i^{(b)}$. Because the realized surrogate images enter both stages of SSFR, the common subject-level multiplier preserves the within-subject dependence between the surrogate-model estimation error and the subsequent SSFR adjustment identified in Theorem \ref{thm:stage2}. Consequently, the empirical bootstrap distribution of $\widehat{\bs\alpha}^{(b)}(\bs z)$ captures the joint sampling variability of the two-stage SSFR estimator, conditional on the realized surrogate images, without requiring a separate covariance-propagation step.

Given the bootstrap replicates $\{\widehat{\bs\alpha}^{(b)}(\bs z)\}_{b=1}^B$, we construct SCCs separately for each coefficient function $\alpha_{0,\ell}(\bs z)$, $\ell=0,\ldots,p$. We first estimate the pointwise bootstrap SE
$\widehat\sigma_\ell(\bs z) = \mathrm{sd}_b \{\widehat\alpha_\ell^{(b)}(\bs z)\}$, $\bs z\in\Omega$. We then compute the studentized supremum statistic, for $b=1,\ldots,B$, $M_\ell^{(b)} = \sup_{\bs z\in\Omega}\{\widehat\sigma_\ell(\bs z)\}^{-1}{|\widehat\alpha_\ell^{(b)}(\bs z)-\widehat\alpha_\ell(\bs z)|}$. Let $\widehat q_\ell$ denote the empirical $(1-\alpha_0)$ quantile of $\{M_\ell^{(b)}\}_{b=1}^B$. The $(1-\alpha_0)$ SCC for $\alpha_{0,\ell}(\bs z)$ is then
\begin{equation}
\left[\widehat\alpha_\ell(\bs z)-\widehat q_\ell\widehat\sigma_\ell(\bs z),\; \widehat\alpha_\ell(\bs z)+\widehat q_\ell\widehat\sigma_\ell(\bs z)\right], \qquad \bs z\in\Omega.
\label{EQ:SCC}
\end{equation}
Algorithm \ref{alg:SCC} summarizes the complete procedure.

\begin{algorithm}[!ht]
\caption{Weighted Bootstrap SCCs for SSFR}
\label{alg:SCC}
\begin{algorithmic}[1]
\STATE \textbf{Input:} SSFR inputs as in Algorithm \ref{alg:SSFR}, the SSFR estimator $\widehat{\bs\alpha}(\bs z)$, the number of bootstrap replicates $B$, and the nominal level $1-\alpha_0$.
\STATE Draw independent multipliers $u_i^{(b)} \overset{\text{iid}}{\sim} \mathrm{Exp}(1)$ for $i=1,\ldots,n$ and $b=1,\ldots,B$.
\FOR{$b = 1, \ldots, B$}
    \STATE Refit the Stage-1 SISR model \eqref{model:surrogate} with per-subject weights $u_i^{(b)}$ to obtain $\widehat{\bs\beta}^{(b)}(\bs z)$, as in \eqref{EQ:PLS}.
    \STATE Refit the Stage-2 SSFR-adjusted model \eqref{EQ:LeastSquare} with per-subject weights $w_iu_i^{(b)}$ replacing $w_i$ to obtain $\widehat{\bs\nu}^{(b)}(\bs z)$ and $\widehat U^{(b)}(\bs z,\bs z')$.
    \STATE Form the bootstrap SSFR estimator $\widehat{\bs\alpha}^{(b)}(\bs z) = \widehat{\bs\nu}^{(b)}(\bs z) + \int_{\Omega} \widehat{U}^{(b)}(\bs{z},\bs{z}') \widehat{\bs{\beta}}^{(b)}(\bs{z}') d\bs{z}'$.
\ENDFOR
\FOR{$\ell = 0, \ldots, p$}
    \STATE Compute $\widehat{\sigma}_{\ell}(\bs{z}) = \mathrm{sd}_b\bigl\{\widehat{\alpha}_{\ell}^{(b)}(\bs{z})\bigr\}_{b=1}^B$.
    \STATE Compute $M_{\ell}^{(b)} = \sup_{\bs{z}\in\Omega} \{\widehat{\sigma}_{\ell}(\bs{z})\}^{-1}{\bigl| \widehat{\alpha}_{\ell}^{(b)}(\bs{z}) - \widehat{\alpha}_{\ell}(\bs{z}) \bigr|}$, $b=1,\ldots,B$.
    \STATE Compute the critical value $\widehat{q}_{\ell}$ as the empirical $(1-\alpha_0)$ quantile of $\{M_{\ell}^{(b)}\}_{b=1}^B$.
\ENDFOR
\STATE \textbf{Output:} For each $\ell = 0, \ldots, p$, the $(1-\alpha_0)$ SCCs 
$\bigl[\widehat{\alpha}_{\ell}(\bs{z}) - \widehat{q}_{\ell}\widehat{\sigma}_{\ell}(\bs{z}),~ \widehat{\alpha}_{\ell}(\bs{z}) + \widehat{q}_{\ell}\widehat{\sigma}_{\ell}(\bs{z})\bigr]$, $\quad\bs{z}\in\Omega$.
\end{algorithmic}
\end{algorithm}

The finite-sample performance of the resulting SCCs is evaluated empirically in Section \ref{SEC:simulation}, where the procedure attains coverage close to the nominal level while translating the efficiency gain from an informative surrogate, as measured by $R^2(\bs{z})$ in \eqref{EQ:VR}, into SCCs substantially narrower than those from the observed-data-only estimator.

\vspace{-6pt}
\subsection{Hyperparameter Selection and Implementation Considerations}
\label{SSEC:implementation}

The proposed SSFR method involves several hyperparameters that affect estimation accuracy and computational efficiency, including the triangulation of the domain, the specification of the spline basis, and the selection of roughness and tensor ridge penalty parameters. This section summarizes their roles and outlines the practical choices adopted in our implementation.

The penalty parameters are selected using generalized cross-validation (GCV) following \cite{Yu:etal:21}, with joint selection of the roughness and ridge parameters in Stage 2. The performance of the proposed SSFR framework also depends on the choice of the triangulation and TPST basis. Ideally, tetrahedra should be well shaped and, as suggested in \cite{Li:etal:24}, triangulations should be sufficiently fine to capture the dominant spatial variation, while remaining coarse enough to keep computation feasible. Given a fixed triangulation, the spline degree $d$ and smoothness parameter $r$ control the flexibility and smoothness of the spline approximation. The parameter $r$ governs continuity across adjacent tetrahedra; in our numerical studies, we set $r = 1$, which provides a reasonable balance between smoothness and flexibility. The degree $d$ regulates the flexibility of the estimator: lower values yield smoother fits, whereas higher values preserve local structure but may increase the risk of overfitting. In practice, $d$ may be specified by the user or selected using GCV. Sensitivity analyses of the effects of triangulation resolution and spline degree are reported in the Supplementary Material.

\vspace{-6pt}
\section{Simulation Studies}
\label{SEC:simulation}
 
In this section, we conduct Monte Carlo experiments to evaluate the finite-sample performance of the proposed SSFR methodology. We first study estimation accuracy and simultaneous inference under an analytically generated surrogate image, whose informativeness and degree of misspecification can be controlled directly (Sections \ref{SSEC:sim-design}--\ref{SSEC:sim-scc}). We then examine whether the same conclusions persist when the surrogate is instead produced by modern image-to-image generative models, including a setting in which surrogate quality varies across a clinically relevant subgroup in a way that a pooled diagnostic could otherwise mask (Section \ref{SSEC:sim-ai-summary}; reported in full in Supplement Section A.2).
 
\vspace{-6pt}
\subsection{Simulation Design}
\label{SSEC:sim-design}

We evaluate SSFR under varying missing-data mechanisms, levels of missingness, surrogate-model misspecification, and surrogate informativeness.

Following \cite{Wang:etal:24a}, we consider a ball domain $\Omega = B(0,1/2) = \{\bs{z}\in\mathbb{R}^3:\|\bs{z}\|_2\leq1/2\}$, and simulate a dataset of $n$ subjects with complete scalar covariate information, while allowing imaging responses to be missing for a subset of subjects. For all subjects, we set $X_{i0}=1$ and generate two scalar covariates $\{X_{i1},X_{i2}\}$ independently as $X_{i1}\sim\mathrm{Bernoulli}(0.4)$ and $X_{i2}\sim\mathcal{N}(0,1)$. For every subject, we generate a target image and a surrogate image jointly; the missingness mechanism then determines which target images are observed. The observed image is generated by $Y_i(\bs{z}_j)=\mathbf{X}_i^\top\bs{\alpha}_{0}(\bs{z}_j)+\eta_i^{(O)}(\bs{z}_j)+\varepsilon_{ij}^{(O)}$, whereas the surrogate is generated by $\widetilde{Y}_i(\bs{z}_j)=\mathbf{X}_i^\top\bs{\beta}_0(\bs{z}_j)+\eta_i^{(S)}(\bs{z}_j)+\varepsilon_{ij}^{(S)}$, with measurement errors having marginal distribution $\mathcal{N}(0,0.5^2)$. For each subject and voxel, the target and surrogate measurement errors have correlation $\omega$, the same parameter used for the latent-score correlation below; the paired errors are independent across subjects and voxels. Both subject-specific processes are built on the same three orthonormal eigenfunctions, $\eta_i^{(\cdot)}(\bs{z})=\sum_{k=1}^3\lambda_k^{1/2}\xi_{ik}^{(\cdot)}\psi_k(\bs{z})$; the eigenvalues, the $\psi_k$, and the true coefficient functions $\bs{\alpha}_0$ are given in Supplement Section B.1.1, and Supplement Figure B.2(a)--(c) displays $\bs{\alpha}_0$. Target--surrogate dependence is controlled through both the measurement errors and the latent scores: for each $k=1,2,3$, $(\xi_{ik}^{(O)},\xi_{ik}^{(S)})^{\top}\sim\mathcal{N}(\mathbf{0},\Sigma_{\omega})$, where $\Sigma_{\omega}$ has unit variances and correlation $\omega$, so that a larger $\omega$ yields a more informative surrogate. We report the average estimated image-level partial $\overline{R^2}$ (Section \ref{SSEC:surrogate_measure}) across the simulation settings. We consider both correctly specified surrogates ($\bs\beta_0=\bs\alpha_0$) and misspecified surrogates with a different coefficient surface, given in Supplement Section B.1.1.
Let $\mathcal{I}_O$ and $\mathcal{I}_S$ denote the index sets of subjects with observed and missing imaging responses, respectively. We consider both MCAR and MAR mechanisms; only MAR is reported here, with MCAR in Supplement Section B.2.2. Under MAR the observation indicator $R_{Y_i}$ follows a logistic model in $X_{i1}$ and $X_{i2}$ alone, as in \cite{ciarleglio2022elucidating}, with the intercept calibrated to a marginal missing rate $r_m$ varying from $25\%$ to $75\%$; we set $\mathcal{I}_O=\{i:R_{Y_i}=1\}$ and $\mathcal{I}_S=\{i:R_{Y_i}=0\}$. To account for MAR selection in the Stage-2 estimation of SSFR, we incorporate inverse probability weighting (IPW) using subject-level weights $w_i \propto 1/\Pr(R_{Y_i}=1\mid X_{i1},X_{i2})$ for $i\in\mathcal{I}_O$.
 
We compare the proposed SSFR method with four alternative approaches: (i) ISR without surrogates, (ii) naive SISR, which directly combines the observed images and synthetically generated surrogate images as the response, (iii) inverse probability weighted ISR (IPW), and (iv) augmented inverse probability weighted ISR (AIPW). IPW and AIPW represent standard robust and doubly robust strategies for handling incomplete outcomes under MAR. 
Specifically, IPW reweights complete-case observations using inverse propensity scores, while AIPW combines an outcome regression fitted to the observed target images with an inverse-probability-weighted residual correction. The augmented responses are then used to fit ISR to the full sample, as detailed in Supplement Section B.1.1.
All methods are implemented using the TPST framework. The domain $\Omega$ is partitioned into 176 tetrahedra, and splines of degree $d=2$ with smoothness parameter $r=1$ are employed. Performance is assessed using the MISE of the estimated coefficient functions, defined as the Monte Carlo average of the integrated squared error $\int_{\bs{z} \in \Omega} \{\widehat{\alpha}_{\ell}(\bs{z}) - \alpha_{0,\ell}(\bs{z})\}^2 d\bs{z}$. We present one representative simulation design in the main paper, under both correctly specified and misspecified surrogates; additional results, including MCAR and sensitivity analyses for triangulation resolution and spline degree, are provided in the Supplementary Material.

\vspace{-6pt}
\subsection{Estimation Accuracy under Correctly Specified and Misspecified Surrogates}
\label{SSEC:sim-results}

\begin{table}
\caption{MISEs of the coefficient functions for different methods with a misspecified surrogate model under MAR. \label{TAB:eg1-mar-mise2}}
\centering
\begin{threeparttable}
\scalebox{0.75}{\setlength{\tabcolsep}{3pt}\begin{tabular}
{ccccccccccccccccc} \toprule
\multirow{2}{*}{$r_m$} &\multirow{2}{*}{$\overline{R^2}$} &\multicolumn{3}{c}{ISR} &\multicolumn{3}{c}{IPW} &\multicolumn{3}{c}{AIPW} &\multicolumn{3}{c}{SISR} &\multicolumn{3}{c}{SSFR}\\ \cmidrule(lr){3-5}\cmidrule(lr){6-8}\cmidrule(lr){9-11} \cmidrule(lr){12-14}\cmidrule(lr){15-17}
& &$\alpha_0$ &$\alpha_1$ &$\alpha_2$ &$\alpha_0$ &$\alpha_1$ &$\alpha_2$ &$\alpha_0$ &$\alpha_1$ &$\alpha_2$ &$\alpha_0$ &$\alpha_1$ &$\alpha_2$ &$\alpha_0$ &$\alpha_1$ &$\alpha_2$\\ \midrule
\multicolumn{17}{c}{$n=200$}\\ \midrule
\multirow{4}{*}{75\%} &0.06 &\multirow{4}{*}{0.0116} &\multirow{4}{*}{0.0314} &\multirow{4}{*}{0.0062} &\multirow{4}{*}{0.0113} &\multirow{4}{*}{0.0306} &\multirow{4}{*}{0.0064} &0.0113 &0.0306 &0.0064 &\multirow{4}{*}{0.5749} &\multirow{4}{*}{0.1524} &\multirow{4}{*}{0.9560} &0.0138&0.0364&0.0084\\
&0.25 & & & & & & &0.0113 &0.0308 &0.0064 & & & &0.0114&0.0302&0.0072\\
&0.56 & & & & & & &0.0114 &0.0310 &0.0065 & & & &0.0077&0.0207&0.0050\\
&0.81 & & & & & & &0.0115 &0.0313 &0.0066 & & & &0.0048&0.0131&0.0031\\ \midrule
\multirow{4}{*}{50\%} &0.06 &\multirow{4}{*}{0.0061} &\multirow{4}{*}{0.0156} &\multirow{4}{*}{0.0037} &\multirow{4}{*}{0.0060} &\multirow{4}{*}{0.0159} &\multirow{4}{*}{0.0036} &0.0061 &0.0159 &0.0036 &\multirow{4}{*}{0.2040} &\multirow{4}{*}{0.1054} &\multirow{4}{*}{0.3947} &0.0062&0.0174&0.0038\\
&0.25 & & & & & & &0.0060 &0.0156 &0.0035 & & & &0.0056&0.0159&0.0035\\
&0.56 & & & & & & &0.0060 &0.0154 &0.0034 & & & &0.0045&0.0125&0.0028\\
&0.81 & & & & & & &0.0060 &0.0152 &0.0033 & & & &0.0035&0.0096&0.0022\\ \midrule
\multirow{4}{*}{25\%} &0.06 &\multirow{4}{*}{0.0039} &\multirow{4}{*}{0.0114} &\multirow{4}{*}{0.0029} &\multirow{4}{*}{0.0039} &\multirow{4}{*}{0.0115} &\multirow{4}{*}{0.0029} &0.0039 &0.0115 &0.0029 &\multirow{4}{*}{0.0473} &\multirow{4}{*}{0.0482} &\multirow{4}{*}{0.0924} &0.0039&0.0110&0.0028\\
&0.25 & & & & & & &0.0039 &0.0113 &0.0029 & & & &0.0036&0.0101&0.0026\\
&0.56 & & & & & & &0.0038 &0.0111 &0.0029 & & & &0.0032&0.0087&0.0022\\
&0.81 & & & & & & &0.0038 &0.0109 &0.0028 & & & &0.0029&0.0076&0.0020\\ \midrule
\multicolumn{17}{c}{$n=400$}\\ \midrule
\multirow{4}{*}{75\%} &0.06 &\multirow{4}{*}{0.0057} &\multirow{4}{*}{0.0183} &\multirow{4}{*}{0.0040} &\multirow{4}{*}{0.0057} &\multirow{4}{*}{0.0184} &\multirow{4}{*}{0.0041} &0.0057 &0.0184 &0.0041 &\multirow{4}{*}{0.5491} &\multirow{4}{*}{0.1487} &\multirow{4}{*}{0.9405} &0.0053&0.0177&0.0043\\
&0.25 & & & & & & &0.0057 &0.0186 &0.0041 & & & &0.0043&0.0150&0.0037\\
&0.56 & & & & & & &0.0057 &0.0188 &0.0041 & & & &0.0029&0.0105&0.0026\\
&0.81 & & & & & & &0.0056 &0.0188 &0.0041 & & & &0.0019&0.0067&0.0017\\ \midrule
\multirow{4}{*}{50\%} &0.07 &\multirow{4}{*}{0.0031} &\multirow{4}{*}{0.0076} &\multirow{4}{*}{0.0020} &\multirow{4}{*}{0.0031} &\multirow{4}{*}{0.0077} &\multirow{4}{*}{0.0021} &0.0031 &0.0077 &0.0021 &\multirow{4}{*}{0.2095} &\multirow{4}{*}{0.0834} &\multirow{4}{*}{0.3870} &0.0030&0.0078&0.0020\\
&0.25 & & & & & & &0.0030 &0.0076 &0.0021 & & & &0.0026&0.0069&0.0018\\
&0.56 & & & & & & &0.0029 &0.0076 &0.0021 & & & &0.0020&0.0055&0.0015\\
&0.81 & & & & & & &0.0028 &0.0076 &0.0021 & & & &0.0016&0.0044&0.0012\\ \midrule
\multirow{4}{*}{25\%} &0.06 &\multirow{4}{*}{0.0019} &\multirow{4}{*}{0.0049} &\multirow{4}{*}{0.0013} &\multirow{4}{*}{0.0019} &\multirow{4}{*}{0.0049} &\multirow{4}{*}{0.0013} &0.0019 &0.0049 &0.0013 &\multirow{4}{*}{0.0445} &\multirow{4}{*}{0.0322} &\multirow{4}{*}{0.0918} &0.0020&0.0049&0.0013\\
&0.25 & & & & & & &0.0019 &0.0049 &0.0013 & & & &0.0019&0.0046&0.0012\\
&0.56 & & & & & & &0.0018 &0.0048 &0.0013 & & & &0.0017&0.0042&0.0011\\
&0.81 & & & & & & &0.0017 &0.0048 &0.0013 & & & &0.0015&0.0039&0.0011\\ \bottomrule
\end{tabular}
}
\end{threeparttable}
\end{table}

Table \ref{TAB:eg1-mar-mise2} reports the average MISEs of the estimated coefficient functions under a misspecified surrogate model. The corresponding results for correctly specified surrogates are provided in Supplementary Table B.2. Across different missing rates, levels of surrogate informativeness, as measured by $\overline{R^2}$, and sample sizes, reducing the missing-data rate and increasing the sample size improve estimation accuracy for both ISR and SSFR. IPW and AIPW perform similarly to ISR. This behavior is expected because, under the MAR mechanism considered here, missingness depends only on fully observed covariates and the outcome model is correctly specified, so correcting the missing-data mechanism alone yields limited efficiency gains. In contrast, SSFR achieves substantially lower MISEs whenever the surrogate contains moderate to strong explanatory power, demonstrating that its efficiency gains arise primarily from effectively leveraging informative surrogate images rather than solely correcting for missing-data bias.

When the surrogate model is correctly specified, the detailed results in Supplementary Table B.2 show that SISR remains largely insensitive to missingness because the surrogate images closely approximate the target images under MAR. In contrast, under surrogate misspecification (Table \ref{TAB:eg1-mar-mise2}), the synthetic images are systematically biased, and directly combining them with the observed images, as in SISR, produces substantial error inflation and highly unstable estimates. This instability is most pronounced under the 75\% missing scenario, where SISR produces MISEs that are orders of magnitude larger than those of ISR, IPW, AIPW, and SSFR. It persists even at moderate missing rates, highlighting SISR's vulnerability to surrogate misspecification. These findings agree with Corollary \ref{cor:misspec}: SSFR remains stable because its second stage recalibrates the surrogate against the observed images rather than directly substituting the surrogate for the missing outcome. Moreover, higher levels of surrogate explanatory power, as measured by $\overline{R^2}$, consistently improve the estimation accuracy of SSFR, with the largest gains occurring when missingness is severe and the surrogate is moderately to highly informative.

Overall, since IPW and AIPW remain close to ISR despite explicitly accounting for MAR, the primary advantage of SSFR stems from exploiting surrogate information rather than correcting selection bias. Unlike naive SISR, SSFR remains robust to misspecification by recalibrating against observed outcomes. MCAR results in the Supplementary Material are qualitatively consistent, with SSFR gains increasing with the surrogate explanatory power measured by $\overline{R^2}$, particularly under severe missingness.

\vspace{-6pt}
\subsection{Simultaneous Confidence Corridors (SCCs)}
\label{SSEC:sim-scc}

Beyond point estimation, we evaluate the finite-sample performance of the proposed weighted bootstrap procedure for constructing SCCs. To facilitate comparison with the estimation results above, we adopt the same simulation setting as in the preceding MAR design, with a 50\% missing rate and sample size $n=400$. In this simulation, we consider the more challenging setting of a misspecified surrogate model. To investigate how the explanatory power of the surrogate affects simultaneous inference, we consider data-generating score correlations $\omega\in \{0.10,0.25,0.50,0.75,0.90\}$. These settings yield average estimated image-level explanatory measures $\overline{R^2}=0.01, 0.07, 0.25, 0.56,$ and $0.81$, respectively, representing progressively greater fractions of residual target-image variation explained by the surrogate. For each Monte Carlo replication, SCCs are constructed based on three estimators: \textit{Oracle}, an ISR estimator fitted to the complete primary imaging responses from all 400 subjects, serving as an unattainable full-data benchmark; \textit{Observed}, an ISR estimator fitted using only subjects with observed primary imaging responses and thus without leveraging surrogate information; and \textit{SSFR}, the proposed two-stage surrogate-assisted estimator that incorporates information from the surrogate images.
 
For each setting, we conduct 100 Monte Carlo replications. Within each replication, SCCs for all three estimators are constructed using the proposed weighted multiplier bootstrap procedure with $B=500$ bootstrap samples. Table \ref{TAB:scc} summarizes the resulting empirical coverage probabilities, average corridor widths, and empirical SEs under the misspecified surrogate model.

\begin{table}[ht!]
\centering
\caption{Performance of bootstrap SCCs under MAR with 50\% missingness ($n=400$), across five levels of surrogate explanatory power ($\overline{R^2}$). Evaluation is performed after removing two boundary layers.}
\label{TAB:scc}
\scalebox{0.75}{\begin{tabular}{clccccccccc}
\toprule
\multirow{2}{*}{$\overline{R^2}$} & \multirow{2}{*}{Method} & \multicolumn{3}{c}{Coverage} & \multicolumn{3}{c}{Average Width} & \multicolumn{3}{c}{Empirical SE} \\
\cmidrule(lr){3-5} \cmidrule(lr){6-8} \cmidrule(lr){9-11}
& & $\alpha_0$ & $\alpha_1$ & $\alpha_2$ & $\alpha_0$ & $\alpha_1$ & $\alpha_2$ & $\alpha_0$ & $\alpha_1$ & $\alpha_2$ \\ \midrule
\multirow{3}{*}{0.01} & Oracle & 0.94 & 0.93 & 0.93 & 0.276 & 0.429 & 0.209 & 0.051 & 0.082 & 0.042 \\
& Observed & 0.94 & 0.91 & 0.93 & 0.385 & 0.612 & 0.294 & 0.073 & 0.116 & 0.060 \\
& SSFR & 0.97 & 0.92 & 0.96 & 0.382 & 0.615 & 0.293 & 0.074 & 0.118 & 0.059 \\ \cmidrule(lr){1-11}
\multirow{3}{*}{0.07} & Oracle & 0.95 & 0.92 & 0.93 & 0.276 & 0.428 & 0.209 & 0.051 & 0.082 & 0.042 \\
& Observed & 0.94 & 0.92 & 0.92 & 0.385 & 0.612 & 0.294 & 0.072 & 0.115 & 0.060 \\
& SSFR & 0.96 & 0.92 & 0.96 & 0.377 & 0.607 & 0.290 & 0.072 & 0.116 & 0.059 \\ \cmidrule(lr){1-11}
\multirow{3}{*}{0.25} & Oracle & 0.95 & 0.92 & 0.93 & 0.276 & 0.428 & 0.209 & 0.050 & 0.081 & 0.043 \\
& Observed & 0.94 & 0.95 & 0.92 & 0.385 & 0.612 & 0.295 & 0.071 & 0.114 & 0.060 \\
& SSFR & 0.96 & 0.92 & 0.94 & 0.359 & 0.576 & 0.276 & 0.067 & 0.109 & 0.056 \\ \cmidrule(lr){1-11}
\multirow{3}{*}{0.56} & Oracle & 0.96 & 0.92 & 0.92 & 0.275 & 0.428 & 0.209 & 0.049 & 0.080 & 0.042 \\
& Observed & 0.95 & 0.96 & 0.92 & 0.384 & 0.611 & 0.296 & 0.070 & 0.113 & 0.060 \\
& SSFR & 0.98 & 0.94 & 0.94 & 0.328 & 0.521 & 0.252 & 0.059 & 0.097 & 0.051 \\ \cmidrule(lr){1-11}
\multirow{3}{*}{0.81} & Oracle & 0.96 & 0.94 & 0.93 & 0.274 & 0.429 & 0.210 & 0.048 & 0.080 & 0.042 \\
& Observed & 0.93 & 0.95 & 0.93 & 0.384 & 0.611 & 0.296 & 0.069 & 0.113 & 0.059 \\
& SSFR & 0.97 & 0.96 & 0.94 & 0.301 & 0.475 & 0.231 & 0.052 & 0.087 & 0.047 \\ \bottomrule
\end{tabular}}
\end{table}

Across all five levels of $\overline{R^2}$, the empirical coverage probabilities of all three estimators remain close to the nominal 95\% level, supporting the validity of the proposed weighted bootstrap procedure for simultaneous inference over a broad range of $\overline{R^2}$. In contrast to the relatively stable coverage, the corridor widths exhibit a clear pattern as $\overline{R^2}$ increases. When the surrogate explains little residual target-image variation ($\overline{R^2}=0.01$ or $0.07$), the SSFR and observed-data estimators yield nearly identical corridor widths. As $\overline{R^2}$ increases, however, the SSFR corridors become progressively narrower, reflecting increasingly effective use of the information contained in the surrogate images. At the highest level of surrogate explanatory power ($\overline{R^2}=0.81$), the average SSFR corridor widths for $(\alpha_0,\alpha_1,\alpha_2)$ decrease to $(0.301,0.475,0.231)$, compared with $(0.384,0.611,0.296)$ for the observed-data estimator, and move substantially closer to the oracle benchmark of $(0.274,0.429,0.210)$. In contrast, the corridor widths of the observed-data estimator remain essentially unchanged across levels of $\overline{R^2}$ because this estimator does not incorporate surrogate information.

The narrowing of SSFR corridors with increasing $\overline{R^2}$ indicates that a more explanatory surrogate enables recovery of a greater proportion of the information lost to missingness. This pattern is consistent with the variance approximation in \eqref{EQ:VR}, which predicts that the leading variance constant is determined by the surrogate explanatory measure $R^2(\bs{z})$ together with the observed-image fraction. The results also show that this efficiency improvement does not compromise simultaneous coverage, even when the surrogate model is misspecified.

\vspace{-6pt}
\subsection{Surrogates Generated by AI Image-to-Image Synthesis Methods}
\label{SSEC:sim-ai-summary}
	
The surrogate above is analytic, so its informativeness and specification can be controlled directly. Two further experiments, reported in full in Supplement Section A.2, replace it with a surrogate learned from an auxiliary imaging modality by a modern image-to-image network, on a discretized $26\times26\times26$ voxel grid but otherwise reusing the design of Section \ref{SSEC:sim-design}.
	
The first benchmarks four generators: a compact 3D UNet, a CNN encoder--decoder, a conditional GAN, and a conditional VAE, under $r_m\in\{25\%,50\%,75\%\}$. The estimated image-level explanatory measure $\overline{R^2}$ ranges from $0.371$ to $0.493$ across all twelve architecture-by-missingness cells, and SSFR improves on ISR, IPW, and AIPW in every one. Naive SISR degrades sharply and unevenly: under the CVAE surrogate at $75\%$ missingness its MISE for $\alpha_2$ is roughly two hundred times that of ISR. 

The second varies surrogate quality across a subgroup defined by the binary covariate $X_{i1}$, with subgroup-specific target--auxiliary correlations $(\omega_0,\omega_1)$ ranging from homogeneity $(0.75,0.75)$ to severe heterogeneity $(0.75,0)$. 
The pooled $\overline{R^2}$ values are $0.465$, $0.309$ and $0.185$ under homogeneous, moderately heterogeneous and severely heterogeneous surrogate quality, respectively.
SSFR degrades gracefully towards ISR as heterogeneity increases, its total MISE rising from $0.0194$ to $0.0239$ against an ISR total of $0.0242$, while the naive-SISR total rises from $0.3285$ to $0.3753$. This is the behavior Corollary \ref{cor:misspec} predicts: an uninformative surrogate costs efficiency without introducing the misspecification bias that afflicts direct substitution. 
The subgroup-specific results are reported in Supplement Section A.2.2.

\vspace{-6pt}
\section{ADNI Data Analysis}
\label{SEC:application}

This section investigates spatially varying associations between subject-level covariates and brain imaging outcomes using the ADNI data described in Section \ref{SEC:data}. Section \ref{SSEC:application-1} presents a validation study under controlled PET missingness using the benchmark cohort with complete PET, MRI, and covariate data. Section \ref{SSEC:application-2} analyzes the expanded cohort augmented with MRI-based PET surrogates, and Section \ref{SSEC:application-scc} presents simultaneous inference based on the proposed SCC procedure.

The analysis incorporates a comprehensive set of demographic, clinical, genetic, and cognitive covariates, including sex, age, years of education, Mini-Mental State Examination (MMSE) score, clinical diagnosis indicators for mild cognitive impairment (MCI) and Alzheimer's disease (AD), and indicators for carrying one or two copies of the APOE-$\epsilon 4$ allele, a well-established genetic risk factor for AD. These covariates have been widely used in prior ADNI analyses and shown to be strongly associated with cerebral glucose metabolism and neurodegenerative progression in aging, MCI, and AD populations \citep{Landau:etal:11, Nosheny:etal:15}. Clinical diagnosis is encoded using two binary indicators, with cognitively normal subjects serving as the reference group. Potential collinearity was assessed before analysis. The largest pairwise correlation is between MMSE and AD ($r=-0.66$), and all variance inflation factors are below 3.1.

We focus on FDG-PET as the primary imaging response because it provides a direct measure of cerebral glucose metabolism, a sensitive biomarker of synaptic dysfunction and neurodegeneration that often precedes detectable structural atrophy. Although structural MRI is available for a substantially larger proportion of participants, it primarily reflects anatomical changes that typically emerge later in the disease process and may be less responsive to early metabolic alterations. Consequently, FDG-PET is more closely aligned with the scientific objective of characterizing spatially heterogeneous associations between subject-level covariates and underlying neurobiological processes in AD. To this end, we model the PET signal using the following ISR model with spatially varying coefficients: for $i = 1,\ldots,n$,
\begin{align}
Y_i(\bs{z}_j) = &\ \alpha_0(\bs{z}_j) + \alpha_1(\bs{z}_j)\mathrm{SEX}_i + \alpha_2(\bs{z}_j)\mathrm{AGE}_i + \alpha_3(\bs{z}_j)\mathrm{Education}_i + \alpha_4(\bs{z}_j)\mathrm{APOE}_{1i} \nonumber \\
& + \alpha_5(\bs{z}_j)\mathrm{APOE}_{2i} + \alpha_6(\bs{z}_j)\mathrm{MMSE}_i + \alpha_7(\bs{z}_j)\mathrm{MCI}_i + \alpha_8(\bs{z}_j)\mathrm{AD}_{i} + \varepsilon_i(\bs{z}_j),
\label{EQ:GFR-app}
\end{align}
where $\alpha_{\ell}(\bs{z}_j)$ denotes the spatially varying effect of covariate $X_{i\ell}$ at voxel $\bs{z}_j\in\Omega$.

In principle, fitting this model directly using PET images provides detailed quantitative maps of how demographic, diagnostic, and genetic factors are associated with metabolic changes across the cortex. In practice, however, PET imaging is expensive and sparsely collected, limiting the effective sample size available for high-resolution spatial modeling. To address this limitation, we leverage the fact that structural MRI is available for a much larger subset of the cohort and can be used to generate PET surrogates via cross-modal synthesis, as described in Section \ref{SEC:data}. This motivates the proposed SSFR framework, which is designed to integrate both observed and synthetic imaging data under substantial modality-level missingness.

\vspace{-6pt}
\subsection{Validation Study under Controlled PET Missingness}
\label{SSEC:application-1}

To evaluate the accuracy and robustness of SSFR under controlled missing-data conditions, we first conduct a validation study using a benchmark ADNI cohort with complete imaging and covariate information. This benchmark cohort consists of the 260 participants with complete PET, MRI, and covariate data identified in Section \ref{SSEC:dataset}.

Because the observed and synthetic PET images are aligned through standard intensity normalization and template mapping, they share a common spatial basis and integrate directly into the two-stage SSFR pipeline. In the first stage, we fit model \eqref{EQ:GFR-app} to the synthetic PET surrogates for all 260 participants; in the second stage, we regress the observed PET images for the PET-available subjects on their covariates and synthetic PET images.

For benchmarking purposes, we also fit model \eqref{EQ:GFR-app} using the full set of 260 observed PET images. The resulting estimates of $\alpha_0,\ldots,\alpha_8$ are treated as the reference truth for assessing model accuracy under controlled missingness; because this reference is itself estimated from finite data, the reported MISEs should be read as measuring agreement with the full-data analysis rather than as absolute estimation error. We evaluate performance through 100 replications of a random missingness experiment and report the average MISE of each coefficient function in Table \ref{TAB:app-mise}. We compare the proposed SSFR method with two benchmarks: conventional ISR, which uses only the observed PET images, and naive SISR, which augments the observed data with synthetic PET images imputed for subjects with missing scans.

We consider two triangulations of the brain domain consisting of 317 and 652 tetrahedra, respectively, shown in Supplement Figure B.1(c) and (d). For each triangulation, we evaluate spline degrees from $d=2$ to $d=6$; balancing GCV performance and approximation quality, we adopt the triangulation with 317 tetrahedra and $d=5$ in the analyses reported below. 

Table \ref{TAB:app-mise} shows that SSFR attains the lowest MISE for eight of the nine coefficients at 50\% missingness and the lowest or tied for the lowest at 25\% missingness. Relative to ISR, SSFR consistently reduces estimation error, with improvements averaging about 8\% across coefficients. The gains are broadly uniform, suggesting a systematic variance reduction rather than improvements confined to a few coefficients.

The comparison with naive SISR is more pronounced. Although SISR occasionally achieves slightly lower MISE than ISR for individual coefficients, it exhibits substantially larger errors for others, particularly under 50\% missingness. This coefficient-dependent behavior reflects the tradeoff between reduced variance and surrogate-induced bias when synthetic images are directly substituted for missing responses. In contrast, SSFR consistently avoids this instability through the Stage-2 recalibration, as predicted by Corollary \ref{cor:misspec}. Overall, these results demonstrate that SSFR provides more accurate coefficient estimation than both ISR and SISR, under realistic missing-data conditions, consistent with the theory of Section \ref{SEC:theory}. 

\begin{table}[t]
\caption{MISEs of the varying coefficient functions in (\ref{EQ:GFR-app}) for different methods in the ADNI study.\label{TAB:app-mise}}
\scalebox{0.8}{\begin{tabular}{ccrrrrrrrrr} \toprule
Missing &\multirow{2}{*}{Method} &\multicolumn{9}{c}{MISE$\times 10^4$}\\ \cline{3-11}
Rate & &$\alpha_0$ &$\alpha_1$ &$\alpha_2$ &$\alpha_3$ &$\alpha_4$ &$\alpha_5$ &$\alpha_6$ &$\alpha_7$ &$\alpha_8$\\ \midrule
\multirow{3}{*}{0.50} &ISR &27.4534 &0.3342 &0.0020 &0.0087 &0.4686 &0.8963 &0.0148 &0.4736 &1.2956\\ \cline{2-11}
&SISR &28.6700 &2.5571 &0.0036 &0.0080 &0.4622 &1.2873 &0.0111 &0.5882 &1.2076\\ \cline{2-11}
&SSFR &24.9971 &0.3157 &0.0018 &0.0079 &0.4204 &0.8630 &0.0135 &0.4284 &1.1921\\ \midrule

\multirow{3}{*}{0.25} &ISR &9.1723 &0.1057 &0.0006 &0.0027 &0.1401 &0.3148 &0.0049 &0.1484 &0.4274\\ \cline{2-11}
&SISR &13.8456 &0.7285 &0.0013 &0.0042 &0.2057 &0.6716 &0.0053 &0.2826 &0.5863\\ \cline{2-11}
&SSFR &8.3095 &0.0991 &0.0006 &0.0025 &0.1289 &0.2904 &0.0045 &0.1373 &0.3985\\ \bottomrule
\end{tabular}}
\end{table}

\vspace{-6pt}
\subsection{Expanded-Cohort Analysis with MRI-based PET Surrogates}
\label{SSEC:application-2}

We apply SSFR to an expanded ADNI cohort augmented with MRI-based PET surrogates to investigate spatially varying associations between FDG-PET images and subject-level covariates. Among participants with complete covariate information, PET scans are available for 260 individuals, whereas 325 have MRI but no corresponding PET scan, yielding an empirical PET missing rate of 56\%. Similar to \cite{Lo:etal:12}, we use logistic regression models to examine the associations between PET scan availability and demographic, genetic, and clinical features. Consistent with prior findings, PET missingness is not completely at random and depends on observed subject characteristics. We estimate inclusion probabilities from the fitted logistic models and incorporate the resulting inverse-probability weights into SSFR. Synthetic PET images are generated using the cross-modal synthesis procedure described in Section \ref{SSEC:synthgen}, yielding surrogate PET images for all 585 participants. Surrogates for participants in the training set are in-sample predictions and may be more accurate than those for MRI-only participants; the Stage-2 co-kriging operator is estimated on the former group and applied to the latter. For the MRI-based surrogate in this cohort, $\overline{R^2}\approx 0.24$. Five-fold cross-validation yields a similar out-of-sample estimate, $\overline{R^2}_{\mathrm{CV}}\approx 0.23$. These results indicate that the MRI-based surrogate explains approximately 24\% of the target-image residual variation beyond the observed covariates, with comparable explanatory power for held-out subjects.

\begin{figure}[p]\centering\setlength{\tabcolsep}{2pt}
{\footnotesize
\begin{tabular}{cccccc}
\multicolumn{1}{l}{\textbf{A.}}& Intercept & Sex & Age & Education & APOE$_1$ \\[1pt]
SSFR & \includegraphics[width=.85in]{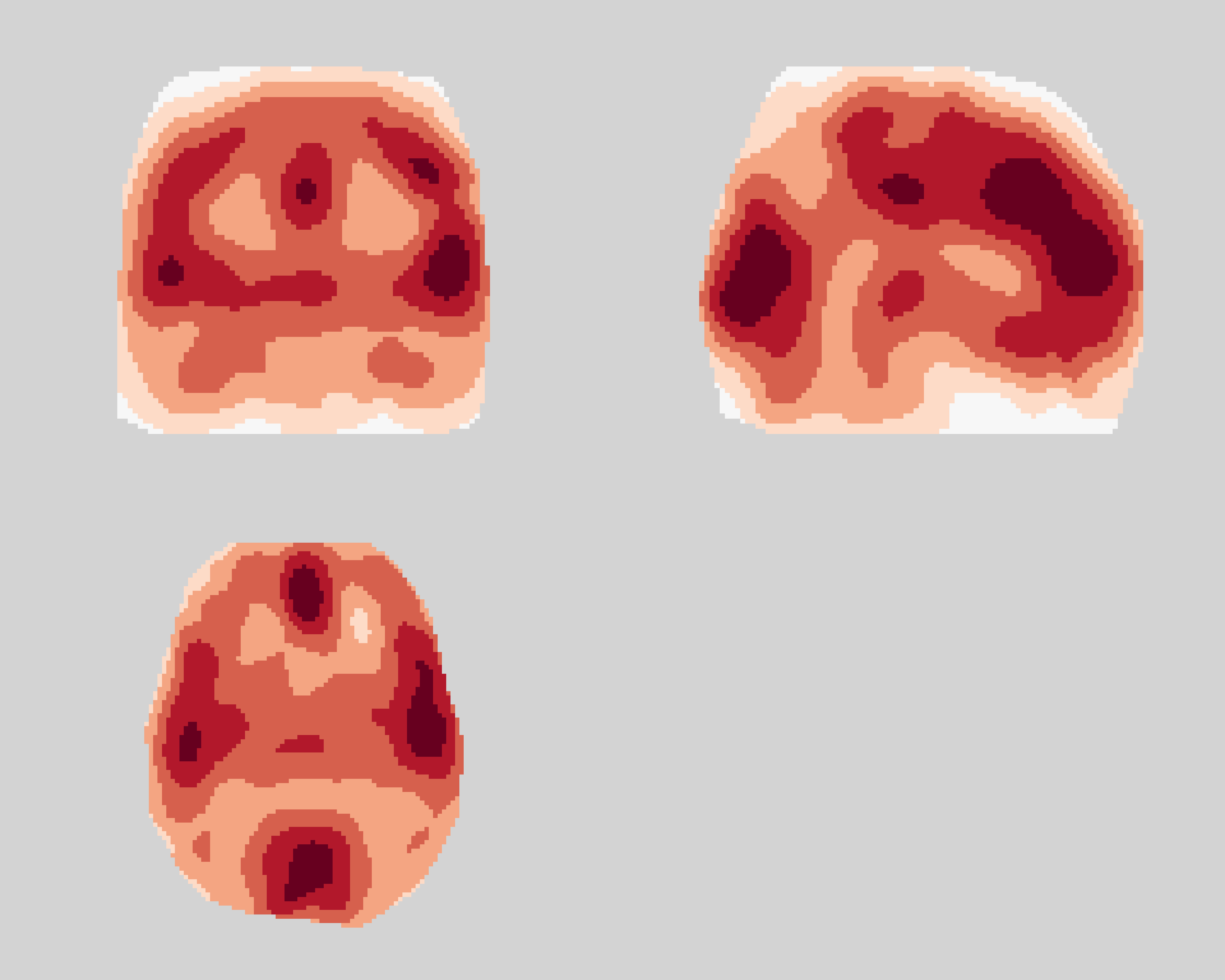} & \includegraphics[width=.85in]{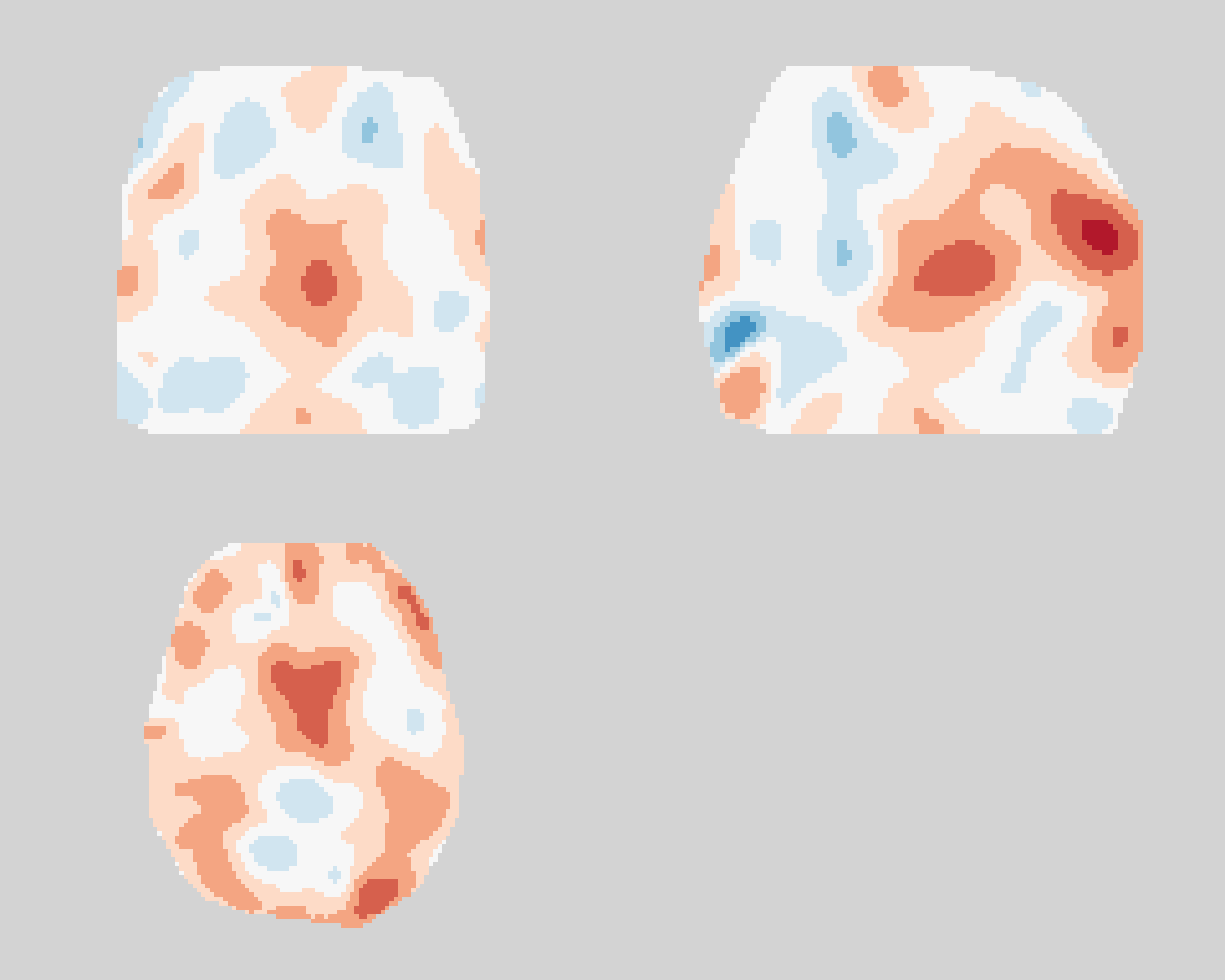} & \includegraphics[width=.85in]{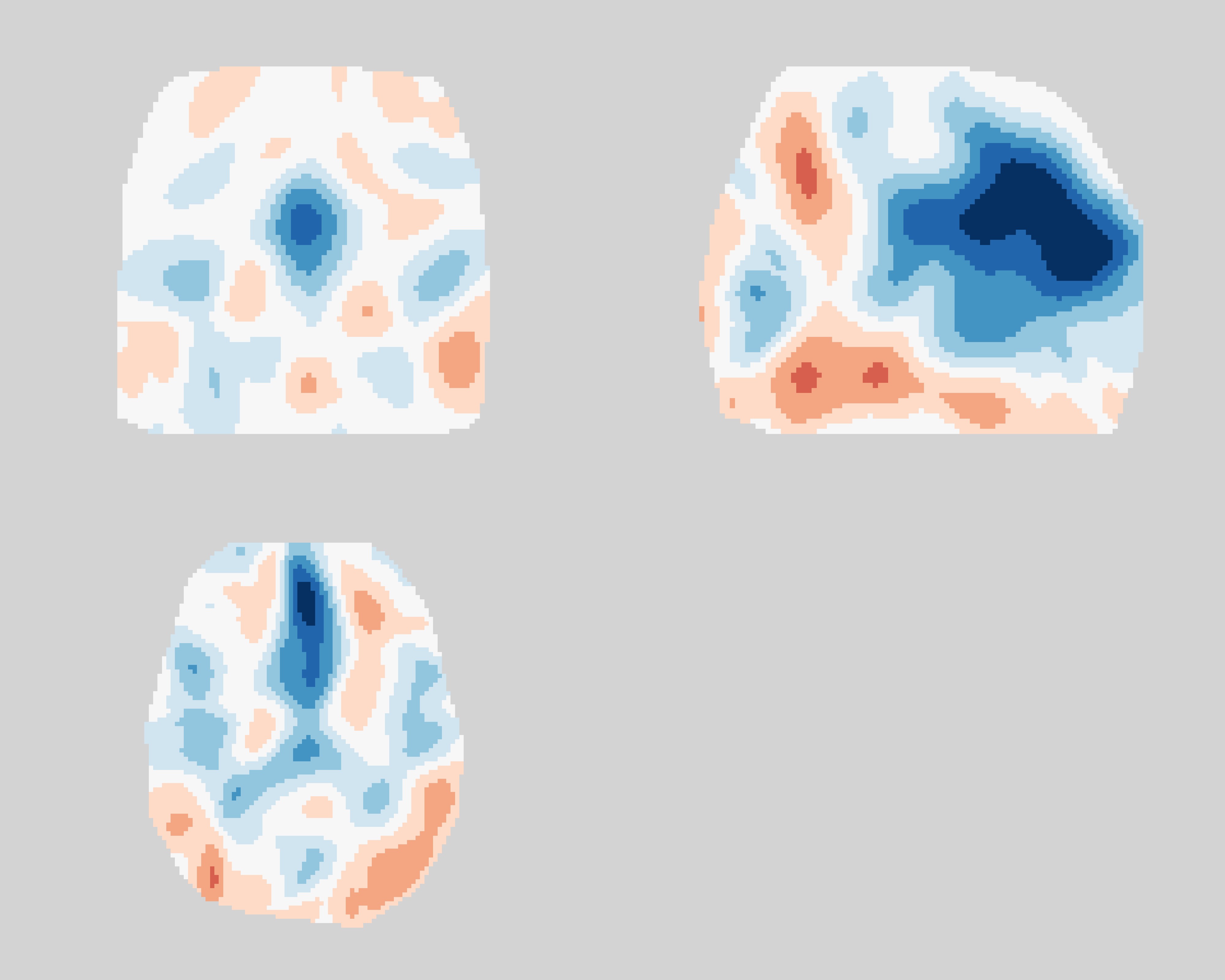} & \includegraphics[width=.85in]{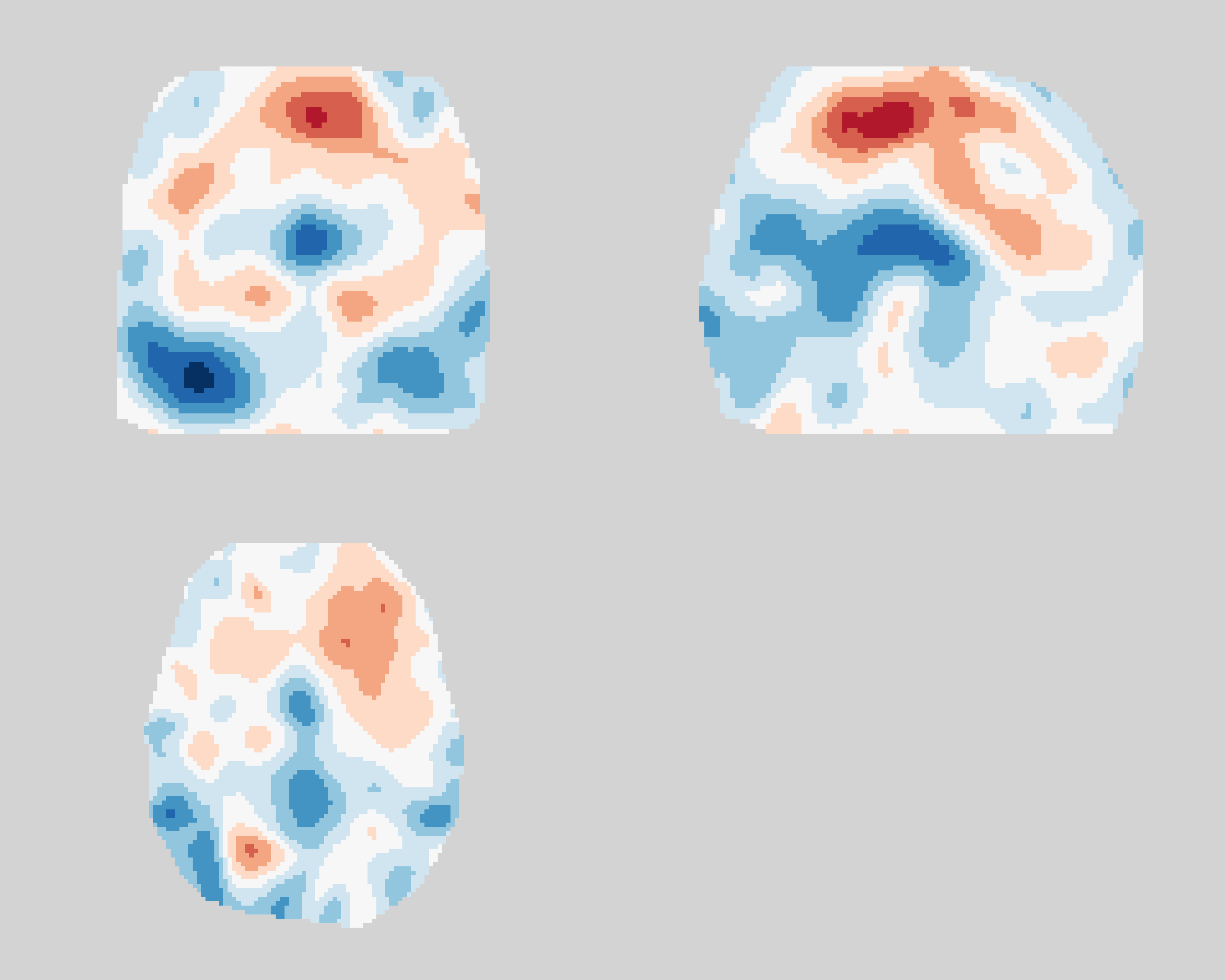} & \includegraphics[width=.85in]{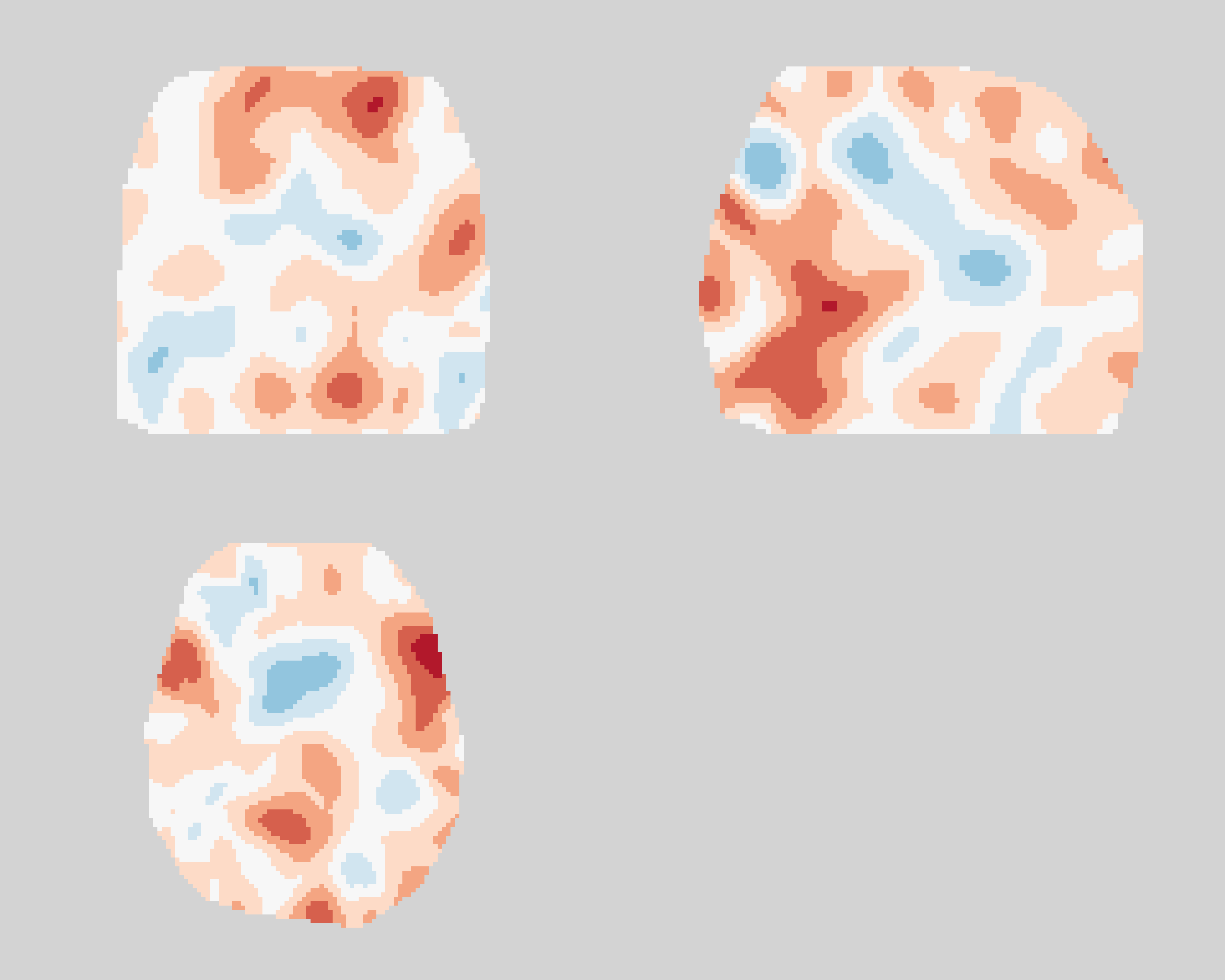} \\[-1pt]
\% SE red. & \includegraphics[width=.85in]{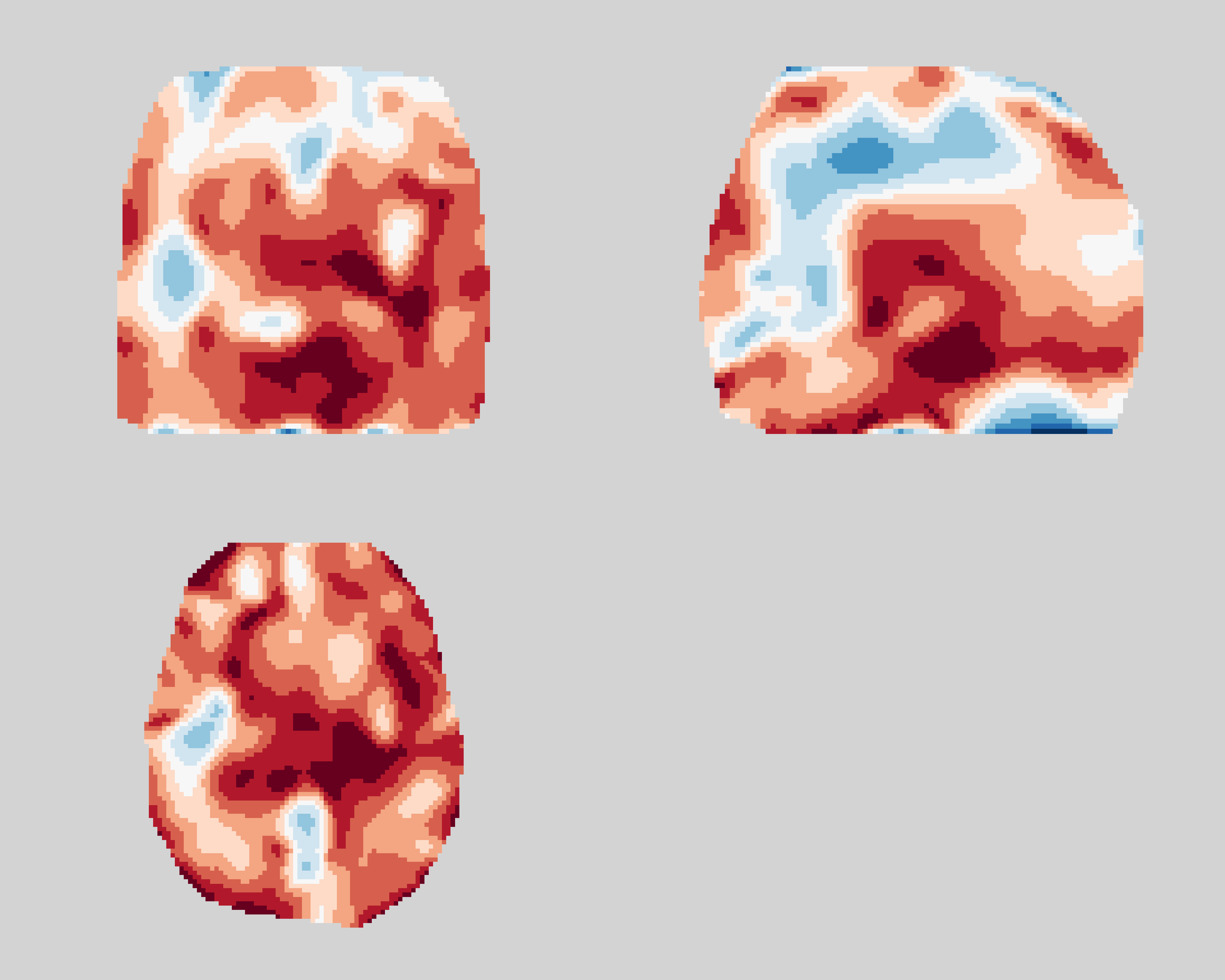} & \includegraphics[width=.85in]{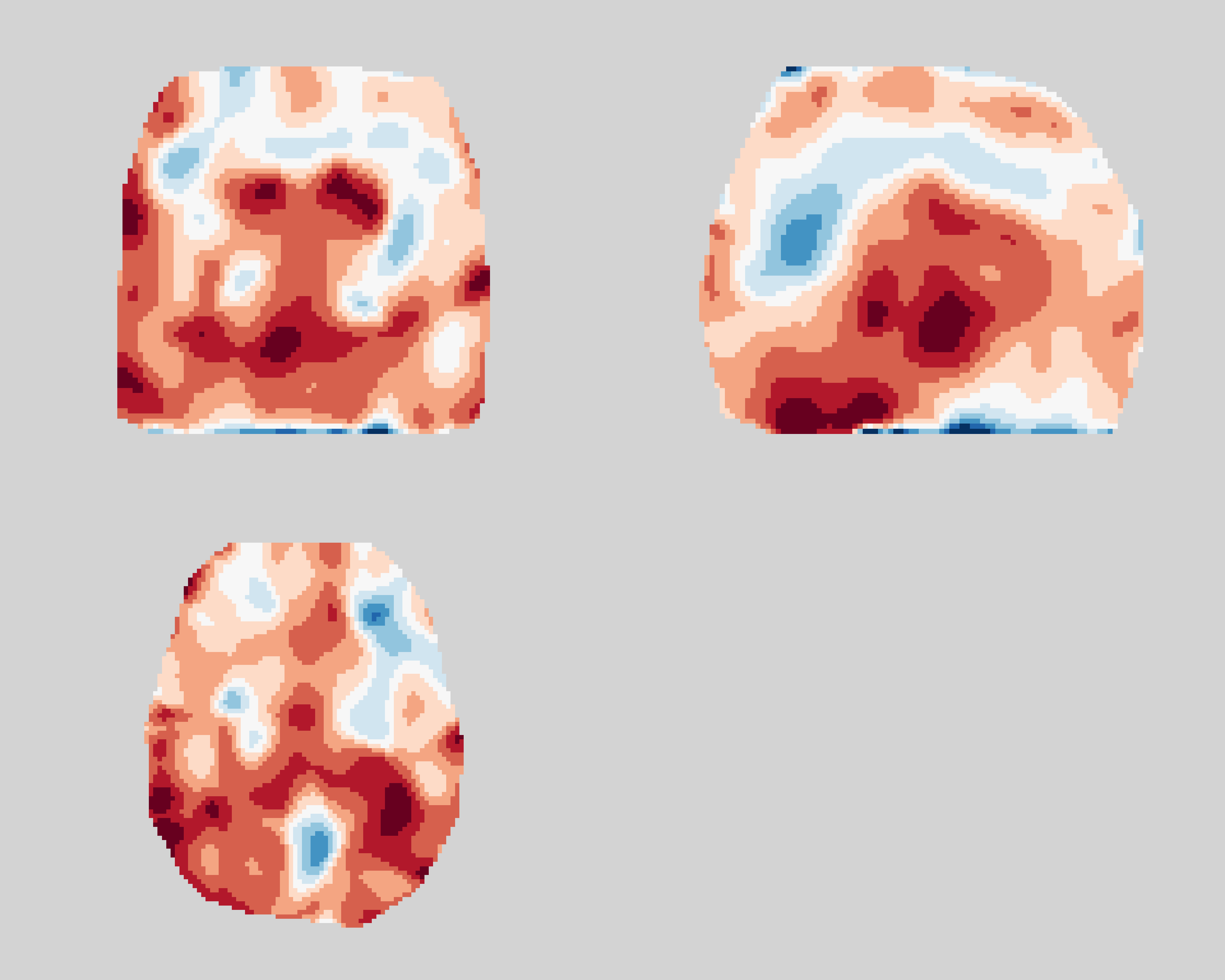} & \includegraphics[width=.85in]{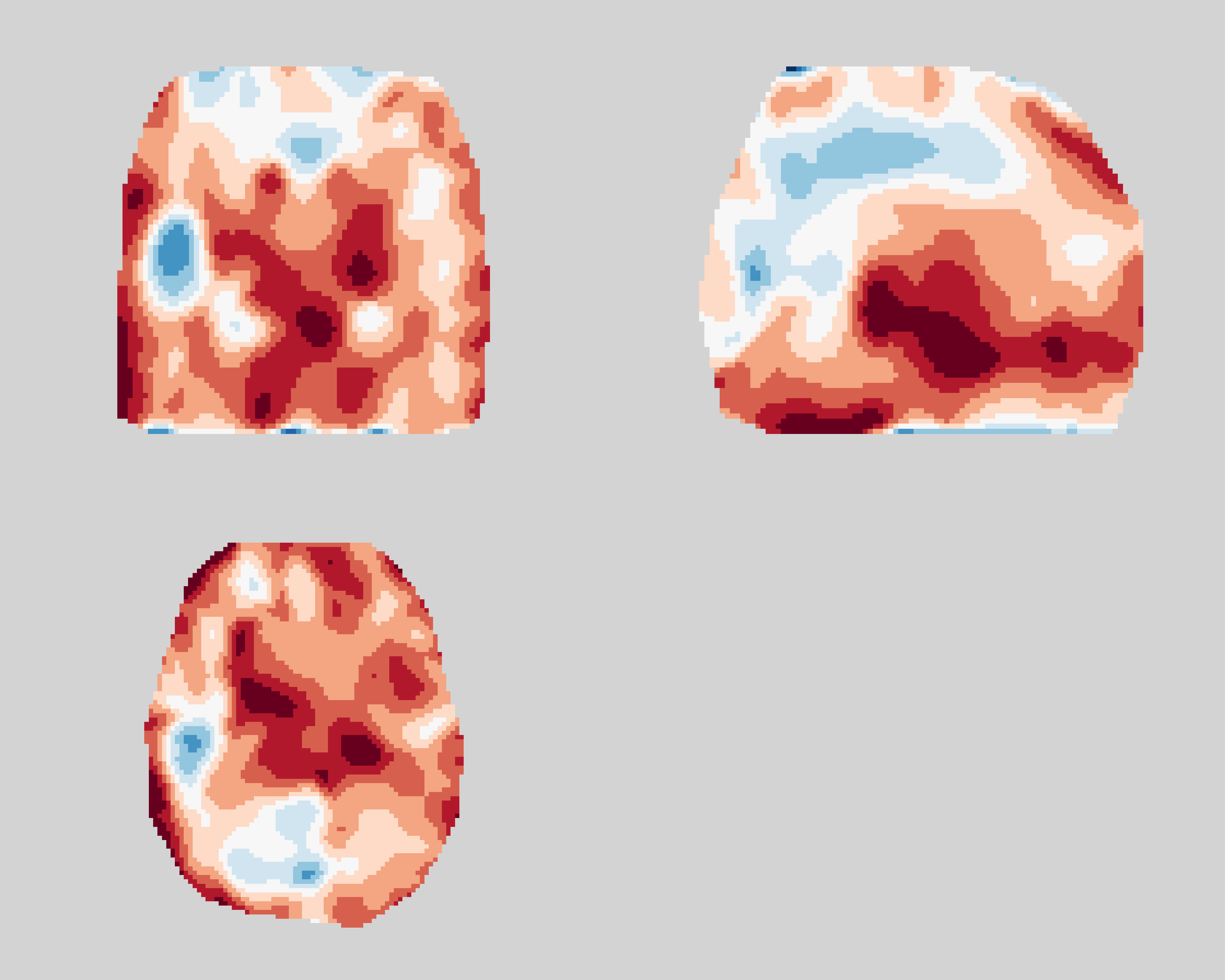} & \includegraphics[width=.85in]{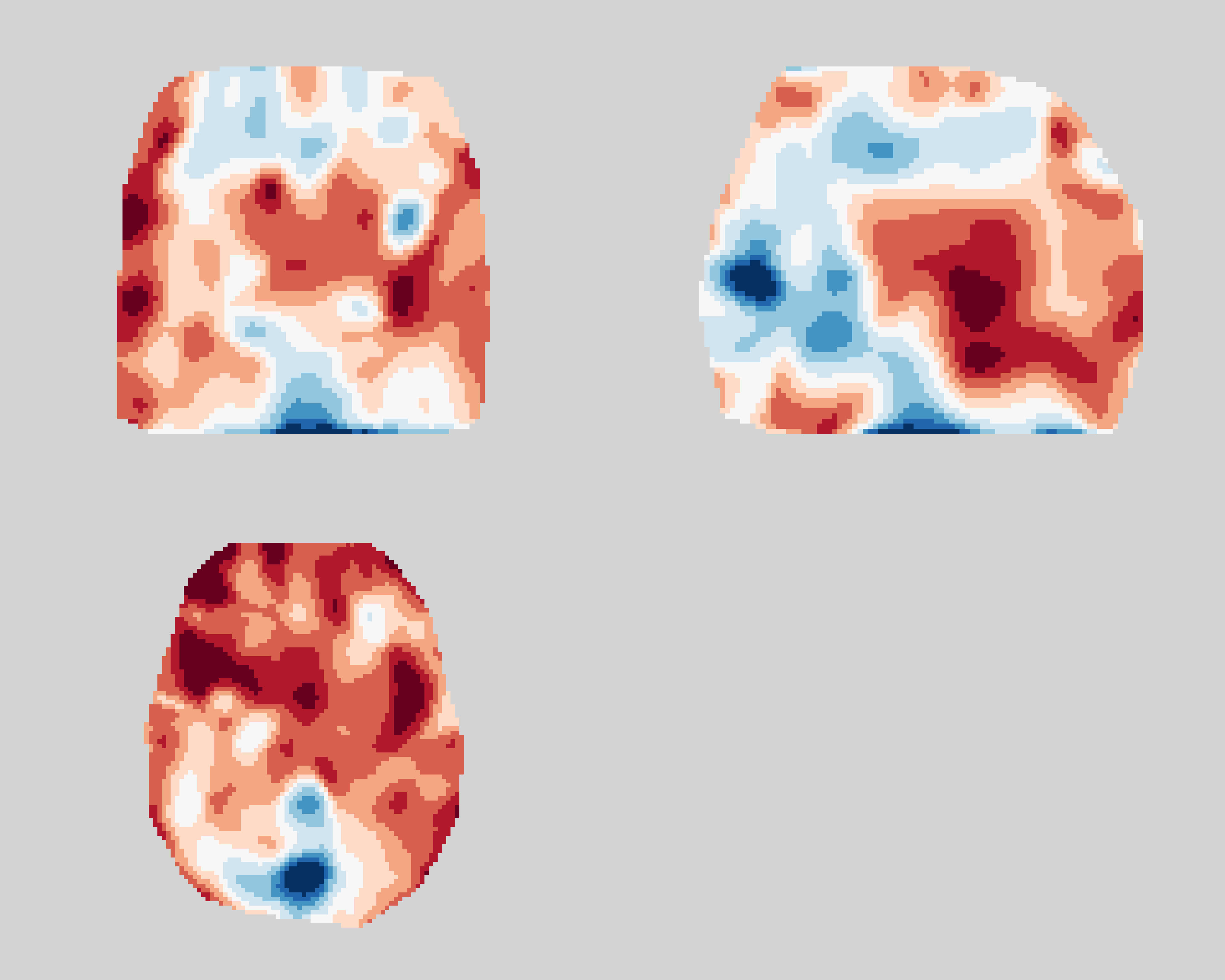} & \includegraphics[width=.85in]{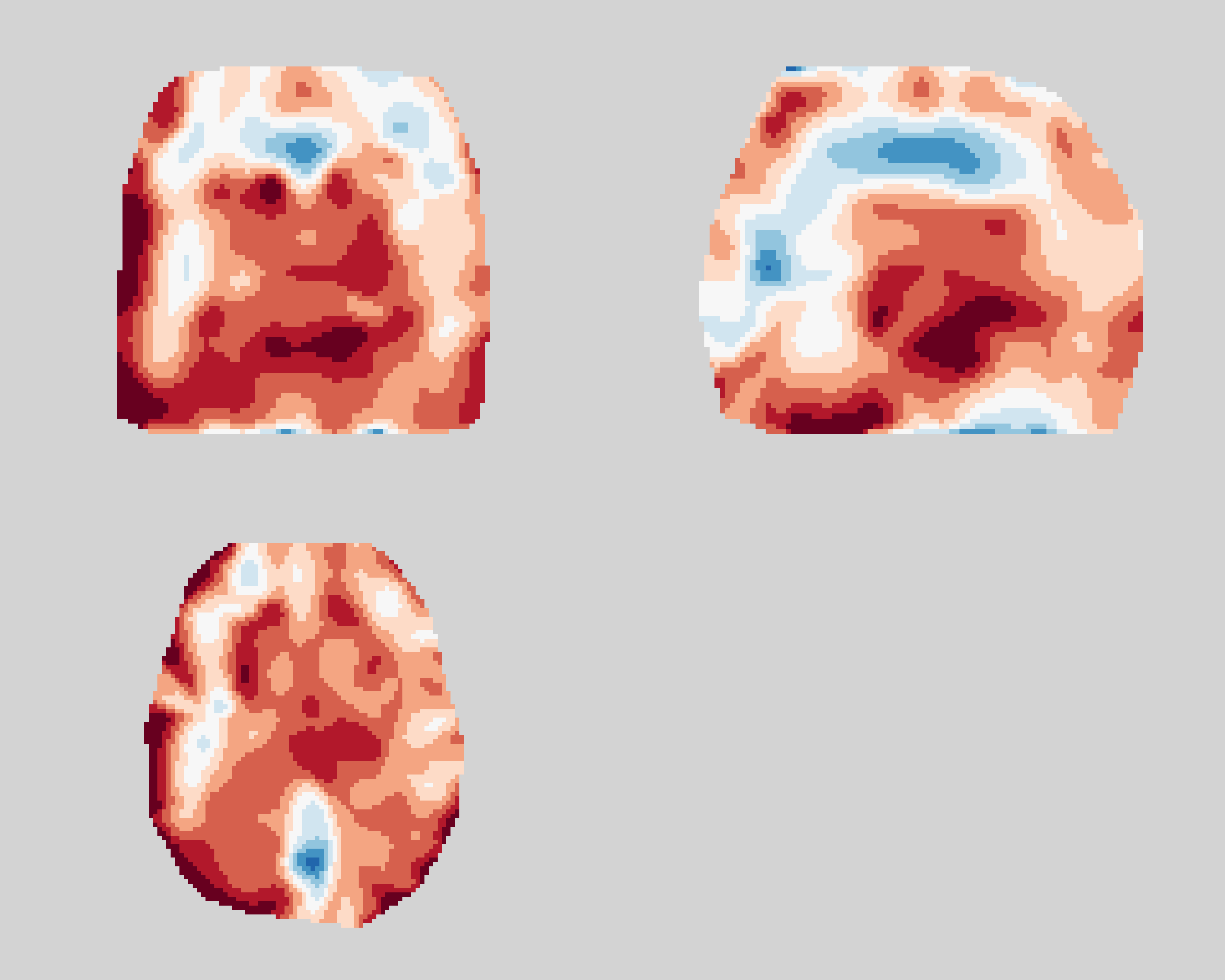} \\[-1pt]
\end{tabular}\\[3pt]
\begin{tabular}{ccccc}
& APOE$_2$ & MMSE & MCI & AD \\[1pt]
SSFR & \includegraphics[width=.85in]{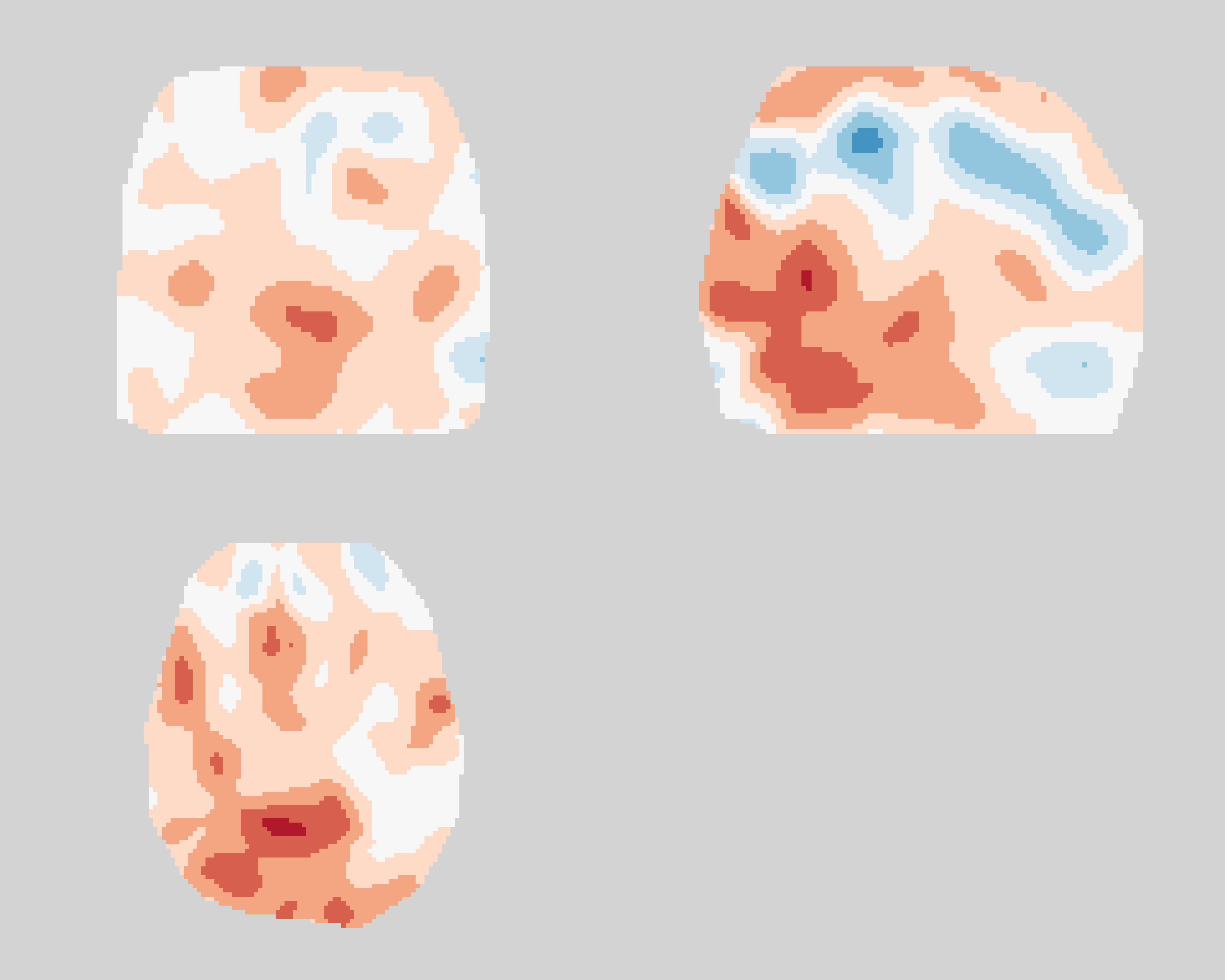} & \includegraphics[width=.85in]{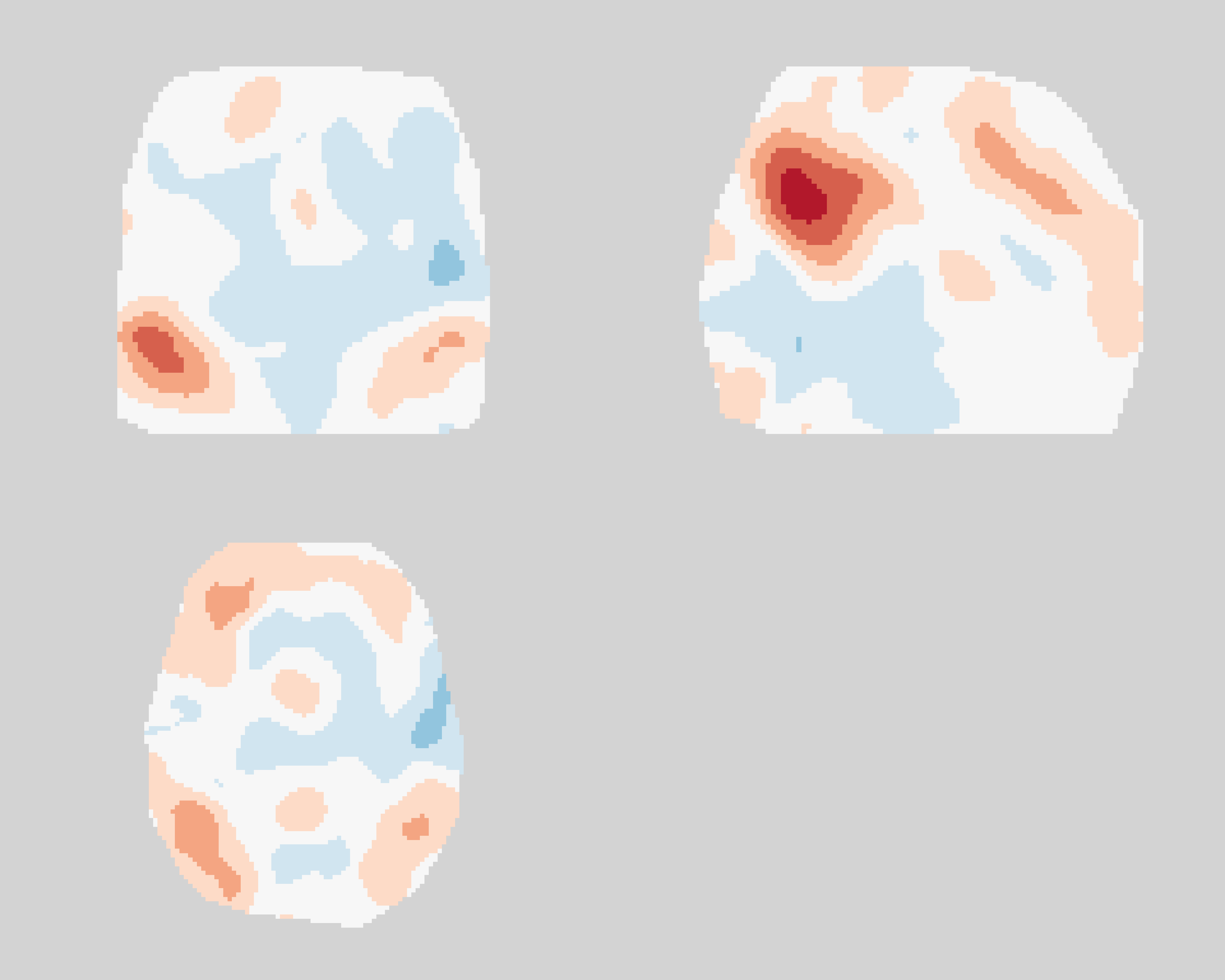} & \includegraphics[width=.85in]{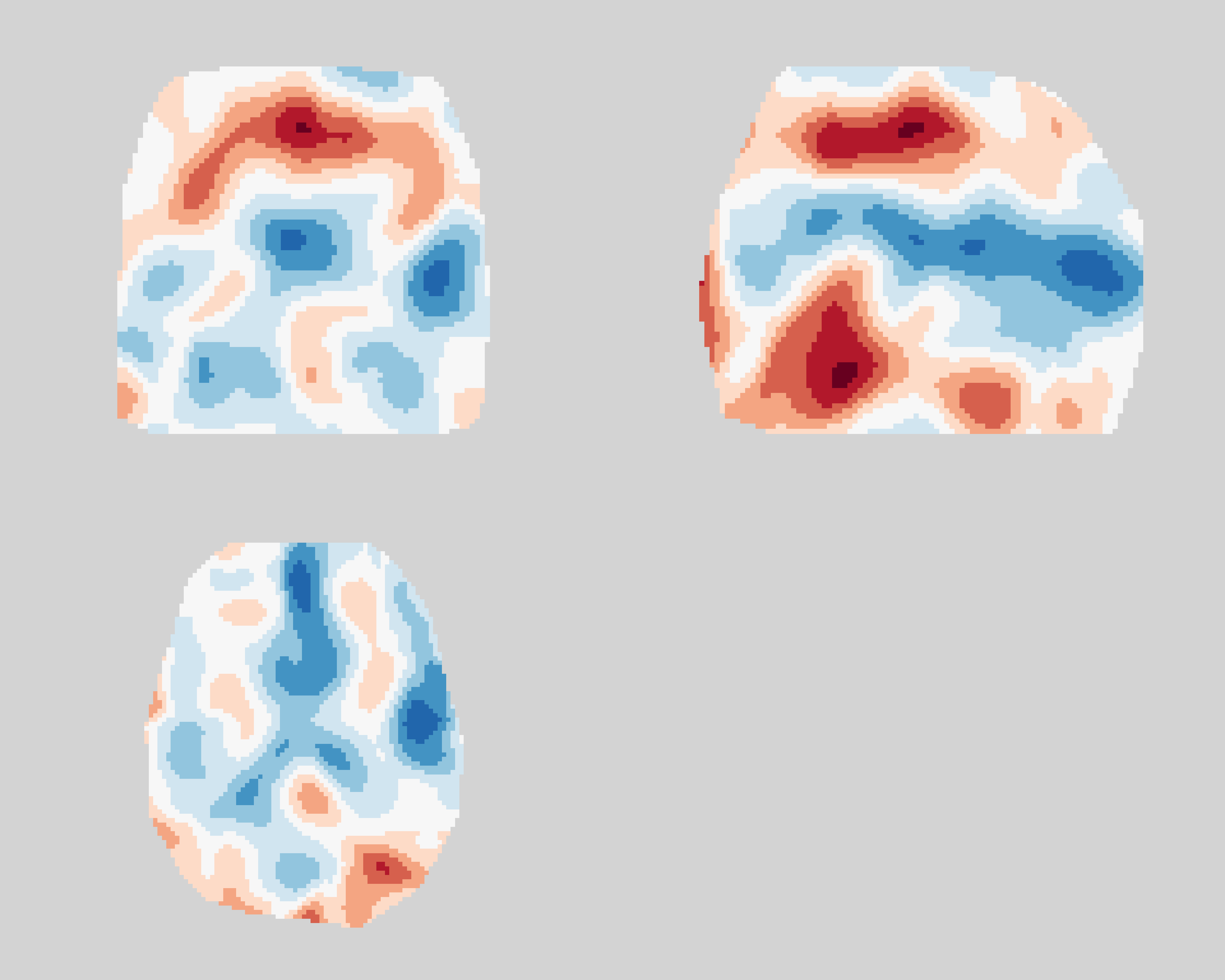} & \includegraphics[width=.85in]{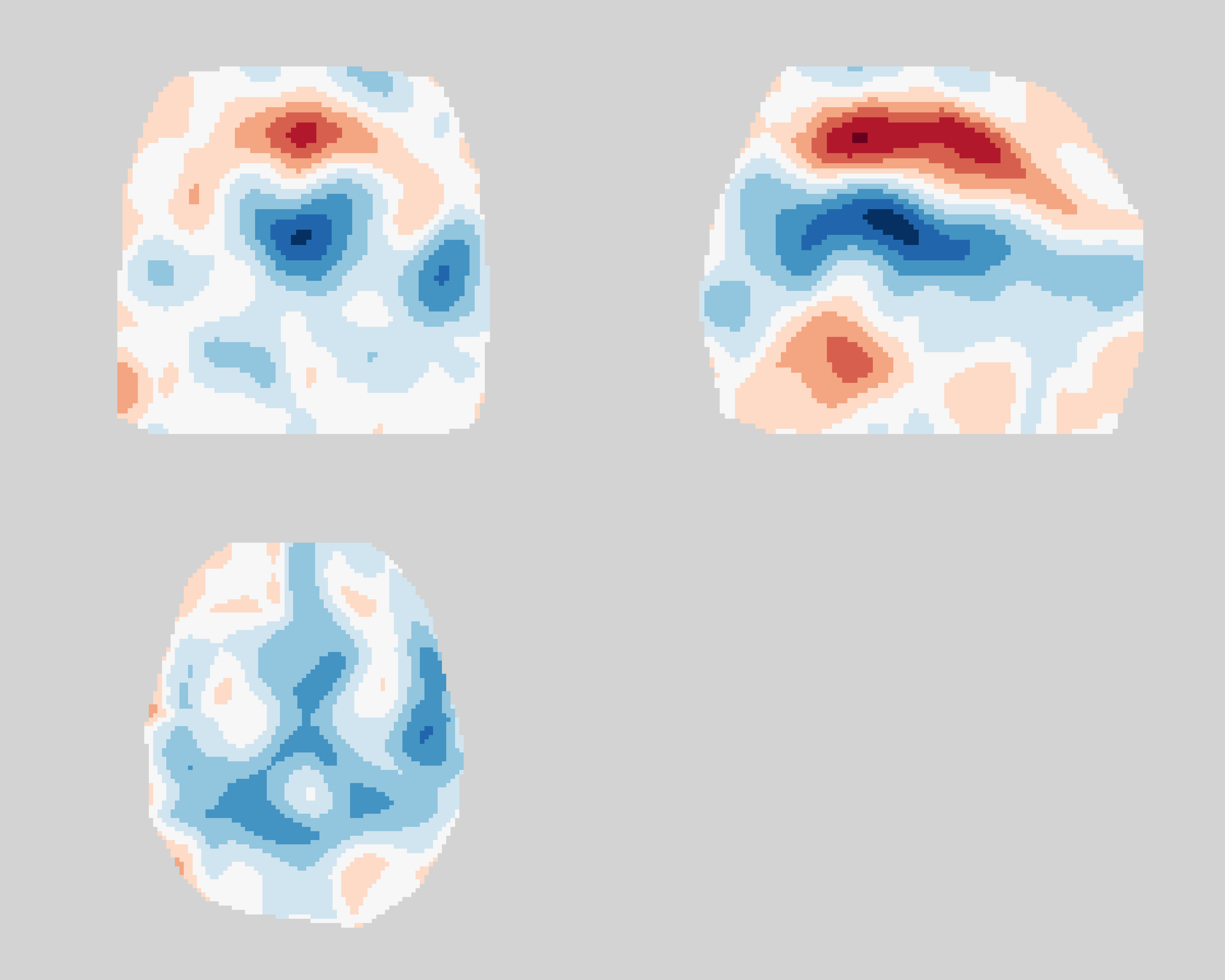} \\[-1pt]
\% SE red. & \includegraphics[width=.85in]{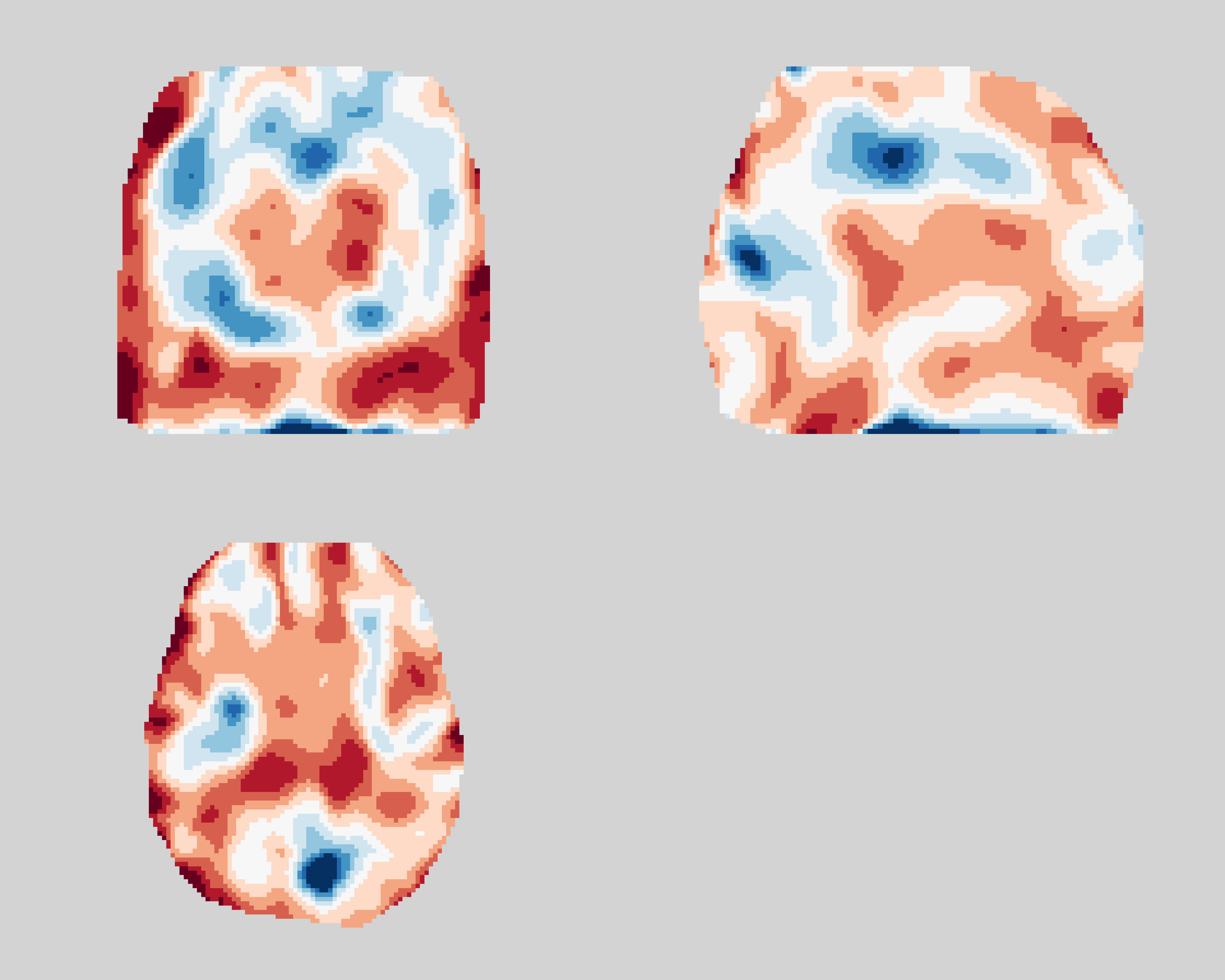} & \includegraphics[width=.85in]{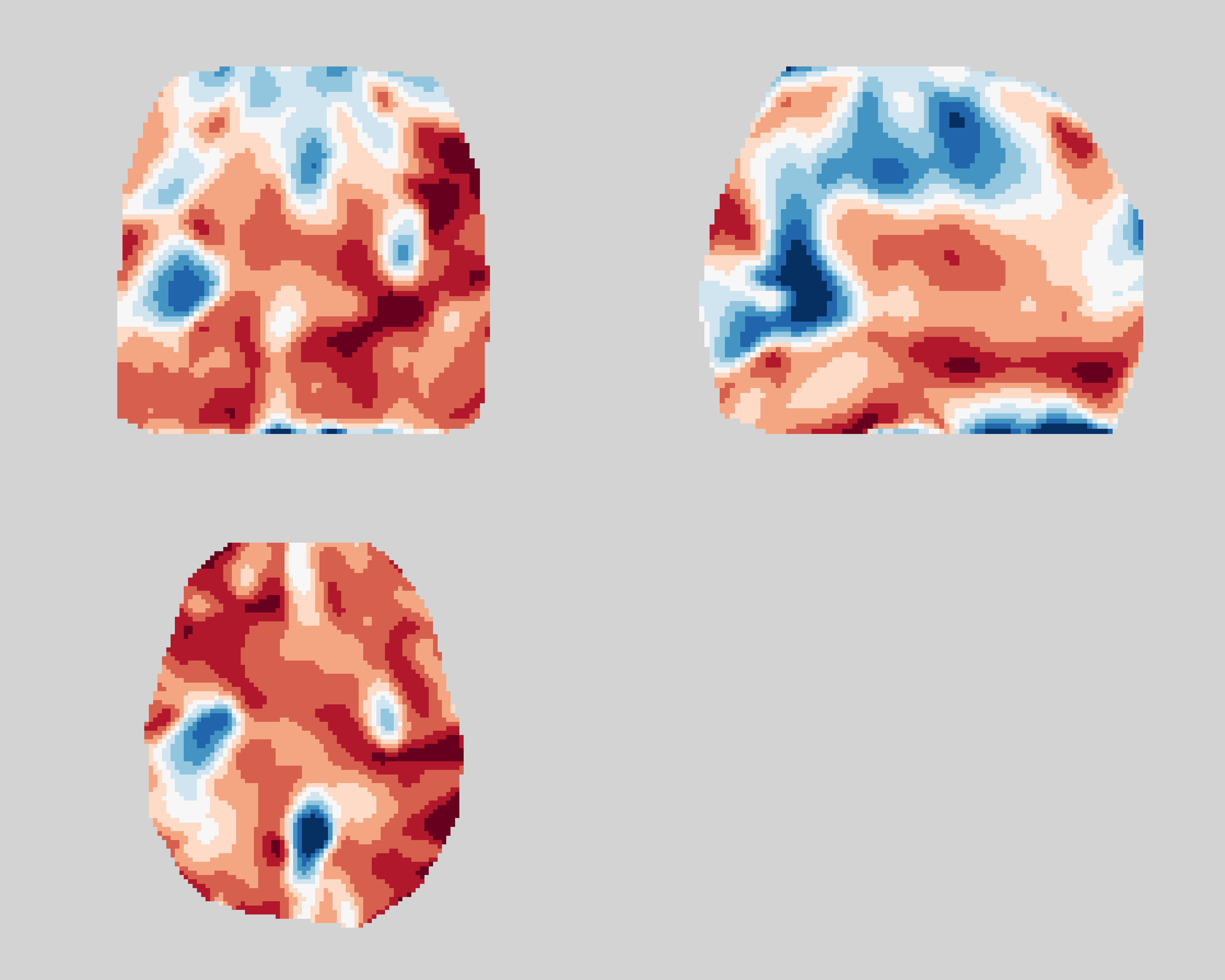} & \includegraphics[width=.85in]{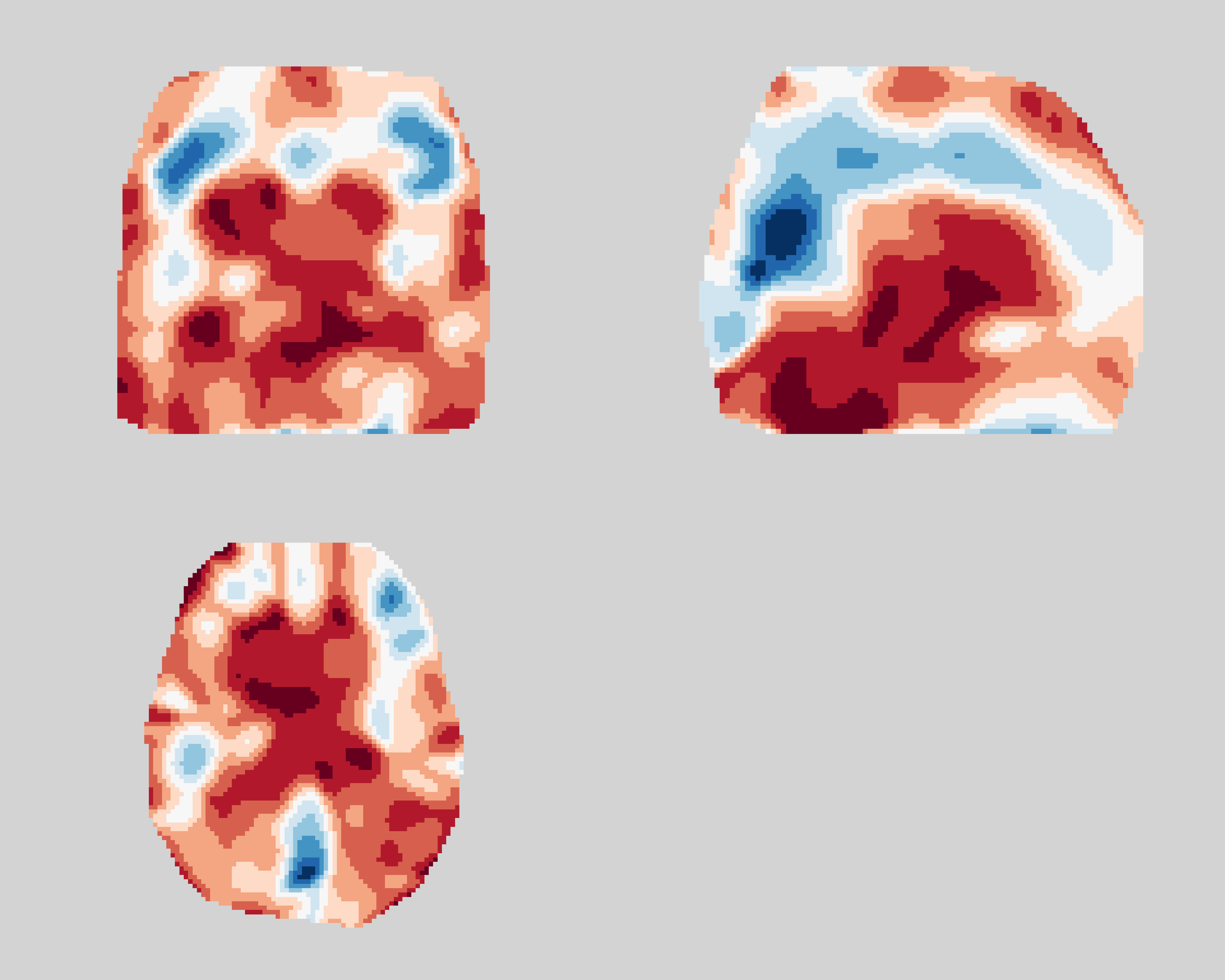} & \includegraphics[width=.85in]{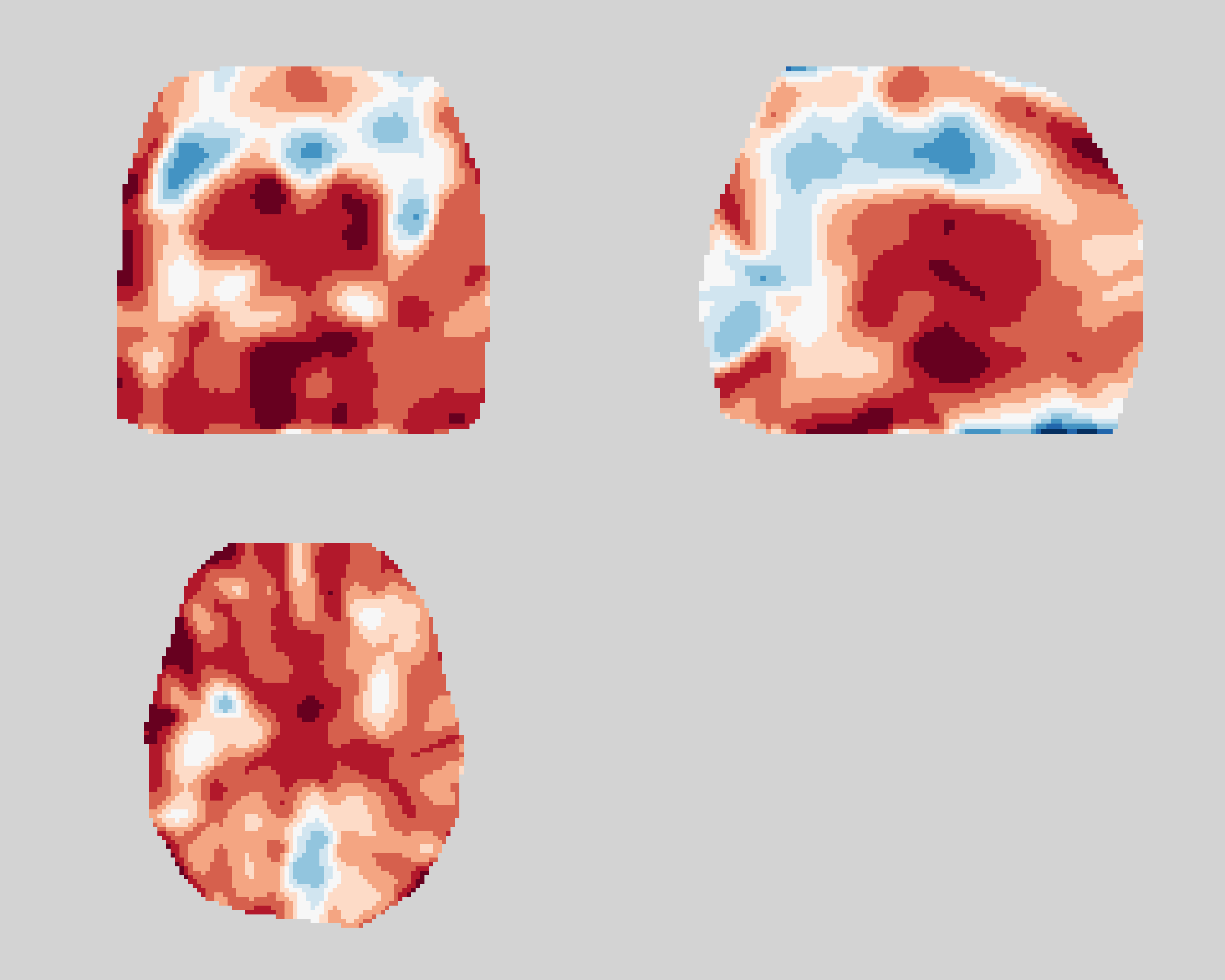} \\[-1pt]
\end{tabular}\\[3pt]
\begin{tabular}{cccccc}
\multicolumn{1}{l}{\textbf{B.}} & Intercept & Sex & Age & APOE$_2$ & MMSE \\[1pt]
Sig.\ ISR & \includegraphics[width=.85in]{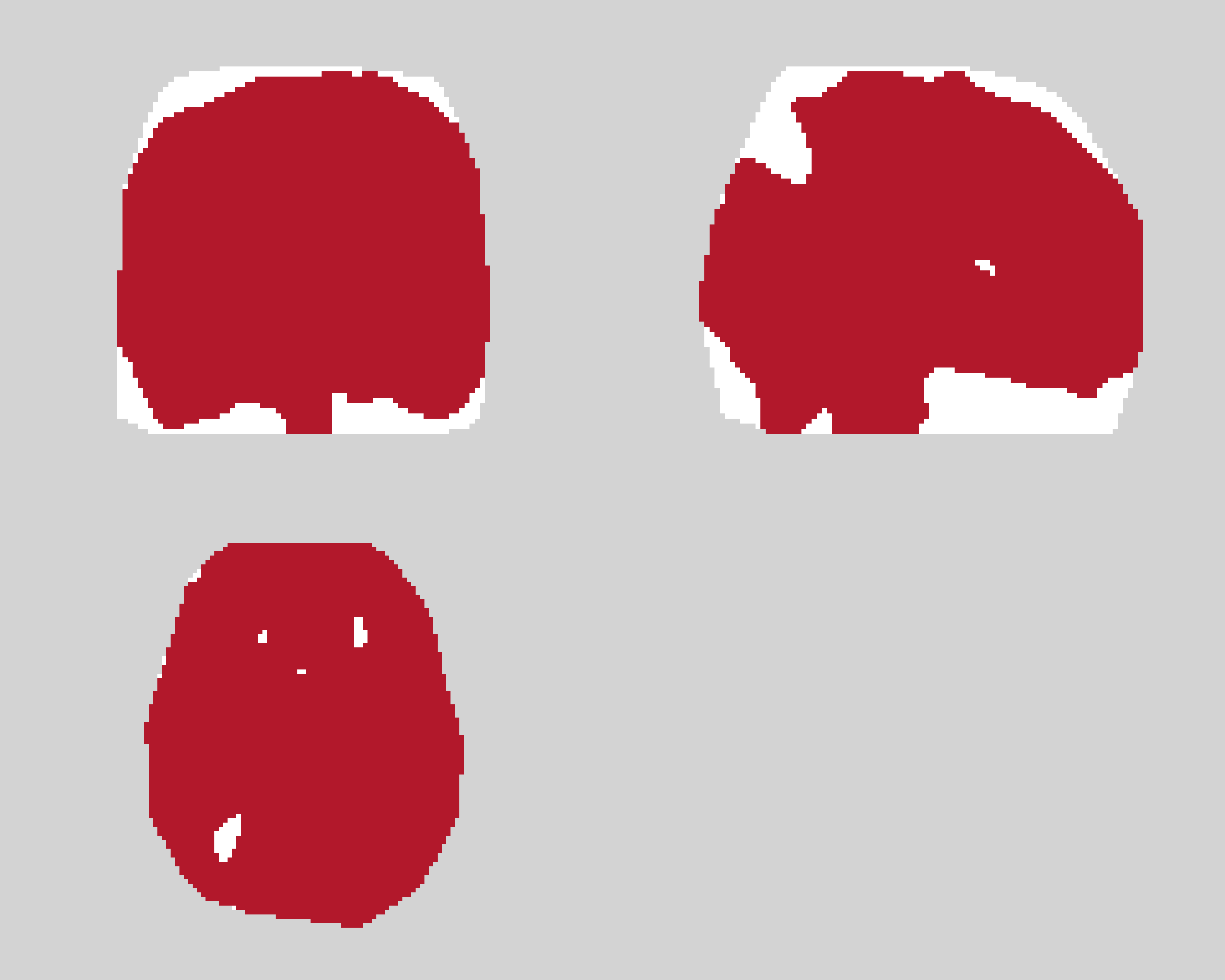} & \includegraphics[width=.85in]{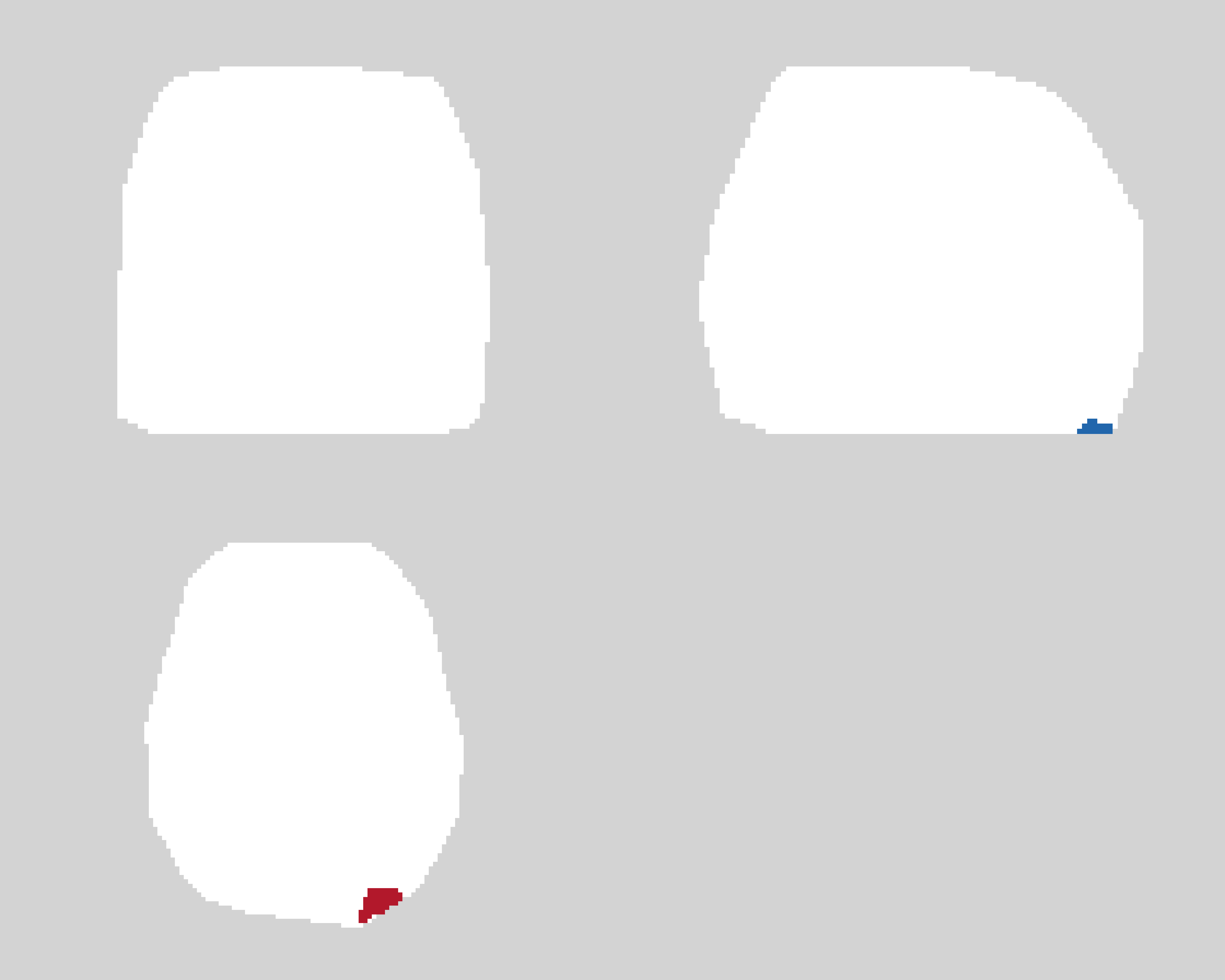} & \includegraphics[width=.85in]{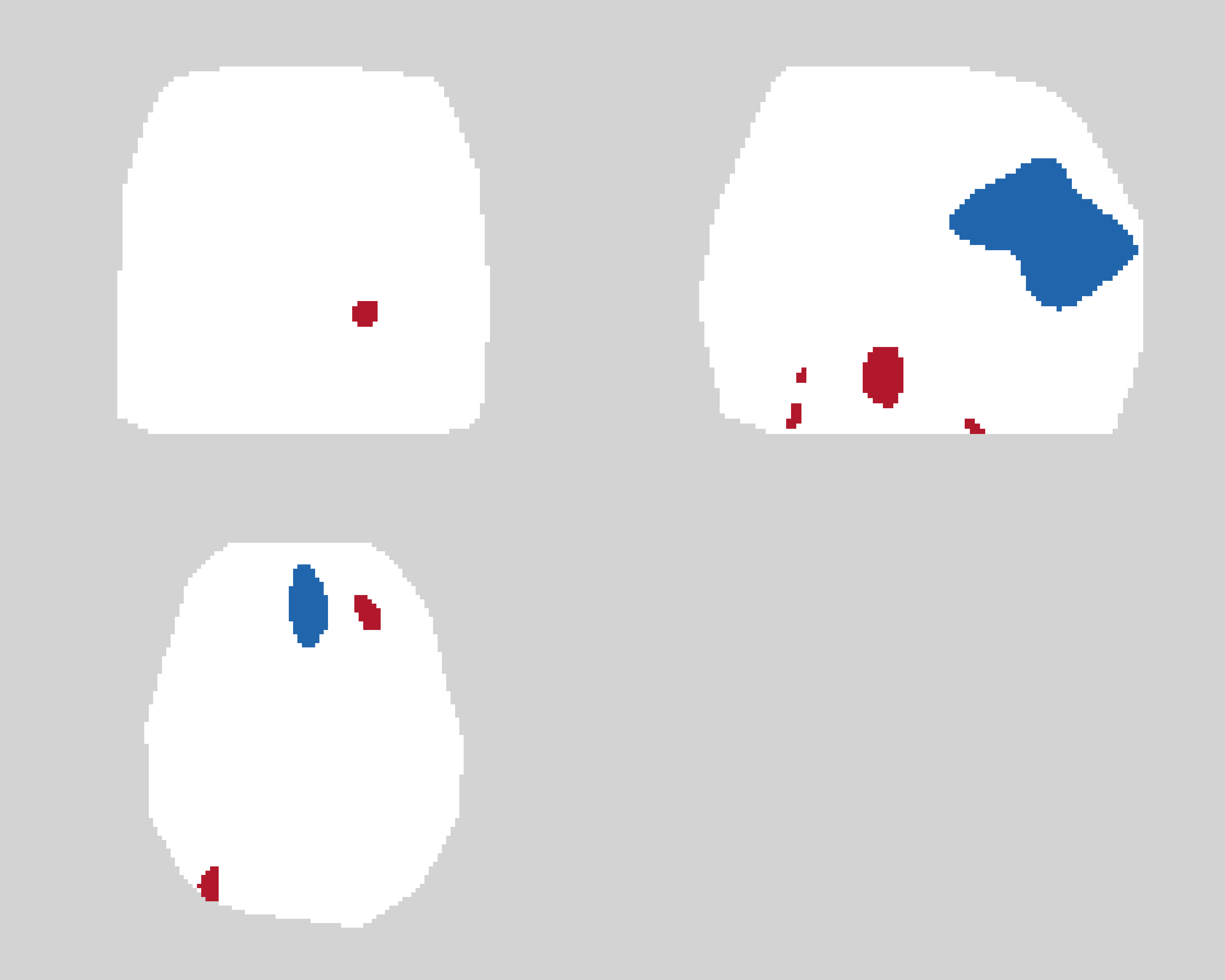} & \includegraphics[width=.85in]{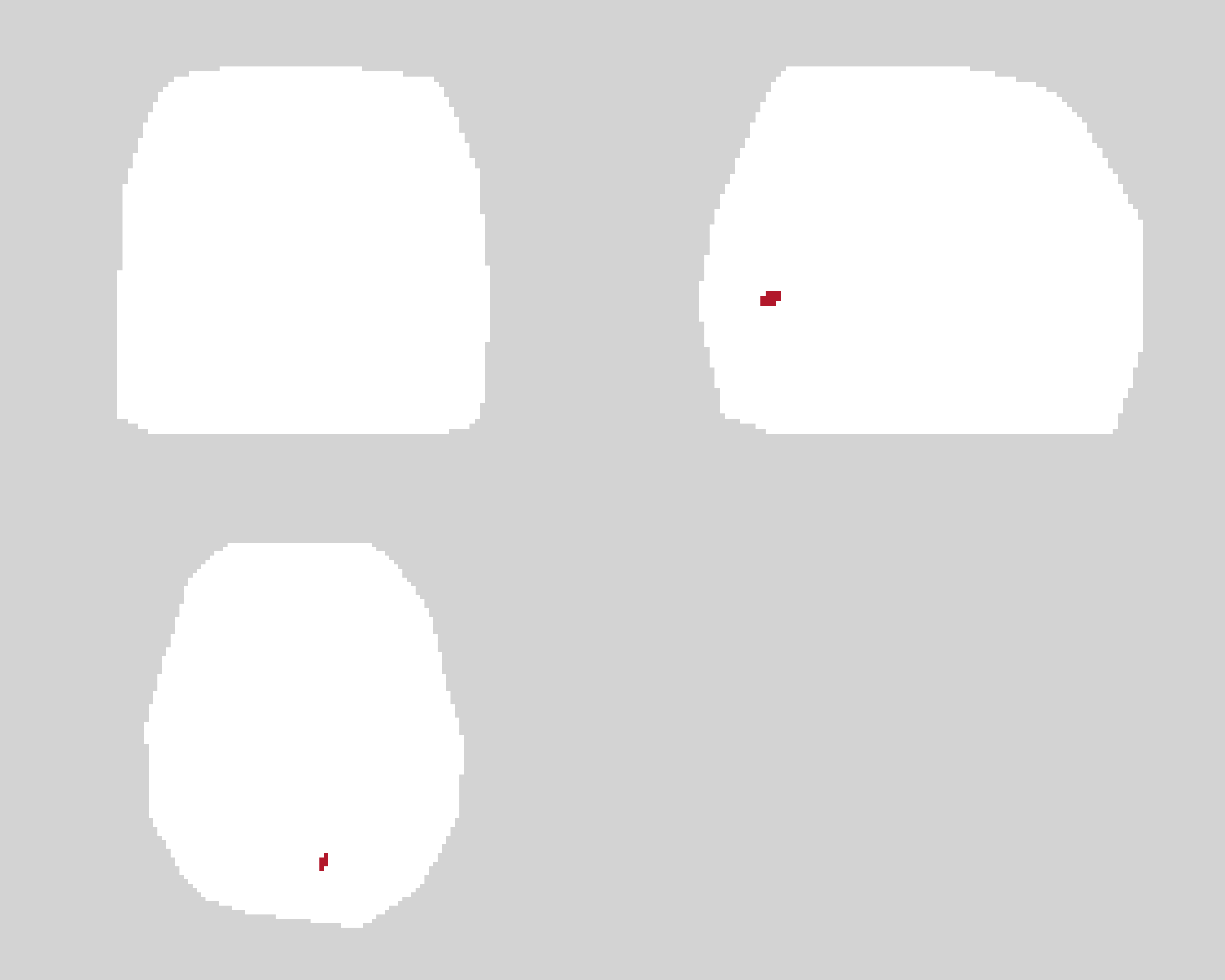} & \includegraphics[width=.85in]{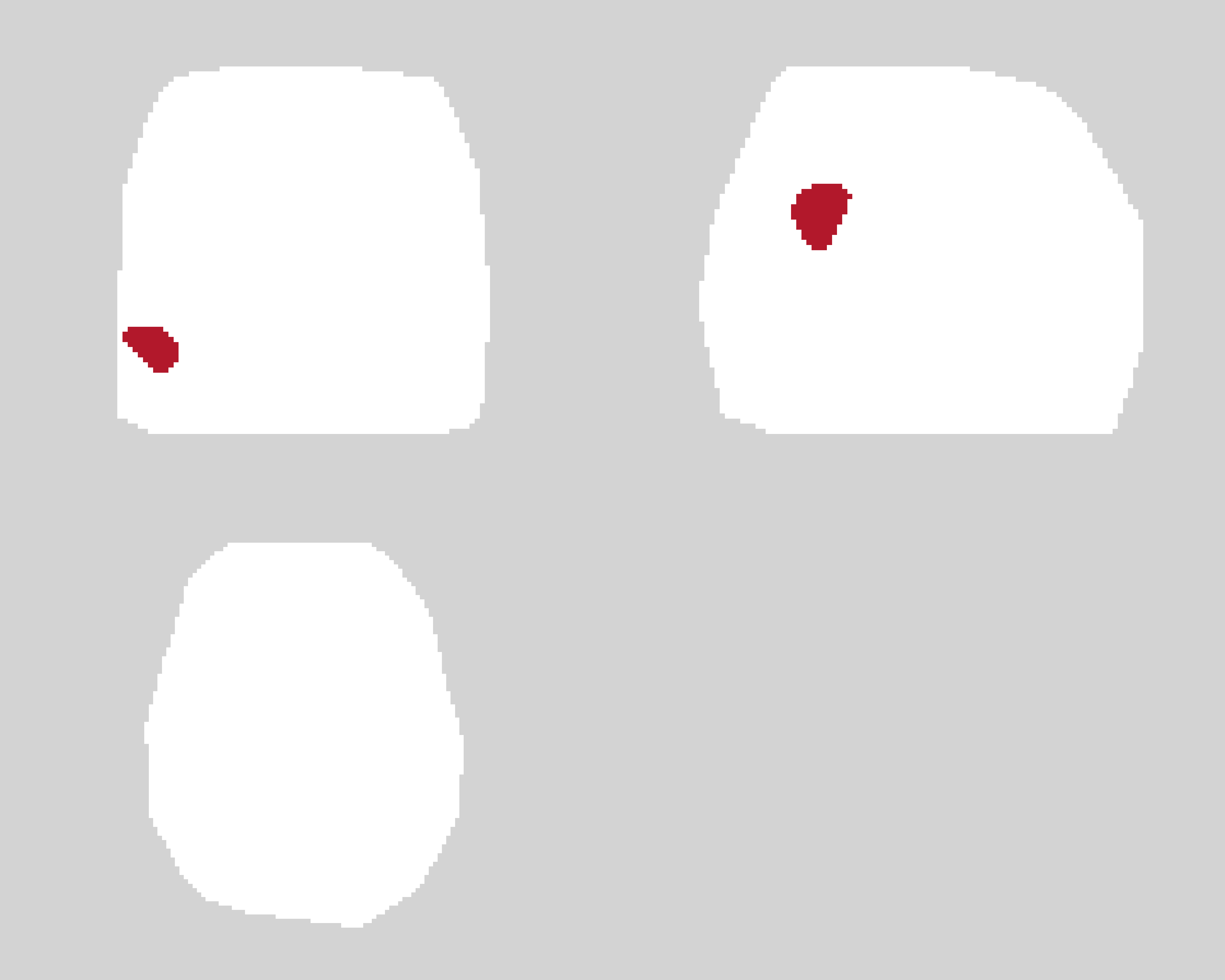} \\[-1pt]
Sig.\ SSFR & \includegraphics[width=.85in]{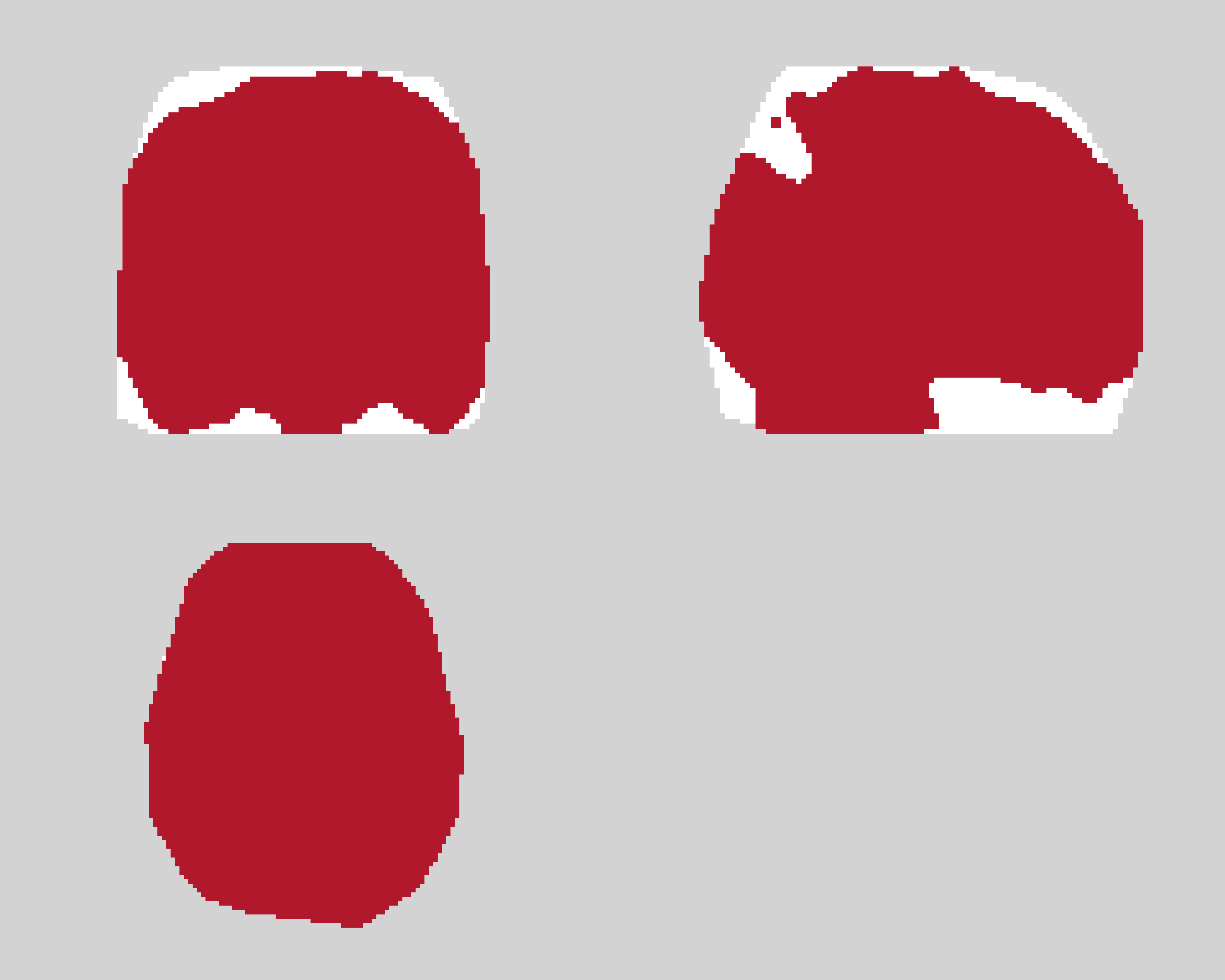} & \includegraphics[width=.85in]{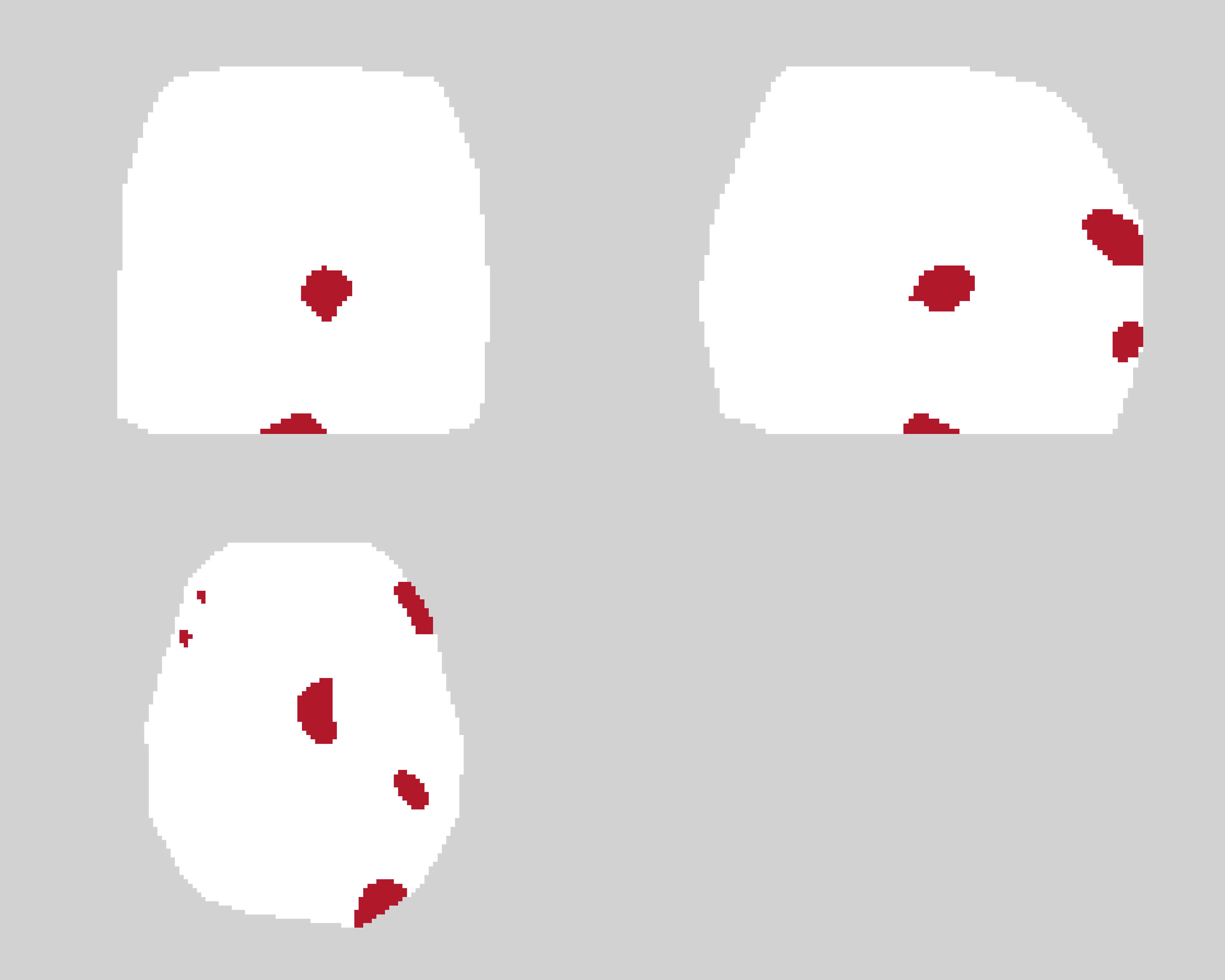} & \includegraphics[width=.85in]{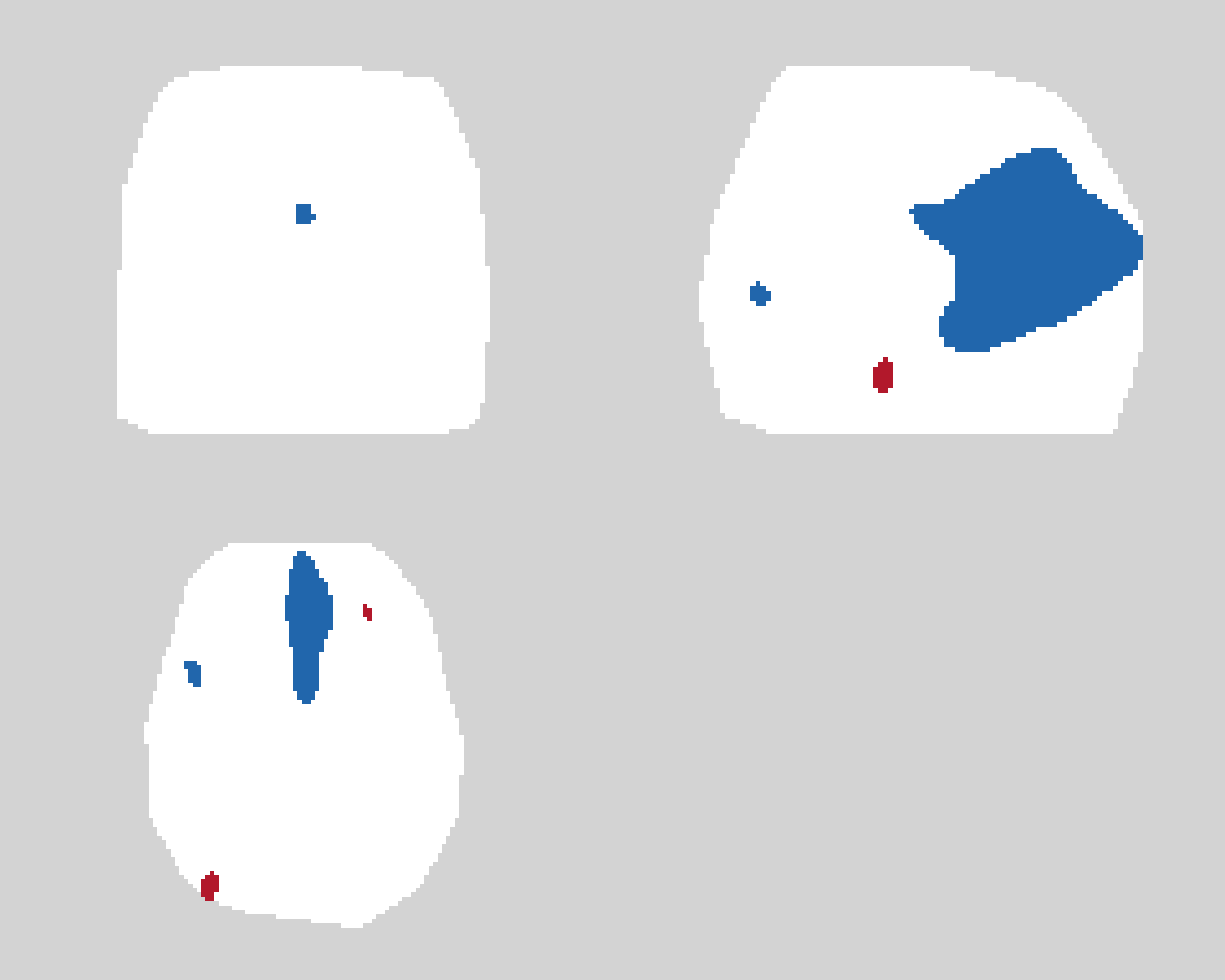} & \includegraphics[width=.85in]{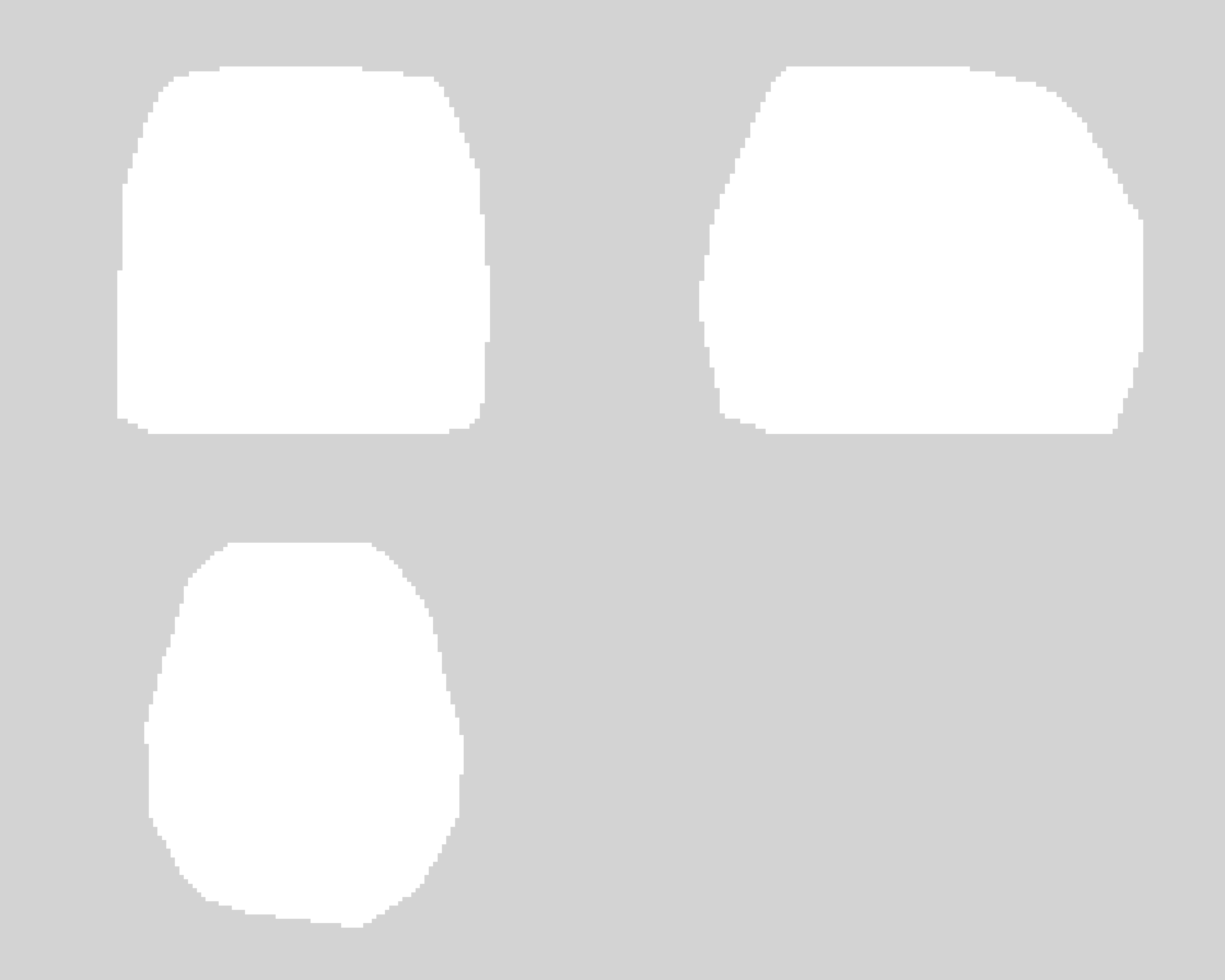} & \includegraphics[width=.85in]{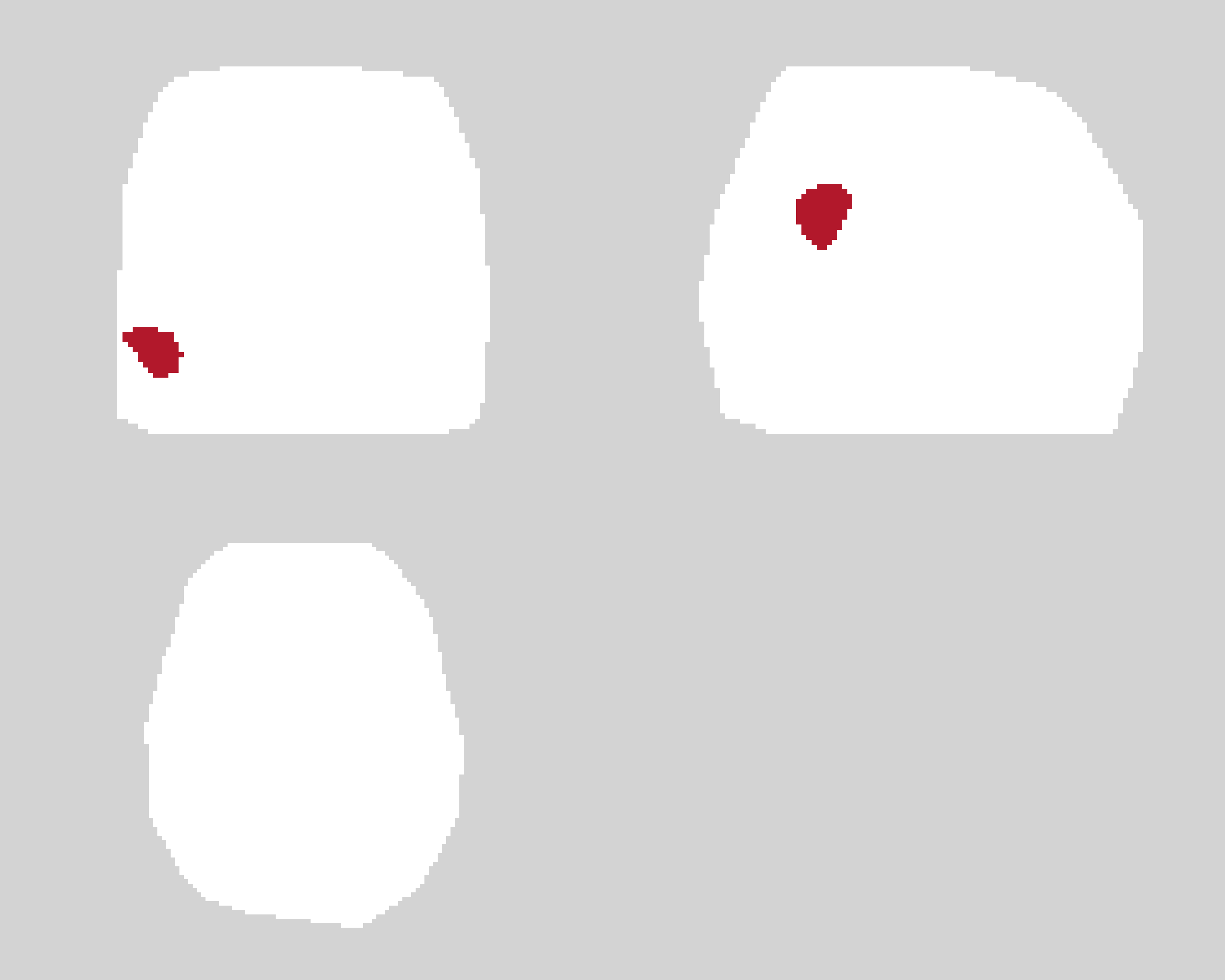} \\[-1pt]
Difference & \includegraphics[width=.85in]{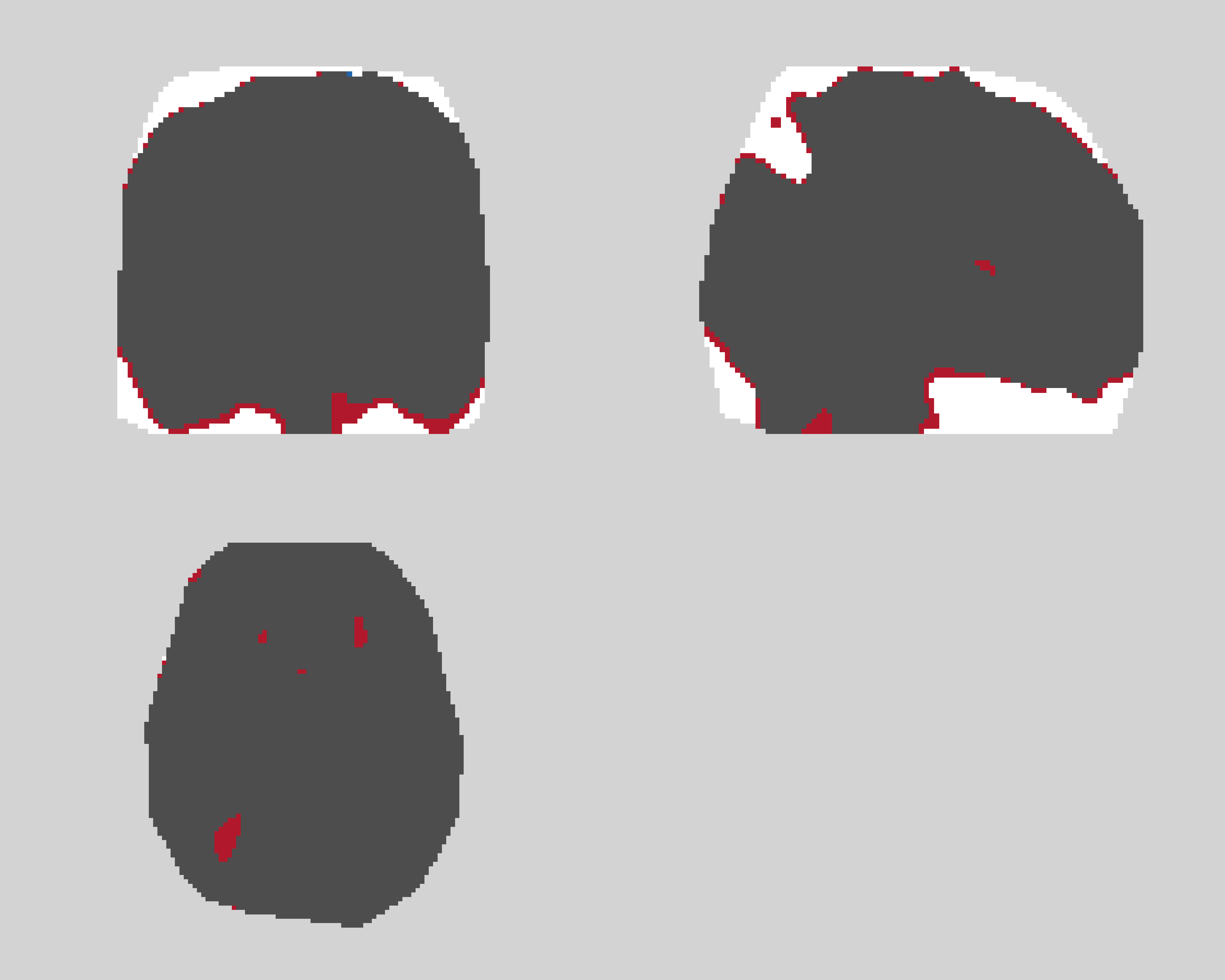} & \includegraphics[width=.85in]{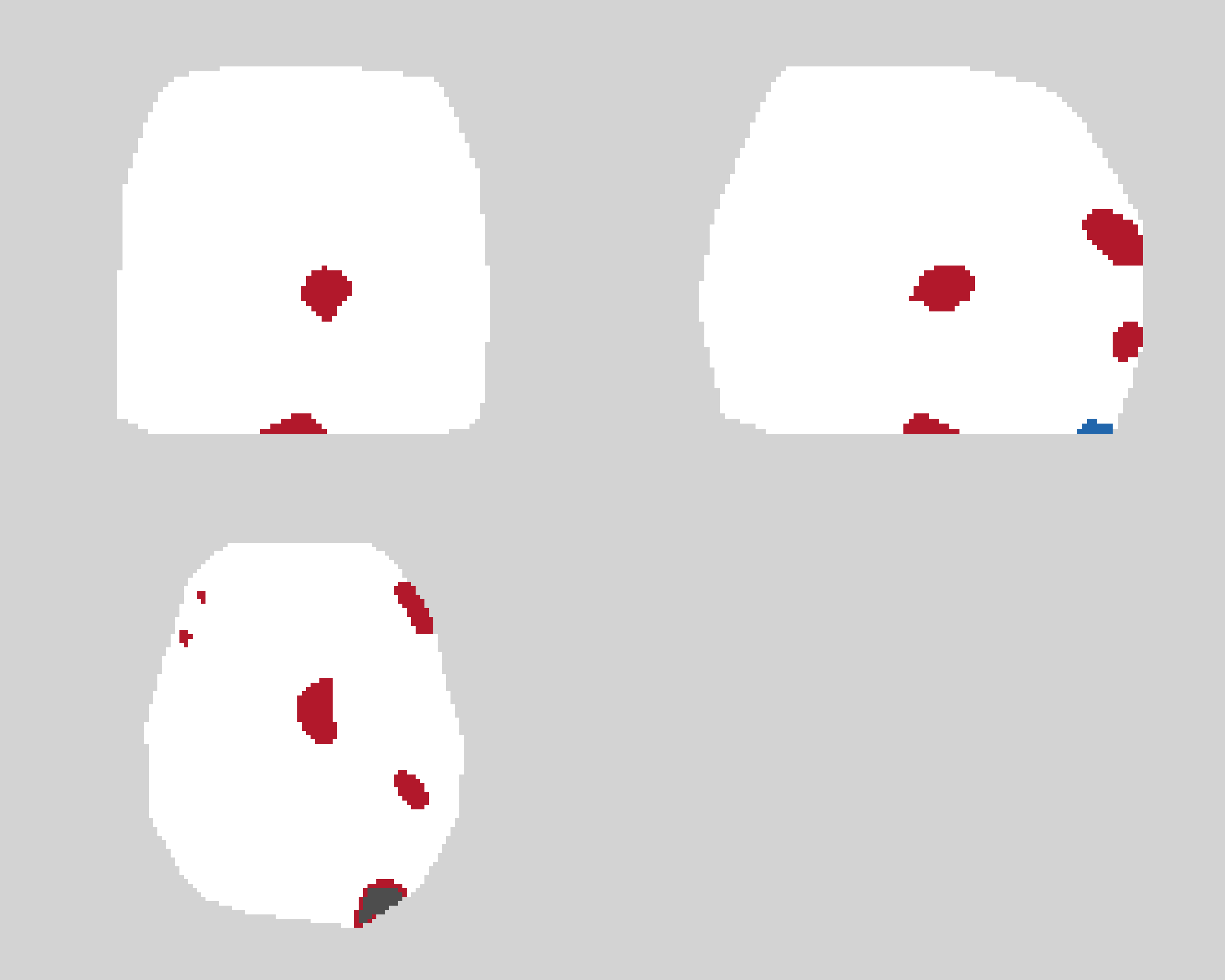} & \includegraphics[width=.85in]{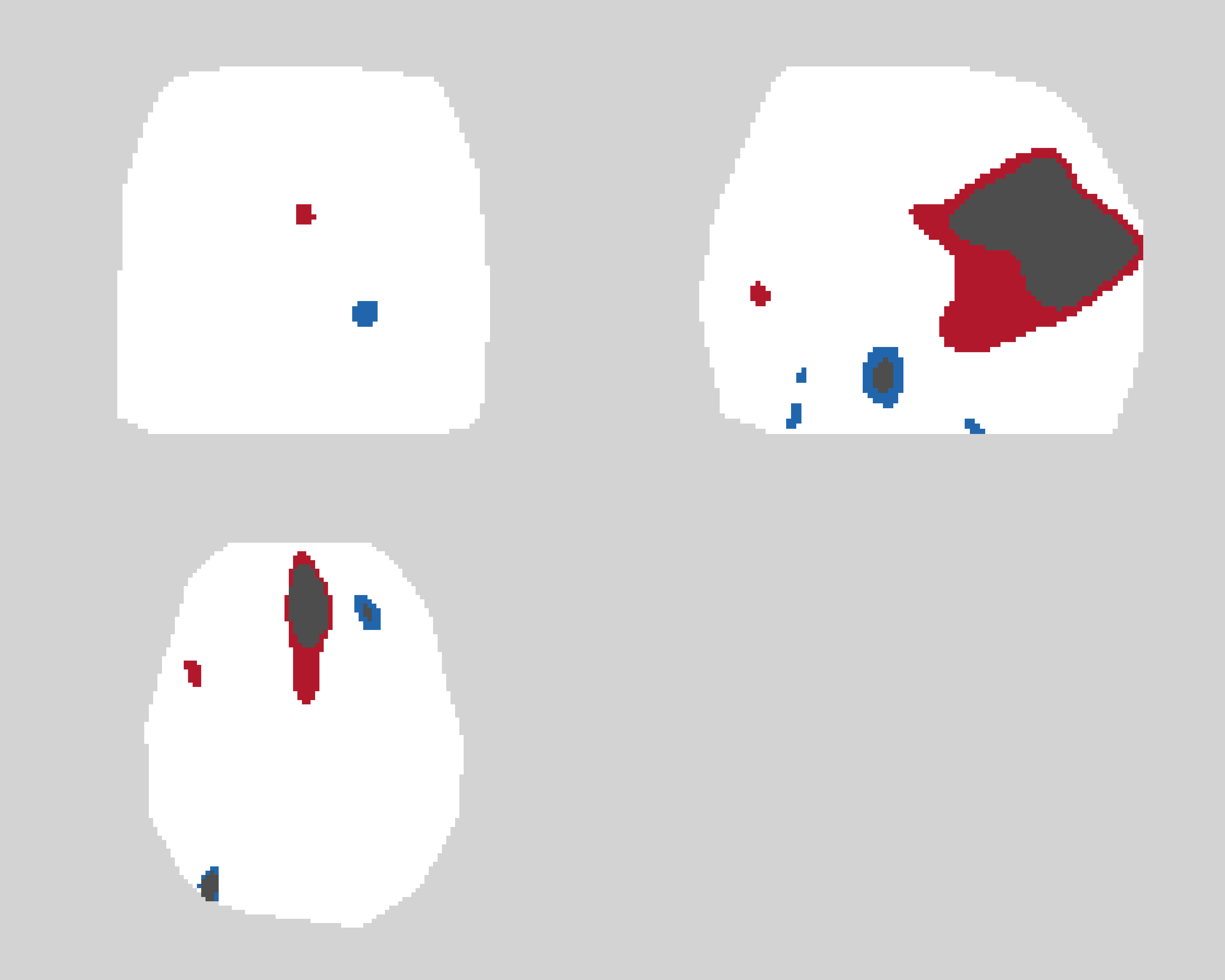} & \includegraphics[width=.85in]{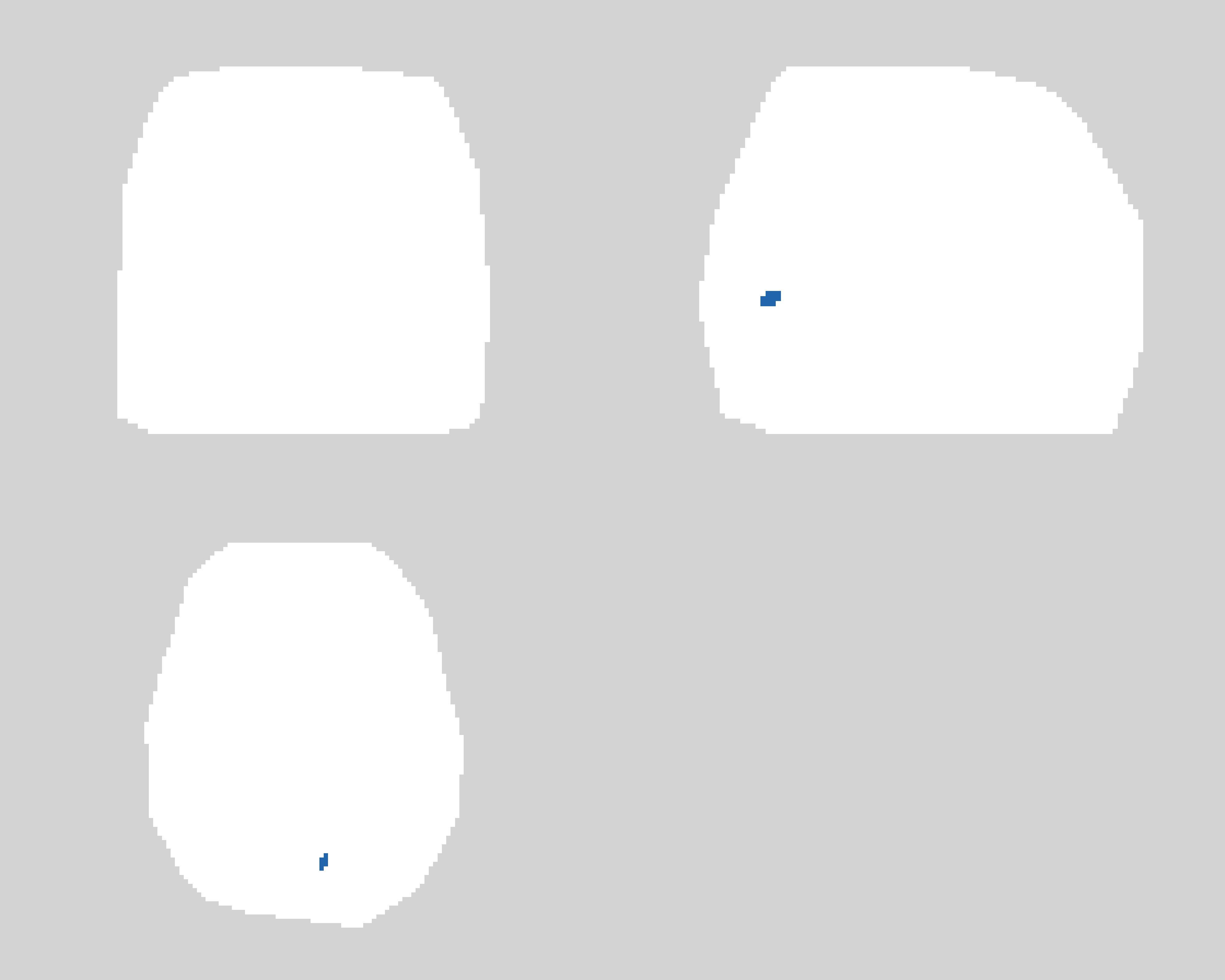} & \includegraphics[width=.85in]{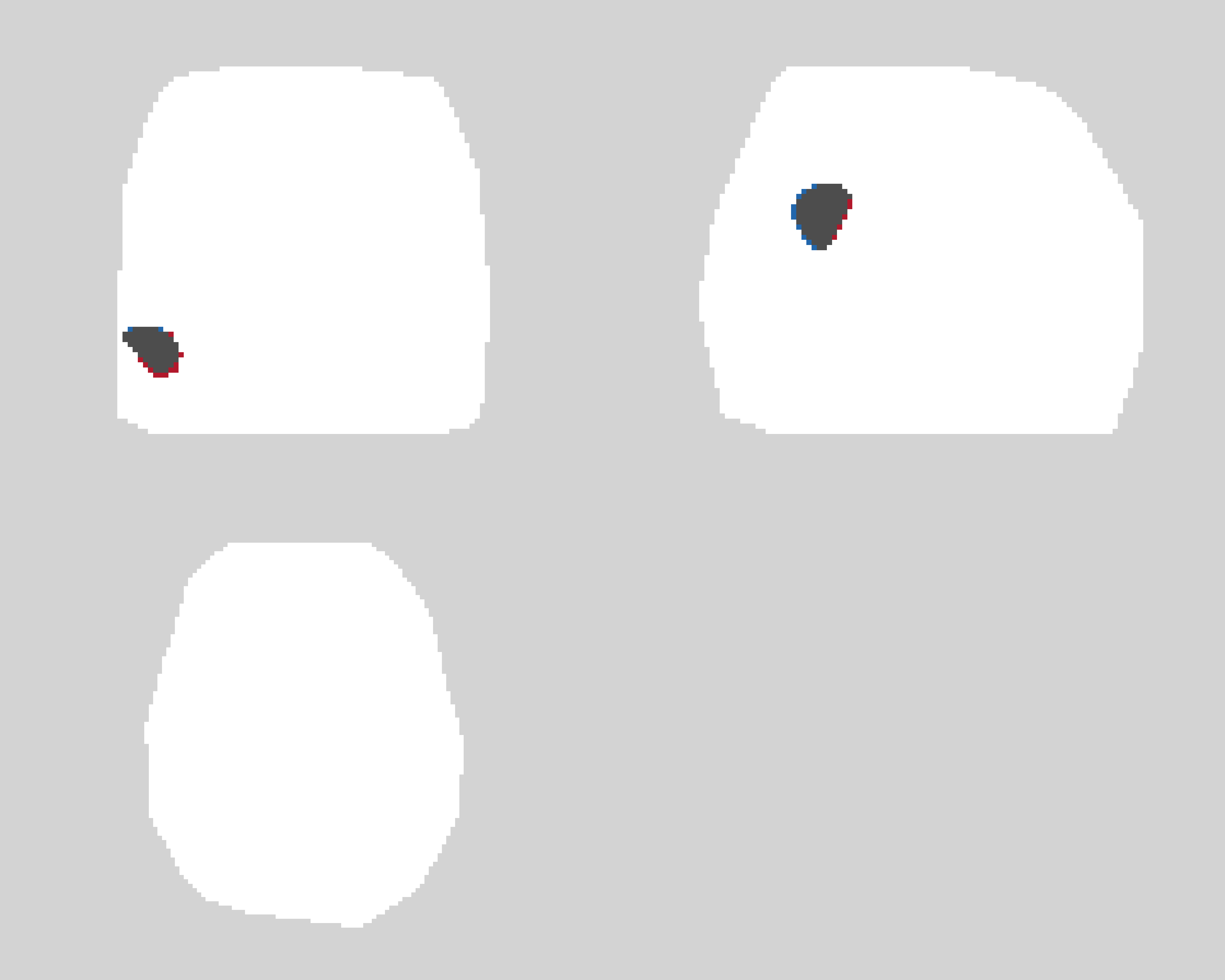} \\[-1pt]
\end{tabular}
}
\caption{ADNI expanded-cohort results. \textbf{A.} SSFR coefficient estimates and percent reduction in pointwise bootstrap SE relative to ISR (red = reduction) for all nine coefficients; \textbf{B.} simultaneously significant regions under ISR, under SSFR, and their difference (grey: both; red: SSFR only; blue: ISR only) for the five displayed coefficients. ISR/SISR maps and full significance maps: Supplement B.3.}
\label{FIG:app-coefs}
\end{figure}

The upper blocks of Figure \ref{FIG:app-coefs} display the SSFR estimates of the nine coefficient functions in sagittal, coronal, and axial views following radiological convention, with darker red (blue) indicating stronger positive (negative) associations between the covariate and the imaging outcome conditional on the other covariates, together with the voxelwise percent reduction in bootstrap SE relative to ISR that is discussed in Section \ref{SSEC:application-scc}. The corresponding ISR and SISR maps for all nine coefficients are given in Supplement Section B.3. The ISR and SSFR maps are closely concordant across all nine coefficient functions in sign, spatial localization, and magnitude. This agreement is expected: ISR is consistent for $\alpha_0$, and Corollary \ref{cor:misspec} predicts that SSFR preserves the target coefficient functions while improving efficiency through surrogate-assisted estimation. The primary benefit of SSFR is therefore reduced sampling variability rather than changes in the estimated coefficient surfaces, as reflected by the MISE results (Table \ref{TAB:app-mise}), the SCC widths (Table \ref{TAB:app-scc}), and the widespread reduction in bootstrap SEs shown in Figure \ref{FIG:app-coefs}. For the AD indicator, the estimated coefficient is negative in canonical AD-vulnerable regions, including the posterior cingulate cortex and precuneus.

The SISR maps (Supplement Section B.3) depart visibly from both ISR and SSFR, and the departures vary across coefficient functions. Some coefficients, such as SEX and MMSE, exhibit attenuated spatial patterns, whereas others, such as APOE$_2$, show exaggerated positive effects relative to ISR and SSFR. This coefficient-specific behavior is consistent with Corollary \ref{cor:misspec}: because naive SISR directly substitutes surrogate images for the missing responses, its estimates are shifted toward $\beta_0$ by an amount determined by the local discrepancy $\beta_{0,\ell}-\alpha_{0,\ell}$. Consequently, SISR may benefit from reduced variance where this discrepancy is small but suffers substantial bias where it is not. In contrast, SSFR recalibrates the surrogate-derived predictor against the observed PET images and therefore avoids this instability.

Overall, the ADNI analysis demonstrates that SSFR preserves the coefficient estimates obtained from ISR while consistently reducing estimation uncertainty, with the gains reflected in the MISE and SCC comparisons.

\vspace{-6pt}
\subsection{SCCs for the ADNI Coefficient Functions}
\label{SSEC:application-scc}

Beyond point estimation, we perform simultaneous inference for the coefficient functions in \eqref{EQ:GFR-app} using the weighted bootstrap procedure of Section \ref{SSEC:bootstrap}. We analyze the expanded cohort described in Section \ref{SSEC:application-2}, comprising 260 subjects with observed PET and 325 additional subjects with MRI-based PET surrogates, using inverse-probability weighting to account for MAR selection. SCCs are constructed using $B=500$ bootstrap replicates at the nominal level $1-\alpha_0=0.95$. For comparison, we also compute SCCs for the observed-data ISR estimator.

For each coefficient function, the \emph{simultaneously significant region} is defined as the set of voxels where the corresponding SCC excludes zero. Because the SCC provides simultaneous coverage over the imaging domain, these regions identify locations where the covariate effect differs significantly from zero while accounting for simultaneous inference across the imaging domain.

Table \ref{TAB:app-scc} reports the mean SCC width for each coefficient under ISR and SSFR together with the relative reduction, $(\bar w_{\mathrm{ISR},\ell} - \bar w_{\mathrm{SSFR},\ell}) / \bar w_{\mathrm{ISR},\ell} \times 100\%$, achieved by SSFR. SSFR produces uniformly narrower SCCs than ISR, with reductions ranging from 6.5\% to 8.3\% and averaging 7.43\% across coefficients.
The consistency of these reductions across all coefficient functions suggests a systematic variance reduction attributable to the shared co-kriging component rather than a coefficient-specific effect.
The percent-reduction maps in Figure \ref{FIG:app-coefs}\,\textbf{A} and Panel B of Table \ref{TAB:app-scc} corroborate this at the pointwise level: the mean reduction in bootstrap SE is 4.52\% and the median 4.71\%, with red regions predominating across all nine coefficient maps, confirming that the variance reduction is spatially broad rather than concentrated at isolated voxels.

Figure \ref{FIG:app-coefs}\,\textbf{B} displays the simultaneously significant regions obtained from ISR and SSFR. The intercept function is significant over the large majority of the brain domain under both estimators, as expected, since it captures the overall metabolic level rather than a covariate contrast; SSFR extends this region modestly. For the AGE coefficient, both estimators identify a focal negative region, and the SSFR region is visibly larger and more spatially contiguous than its ISR counterpart, with the additional voxels predominantly carrying a negative sign. The SEX coefficient yields a small number of scattered significant voxels under both estimators, again somewhat more numerous under SSFR, and the MMSE coefficient yields two small positive foci detected by both. For the $APOE_2$ coefficient, ISR identifies a small number of significant voxels, whereas SSFR identifies only 15. This difference is compatible with the generally narrower SSFR SCCs, since significance depends jointly on the estimated coefficient and the corresponding corridor width. Thus, modest differences between the ISR and SSFR estimates may lead to slightly different detection patterns. 
For EDUCATION, APOE$_1$, and MCI coefficients, neither estimator identifies any simultaneously significant voxel at the 95\% level. AD has 94 significant voxels under ISR and 164 under SSFR. Maps for these four coefficients appear in Supplement Section B.3; the lower block of Figure \ref{FIG:app-coefs} displays the other five coefficients.
The Difference row of Figure \ref{FIG:app-coefs} makes this gain concrete: dark grey voxels are declared significant by both estimators, red voxels only by SSFR, and blue voxels only by ISR. SSFR-only detections are concentrated in the intercept and AGE coefficients, consistent with those being the coefficients nearest the detection threshold under ISR.

Overall, the ADNI analysis demonstrates that the efficiency gains associated with surrogate-assisted estimation translate into uniformly tighter simultaneous uncertainty quantification in the real-data setting, and that the resulting gain in detection is concentrated primarily in the AGE coefficient, for which ISR was already near the detection threshold, with the intercept gaining coverage area rather than new sign-specific detections.

\begin{table}[p]\centering
\caption{Simultaneous and pointwise uncertainty for the observed-data ISR and SSFR estimators in the ADNI application. Panel A reports the average SCC width across voxels; Panel B reports the average pointwise bootstrap standard error. Reductions are relative to ISR.}
\label{TAB:app-scc}
\resizebox{\linewidth}{!}{
\begin{tabular}{lccccccccc} \toprule
& Intercept & Sex & Age & Education & APOE$_1$ & APOE$_2$ & MMSE & MCI & AD \\ \midrule
\multicolumn{10}{l}{\textit{Panel A: average simultaneous confidence corridor (SCC) width}}\\
ISR & 0.6004 & 0.0695 & 0.0054 & 0.0115 & 0.0787 & 0.1121 & 0.0143 & 0.0856 & 0.1378 \\
SSFR & 0.5612 & 0.0642 & 0.0049 & 0.0106 & 0.0735 & 0.1033 & 0.0133 & 0.0795 & 0.1263 \\
Mean reduction (\%) & 6.5 & 7.6 & 7.9 & 8.0 & 6.5 & 7.8 & 7.1 & 7.1 & 8.3 \\
Median reduction (\%) & 7.1 & 7.4 & 8.2 & 8.3 & 6.4 & 7.7 & 7.6 & 7.3 & 8.6 \\ \midrule
\multicolumn{10}{l}{\textit{Panel B: average pointwise bootstrap standard error}}\\
ISR & 0.0640 & 0.0074 & 0.0006 & 0.0012 & 0.0084 & 0.0117 & 0.0015 & 0.0089 & 0.0146 \\
SSFR & 0.0608 & 0.0071 & 0.0005 & 0.0012 & 0.0079 & 0.0112 & 0.0014 & 0.0085 & 0.0138 \\
Mean reduction (\%) & 5.0 & 3.3 & 5.2 & 4.2 & 5.2 & 4.2 & 4.5 & 4.0 & 5.2 \\
Median reduction (\%) & 5.5 & 3.1 & 5.5 & 4.5 & 5.0 & 4.2 & 5.0 & 4.1 & 5.5 \\ \bottomrule
\end{tabular}}
\end{table}

\section{Conclusion and Discussion}
\label{SEC:conclusion}

In this paper, we introduce the SSFR framework for analyzing brain imaging data subject to modality-level missingness. SSFR acts as a statistical safeguard for integrating observed and synthetic surrogate images within a unified two-stage functional regression framework, enabling reliable estimation of spatially varying covariate effects when the primary imaging modality is only partially observed.

The proposed framework offers three main methodological innovations. First, unlike conventional imputation or direct surrogate substitution strategies, SSFR employs a two-stage design specifically to de-bias synthetic surrogates. This mechanism offers principled protection against inaccuracies in generative imaging models, allowing surrogate information to be harnessed without requiring perfect synthesis quality: Theorem \ref{thm:stage2} and Corollary \ref{cor:misspec} formalize this robustness, showing that $\widehat{\bs\alpha}$ remains consistent for $\bs\alpha_0$ at the same rate as an oracle estimator, even when the surrogate model is misspecified. Second,the local partial $R^2(\bs{z})$ introduced in Section \ref{SSEC:surrogate_measure} quantifies the fraction of residual target-image variation explained by the surrogate and characterizes the potential efficiency gain through \eqref{EQ:VR}. In particular, the image-level $R^2$, $\overline{R^2}$, together with the observed-image fraction, provides a data-driven criterion for screening and ranking candidate surrogates before they are incorporated into the analysis. Third, the weighted bootstrap procedure of Section \ref{SSEC:bootstrap} furnishes SCCs that achieve empirical coverage close to the nominal level in our simulations (Section \ref{SEC:simulation}), enabling identification of spatial regions with significant covariate effects rather than relying on visual comparison of coefficient maps alone; a formal proof of asymptotic validity for the bootstrap procedure is left for future work.

A natural question raised in the course of this work is whether the efficiency gains achieved by SSFR could instead be replicated by a simpler group-level regularization or prior toward a benchmark image, rather than genuine subject-specific information transfer. Our framework gives a sharp answer. The surrogate enters the second stage only through the co-kriging term $C(\bs z,\widetilde Y_i)$, and by Assumption A5 and Remark \ref{rem:surrogate-efficiency}, the efficiency gain is governed by the amount of residual target-image variation explained by the surrogate, as quantified by $R^2(\bs{z})$. If a surrogate is constructed from a group-specific benchmark without individual-level auxiliary data (e.g., a stratum mean of PET by diagnosis, age band, and sex, using no MRI), then when the group is defined by covariates already in the model, the surrogate-derived predictor lies in the span of $\mathbf X_i$, Assumption A5 fails, and SSFR reduces to ISR because the surrogate contributes no subject-specific residual information beyond the covariates, corresponding to $R^2(\bs{z})=0$; when the group is defined outside the covariate set, the surrogate contributes only a group-level fixed effect. In neither case does a group-benchmark surrogate transfer subject-specific information, and in neither case does SSFR behave like a shrinkage-toward-benchmark procedure, since by Corollary~\ref{cor:misspec} $\widehat{\bs\alpha}$ remains consistent for $\bs\alpha_0$ regardless of surrogate quality, whereas a prior or penalty toward benchmark PET would introduce bias toward that benchmark. This distinguishes the mechanism underlying SSFR's efficiency gains from ordinary regularization.

Extensive simulation studies show that SSFR yields stable and accurate estimation across a range of realistic settings, including varying degrees of imaging missingness, surrogate-model misspecification, and surrogates from several distinct AI architectures. Estimation accuracy improves as the surrogate explanatory power increases, consistent with the efficiency discussion accompanying the asymptotic theory of Section \ref{SEC:theory}. The ADNI application further illustrates its practical utility in a real neuroimaging setting, where PET data are costly and incompletely observed. By leveraging MRI-based PET surrogates, SSFR recovers biologically meaningful and spatially heterogeneous associations between subject-level covariates and cerebral glucose metabolism that are consistent with established Alzheimer's disease pathology, and the accompanying SCCs provide formal statistical support for the localization of these associations.

Our framework also has limitations that merit discussion. First, valid estimation in the second stage relies on the missing-at-random assumption; when PET acquisition depends on unobserved metabolic status, the missingness is more accurately described as missing-not-at-random (MNAR), and inverse-probability weighting no longer fully removes selection bias. Because SSFR draws its efficiency from surrogate information that is largely orthogonal to the missingness mechanism, it is expected to remain more accurate than observed-data-only analyses under moderate departures from MAR; nevertheless, robustness to the missingness mechanism is distinct from the robustness to surrogate misspecification that our two-stage design guarantees (Corollary \ref{cor:misspec}). Second, the efficiency gain depends on how much residual target-image variation the surrogate explains, as quantified by the local and global partial $R^2$ measures. When a candidate surrogate is essentially uninformative, incorporating it yields little benefit and, under severe missingness, may slightly reduce efficiency relative to the observed-data estimator, because degrees of freedom are spent estimating an uninformative surrogate operator. We therefore recommend computing $\overline{R^2}$ and the approximate ratio \eqref{EQ:VR} for each candidate surrogate, and augmenting only when the predicted gain clearly exceeds the finite-sample cost reflected in $c_L$.

Several important directions remain for future research. Scaling the TPST estimation procedure of Section \ref{SSEC:estimation} to whole-brain voxel grids with hundreds of thousands of locations introduces substantial computational challenges; a distributed extension of SSFR based on triangulated domain decomposition, together with a communication-efficient bootstrap procedure adapted to that setting, is developed in a companion paper. In addition, extending SSFR to imaging genetics is a challenging but important avenue for future work. Imaging genetics studies often face the ``big data square'' problem \citep{nathoo2019review, zhu2023statistical, zhao2023heart}, which arises from the combination of high-dimensional imaging outcomes, large numbers of genetic predictors, and limited effective sample sizes due to the scarcity of subjects with both imaging and genetic data. The present study characterizes how subject-level covariates relate to imaging phenotypes; a natural complementary direction is to reverse the modeling direction and use the surrogate-augmented imaging outcomes to predict cognitive and clinical endpoints, a setting in which the larger effective sample size afforded by SSFR may be especially advantageous. Principled selection or combination of multiple candidate surrogates using their estimated explanatory power, as well as accommodating MNAR missingness through selection-model or pattern-mixture formulations, are further important directions.

\section*{Acknowledgments}
Data used in the preparation of this paper were obtained from the ADNI database (adni.loni.usc.edu). The investigators within ADNI contributed to its design and implementation and/or provided data but did not participate in the analysis or writing of this report. A complete listing of ADNI investigators can be found at \url{http://adni.loni.usc.edu/wp-content/uploads/how_to_apply/ADNI_Acknowledgement_List.pdf}. 

\section*{Funding}
This research is partially supported by NIH award 1R01AG085616, NSF awards DMS-2426173 and DMS-2515788, and Simons Foundation Award 963447.

\section*{Statement on AI use}
The authors used generative AI tools (Claude Opus 5.5, Anthropic; and GPT-6.1 Sol, OpenAI) to polish the language and improve the readability of the text, and to assist with writing and debugging code for the simulations and figures. The authors conceived the research question, developed the methodology, formulated the identification conditions and theoretical results, designed the proof strategies, and verified all proofs, numerical results, and references. The authors reviewed and edited all AI-generated text and code and take full responsibility for the content of the manuscript.

\clearpage
\setcounter{page}{1}
\begin{center}
{\large\bfseries SUPPLEMENT TO ``SYNTHETIC SURROGATE FUNCTIONAL REGRESSION (SSFR): A ROBUST FRAMEWORK FOR ESTIMATING SPATIALLY VARYING COVARIATE EFFECTS IN SURROGATE-AUGMENTED NEUROIMAGING''}
\end{center}

\setcounter{section}{0}
\renewcommand{\thesection}{\Alph{section}}

\renewcommand{\thesubsection}{A.\arabic{subsection}}
\renewcommand{\thesubsubsection}{A.\arabic{subsection}.\arabic{subsubsection}}
\renewcommand{\thetheorem}{{\sc A.\arabic{theorem}}}
\renewcommand{\thelemma}{{\sc A.\arabic{lemma}}}
\renewcommand{\theproposition}{{\sc A.\arabic{proposition}}}
\renewcommand{\thecorollary}{\sc A.\arabic{corollary}}
\renewcommand{\thefigure}{A.\arabic{figure}}
\renewcommand{\thetable}{A.\arabic{table}}
\renewcommand{\theequation}{A.\arabic{equation}}
\renewcommand{\theremark}{A.\arabic{remark}}
\renewcommand{\thealgorithm}{A.\arabic{algorithm}}
\setcounter{equation}{0}
\setcounter{theorem}{0}
\setcounter{figure}{0}
\setcounter{table}{0}
\setcounter{subsection}{0}
\setcounter{subsubsection}{0}

\section{AI Surrogate Generation and Surrogate Explanatory Power}
\label{SEC:supp-ai-architectures}
This section provides implementation details for the AI surrogate generators used in the simulation studies summarized in Section 6.4 of the main paper and presents additional experiments illustrating how the estimated $\overline{R^2}$ can be used to assess surrogate quality. We first describe the AI architectures and training procedures. The subsequent subsections compare different image-to-image synthesis methods and investigate robustness when surrogate quality varies across clinically relevant subgroups.

\subsection{Architectures and Training Details}
\label{ssec:supp-ai-arch}

This subsection gives implementation details for the AI surrogate generators used in the simulation study with additional AI methods. The benchmark includes a compact 3D UNet \citep{Ronneberger2015,Cicek2016}, a deterministic convolutional encoder--decoder baseline related to fully convolutional CNNs \citep{LeCun1998,Long2015}, a conditional variational autoencoder (CVAE) \citep{KingmaWelling2014,sohn_learning_2015}, and a pix2pix-style conditional GAN \citep{Goodfellow2014,Isola2017}. All architectures are designed to map an auxiliary image $M_i$ to a predicted target image $\widetilde Y_i$. During training, the observed target image $Y_i$ serves as the supervision signal, while after training the fitted network is applied to every subject to generate an AI surrogate image $\widetilde Y_i$. The downstream statistical analysis is identical across architectures, and surrogate quality is evaluated using the estimated image-level explanatory measure $\overline{R^2}$.

The simulated images are represented on a $26\times26\times26$ voxel lattice, with losses and reported images restricted to the 7290 voxels inside the ball domain. Before training, both $M_i$ and $Y_i$ are standardized using only the complete-pair training subjects, and all losses are evaluated on the inside-ball mask. Within each Monte Carlo replication, complete pairs are randomly divided into 80\% training and 20\% validation sets. Unless otherwise stated, every network is trained for 200 epochs using the AdamW optimizer with learning rate $10^{-3}$, weight decay $10^{-5}$, batch size 8, ReduceLROnPlateau learning-rate scheduling, gradient clipping at norm 5, and checkpoint selection based on the smallest validation loss.

\begin{table}[t]
\caption{Architecture and training settings for AI surrogate generators.}
\label{tab:supp-ai-training-settings}
\centering
\small
\begin{tabularx}{\linewidth}{
  @{}p{0.12\linewidth}
  X
  p{0.34\linewidth}@{}
} \toprule
Model & Architecture & Objective and method-specific settings \\
\midrule
UNet &Two-level 3D UNet \citep{Ronneberger2015,Cicek2016}: \(3^3\) convolutional blocks, stride-2 downsampling, transposed-convolution upsampling, one skip connection, and \(1^3\) output convolution. &Masked \(L_1\) loss; base channels \(b=4\). \\
\midrule
CNN &3D encoder-decoder CNN \citep{LeCun1998,Long2015} with the same channel widths as UNet but no skip connection. &Masked \(L_1\) loss; base channels \(b=4\). \\
\midrule
CVAE &Conditional 3D VAE \citep{KingmaWelling2014,sohn_learning_2015}: condition encoder for \(M_i\), posterior encoder for \((M_i,Y_i)\), latent code concatenated with condition features, and convolutional decoder. Prediction uses $z=0$. &Masked \(L_1\) plus KL penalty; latent dimension 8; KL weight \(10^{-4}\); \(b=4\). \\
\midrule
GAN &Pix2pix-style 3D conditional GAN \citep{Goodfellow2014,Isola2017}: UNet generator and PatchGAN discriminator applied to \((M_i,Y)\). &Generator BCE adversarial loss plus \(50\times\) masked \(L_1\); discriminator BCE loss; \(b=4\); AdamW betas \((0.5,0.999)\). \\
\bottomrule
\end{tabularx}
\begin{flushleft}
\footnotesize
Note: All main architecture comparisons use \(n=200\), MAR missingness with IPW in the downstream stage, missing rates \(25\%\), \(50\%\), and \(75\%\), and 100 Monte Carlo replications per missingness setting. The same complete-pair split, image standardization, masked loss, and downstream SSFR evaluation pipeline are used for every architecture.
\end{flushleft}
\end{table}

Table \ref{tab:supp-ai-training-settings} summarizes the architecture-specific configurations. To ensure a fair comparison, all architectures use the same training/validation split, image standardization, masked-loss evaluation, optimization strategy, and downstream SSFR pipeline. Consequently, the only difference among the competing methods is the image-to-image synthesis model used to generate the surrogate images.

\subsection{Surrogates Generated by AI Image-to-image Synthesis Methods}
\label{ssec:supp-ai-sim}

The simulations in the main paper use analytically generated surrogate images so that surrogate informativeness and model misspecification can be controlled directly. This subsection investigates whether the same conclusions continue to hold when surrogate images are generated by modern AI image-to-image synthesis methods. We first compare several neural-network architectures for surrogate generation and then examine a setting in which the AI-generated surrogate quality varies across subject subgroups.

\subsubsection{Benchmark across Generator Architectures}
\label{SSSEC:sim-ai-benchmark}

We compare SSFR using surrogate images generated by a compact 3D UNet \citep{Ronneberger2015,Cicek2016}, a CNN encoder--decoder baseline \citep{LeCun1998,Long2015}, a conditional variational autoencoder (CVAE) \citep{KingmaWelling2014,sohn_learning_2015}, and a pix2pix-style conditional GAN \citep{Goodfellow2014,Isola2017}. All architectures satisfy the same input-output interface and therefore can be incorporated into an identical SSFR pipeline. Architectural specifications and training details are provided in Supplementary Section \ref{ssec:supp-ai-arch}.

The synthetic multimodal data reuse the domain $\Omega$, covariates $\mathbf X_i$, coefficient functions $\bs\alpha_0$, latent-score construction $\{\lambda_k,\psi_k\}_{k=1}^3$, and noise level described in the main article. Because the image-to-image networks require regular voxel grids, the images are generated on a $26\times26\times26$ lattice defined on $\{0,0.04,\ldots,1\}^3$ and restricted to the 7290 voxels inside the ball. The target image is generated according to
\[
Y_i(\bs{z}_j) = \mathbf X_i^\top\bs\alpha_0(\bs{z}_j) + \eta_i^{(O)}(\bs{z}_j) + \varepsilon_{ij}^{(O)}.
\]
In addition to the target image, an auxiliary modality is generated as
\[
M_i(\bs{z}_j) = \mathbf X_i^\top \bs\beta_0^{M}(\bs{z}_j) + \eta_i^{(M)}(\bs{z}_j) + \varepsilon_{ij}^{(M)},
\]
where $\bs\beta_0^{M}$ follows the same misspecified functional form as the misspecified surrogate model in main article. Thus, $M_i$ remains informative for the target image while intentionally violating the target regression model.

The target and auxiliary images are jointly generated by correlating both the latent scores $(\xi_{ik}^{(O)},\xi_{ik}^{(M)})^\top$ and the voxel-level noise $(\varepsilon_{ij}^{(O)},\varepsilon_{ij}^{(M)})^\top$, using correlation $\omega_M=0.75$. Importantly, the target image is not generated from the auxiliary image. Instead, $M_i$ serves as an informative but imperfect auxiliary imaging modality.

Target images are then made missing according to the same MAR mechanism described in Section 6.1, with the intercept calibrated to attain missing rates $r_m\in\{25\%,50\%,75\%\}$. The auxiliary images and scalar covariates remain fully observed.

Within each Monte Carlo replication, every AI model is trained only on the complete image pairs $(M_i,Y_i)$ and then used to generate an AI surrogate image $\widetilde Y_i$ for every subject.

The resulting surrogate images are analyzed using the same downstream methods considered throughout the paper, namely ISR, IPW, AIPW, naive SISR, and SSFR. For naive SISR, missing target images are replaced directly by the generated surrogate images before fitting ISR. In contrast, SSFR treats $\widetilde Y_i$ only as a surrogate outcome and recalibrates its information through the proposed two-stage estimation procedure. Performance is summarized using the coefficient-wise MISE together with the image-level surrogate explanatory measure $\overline{R^2}$ defined in the main paper.

\begin{table}[htbp]
\caption{MISEs of the coefficient functions across AI-surrogate benchmarks under MAR. \label{TAB:ai-unet-mise}}
\centering
\resizebox{\linewidth}{!}{
\begin{threeparttable}
{\setlength{\tabcolsep}{2.5pt}
\begin{tabular}{ccccccccccccccccc} 
\toprule
\multirow{2}{*}{Architecture} &\multirow{2}{*}{$\overline{R^2}$} &\multicolumn{3}{c}{ISR} &\multicolumn{3}{c}{IPW} &\multicolumn{3}{c}{AIPW} &\multicolumn{3}{c}{SISR} &\multicolumn{3}{c}{SSFR}\\ \cmidrule(lr){3-5}\cmidrule(lr){6-8}\cmidrule(lr){9-11} \cmidrule(lr){12-14}\cmidrule(lr){15-17}
& &$\alpha_0$ &$\alpha_1$ &$\alpha_2$ &$\alpha_0$ &$\alpha_1$ &$\alpha_2$ &$\alpha_0$ &$\alpha_1$ &$\alpha_2$ &$\alpha_0$ &$\alpha_1$ &$\alpha_2$ &$\alpha_0$ &$\alpha_1$ &$\alpha_2$\\ \midrule
\multicolumn{17}{c}{\textbf{Missingness Rate $r_m = 25\%$}}\\ \midrule
UNet &0.493 &\multirow{4}{*}{0.0038} &\multirow{4}{*}{0.0110} &\multirow{4}{*}{0.0029} &\multirow{4}{*}{0.0038} &\multirow{4}{*}{0.0111} &\multirow{4}{*}{0.0029} &\multirow{4}{*}{0.0038} &\multirow{4}{*}{0.0111} &\multirow{4}{*}{0.0029} &0.0108 &0.0309 &0.0293 &0.0033 &0.0088 &0.0023\\
CNN  &0.430 & & & & & & & & & &0.0170 &0.0331 &0.0516 &0.0033 &0.0092 &0.0024\\
GAN  &0.484 & & & & & & & & & &0.0123 &0.0305 &0.0378 &0.0033 &0.0088 &0.0024\\
CVAE &0.454 & & & & & & & & & &0.1285 &0.0528 &0.1278 &0.0033 &0.0088 &0.0023\\ \midrule
\multicolumn{17}{c}{\textbf{Missingness Rate $r_m = 50\%$}}\\ \midrule
UNet &0.484 &\multirow{4}{*}{0.0060} &\multirow{4}{*}{0.0151} &\multirow{4}{*}{0.0035} &\multirow{4}{*}{0.0060} &\multirow{4}{*}{0.0154} &\multirow{4}{*}{0.0034} &\multirow{4}{*}{0.0060} &\multirow{4}{*}{0.0154} &\multirow{4}{*}{0.0034} &0.0397 &0.0835 &0.1560 &0.0048 &0.0121 &0.0031\\
CNN  &0.422 & & & & & & & & & &0.0572 &0.0962 &0.2548 &0.0047 &0.0130 &0.0031\\
GAN  &0.486 & & & & & & & & & &0.0398 &0.0804 &0.1700 &0.0046 &0.0121 &0.0031\\
CVAE &0.421 & & & & & & & & & &0.3478 &0.1226 &0.5305 &0.0050 &0.0127 &0.0029\\ \midrule
\multicolumn{17}{c}{\textbf{Missingness Rate $r_m = 75\%$}}\\ \midrule
UNet &0.468 &\multirow{4}{*}{0.0116} &\multirow{4}{*}{0.0316} &\multirow{4}{*}{0.0063} &\multirow{4}{*}{0.0113} &\multirow{4}{*}{0.0310} &\multirow{4}{*}{0.0065} &\multirow{4}{*}{0.0114} &\multirow{4}{*}{0.0310} &\multirow{4}{*}{0.0065} &0.1910 &0.1770 &0.7589 &0.0074 &0.0194 &0.0055\\
CNN  &0.409 & & & & & & & & & &0.1762 &0.1973 &0.8389 &0.0083 &0.0244 &0.0062\\
GAN  &0.482 & & & & & & & & & &0.1744 &0.1789 &0.6885 &0.0080 &0.0206 &0.0052\\
CVAE &0.371 & & & & & & & & & &0.5255 &0.2271 &1.2560 &0.0085 &0.0232 &0.0060\\ \bottomrule
\end{tabular}}
\begin{tablenotes}
\small
\item Note: All surrogate models were trained for 200 epochs across 100 simulation replications. For CVAE, KL weight was set to $10^{-4}$. The $\overline{R^2}$ values use all subjects with observed target images. 
\end{tablenotes}
\end{threeparttable}}
\end{table}

Based on Table \ref{TAB:ai-unet-mise}, we can see that across all architectures and missingness rates, SSFR consistently outperforms the missing-data estimators IPW and AIPW, whereas naive SISR is considerably less stable because it directly substitutes AI-generated images for the missing targets. Although the AI generators exhibit noticeably different image-generation quality, the resulting SSFR performance remains remarkably stable across architectures. These results indicate that the efficiency gain of SSFR depends primarily on the transferable information captured by the generated surrogate, as quantified by the $\overline{R^2}$, rather than on the particular deep-learning architecture used to generate the surrogate images.

\subsubsection{Robustness to Subgroup-varying Surrogate Quality}
\label{SSEC:ai-subgroup-quality}

We next consider a focused robustness experiment in which surrogate quality varies across a prespecified subject subgroup. Whereas the preceding experiment examines robustness to the choice of AI architecture, the present experiment asks whether a pooled measure of surrogate quality can conceal poor performance within a clinically relevant subgroup.

We use the binary covariate $X_{i1}$ to define the two subgroups while retaining the target- and auxiliary-image models, coefficient functions, spatial grid, and noise levels described above. The only modification is that the correlation linking the target and auxiliary images is allowed to differ between the two subgroups. Specifically, both the latent scores $(\xi_{ik}^{(O)},\xi_{ik}^{(M)})$ and the voxel-level noise $(\varepsilon_{ij}^{(O)},\varepsilon_{ij}^{(M)})$ are generated using subgroup-specific correlations $\omega_g$, where
$g\in\{0,1\}$.

Three correlation regimes are considered, $(\omega_0,\omega_1) \in \{(0.75,0.75), (0.75,0.40), (0.75,0)\}$, representing homogeneous surrogate quality, moderate subgroup heterogeneity, and severe subgroup heterogeneity, respectively. Since $\Pr(X_{i1}=1)=0.4$, the severe regime is intentionally designed so that a pooled quality measure remains moderately favorable despite the surrogate carrying essentially no residual target information within subgroup-1.

To isolate the subgroup-quality question, only the compact 3D UNet is used in this experiment. Throughout, we fix $n=200$, generate target images under MAR with an overall missing rate of 50\%, and perform 100 Monte Carlo replications. Within each replication, the intercept in the missingness model is recalibrated to maintain the desired overall missing rate, while subgroup-specific observation rates are recorded. 
Among complete pairs, subjects are split into equal-sized training and validation sets stratified by subgroup membership, rather than the 80\%/20\% split used in the architecture benchmark. The remaining training procedures follow Supplementary Section \ref{ssec:supp-ai-arch}; subgroup membership is never supplied to the image generator.

We compare ISR, IPW, AIPW, naive SISR, and SSFR using the coefficient-wise MISE defined in Section 6. Because the regression model is estimated from the pooled sample, only one coefficient-wise MISE is available for each correlation regime. Alongside the MISE, Table \ref{TAB:ai-subgroup-quality} reports the pooled and subgroup-specific surrogate explanatory measures, $\overline{R^2}$, $\overline{R^2}_0$, and $\overline{R^2}_1$, respectively.

\begin{table}[t]
\caption{Coefficient-wise MISE and pooled and subgroup-specific surrogate explanatory power under different surrogate-quality regimes for the UNet surrogate. \label{TAB:ai-subgroup-quality}}
\centering
\scalebox{0.74}{\setlength{\tabcolsep}{4pt}
\begin{tabular}{lcccccccccccc} \toprule
\multirow{2}{*}{Regime} &\multirow{2}{*}{$(\omega_0,\omega_1)$} &\multirow{2}{*}{$(\overline{R^2}_0,\overline{R^2}_1)$} &\multirow{2}{*}{$\overline{R^2}$} &\multicolumn{4}{c}{SISR} &\multicolumn{4}{c}{SSFR}\\ \cmidrule(lr){5-8}\cmidrule(lr){9-12}
& & & &$\alpha_0$ &$\alpha_1$ &$\alpha_2$ &total &$\alpha_0$ &$\alpha_1$ &$\alpha_2$ &total\\ \midrule
Homogeneous &$(0.75,0.75)$ &$(0.451,0.466)$ &0.465 &0.0448 &0.0838 &0.1999 &0.3285 &0.0048 &0.0119 &0.0027 &0.0194\\
Moderate &$(0.75,0.40)$ &$(0.438,0.121)$ &0.309 &0.0459 &0.0830 &0.2195 &0.3484 &0.0050 &0.0147 &0.0031 &0.0228\\
Severe &$(0.75,0.00)$ &$(0.434,0.000)$ &0.185 &0.0522 &0.0897 &0.2335 &0.3753 &0.0049 &0.0158 &0.0032 &0.0239\\ \bottomrule
\end{tabular}} \smallskip

{\footnotesize\raggedright Note: $\omega_g$ denotes the target--auxiliary correlation for subjects with $X_{i1}=g$. The quantities $\overline{R^2}_0$ and $\overline{R^2}_1$ denote the image-level surrogate explanatory measures computed within subgroups $X_{i1}=0$ and $X_{i1}=1$, respectively, while $\overline{R^2}$ denotes the pooled image-level measure. ISR, IPW, and AIPW do not use the surrogate images and are therefore identical across the three regimes: ISR $(0.0060,0.0149,0.0033)$, total $0.0242$; IPW $(0.0059,0.0152,0.0033)$, total $0.0244$; and AIPW $(0.0060,0.0152,0.0033)$, total $0.0244$. The evaluation samples are the same as those used in Table \ref{TAB:ai-unet-mise}.\par}
\end{table}
Table \ref{TAB:ai-subgroup-quality} shows that the pooled image-level surrogate explanatory measure $\overline{R^2}$ decreases from $0.465$ under homogeneous surrogate quality to $0.309$ under moderate heterogeneity and $0.185$ under severe heterogeneity. Although the homogeneous regime uses the same target--auxiliary correlation as the UNet benchmark with 50\% missingness in Table \ref{TAB:ai-unet-mise}, its pooled $\overline{R^2}$ need not equal the benchmark value of $0.484$, as the subgroup-stratified 50\%/50\% split allocates fewer complete pairs to training than the benchmark's 80\%/20\% split. The resulting fitted UNet surrogates therefore differ, yielding different estimated explanatory powers despite the common underlying correlation. The subgroup-specific measures provide additional insight: $\overline{R^2}_0$ remains relatively stable across the three regimes, whereas $\overline{R^2}_1$ decreases from $0.466$ under homogeneous surrogate quality to $0.121$ under moderate heterogeneity and $0$ under severe heterogeneity. Thus, the pooled measure reflects the overall decline in surrogate explanatory power.

The coefficient-wise MISEs show the expected reduction in SSFR efficiency as surrogate explanatory power deteriorates in subgroup 1. Nevertheless, the total SSFR MISE remains close to that of ISR even under severe subgroup heterogeneity, whereas naive SISR continues to exhibit substantial bias. These findings are consistent with Corollary 4.4: recalibration against the observed images allows an uninformative surrogate in a substantial subgroup to reduce efficiency without introducing the large bias associated with direct substitution.

Overall, the results demonstrate that $\overline{R^2}$ provides a useful summary of overall surrogate explanatory power, while $(\overline{R^2}_0,\overline{R^2}_1)$ identifies clinically relevant heterogeneity in surrogate performance. As the explanatory power decreases in subgroup 1, the efficiency gains of SSFR diminish and its total MISE approaches that of ISR. This subgroup-level pattern illustrates the value of reporting both pooled and subgroup-specific $R^2$ measures when assessing candidate surrogates.

\renewcommand{\thesubsection}{B.\arabic{subsection}}
\renewcommand{\thesubsubsection}{B.\arabic{subsection}.\arabic{subsubsection}}
\renewcommand{\thetheorem}{{\sc B.\arabic{theorem}}}
\renewcommand{\thelemma}{{\sc B.\arabic{lemma}}}
\renewcommand{\theproposition}{{\sc B.\arabic{proposition}}}
\renewcommand{\thecorollary}{\sc B.\arabic{corollary}}
\renewcommand{\thefigure}{B.\arabic{figure}}
\renewcommand{\thetable}{B.\arabic{table}}
\renewcommand{\theequation}{B.\arabic{equation}}
\renewcommand{\theremark}{B.\arabic{remark}}
\renewcommand{\thealgorithm}{B.\arabic{algorithm}}
\setcounter{equation}{0}
\setcounter{theorem}{0}
\setcounter{figure}{0}
\setcounter{table}{0}
\setcounter{subsection}{0}
\setcounter{subsubsection}{0}

\section{Additional Simulation and Application Results}
\label{SEC:simulation-supp}
\setlength\abovedisplayskip{4pt plus 4pt}
\setlength\belowdisplayskip{4pt plus 4pt}

This section collects additional simulation and application results that supplement the findings reported in the main paper. We first provide the complete data-generating mechanism and additional sensitivity analyses for the simulation studies. We then present simulation results under a missing completely at random (MCAR) mechanism, followed by additional figures from the ADNI application that complement the results shown in the main text.

\subsection{Simulation Design Details and Sensitivity Analysis}
\label{ssec:sensitivity}

This subsection provides the complete specification of the simulation design summarized in Section 6.1 of the main paper and reports additional sensitivity analyses that assess the robustness of the proposed estimation procedure with respect to several implementation choices.

\subsubsection{Data-Generating Details}
\label{ssec:supp-simdesign}

This subsection provides the complete specification of the simulation design summarized in Section 6.1 of the main paper, including the image-generating models, latent score construction, true coefficient functions, and missingness mechanism.

The subject-specific stochastic processes are $\eta_i^{(\cdot)}(\bs{z}) = \sum_{k=1}^{3} \lambda_k^{1/2} \xi_{ik}^{(\cdot)} \psi_k(\bs{z})$, with eigenvalues $(\lambda_1,\lambda_2,\lambda_3) = (0.2,0.1,0.05)$, and basis functions $\psi_1(\bs{z}) = \sin(\pi z_1)+c_1$, $\psi_2(\bs{z}) = c_2\cos(\pi z_2)$, and $\psi_3(\bs{z}) = c_3(z_3-1/2)$, where the constants $c_1,c_2,c_3$ are chosen so that $\{\psi_1,\psi_2,\psi_3\}$ forms an orthonormal basis on $\Omega$. The measurement errors $\varepsilon_{ij}^{(O)}$ and $\varepsilon_{ij}^{(S)}$ each have marginal distribution $\mathcal N(0,0.5^2)$ and have correlation $\omega$ within a subject and voxel, using the same correlation parameter as the latent scores in the main simulation design. The paired errors are independent across subjects and voxels.

The true coefficient functions are
\begin{align*}
\alpha_{0,0}(\bs{z}) &= 10\left\{ (z_1-0.5)^2 + (z_2-0.5)^2 + (z_3-0.5)^2 \right\}, \\
\alpha_{0,1}(\bs{z}) &= 0.2(z_1+z_2+z_3)^3, ~\mathrm{and}~\\
\alpha_{0,2}(\bs{z}) &= 2\exp\left\{ (z_1-0.5)^2 + (z_2-0.5)^2 + (z_3-0.5)^2 \right\},
\end{align*}
and Figure \ref{FIG:supp-eg1-coefs}(a)--(c) shows 17 equally spaced horizontal slices of the three coefficient functions.

Under the correctly specified surrogate model, $\bs\beta_0=\bs\alpha_0$. Under the misspecified surrogate model,
\begin{align*}
\beta_{0,0}(\bs{z}) &= z_1^2+z_2^2-z_3, \\
\beta_{0,1}(\bs{z}) &= z_1^2-z_2^2+z_3, ~\mathrm{and}~\\
\beta_{0,2}(\bs{z}) &= -z_1^2+z_2^2+z_3.
\end{align*}
Under the MAR setting, the observation indicator $R_{Y_i}\in\{0,1\}$ is generated according to
\begin{equation}
\operatorname{logit} \{ \Pr(R_{Y_i}=1 \mid X_{i1},X_{i2}) \} = b_0 - \{1+\exp(-X_{i1})\}^{-1} + \{1+\exp(-X_{i2})\}^{-1},
\label{model:missing}
\end{equation}
following \cite{ciarleglio2022elucidating}, where the intercept $b_0$ is calibrated to attain the desired marginal missingness rate.

The observation probabilities in \eqref{model:missing} also enter the AIPW comparator described in Section 6.1 of the main paper. We first fit ISR to the observed target images and use the fitted model to obtain a prediction $\widehat m_i(\bs z)$ for every subject.
Let $\pi_i=\Pr(R_{Y_i}=1\mid\mathbf X_i)$ be the observation probability specified by the simulation design. The augmented response is
\[
Y_i^{\mathrm{AIPW}}(\bs z)
=
\begin{cases}
\widehat m_i(\bs z)
  + \{Y_i(\bs z)-\widehat m_i(\bs z)\}/\pi_i,
    & R_{Y_i}=1,\\
\widehat m_i(\bs z), & R_{Y_i}=0.
\end{cases}
\]
Every subject contributes a prediction, and subjects with observed target images also contribute a weighted residual. We fit ISR to these augmented responses for all $n$ subjects. The construction uses only the scalar covariates and observed target images, so AIPW gives the same results in paired simulations that differ only in the surrogate model.

\subsubsection{Sensitivity Analysis}
\label{ssec:supp-sensitivity}

To assess the robustness of the proposed estimation procedure with respect to implementation choices, we conducted additional sensitivity analyses that vary several key design factors, including sample size, signal-to-noise ratio (SNR), spline degree, and triangulation fineness. These experiments are performed under ISR, which is the regression model used in both stages of SSFR. Since SISR differs only in the response images used for estimation, the tuning conclusions apply equally to both procedures.

The sensitivity study uses the same coefficient functions and spatial error basis, with no missing responses. Its two non-intercept covariates are jointly Gaussian with correlation $0.5$, standardized within each sample and clipped to $[-3,3]$. Sample sizes are varied over $n\in\{20,50,100\}$. The signal-to-noise ratio is defined as
\[
\mathrm{SNR} = \frac{ \sum_{j=1}^{N} \mathrm{Var} \left\{ \sum_{\ell=1}^{p} X_{i\ell}\beta_\ell(\bs{z}_j) \right\} + \mathrm{Var} \{\beta_0(\bs{z}_j)\} }{ \sum_{j=1}^{N} \mathrm{Var} \{ \eta_i(\bs{z}_j) + \sigma\varepsilon_{ij} \} }.
\]

\begin{figure}[ht!]
\centering
\begin{tabular}{cccc}
\includegraphics[height = 0.8in]{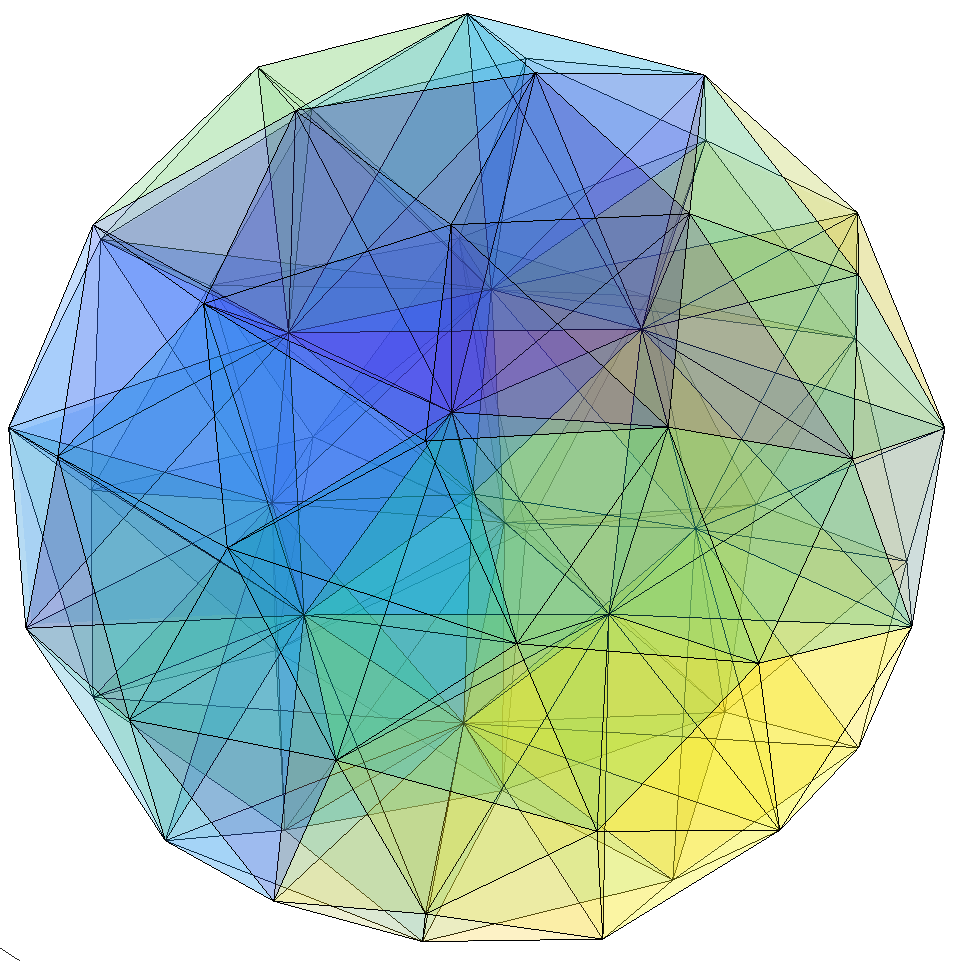} &\includegraphics[height = 0.8in]{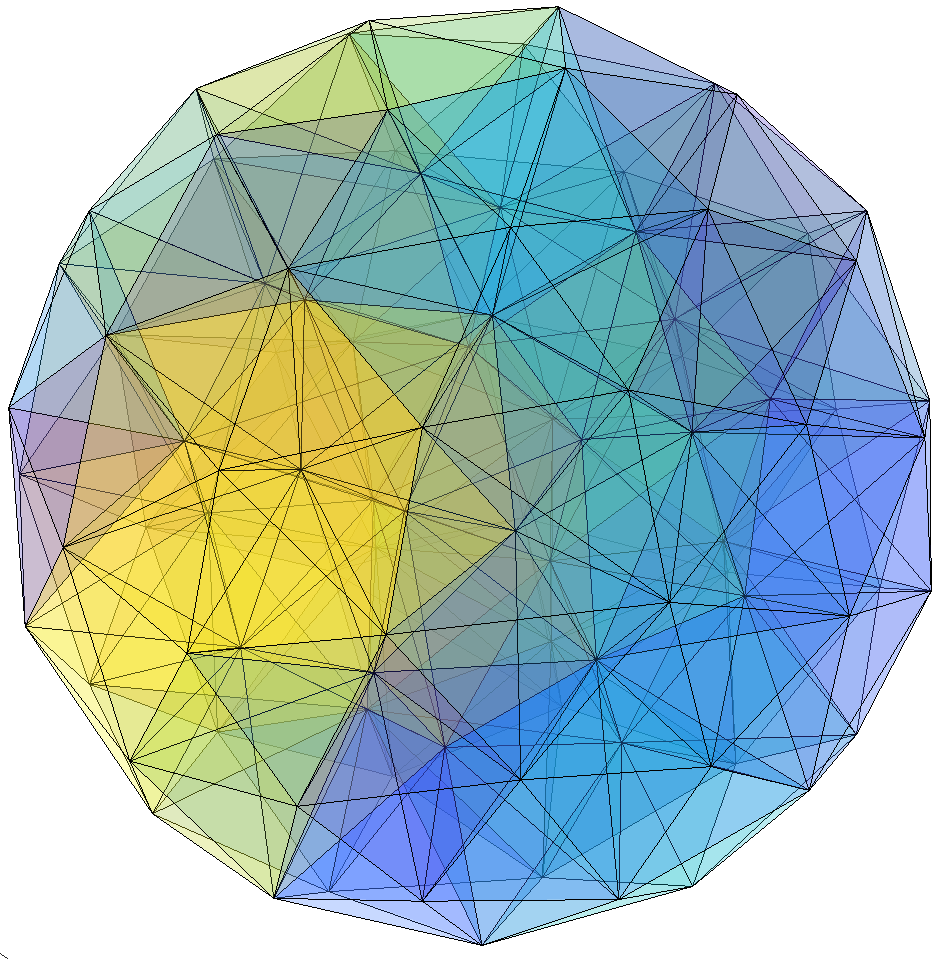} &\includegraphics[height = 0.8in]{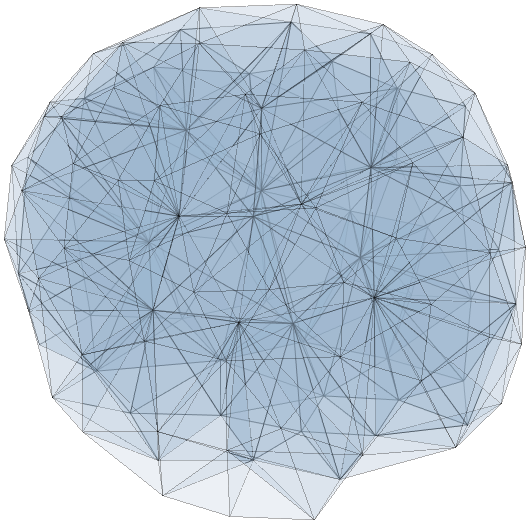} &\includegraphics[height = 0.8in]{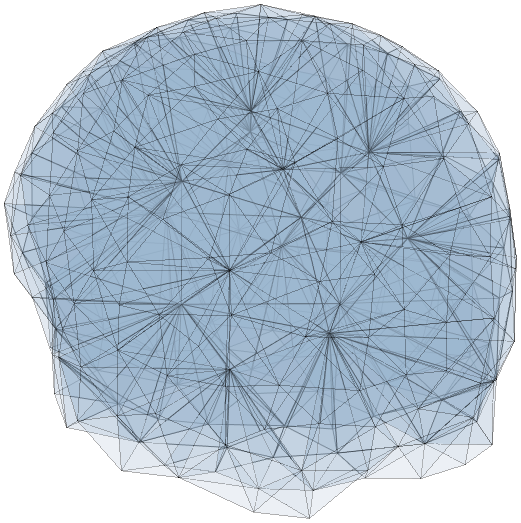}\\
(a) $\triangle_1$ & (b) $\triangle_2$ & (c)  $\triangle_3$ & (d) $\triangle_4$
\end{tabular} \vspace{-8pt}
\caption{\small Triangulations used in  numerical studies: (a)-(b) for simulation studies; and (c)-(d) for ADNI data analysis.} 
\label{FIG:tri} 
\end{figure}

To investigate the effect of mesh complexity, two tetrahedral triangulations are considered: $\triangle_1$ with 176 tetrahedra and 62 vertices, and $\triangle_2$ with 332 tetrahedra and 96 vertices. The two triangulations are shown in Figure \ref{FIG:tri}(a)--(b). In addition, spline degrees $d=2,3,$ and $4$ are examined to evaluate the trade-off between approximation flexibility and smoothness.

\begin{table}[ht!]
\caption{Averaged MISEs of varying-coefficient estimates over 100 replications. Spline degree $d$ varies within each coefficient block; the kernel estimator does not depend on $d$ or $\triangle$.\label{TAB:supp-eg1-mise}}
\centering
\scalebox{0.72}{\setlength{\tabcolsep}{4pt}
\begin{tabular}{ccc ccc ccc ccc ccc} \toprule
& & & \multicolumn{9}{c}{TPST} & \multicolumn{3}{c}{Kernel} \\ \cmidrule(lr){4-12}\cmidrule(lr){13-15}
& & & \multicolumn{3}{c}{$\beta_0$} & \multicolumn{3}{c}{$\beta_1$} & \multicolumn{3}{c}{$\beta_2$} & & & \\
\cmidrule(lr){4-6}\cmidrule(lr){7-9}\cmidrule(lr){10-12}
$n$ & SNR & $\triangle$ & $d{=}2$ & $d{=}3$ & $d{=}4$ & $d{=}2$ & $d{=}3$ & $d{=}4$ & $d{=}2$ & $d{=}3$ & $d{=}4$ & $\beta_0$ & $\beta_1$ & $\beta_2$ \\ \midrule
\multirow{4}{*}{20}
& \multirow{2}{*}{9.0} & $\triangle_1$ & 0.018 & 0.018 & 0.018 & 0.030 & 0.030 & 0.030 & 0.023 & 0.023 & 0.023 & \multirow{2}{*}{.025} & \multirow{2}{*}{.040} & \multirow{2}{*}{.032}\\
& & $\triangle_2$ & 0.018 & 0.018 & 0.018 & 0.030 & 0.030 & 0.030 & 0.023 & 0.023 & 0.023 & & & \\ \cmidrule(lr){2-15}
& \multirow{2}{*}{4.5} & $\triangle_1$ & 0.042 & 0.042 & 0.042 & 0.070 & 0.070 & 0.070 & 0.052 & 0.052 & 0.052 & \multirow{2}{*}{.049} & \multirow{2}{*}{.080} & \multirow{2}{*}{.062}\\
& & $\triangle_2$ & 0.042 & 0.043 & 0.043 & 0.070 & 0.070 & 0.070 & 0.052 & 0.052 & 0.052 & & & \\ \midrule
\multirow{4}{*}{50} & \multirow{2}{*}{9.0} & $\triangle_1$ & 0.007 & 0.007 & 0.007 & 0.010 & 0.010 & 0.010 & 0.010 & 0.010 & 0.010 & \multirow{2}{*}{0.010} & \multirow{2}{*}{0.014} & \multirow{2}{*}{0.014}\\
& & $\triangle_2$ & 0.007 & 0.007 & 0.007 & 0.010 & 0.010 & 0.010 & 0.010 & 0.010 & 0.010 & & & \\ \cmidrule(lr){2-15}
& \multirow{2}{*}{4.5} & $\triangle_1$ & 0.016 & 0.016 & 0.016 & 0.024 & 0.024 & 0.024 & 0.023 & 0.023 & 0.023 & \multirow{2}{*}{0.019} & \multirow{2}{*}{0.027} & \multirow{2}{*}{0.027}\\
& & $\triangle_2$ & 0.016 & 0.016 & 0.017 & 0.024 & 0.024 & 0.024 & 0.023 & 0.023 & 0.023 & & & \\ \midrule
\multirow{4}{*}{100} & \multirow{2}{*}{9.0} & $\triangle_1$ & 0.003 & 0.003 & 0.003 & 0.005 & 0.005 & 0.005 & 0.005 & 0.005 & 0.005 & \multirow{2}{*}{0.004} & \multirow{2}{*}{0.006} & \multirow{2}{*}{0.006}\\
& & $\triangle_2$ & 0.003 & 0.003 & 0.003 & 0.005 & 0.005 & 0.005 & 0.005 & 0.005 & 0.005 & & & \\ \cmidrule(lr){2-15}
& \multirow{2}{*}{4.5} & $\triangle_1$ & 0.006 & 0.006 & 0.006 & 0.010 & 0.010 & 0.010 & 0.010 & 0.010 & 0.010 & \multirow{2}{*}{0.008} & \multirow{2}{*}{0.012} & \multirow{2}{*}{0.012}\\
& & $\triangle_2$ & 0.006 & 0.006 & 0.006 & 0.010 & 0.010 & 0.010 & 0.010 & 0.010 & 0.010 & & & \\
\bottomrule
\end{tabular}}
\end{table}

Table \ref{TAB:supp-eg1-mise} summarizes the mean integrated squared errors over 100 Monte Carlo replications. The TPST estimator consistently outperforms the kernel estimator of \cite{Zhu:Fan:Kong:14}. As expected, estimation accuracy improves with increasing sample size and higher signal-to-noise ratio. In contrast, spline degree and triangulation fineness have only minor effects on estimation performance, with little variation observed across the three spline degrees or between the two triangulations.

\begin{figure}[ht!]
\begin{center}
{\footnotesize
\begin{tabular}{ccc}
\fbox{\includegraphics[width = 1.25in]{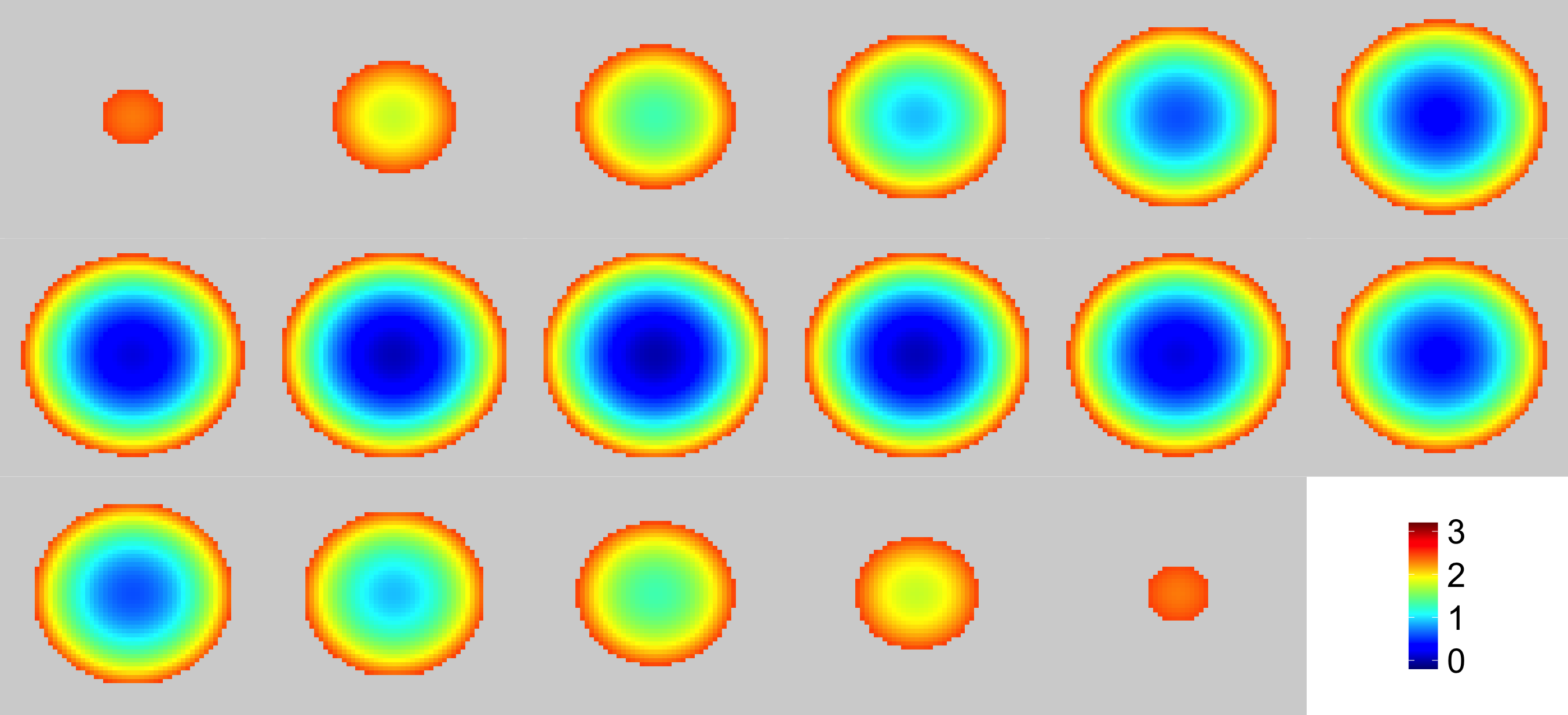}} & \fbox{\includegraphics[width = 1.25in]{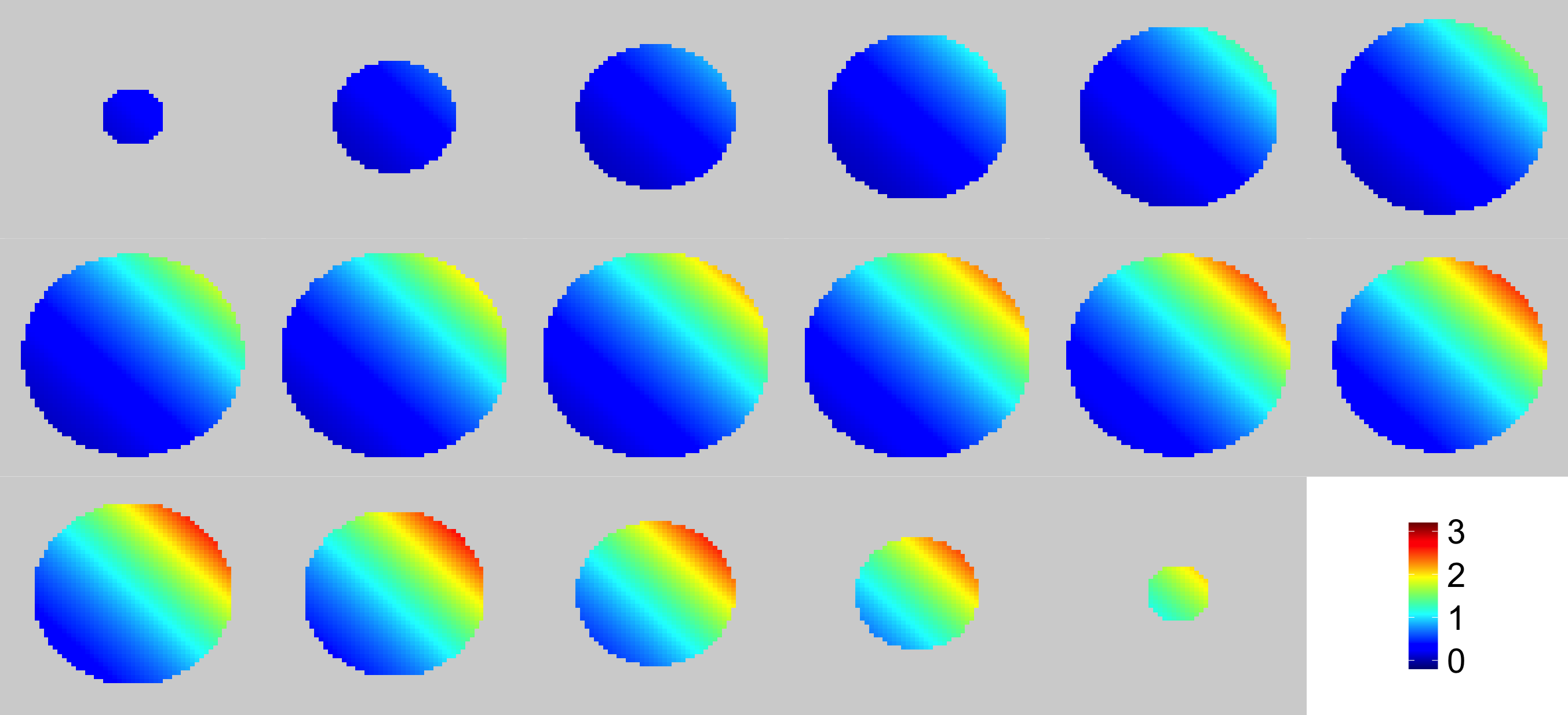}} &\fbox{\includegraphics[width = 1.25in]{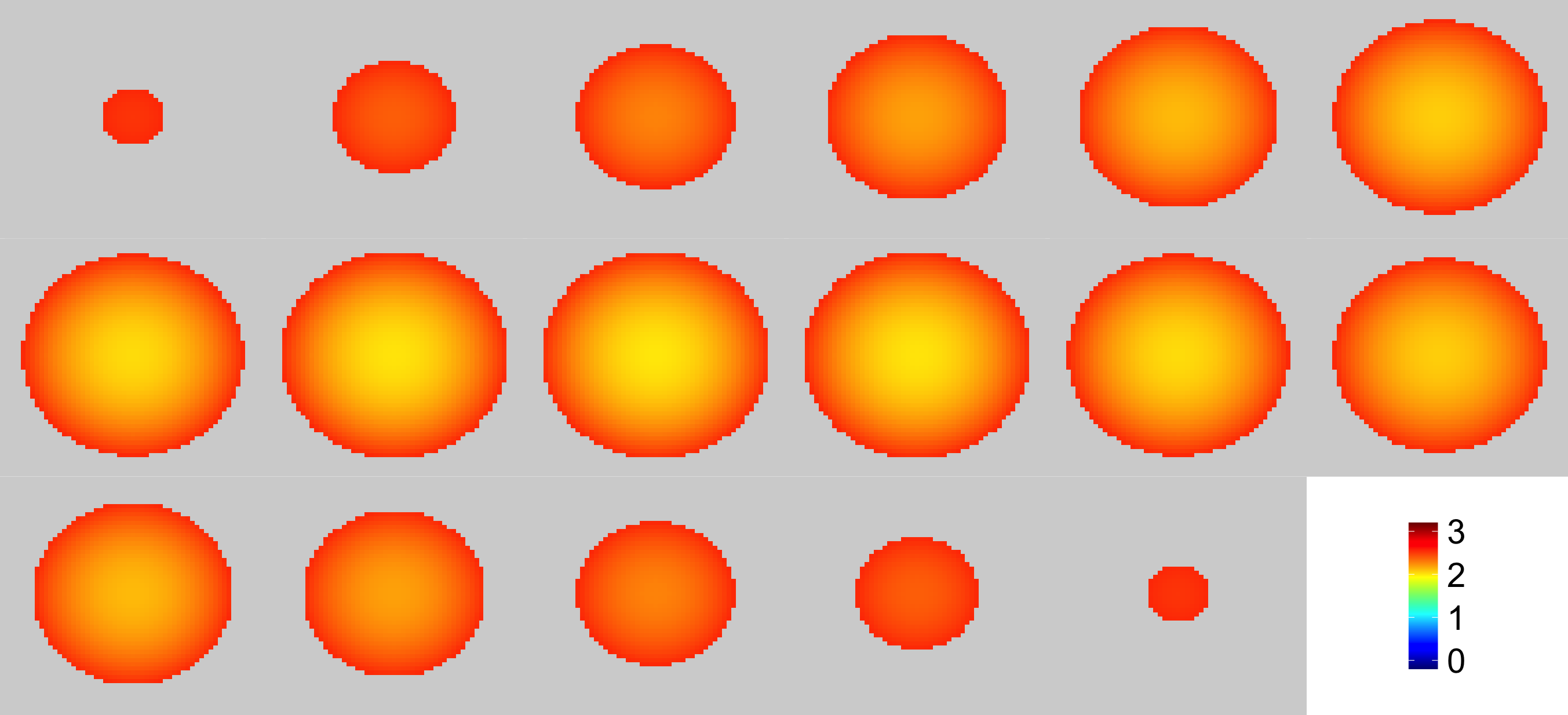}}\\[3pt]
(a) True $\beta_0(\cdot)$ &(b) True $\beta_1(\cdot)$ &(c) True $\beta_2(\cdot)$\\
\fbox{\includegraphics[width = 1.25in]{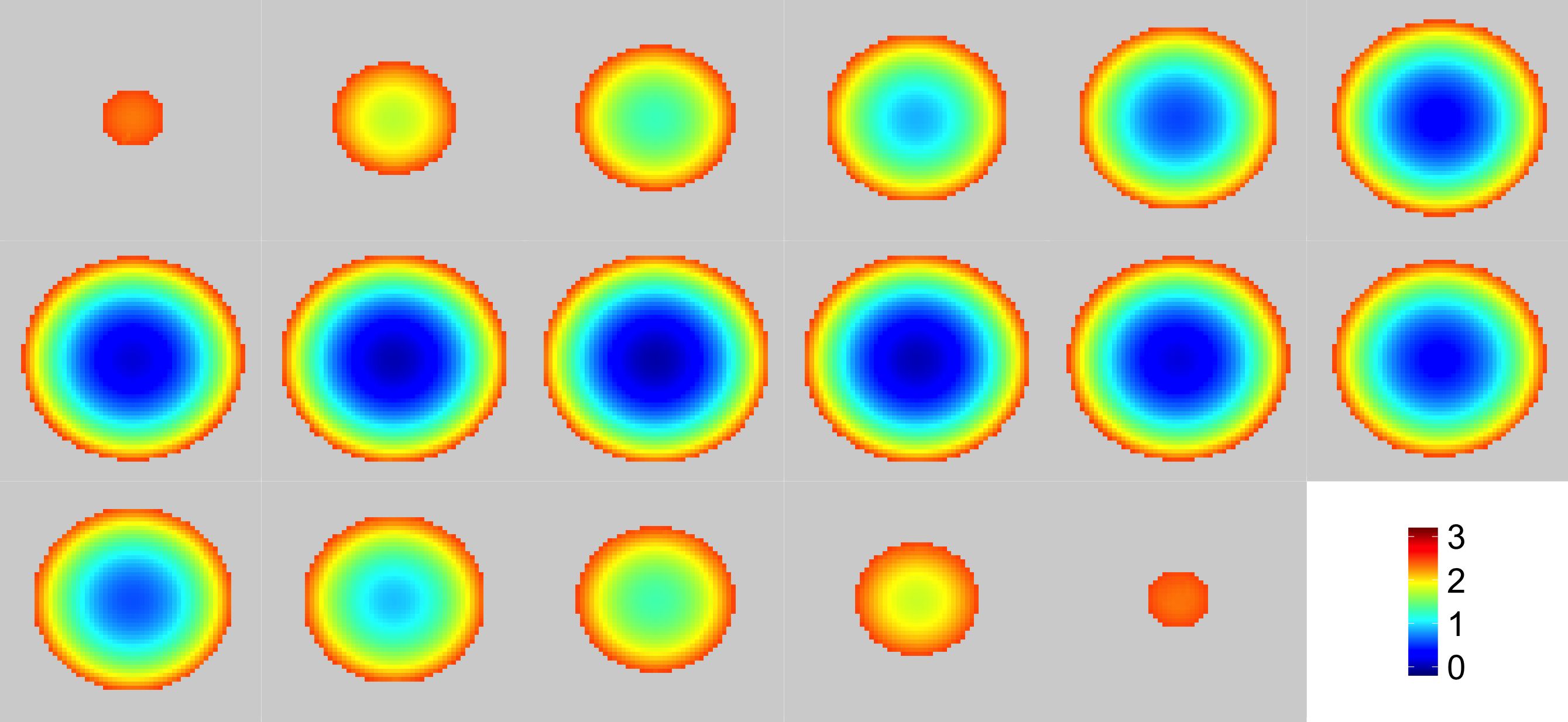}} & \fbox{\includegraphics[width = 1.25in]{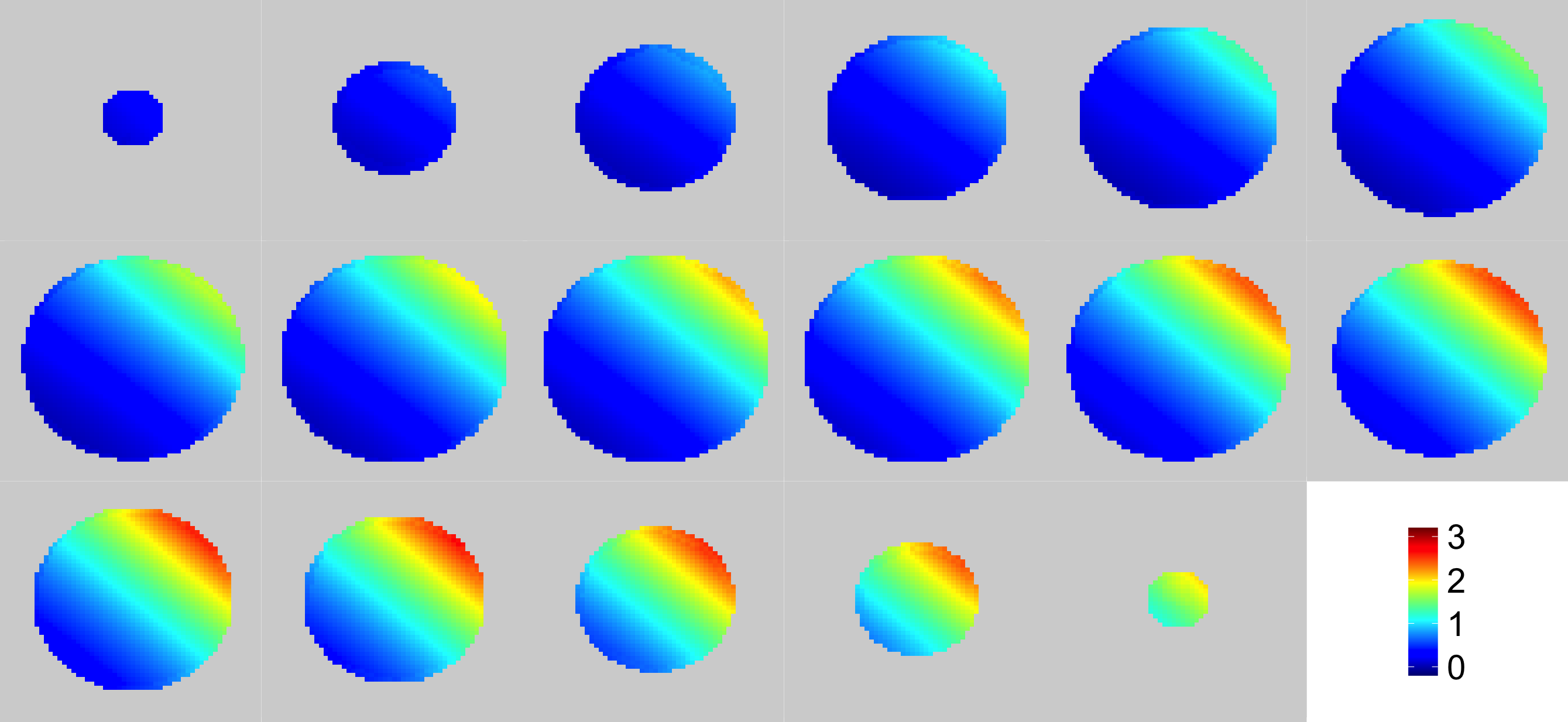}} & \fbox{\includegraphics[width = 1.225in]{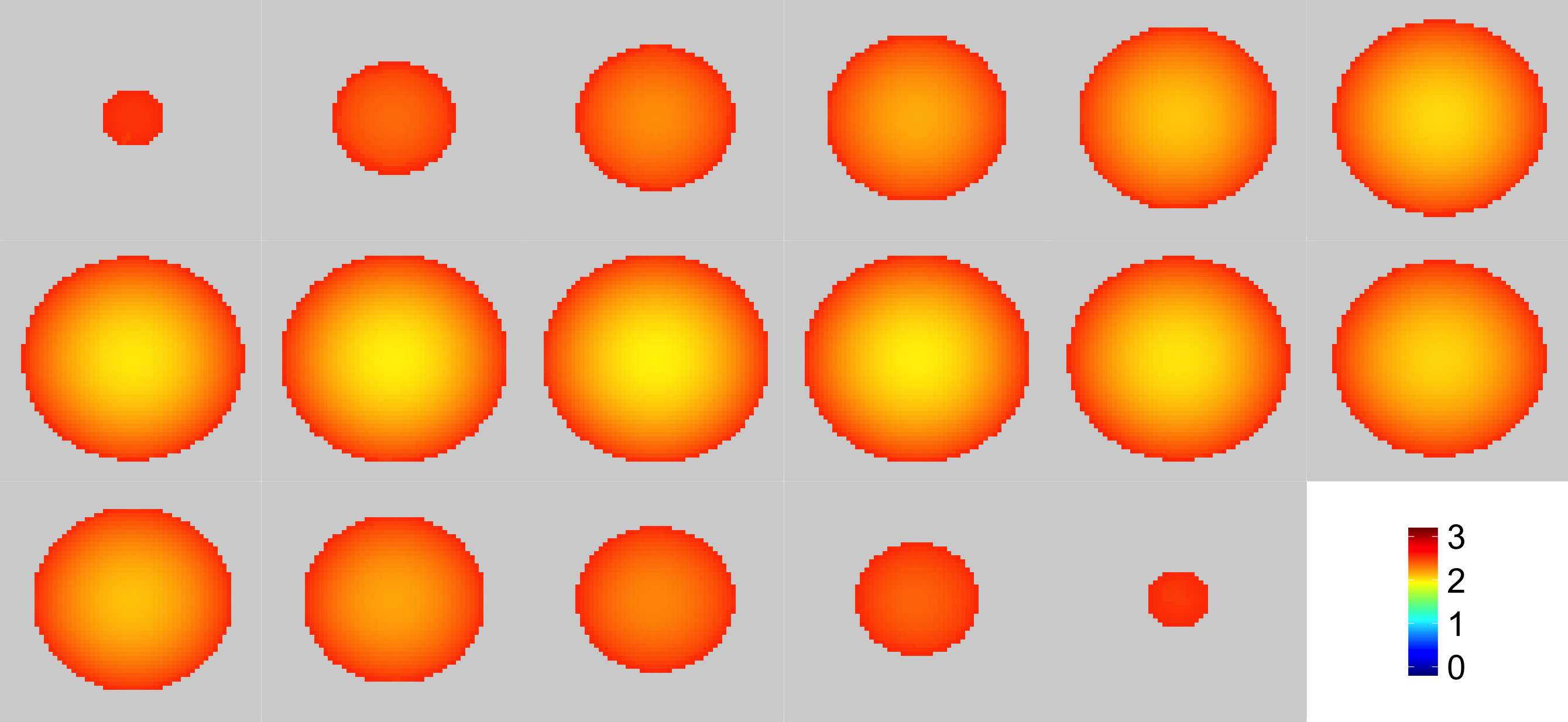}}\\[3pt]
(d) $\widehat{\beta}_0(\cdot)$ via TPST &(e) $\widehat{\beta}_1(\cdot)$ via TPST &(f) $\widehat{\beta}_2(\cdot)$ via TPST\\
\fbox{\includegraphics[width = 1.25in]{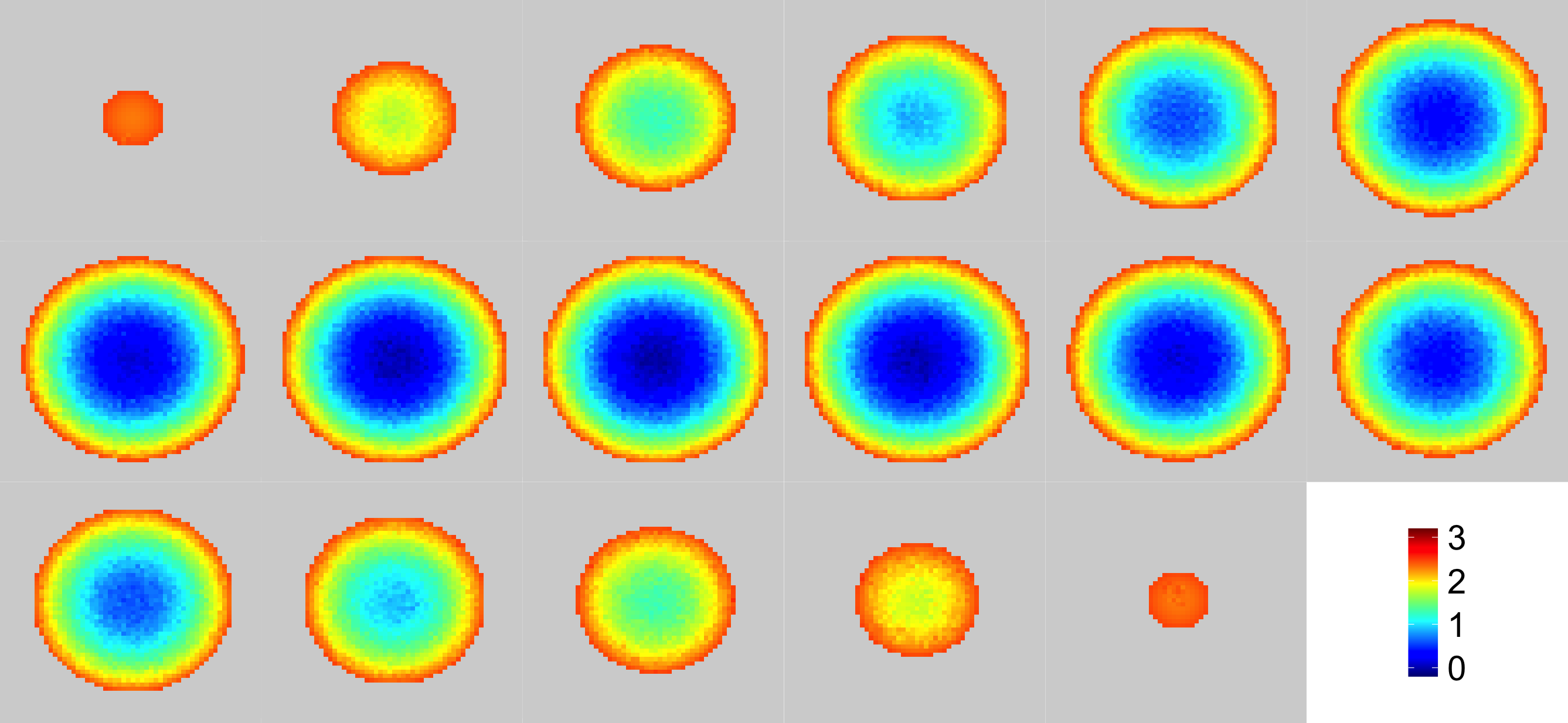}} & \fbox{\includegraphics[width = 1.25in]{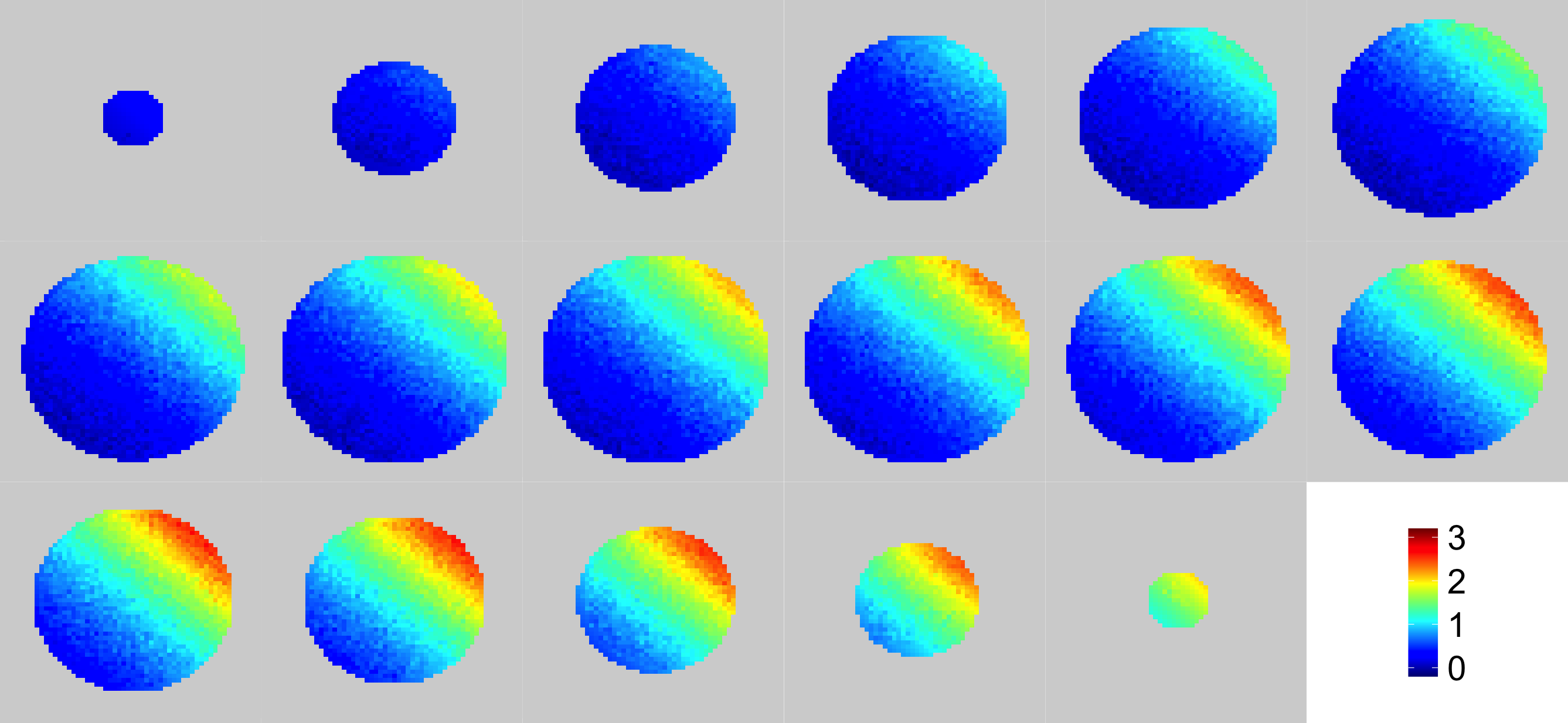}} & \fbox{\includegraphics[width = 1.25in]{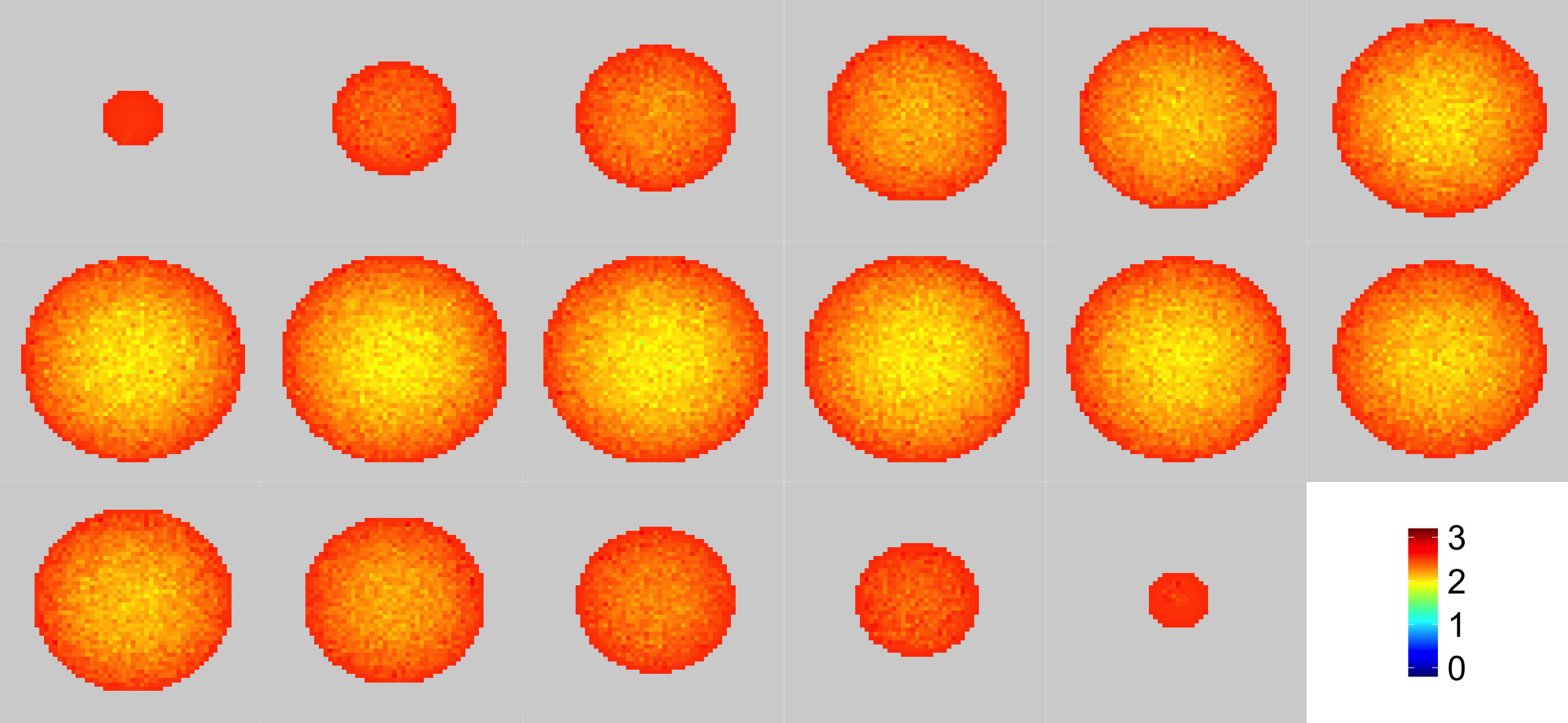}}\\[3pt]
(g) $\widehat{\beta}_0(\cdot)$ via Kernel &(h) $\widehat{\beta}_1(\cdot)$ via Kernel &(i) $\widehat{\beta}_2(\cdot)$ via Kernel
\end{tabular}} \vspace{-.2in}
\end{center}
\caption{True coefficient functions and their estimates via different methods.}
\label{FIG:supp-eg1-coefs}
\end{figure}

Figure \ref{FIG:supp-eg1-coefs} provides representative coefficient estimates for $n=100$, illustrating the improved estimation accuracy achieved by TPST relative to the kernel method.
\subsection{Additional Simulation Results}
\label{ssec:addsimu}

\subsubsection{Correctly Specified Surrogate Model under MAR}

The main paper focuses on the misspecified surrogate setting because it provides the most direct assessment of the robustness of SSFR. For completeness, Table \ref{TAB:eg1-mar-mise1} reports the corresponding results under MAR when the surrogate model is correctly specified.

Table \ref{TAB:eg1-mar-mise1} shows that ISR, IPW, and AIPW yield the same MISEs as in the paired simulations under surrogate misspecification
reported in the main paper, because these methods do not use surrogate images. Consequently, their estimation procedures are completely unaffected by whether the surrogate model is correctly specified.

In contrast, SISR and SSFR explicitly incorporate surrogate information and therefore respond differently to surrogate misspecification. Under correct specification, the generated surrogate images closely approximate the true target images, allowing SISR to achieve substantially improved estimation accuracy. However, as demonstrated in the main paper, SISR becomes highly unstable when the surrogate model is misspecified because the biased surrogate images are directly substituted for the missing target images. By comparison, SSFR exhibits nearly identical performance under correctly specified and misspecified surrogate models. This robustness is a consequence of its two-stage estimation procedure, which recalibrates the surrogate information using the observed target images rather than directly replacing the missing outcomes with surrogate images. As the surrogate becomes more informative (larger $\overline{R^2}$), SSFR consistently achieves lower MISEs than ISR, IPW, and AIPW, with the greatest improvements observed under higher missing-data rates.

\begin{table}
\caption{MISEs of the coefficient functions for different methods with a correctly specified surrogate model under MAR. \label{TAB:eg1-mar-mise1}}
\begin{center}
\begin{threeparttable}
\scalebox{0.75}{\setlength{\tabcolsep}{3pt}\begin{tabular}
{ccccccccccccccccc} \toprule
\multirow{2}{*}{$r_m$} &\multirow{2}{*}{$\overline{R^2}$} &\multicolumn{3}{c}{ISR} &\multicolumn{3}{c}{IPW} &\multicolumn{3}{c}{AIPW} &\multicolumn{3}{c}{SISR} &\multicolumn{3}{c}{SSFR}\\ \cmidrule(lr){3-5}\cmidrule(lr){6-8}\cmidrule(lr){9-11} \cmidrule(lr){12-14}\cmidrule(lr){15-17}
& &$\alpha_0$ &$\alpha_1$ &$\alpha_2$ &$\alpha_0$ &$\alpha_1$ &$\alpha_2$ &$\alpha_0$ &$\alpha_1$ &$\alpha_2$ &$\alpha_0$ &$\alpha_1$ &$\alpha_2$ &$\alpha_0$ &$\alpha_1$ &$\alpha_2$\\ \midrule
\multicolumn{17}{c}{$n=200$}\\ \midrule
\multirow{4}{*}{75\%} &0.06 &\multirow{4}{*}{0.0116} &\multirow{4}{*}{0.0314} &\multirow{4}{*}{0.0062} &\multirow{4}{*}{0.0113} &\multirow{4}{*}{0.0306} &\multirow{4}{*}{0.0064} &0.0113 &0.0306 &0.0064 &\multirow{4}{*}{0.0030} &\multirow{4}{*}{0.0074} &\multirow{4}{*}{0.0015} &0.0138&0.0364&0.0084\\
&0.25 & & & & & & &0.0113 &0.0308 &0.0064 & & & &0.0114&0.0302&0.0072\\
&0.56 & & & & & & &0.0114 &0.0310 &0.0065 & & & &0.0077&0.0207&0.0050\\
&0.81 & & & & & & &0.0115 &0.0313 &0.0066 & & & &0.0048&0.0131&0.0031\\ \midrule
\multirow{4}{*}{50\%} &0.06 &\multirow{4}{*}{0.0061} &\multirow{4}{*}{0.0156} &\multirow{4}{*}{0.0037} &\multirow{4}{*}{0.0060} &\multirow{4}{*}{0.0159} &\multirow{4}{*}{0.0036} &0.0061 &0.0159 &0.0036 &\multirow{4}{*}{0.0030} &\multirow{4}{*}{0.0078} &\multirow{4}{*}{0.0016} &0.0062&0.0174&0.0038\\
&0.25 & & & & & & &0.0060 &0.0156 &0.0035 & & & &0.0056&0.0159&0.0035\\
&0.56 & & & & & & &0.0060 &0.0154 &0.0034 & & & &0.0045&0.0125&0.0028\\
&0.81 & & & & & & &0.0060 &0.0152 &0.0033 & & & &0.0035&0.0096&0.0022\\ \midrule
\multirow{4}{*}{25\%} &0.06 &\multirow{4}{*}{0.0039} &\multirow{4}{*}{0.0114} &\multirow{4}{*}{0.0029} &\multirow{4}{*}{0.0039} &\multirow{4}{*}{0.0115} &\multirow{4}{*}{0.0029} &0.0039 &0.0115 &0.0029 &\multirow{4}{*}{0.0030} &\multirow{4}{*}{0.0076} &\multirow{4}{*}{0.0019} &0.0039&0.0111&0.0028\\
&0.25 & & & & & & &0.0039 &0.0113 &0.0029 & & & &0.0036&0.0101&0.0026\\
&0.57 & & & & & & &0.0038 &0.0111 &0.0029 & & & &0.0032&0.0087&0.0022\\
&0.81 & & & & & & &0.0038 &0.0109 &0.0028 & & & &0.0029&0.0077&0.0020\\ \midrule
\multicolumn{17}{c}{$n=400$}\\ \midrule
\multirow{4}{*}{75\%} &0.06 &\multirow{4}{*}{0.0057} &\multirow{4}{*}{0.0183} &\multirow{4}{*}{0.0040} &\multirow{4}{*}{0.0057} &\multirow{4}{*}{0.0184} &\multirow{4}{*}{0.0041} &0.0057 &0.0184 &0.0041 &\multirow{4}{*}{0.0012} &\multirow{4}{*}{0.0039} &\multirow{4}{*}{0.0009} &0.0053&0.0177&0.0043\\
&0.25 & & & & & & &0.0057 &0.0186 &0.0041 & & & &0.0043&0.0150&0.0037\\
&0.56 & & & & & & &0.0057 &0.0188 &0.0041 & & & &0.0029&0.0105&0.0026\\
&0.81 & & & & & & &0.0056 &0.0188 &0.0041 & & & &0.0019&0.0067&0.0017\\ \midrule
\multirow{4}{*}{50\%} &0.07 &\multirow{4}{*}{0.0031} &\multirow{4}{*}{0.0076} &\multirow{4}{*}{0.0020} &\multirow{4}{*}{0.0031} &\multirow{4}{*}{0.0077} &\multirow{4}{*}{0.0021} &0.0031 &0.0077 &0.0021 &\multirow{4}{*}{0.0013} &\multirow{4}{*}{0.0036} &\multirow{4}{*}{0.0010} &0.0030&0.0078&0.0020\\
&0.25 & & & & & & &0.0030 &0.0076 &0.0021 & & & &0.0026&0.0070&0.0018\\
&0.56 & & & & & & &0.0029 &0.0076 &0.0021 & & & &0.0020&0.0055&0.0015\\
&0.81 & & & & & & &0.0028 &0.0076 &0.0021 & & & &0.0016&0.0044&0.0012\\ \midrule
\multirow{4}{*}{25\%} &0.06 &\multirow{4}{*}{0.0019} &\multirow{4}{*}{0.0049} &\multirow{4}{*}{0.0013} &\multirow{4}{*}{0.0019} &\multirow{4}{*}{0.0049} &\multirow{4}{*}{0.0013} &0.0019 &0.0049 &0.0013 &\multirow{4}{*}{0.0015} &\multirow{4}{*}{0.0038} &\multirow{4}{*}{0.0010} &0.0020&0.0049&0.0013\\
&0.25 & & & & & & &0.0019 &0.0049 &0.0013 & & & &0.0019&0.0046&0.0012\\
&0.56 & & & & & & &0.0018 &0.0048 &0.0013 & & & &0.0017&0.0042&0.0011\\
&0.81 & & & & & & &0.0017 &0.0048 &0.0013 & & & &0.0015&0.0039&0.0011\\ \bottomrule
\end{tabular}
}
\end{threeparttable}
\end{center}
\end{table}

\subsubsection{Simulation Results with Responses Missing Completely at Random}
\label{ssec:mcar}

The primary simulation studies in the main paper consider a missing at random (MAR) mechanism, in which the probability that a target image is observed depends on the fully observed covariates. To examine whether the main conclusions depend on this assumption, we conduct an additional set of experiments under a missing completely at random (MCAR) mechanism. Under MCAR, imaging availability is independent of both the covariates and the imaging outcomes, so the observed sample is formed by selecting subjects uniformly at random according to the desired missingness rate.

Under MCAR, every subject has the same observation probability. Consequently, the inverse-probability weights are constant up to a multiplicative factor, and IPW is theoretically equivalent to ISR. Similarly, because both the missingness mechanism and the outcome model are correctly specified, AIPW is expected to provide little additional efficiency beyond ISR. Therefore, to avoid redundant comparisons, we report only ISR, naive SISR, and SSFR. All three use TPST, with kernel-based ISR and SISR from \cite{Zhu:Fan:Kong:14} included as additional benchmarks.

\begin{table}
\caption{MISEs of the coefficient functions for different methods with a correctly specified surrogate model under MCAR. \label{TAB:eg1-mcar-mise1}}
\centering
\begin{threeparttable}
\scalebox{0.65}{\setlength{\tabcolsep}{3pt}\begin{tabular}
{ccccc|ccc|ccc|ccc|ccc} \toprule
\multirow{3}{*}{\shortstack{Missing\\Rate}} &\multirow{3}{*}{$\overline{R^2}$} & \multicolumn{6}{c}{ISR} &\multicolumn{6}{c}{SISR}  & \multicolumn{3}{c}{SSFR}\\ \cmidrule(lr){3-8} \cmidrule(lr){9-14}
& &\multicolumn{3}{c}{TPST} &\multicolumn{3}{c}{Kernel} &\multicolumn{3}{c}{TPST} &\multicolumn{3}{c}{Kernel} & \multicolumn{3}{c}{TPST} \\ \cmidrule(lr){3-5} \cmidrule(lr){6-8} \cmidrule(lr){9-11} \cmidrule(lr){12-14} \cmidrule(lr){15-17}
& &$\alpha_0$ &$\alpha_1$ &$\alpha_2$ &$\alpha_0$ &$\alpha_1$ &$\alpha_2$ &$\alpha_0$ &$\alpha_1$ &$\alpha_2$ &$\alpha_0$ &$\alpha_1$ &$\alpha_2$ &$\alpha_0$ &$\alpha_1$ &$\alpha_2$ \\ \midrule
\multicolumn{17}{c}{$n=200$}\\ \midrule
\multirow{4}{*}{75\%} &0.05 &\multirow{4}{*}{0.0129} &\multirow{4}{*}{0.0297} &\multirow{4}{*}{0.0083} &
\multirow{4}{*}{0.0174} &\multirow{4}{*}{0.0410} &\multirow{4}{*}{0.0111} &\multirow{4}{*}{0.0029} &
\multirow{4}{*}{0.0074} &\multirow{4}{*}{0.0017} &\multirow{4}{*}{0.0040} &\multirow{4}{*}{0.0101} &
\multirow{4}{*}{0.0024} &0.0152 &0.0397 &0.0095 \\
&0.24 & & & & & & & & & & & & &0.0126 &0.0333 &0.0080 \\
&0.55 & & & & & & & & & & & & &0.0084 &0.0223 &0.0054 \\
&0.80 & & & & & & & & & & & & &0.0052 &0.0137 &0.0034 \\ \midrule
\multirow{4}{*}{50\%} &0.06 &\multirow{4}{*}{0.0056} &\multirow{4}{*}{0.0142} &\multirow{4}{*}{0.0037} &
\multirow{4}{*}{0.0079} &\multirow{4}{*}{0.0197} &\multirow{4}{*}{0.0050} &\multirow{4}{*}{0.0031} &
\multirow{4}{*}{0.0070} &\multirow{4}{*}{0.0018} &\multirow{4}{*}{0.0042} &\multirow{4}{*}{0.0097} &
\multirow{4}{*}{0.0025} &0.0060 &0.0157 &0.0038 \\
&0.25 & & & & & & & & & & & & &0.0055 &0.0139 &0.0034 \\
&0.56 & & & & & & & & & & & & &0.0043 &0.0110 &0.0028 \\
&0.81 & & & & & & & & & & & & &0.0034 &0.0087 &0.0022 \\ \midrule
\multirow{4}{*}{25\%} &0.06 &\multirow{4}{*}{0.0038} &\multirow{4}{*}{0.0099} &\multirow{4}{*}{0.0025} &
\multirow{4}{*}{0.0053} &\multirow{4}{*}{0.0135} &\multirow{4}{*}{0.0034} &\multirow{4}{*}{0.0029} &
\multirow{4}{*}{0.0069} &\multirow{4}{*}{0.0019} &\multirow{4}{*}{0.0040} &\multirow{4}{*}{0.0095} &
\multirow{4}{*}{0.0025} &0.0039 &0.0101 &0.0024 \\
&0.25 & & & & & & & & & & & & &0.0037 &0.0094 &0.0023 \\
&0.56 & & & & & & & & & & & & &0.0033 &0.0084 &0.0021 \\
&0.81 & & & & & & & & & & & & &0.0030 &0.0076 &0.0019 \\ \midrule
\multicolumn{17}{c}{$n=400$}\\ \midrule
\multirow{4}{*}{75\%} &0.07 &\multirow{4}{*}{0.0060} &\multirow{4}{*}{0.0151} &\multirow{4}{*}{0.0037} &
\multirow{4}{*}{0.0083} &\multirow{4}{*}{0.0205} &\multirow{4}{*}{0.0050} &\multirow{4}{*}{0.0014} &
\multirow{4}{*}{0.0036} &\multirow{4}{*}{0.0008} &\multirow{4}{*}{0.0019} &\multirow{4}{*}{0.0049} &
\multirow{4}{*}{0.0011} &0.0062 &0.0151 &0.0038 \\
&0.25 & & & & & & & & & & & & &0.0052 &0.0126 &0.0033 \\
&0.56 & & & & & & & & & & & & &0.0035 &0.0086 &0.0023 \\
&0.81 & & & & & & & & & & & & &0.0023 &0.0057 &0.0015 \\ \midrule
\multirow{4}{*}{50\%} &0.07 &\multirow{4}{*}{0.0032} &\multirow{4}{*}{0.0072} &\multirow{4}{*}{0.0021} &
\multirow{4}{*}{0.0043} &\multirow{4}{*}{0.0098} &\multirow{4}{*}{0.0028} &\multirow{4}{*}{0.0014} &
\multirow{4}{*}{0.0036} &\multirow{4}{*}{0.0009} &\multirow{4}{*}{0.0019} &\multirow{4}{*}{0.0049} &
\multirow{4}{*}{0.0012} &0.0032 &0.0070 &0.0020 \\
&0.25 & & & & & & & & & & & & &0.0029 &0.0064 &0.0019 \\
&0.56 & & & & & & & & & & & & &0.0022 &0.0053 &0.0015 \\
&0.81 & & & & & & & & & & & & &0.0017 &0.0044 &0.0012 \\ \midrule
\multirow{4}{*}{25\%} &0.06 &\multirow{4}{*}{0.0022} &\multirow{4}{*}{0.0051} &\multirow{4}{*}{0.0015} &
\multirow{4}{*}{0.0030} &\multirow{4}{*}{0.0068} &\multirow{4}{*}{0.0019} &\multirow{4}{*}{0.0016} &
\multirow{4}{*}{0.0043} &\multirow{4}{*}{0.0010} &\multirow{4}{*}{0.0022} &\multirow{4}{*}{0.0055} &
\multirow{4}{*}{0.0013} &0.0022 &0.0051 &0.0015 \\
&0.25 & & & & & & & & & & & & &0.0021 &0.0049 &0.0014 \\
&0.56 & & & & & & & & & & & & &0.0018 &0.0044 &0.0013 \\
&0.81 & & & & & & & & & & & & &0.0015 &0.0040 &0.0011 \\ \bottomrule
\end{tabular}\vspace{3pt}}
\end{threeparttable}
\end{table}

\begin{table}
\caption{MISEs of the coefficient functions for different methods with a misspecified surrogate model under MCAR. \label{TAB:eg1-mcar-mise2}}
\centering
\begin{threeparttable}
\scalebox{0.65}{\setlength{\tabcolsep}{3pt}\begin{tabular}
{ccccc|ccc|ccc|ccc|ccc} \toprule
\multirow{3}{*}{\shortstack{Missing\\Rate}} &\multirow{3}{*}{$\overline{R^2}$} & \multicolumn{6}{c}{ISR} &\multicolumn{6}{c}{SISR} & \multicolumn{3}{c}{SSFR}\\ \cmidrule(lr){3-8} \cmidrule(lr){9-14}
& &\multicolumn{3}{c}{TPST} &\multicolumn{3}{c}{Kernel} &\multicolumn{3}{c}{TPST} &\multicolumn{3}{c}{Kernel} & \multicolumn{3}{c}{TPST} \\ \cmidrule(lr){3-5} \cmidrule(lr){6-8} \cmidrule(lr){9-11} \cmidrule(lr){12-14} \cmidrule(lr){15-17}
& &$\alpha_0$ &$\alpha_1$ &$\alpha_2$ &$\alpha_0$ &$\alpha_1$ &$\alpha_2$ &$\alpha_0$ &$\alpha_1$ &$\alpha_2$ &$\alpha_0$ &$\alpha_1$ &$\alpha_2$ &$\alpha_0$ &$\alpha_1$ &$\alpha_2$ \\ \midrule
\multicolumn{17}{c}{$n=200$}\\ \midrule
\multirow{4}{*}{75\%} &0.05 &\multirow{4}{*}{0.0129} &\multirow{4}{*}{0.0297} &\multirow{4}{*}{0.0083} &
\multirow{4}{*}{0.0174} &\multirow{4}{*}{0.0410} &\multirow{4}{*}{0.0111} &\multirow{4}{*}{0.6528} &
\multirow{4}{*}{0.1323} &\multirow{4}{*}{1.0034} &\multirow{4}{*}{0.6458} &\multirow{4}{*}{0.1341} &
\multirow{4}{*}{1.0015} &0.0152 &0.0396 &0.0095 \\
&0.24 & & & & & & & & & & & & &0.0126 &0.0332 &0.0080 \\
&0.55 & & & & & & & & & & & & &0.0084 &0.0222 &0.0054 \\
&0.80 & & & & & & & & & & & & &0.0052 &0.0136 &0.0034 \\ \midrule
\multirow{4}{*}{50\%} &0.06 &\multirow{4}{*}{0.0056} &\multirow{4}{*}{0.0142} &\multirow{4}{*}{0.0037} &
\multirow{4}{*}{0.0079} &\multirow{4}{*}{0.0197} &\multirow{4}{*}{0.0050} &\multirow{4}{*}{0.2928} &
\multirow{4}{*}{0.0764} &\multirow{4}{*}{0.4406} &\multirow{4}{*}{0.2903} &\multirow{4}{*}{0.0786} &
\multirow{4}{*}{0.4401} &0.0060 &0.0156 &0.0038 \\
&0.25 & & & & & & & & & & & & &0.0054 &0.0138 &0.0034 \\
&0.56 & & & & & & & & & & & & &0.0043 &0.0110 &0.0028 \\
&0.81 & & & & & & & & & & & & &0.0034 &0.0087 &0.0022 \\ \midrule
\multirow{4}{*}{25\%} &0.06 &\multirow{4}{*}{0.0038} &\multirow{4}{*}{0.0099} &\multirow{4}{*}{0.0025} &
\multirow{4}{*}{0.0053} &\multirow{4}{*}{0.0135} &\multirow{4}{*}{0.0034} &\multirow{4}{*}{0.0797} &
\multirow{4}{*}{0.0303} &\multirow{4}{*}{0.1116} &\multirow{4}{*}{0.0799} &\multirow{4}{*}{0.0328} &
\multirow{4}{*}{0.1119} &0.0039 &0.0101 &0.0024 \\
&0.25 & & & & & & & & & & & & &0.0037 &0.0094 &0.0023 \\
&0.56 & & & & & & & & & & & & &0.0033 &0.0083 &0.0021 \\
&0.81 & & & & & & & & & & & & &0.0030 &0.0076 &0.0019 \\ \midrule
\multicolumn{17}{c}{$n=400$}\\ \midrule
\multirow{4}{*}{75\%} &0.07 &\multirow{4}{*}{0.0060} &\multirow{4}{*}{0.0151} &\multirow{4}{*}{0.0037} &
\multirow{4}{*}{0.0083} &\multirow{4}{*}{0.0205} &\multirow{4}{*}{0.0050} &\multirow{4}{*}{0.6511} &
\multirow{4}{*}{0.1243} &\multirow{4}{*}{1.0140} &\multirow{4}{*}{0.6436} &\multirow{4}{*}{0.1248} &
\multirow{4}{*}{1.0117} &0.0062 &0.0151 &0.0038 \\
&0.25 & & & & & & & & & & & & &0.0052 &0.0126 &0.0033 \\
&0.56 & & & & & & & & & & & & &0.0035 &0.0086 &0.0023 \\
&0.81 & & & & & & & & & & & & &0.0023 &0.0057 &0.0015 \\ \midrule
\multirow{4}{*}{50\%} &0.07 &\multirow{4}{*}{0.0032} &\multirow{4}{*}{0.0072} &\multirow{4}{*}{0.0021} &
\multirow{4}{*}{0.0043} &\multirow{4}{*}{0.0098} &\multirow{4}{*}{0.0028} &\multirow{4}{*}{0.2872} &
\multirow{4}{*}{0.0633} &\multirow{4}{*}{0.4552} &\multirow{4}{*}{0.2842} &\multirow{4}{*}{0.0642} &
\multirow{4}{*}{0.4543} &0.0032 &0.0070 &0.0020 \\
&0.25 & & & & & & & & & & & & &0.0029 &0.0064 &0.0019 \\
&0.56 & & & & & & & & & & & & &0.0022 &0.0053 &0.0015 \\
&0.81 & & & & & & & & & & & & &0.0017 &0.0044 &0.0012 \\ \midrule
\multirow{4}{*}{25\%} &0.06 &\multirow{4}{*}{0.0022} &\multirow{4}{*}{0.0051} &\multirow{4}{*}{0.0015} &
\multirow{4}{*}{0.0030} &\multirow{4}{*}{0.0068} &\multirow{4}{*}{0.0019} &\multirow{4}{*}{0.0752} &
\multirow{4}{*}{0.0218} &\multirow{4}{*}{0.1220} &\multirow{4}{*}{0.0749} &\multirow{4}{*}{0.0230} &
\multirow{4}{*}{0.1220} &0.0022 &0.0051 &0.0015 \\
&0.25 & & & & & & & & & & & & &0.0021 &0.0049 &0.0014 \\
&0.56 & & & & & & & & & & & & &0.0018 &0.0044 &0.0013 \\
&0.81 & & & & & & & & & & & & &0.0015 &0.0040 &0.0011 \\ \bottomrule
\end{tabular}\vspace{3pt}}
\end{threeparttable}
\end{table}

Table \ref{TAB:eg1-mcar-mise1} reports the results when the surrogate model is correctly specified. As under the MAR experiments, SSFR consistently improves upon ISR by effectively incorporating the surrogate information, with the efficiency gain becoming more pronounced as both the surrogate--target correlation and the missingness rate increase. In this setting, naive SISR also performs competitively because the generated surrogate images closely approximate the true target images.

Table \ref{TAB:eg1-mcar-mise2} summarizes the results under surrogate-model misspecification. The qualitative conclusions remain unchanged. Directly replacing missing target images with misspecified surrogate images causes naive SISR to accumulate substantial bias, resulting in large MISE inflation across all missingness rates and surrogate correlations. In contrast, SSFR remains remarkably stable because the surrogate images are incorporated only through the proposed calibration stage rather than treated as replacements for the missing outcomes. Consequently, SSFR achieves estimation accuracy that remains close to ISR while avoiding the severe bias exhibited by naive SISR.

Across both correctly specified and misspecified surrogate models, the TPST implementation consistently outperforms the kernel estimator, demonstrating the advantage of triangulation-based smoothing for functional imaging data defined on irregular three-dimensional domains. The improvement is observed uniformly across all sample sizes, missingness rates, and surrogate correlation levels.

Overall, the MCAR experiments reinforce the main findings of the paper. Although the missing-data mechanism differs substantially from the MAR setting considered in the primary simulations, the relative performance of the competing estimators remains essentially unchanged. The principal advantage of SSFR arises from its ability to exploit informative surrogate outcomes while remaining robust to surrogate-model misspecification, rather than from the specific form of the missing-data mechanism itself.

\subsection{Additional Results for the ADNI Application}
\label{ssec:app-extra-coefs}

This subsection presents additional results for the ADNI application that complement the analyses reported in Sections 7.2 and 7.3 of the main paper. The main text focuses on the most scientifically relevant findings, whereas the complete coefficient maps and significance maps for all regression coefficients are provided here for completeness.

\begin{figure}[p]\centering\setlength{\tabcolsep}{2pt}
{\footnotesize
\begin{tabular}{cccccc}
& Intercept & Sex & Age & Education \\[1pt]
ISR &\includegraphics[width=.95in]{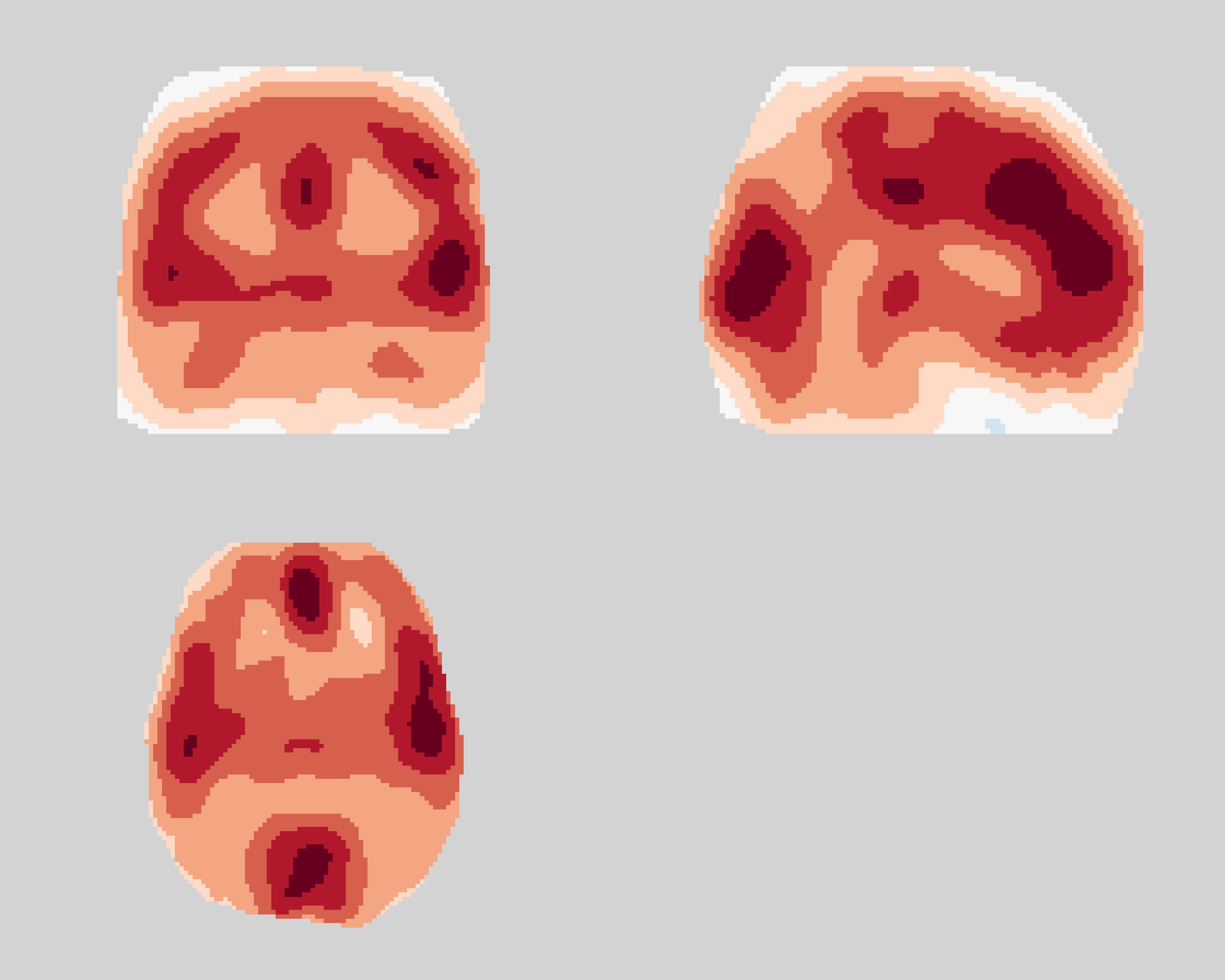} &\includegraphics[width=.95in]{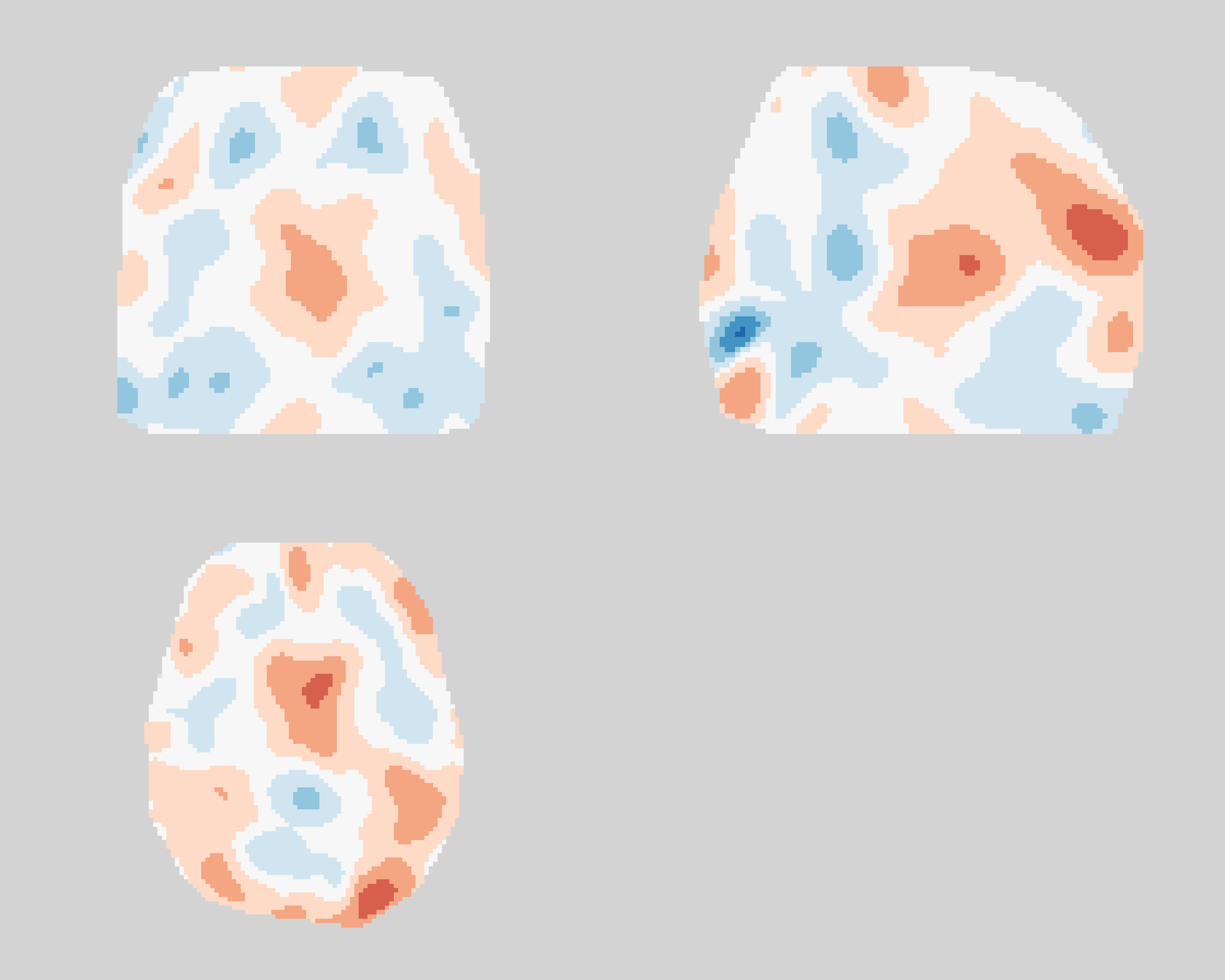} &\includegraphics[width=.95in]{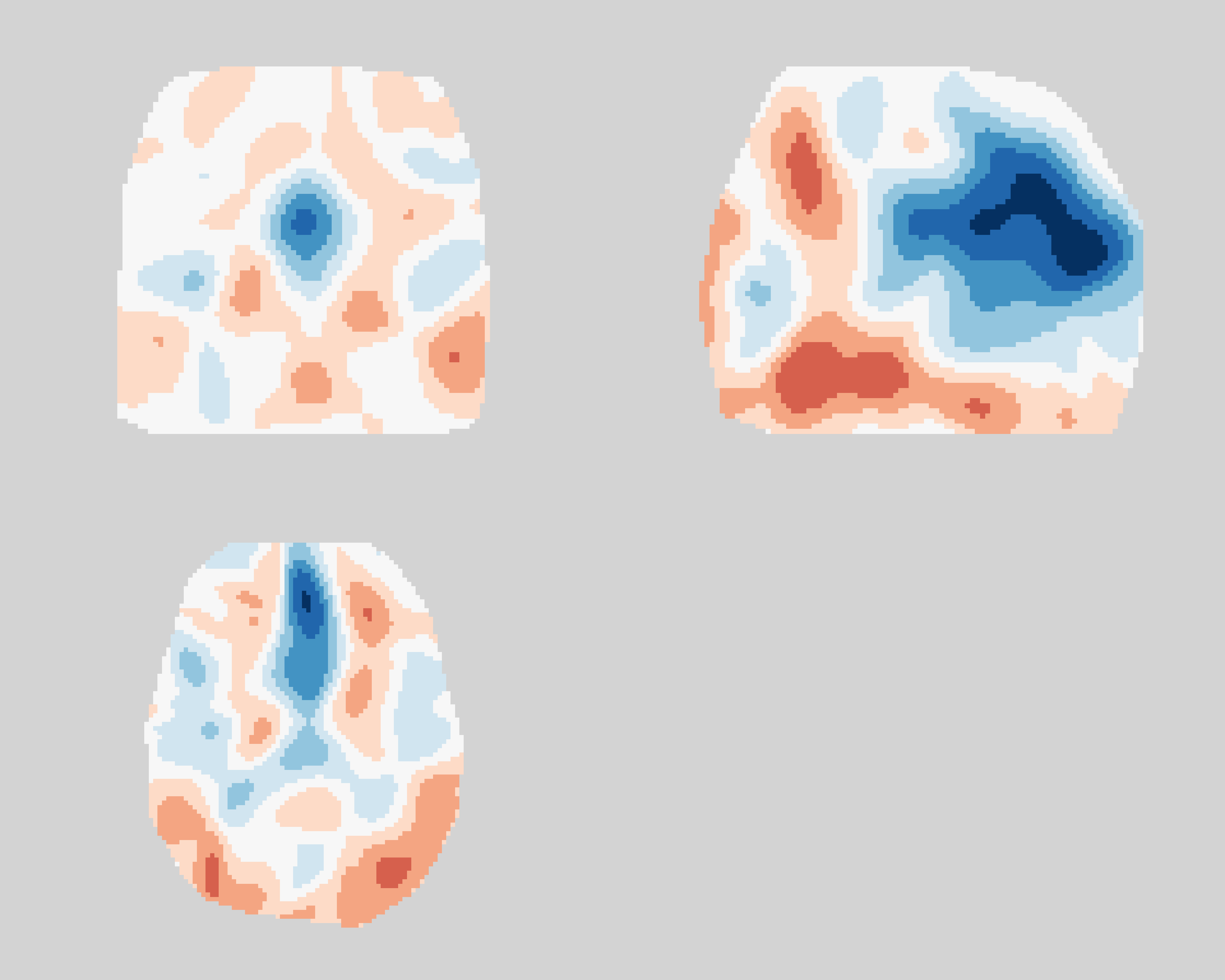} &\includegraphics[width=.95in]{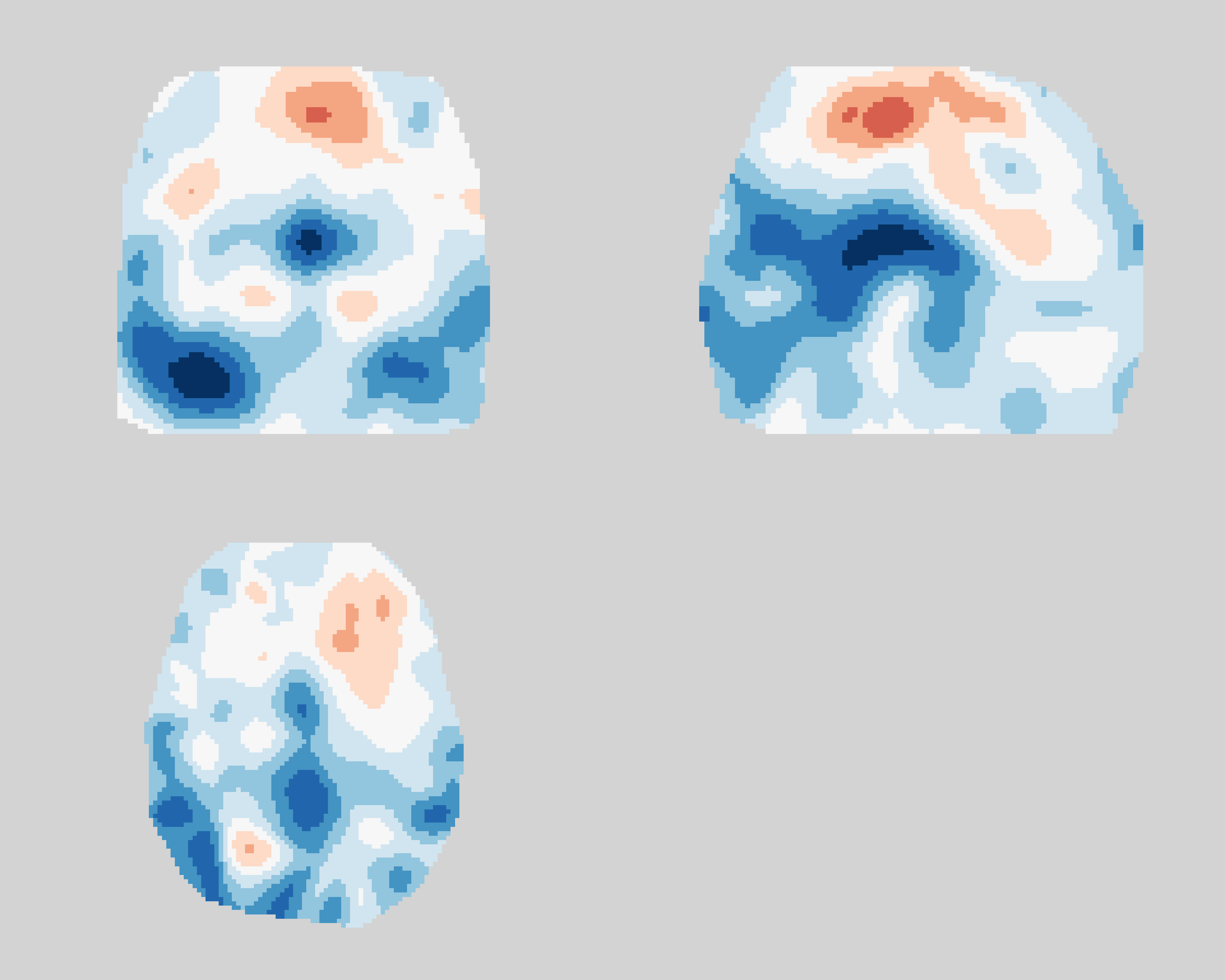} \\[-1pt]
SISR &\includegraphics[width=.95in]{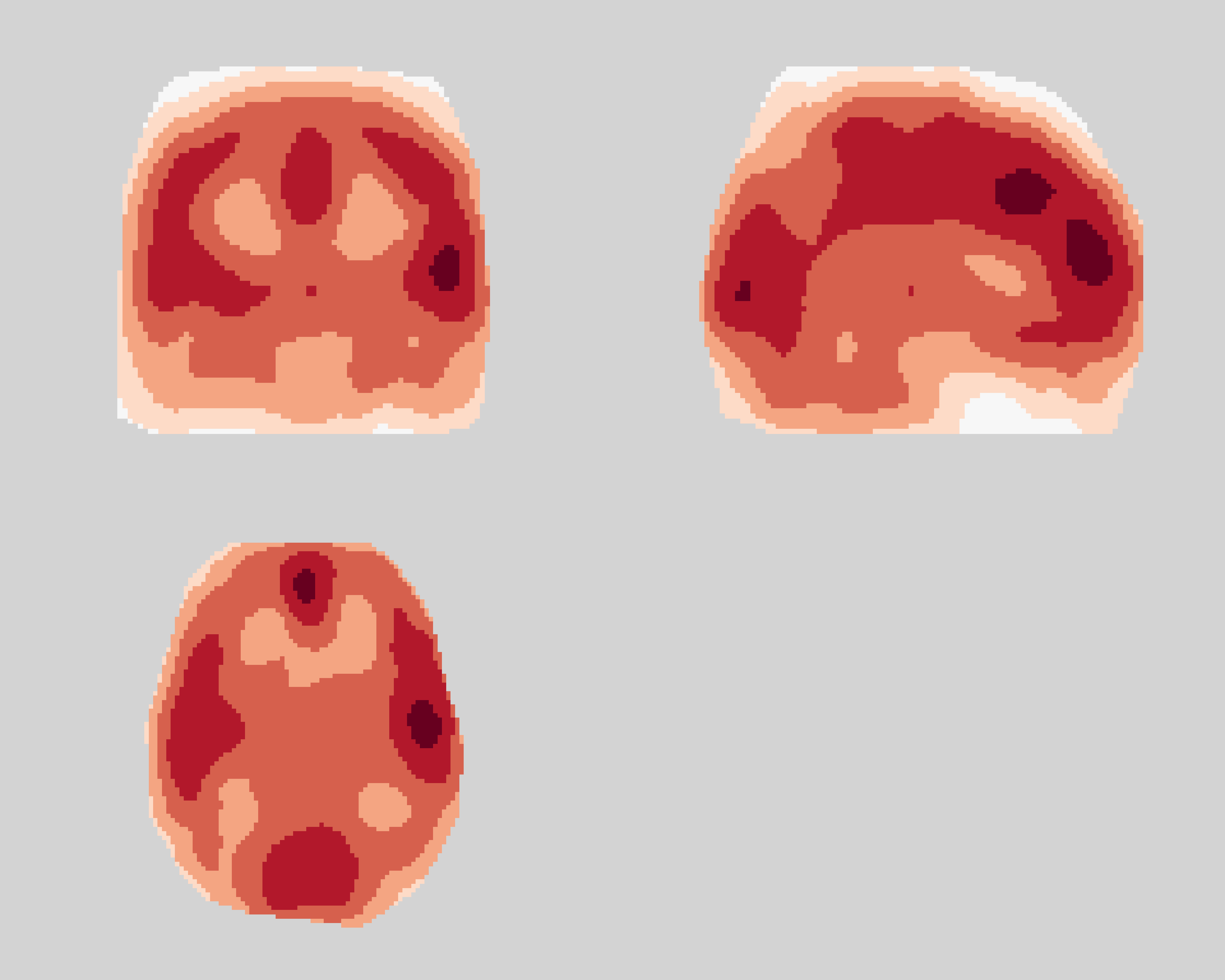} &\includegraphics[width=.95in]{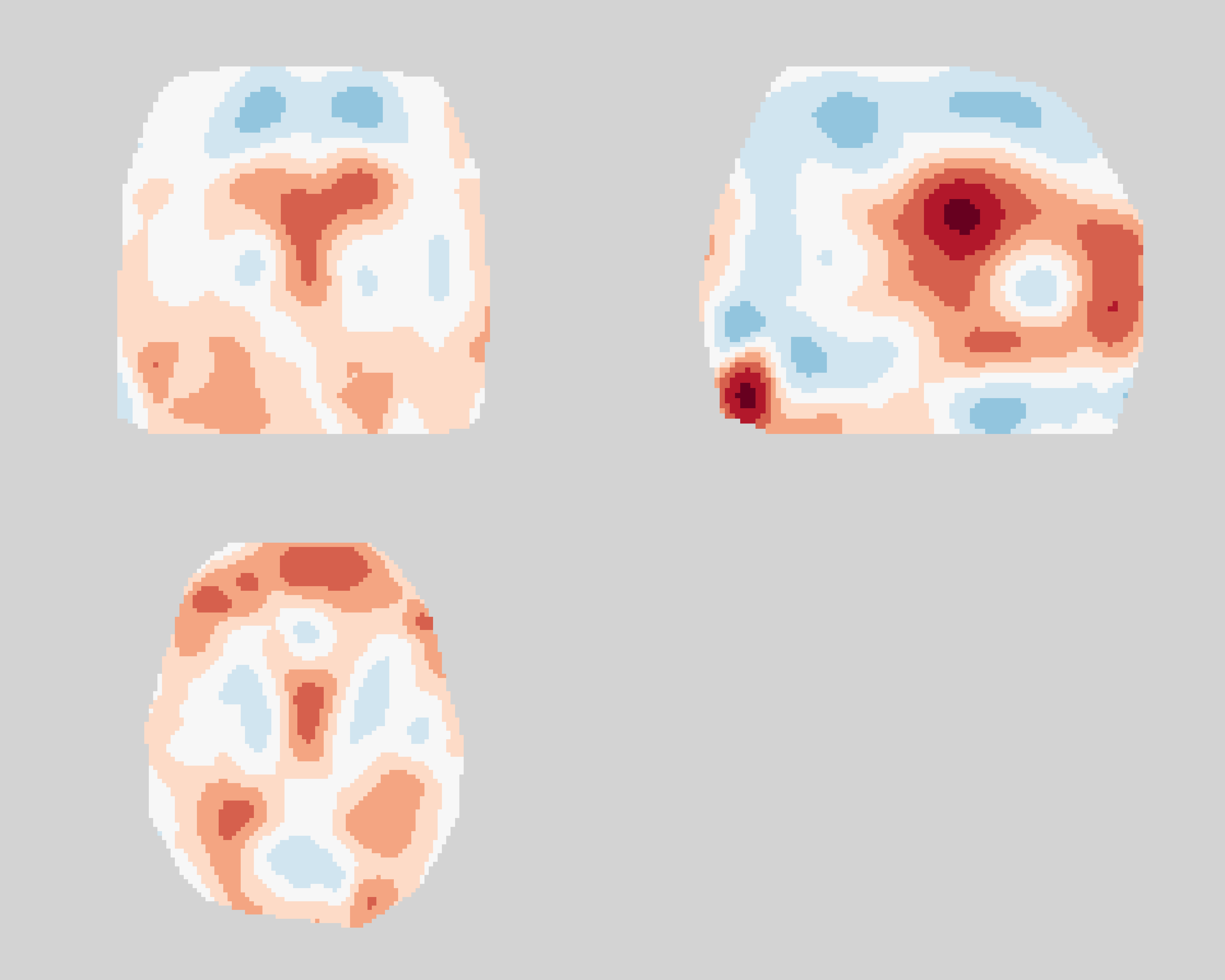} &\includegraphics[width=.95in]{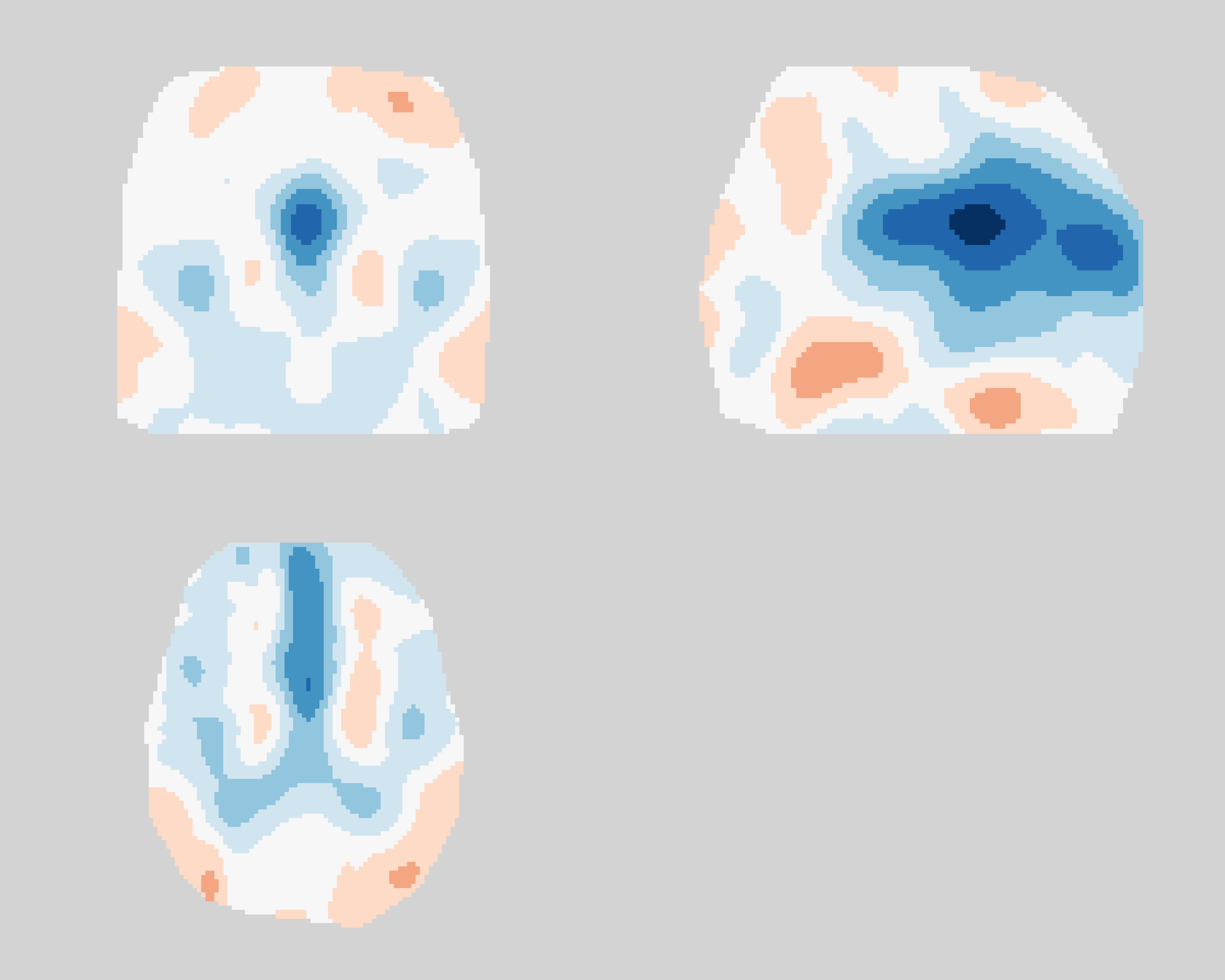} &\includegraphics[width=.95in]{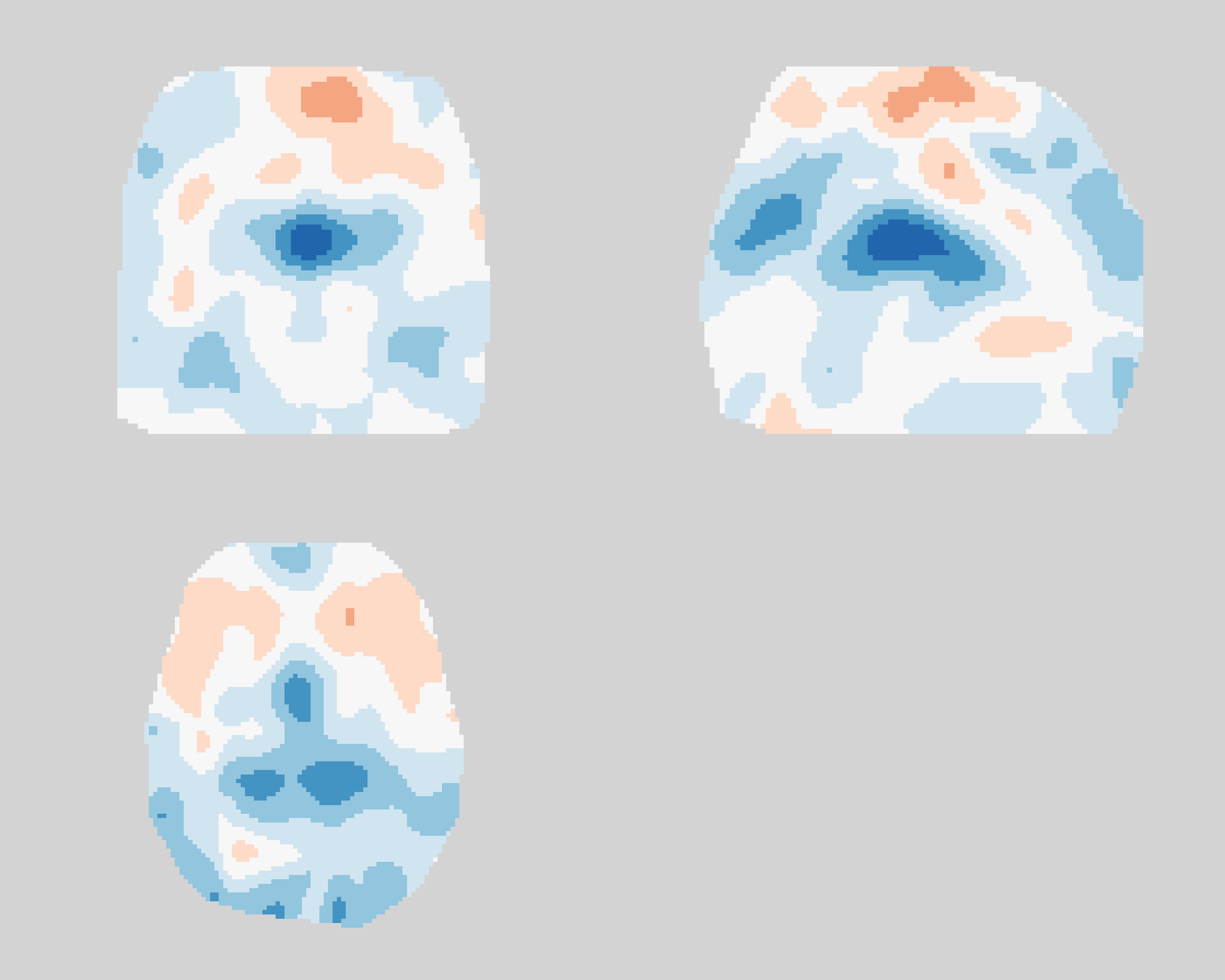} \\[-1pt]
SSFR &\includegraphics[width=.95in]{plots/app_e10/INT_wSSFR_e10.png} &\includegraphics[width=.95in]{plots/app_e10/SEX_wSSFR_e10.png} &\includegraphics[width=.95in]{plots/app_e10/AGE_wSSFR_e10.png} &\includegraphics[width=.95in]{plots/app_e10/EDU_wSSFR_e10.png} \\
& APOE$_1$ & APOE$_2$ & MMSE & MCI & AD \\[1pt]
ISR &\includegraphics[width=.95in]{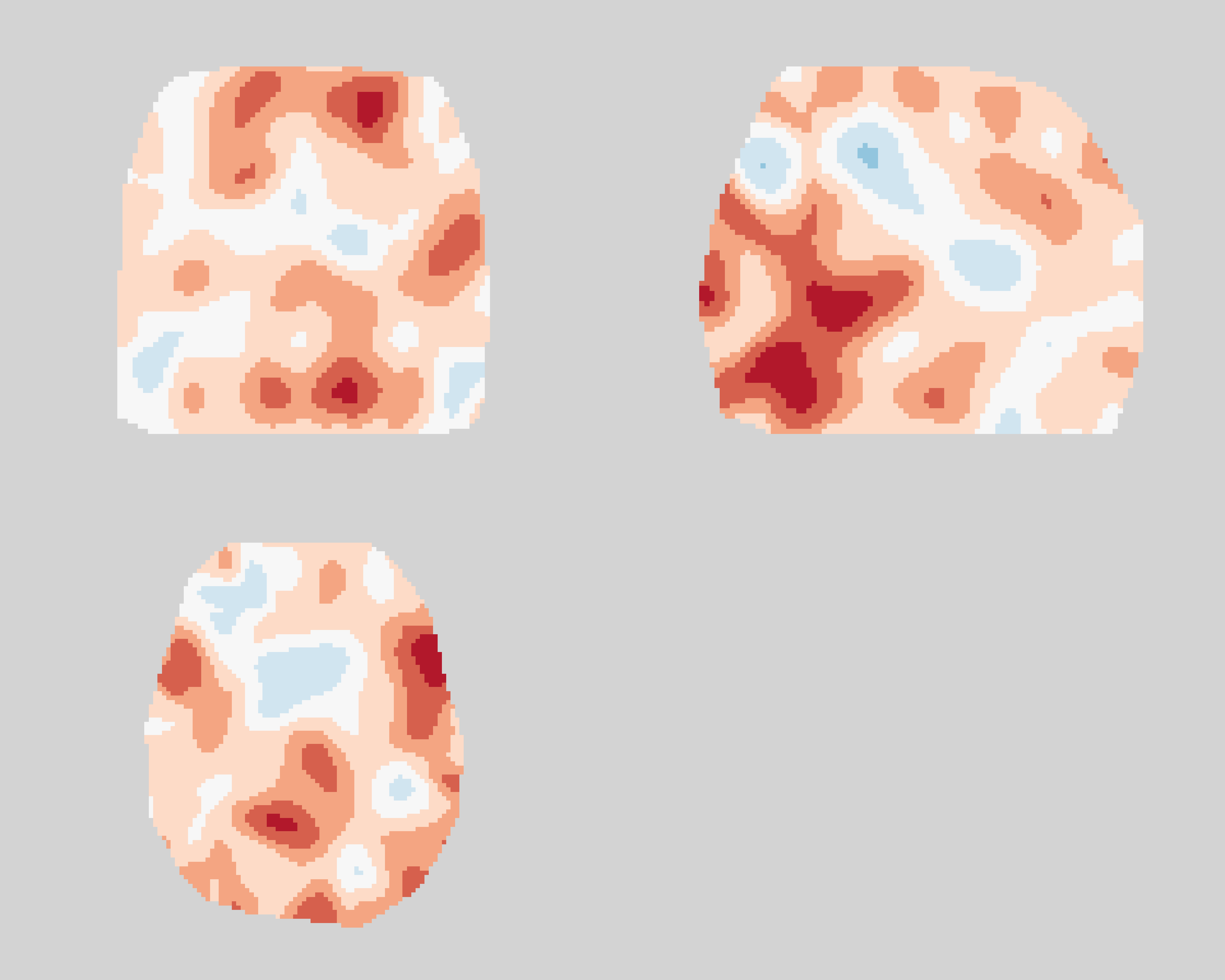} &\includegraphics[width=.95in]{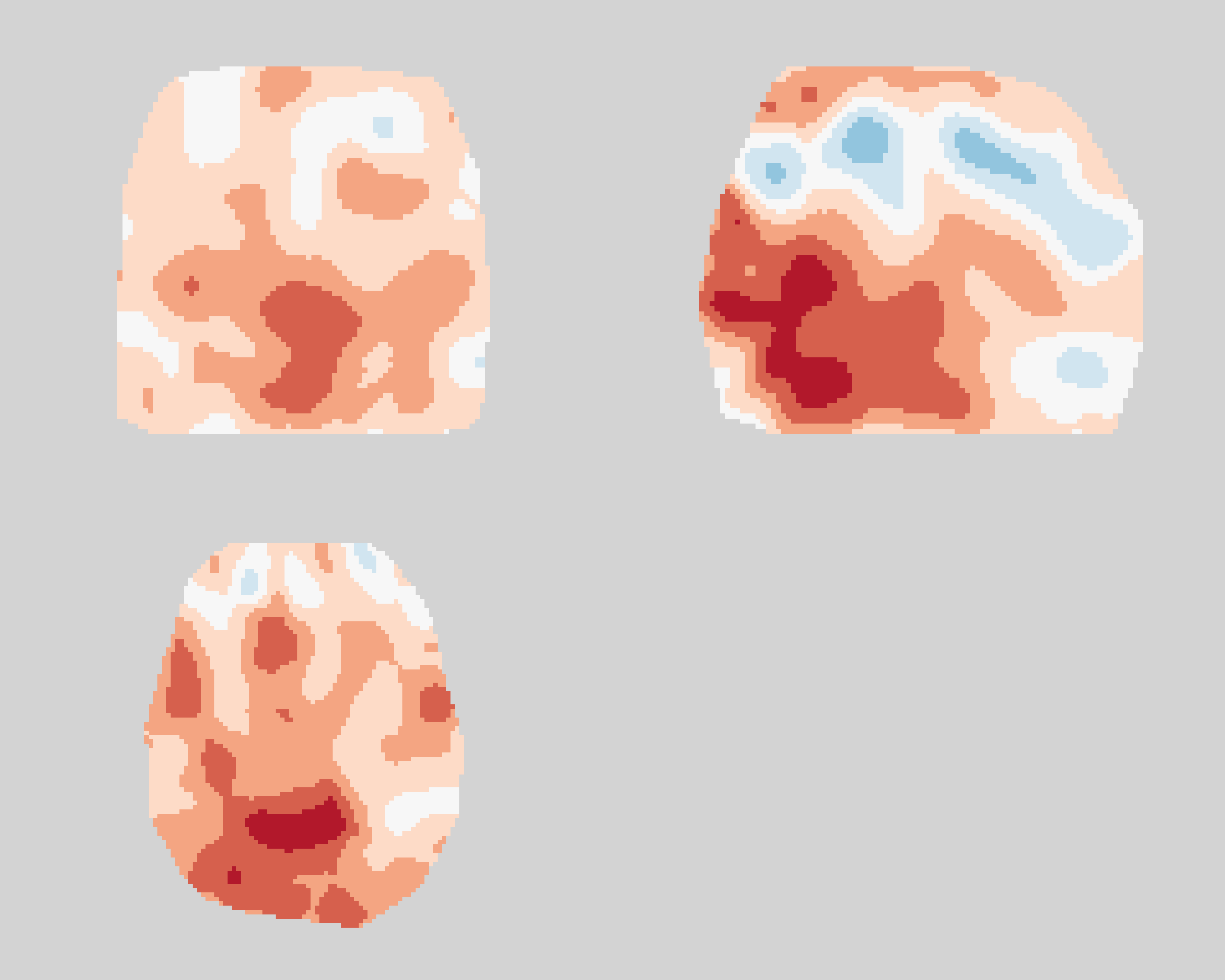} &\includegraphics[width=.95in]{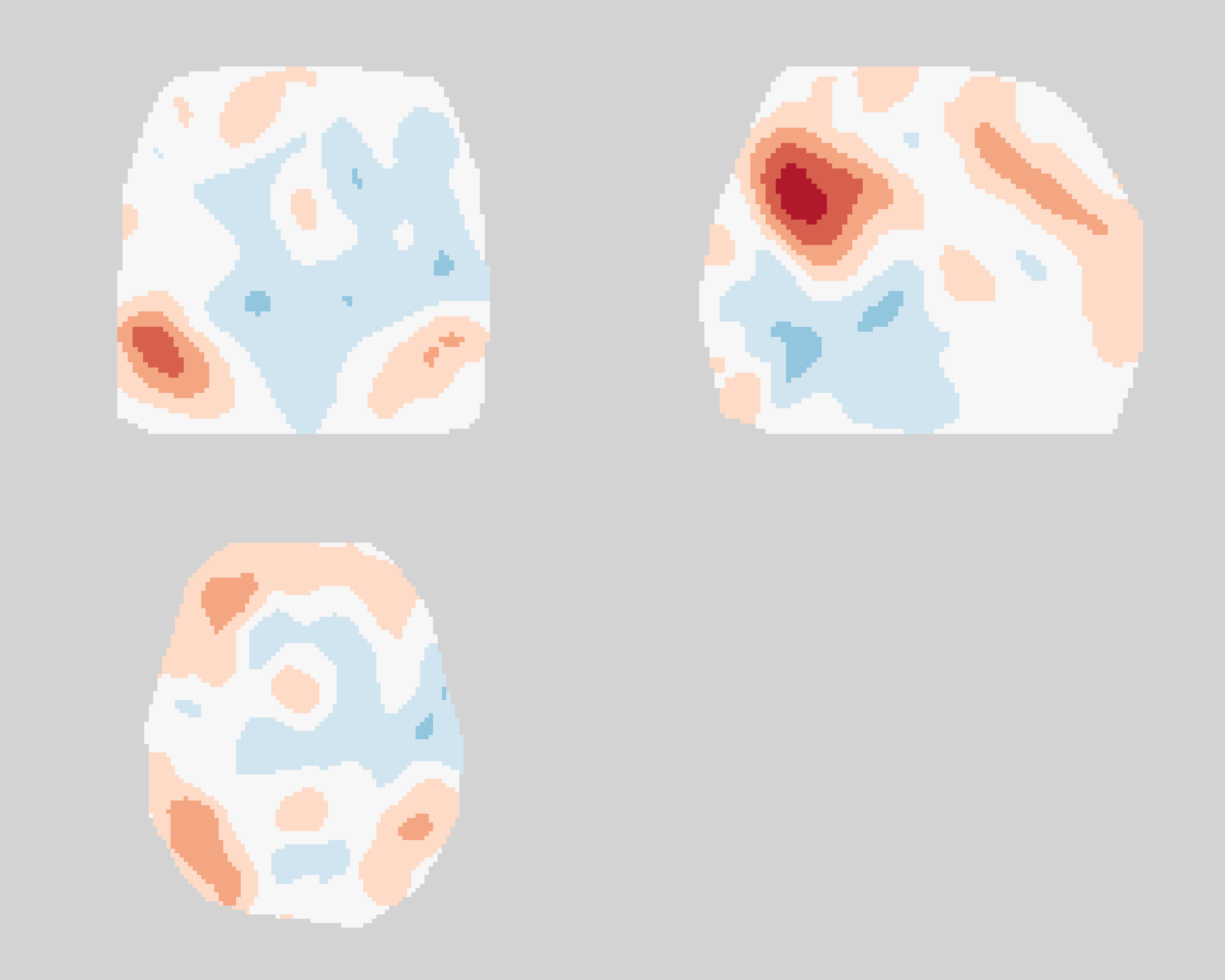} &\includegraphics[width=.95in]{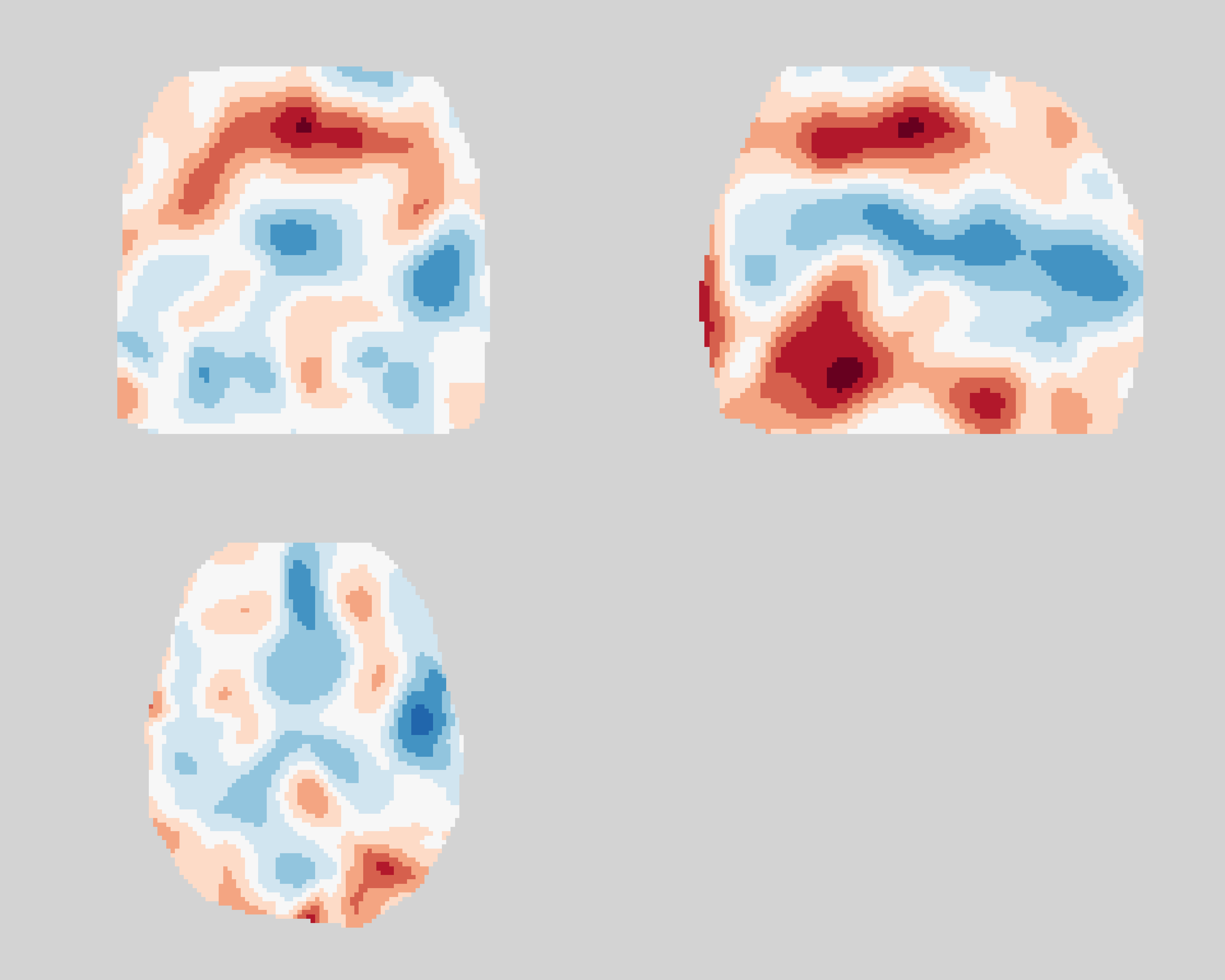} &\includegraphics[width=.95in]{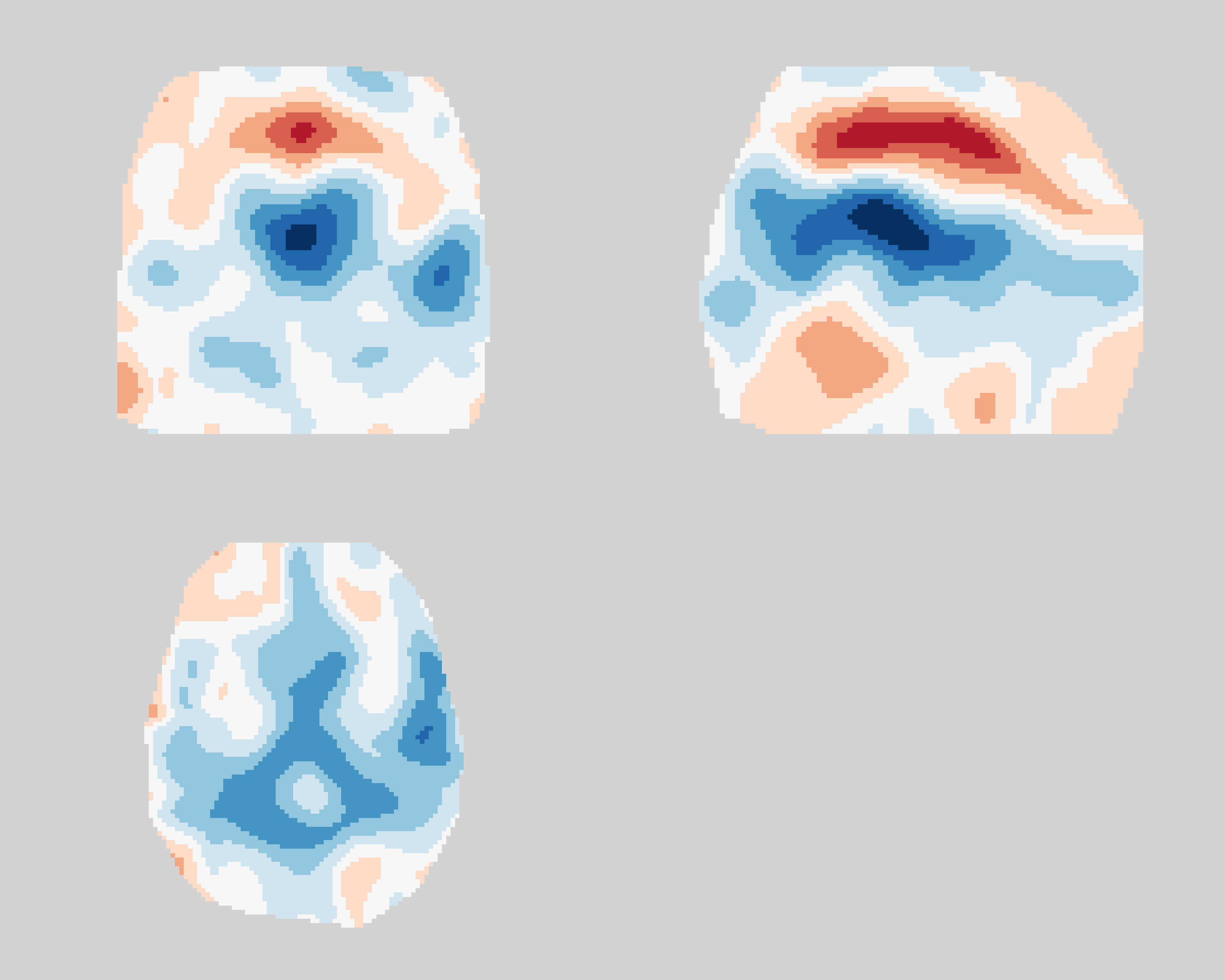} \\[-1pt]
SISR &\includegraphics[width=.95in]{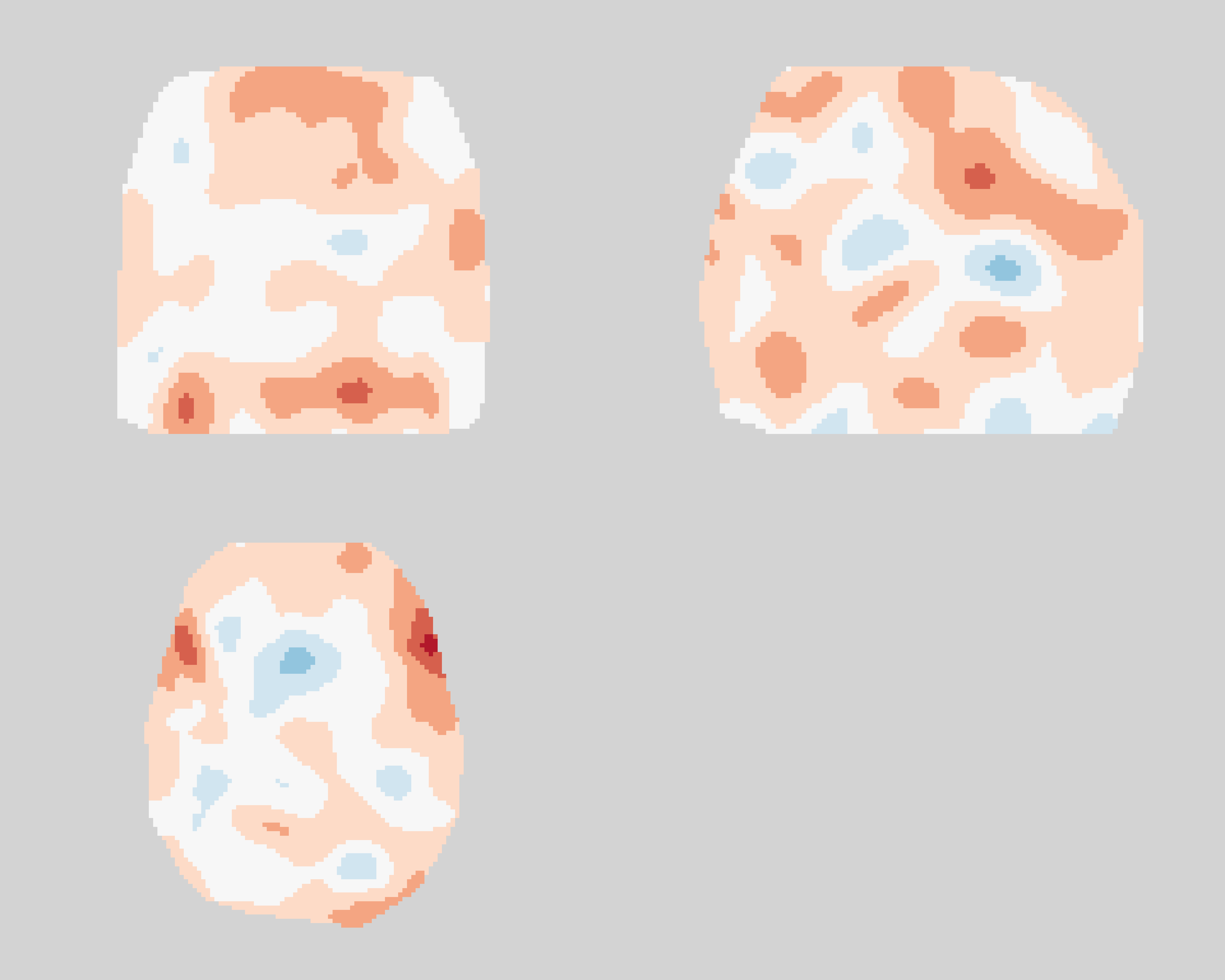} &\includegraphics[width=.95in]{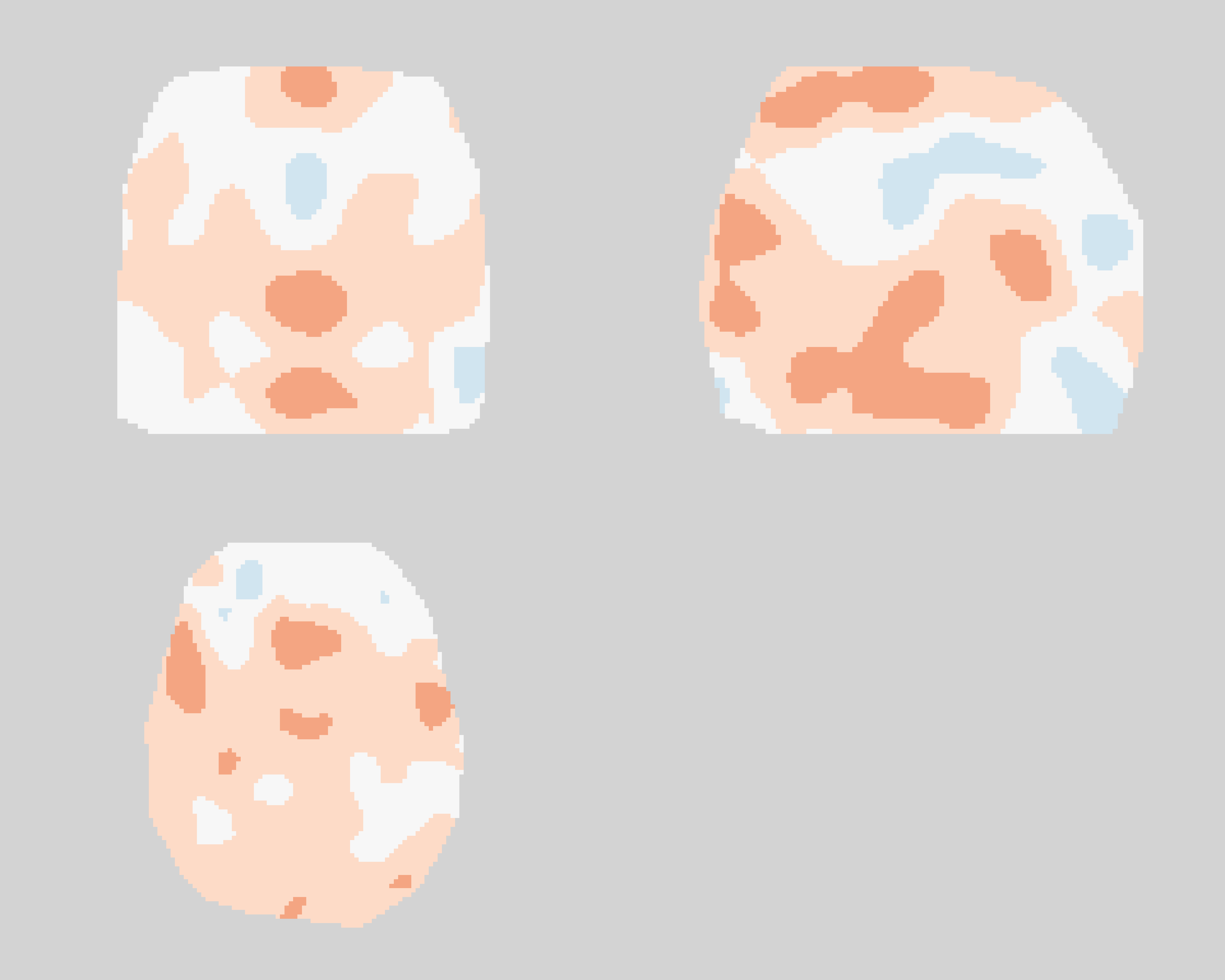} &\includegraphics[width=.95in]{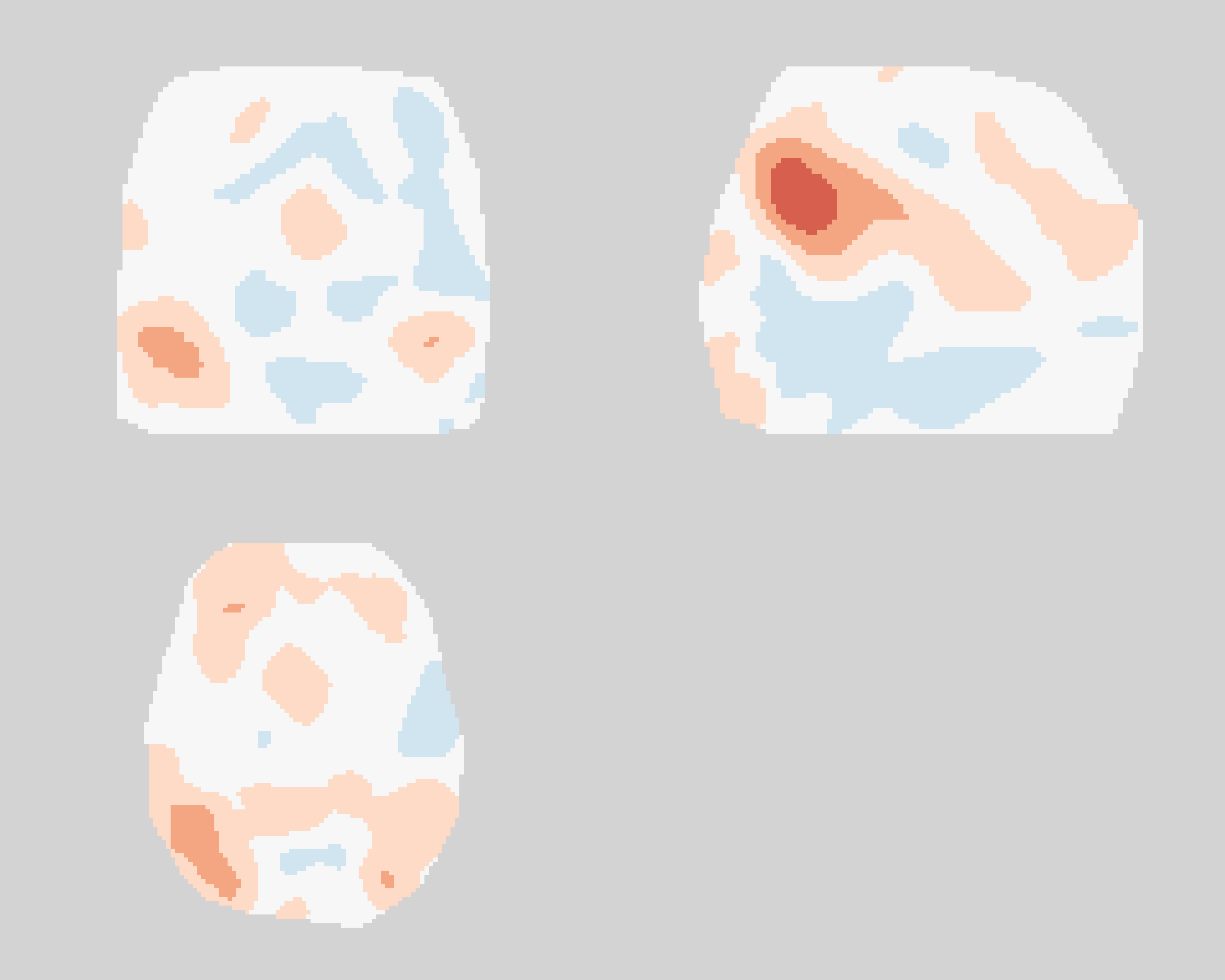} &\includegraphics[width=.95in]{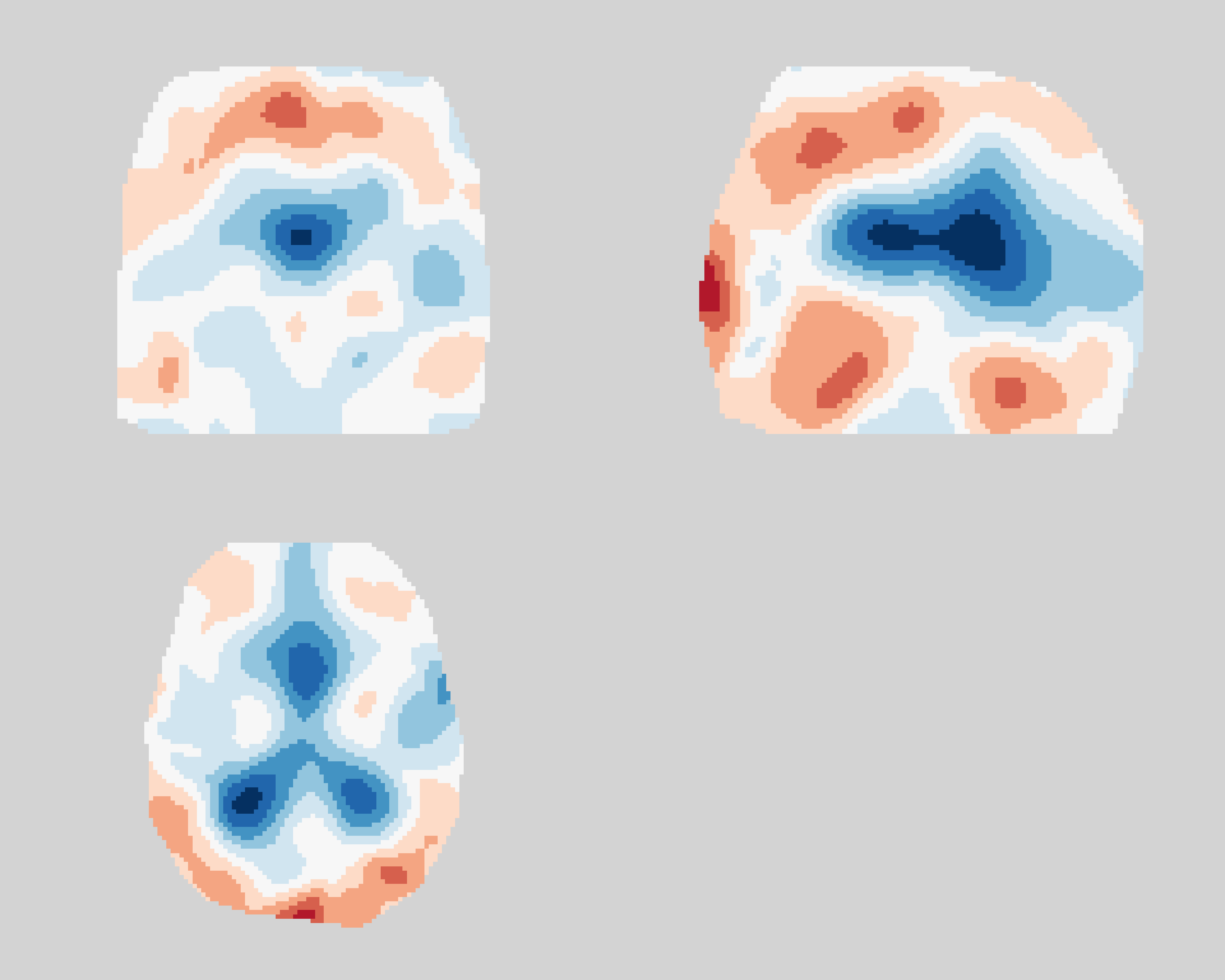} &\includegraphics[width=.95in]{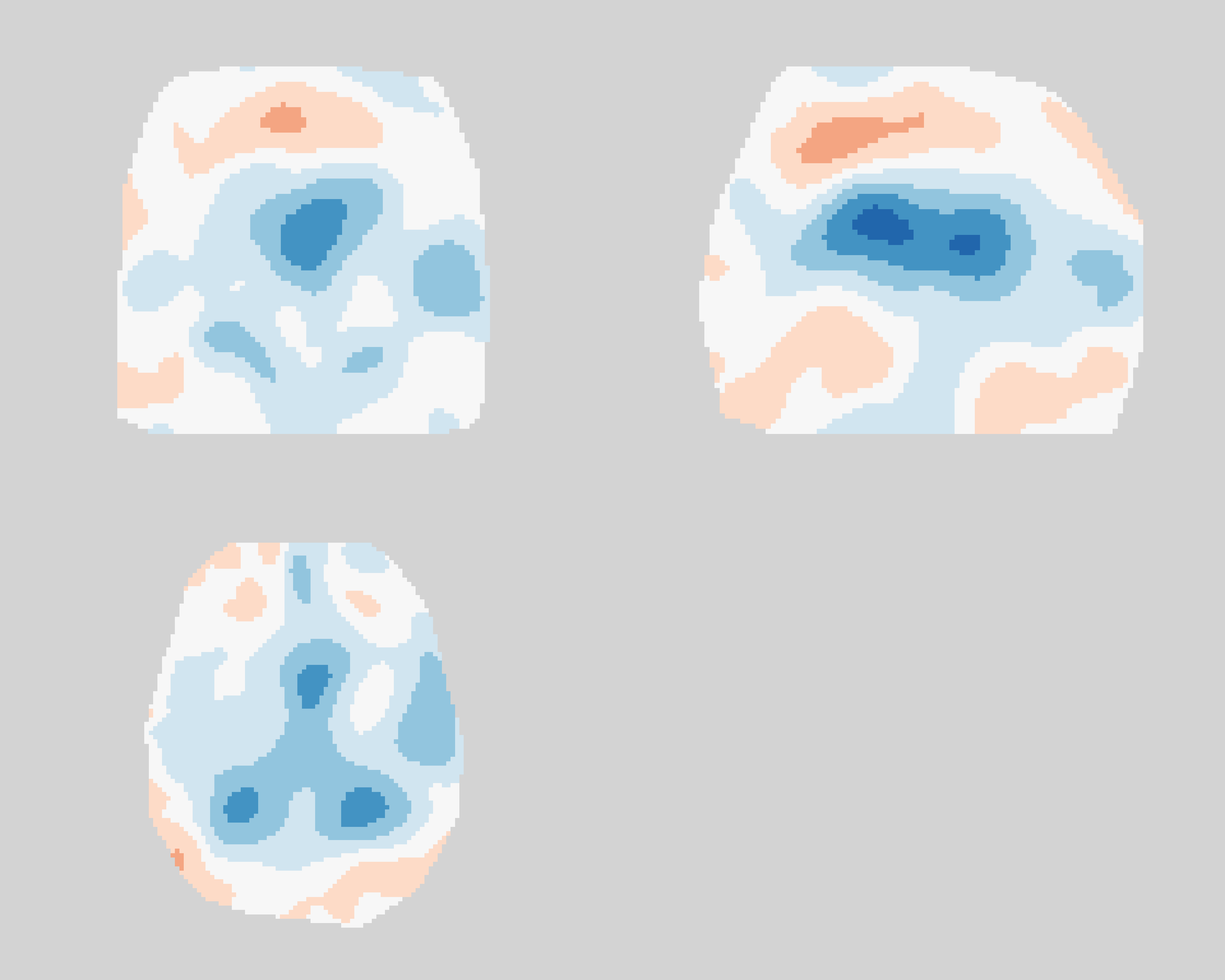} \\[-1pt]
SSFR &\includegraphics[width=.95in]{plots/app_e10/APOE1_wSSFR_e10.png} &\includegraphics[width=.95in]{plots/app_e10/APOE2_wSSFR_e10.png} &\includegraphics[width=.95in]{plots/app_e10/MMSE_wSSFR_e10.png} &\includegraphics[width=.95in]{plots/app_e10/MCI_wSSFR_e10.png} &\includegraphics[width=.95in]{plots/app_e10/AD_wSSFR_e10.png} \\[-4pt]
\end{tabular}}
\caption{Estimated coefficient maps from ISR, SISR, and SSFR in sagittal, coronal, and axial views (left, right, and bottom), with darker blue/red indicating more negative/positive values for the coefficient functions.}
\label{FIG:supp-app-coefs}
\end{figure}

Figure \ref{FIG:supp-app-coefs} displays the estimated coefficient functions obtained from ISR, naive SISR, and SSFR for all nine regression coefficients in the expanded-cohort analysis. The SSFR coefficient maps corresponding to the principal findings are reproduced in Figure 3 of the main paper.

\begin{figure}[p]\centering\setlength{\tabcolsep}{2pt}
{\footnotesize
\begin{tabular}{cccccc}
& Intercept & Sex & Age & Education \\[1pt]
ISR &\includegraphics[width=.95in]{plots/app_e10/SIG_INT_ISR_e10.png} &\includegraphics[width=.95in]{plots/app_e10/SIG_SEX_ISR_e10.png} &\includegraphics[width=.95in]{plots/app_e10/SIG_AGE_ISR_e10.png} &\includegraphics[width=.95in]{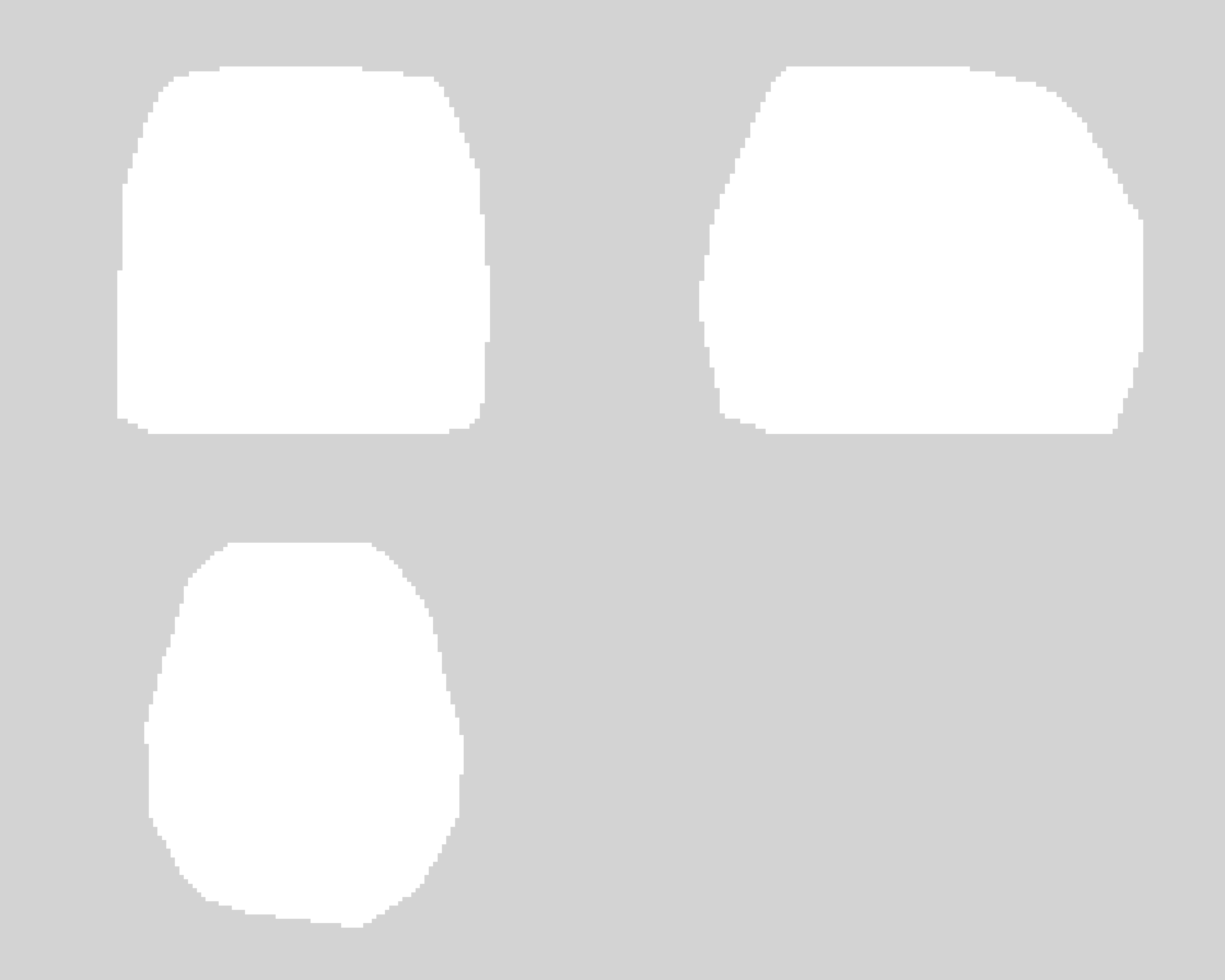} \\[-1pt]
SSFR &\includegraphics[width=.95in]{plots/app_e10/SIG_INT_wSSFR_e10.png} &\includegraphics[width=.95in]{plots/app_e10/SIG_SEX_wSSFR_e10.png} &\includegraphics[width=.95in]{plots/app_e10/SIG_AGE_wSSFR_e10.png} &\includegraphics[width=.95in]{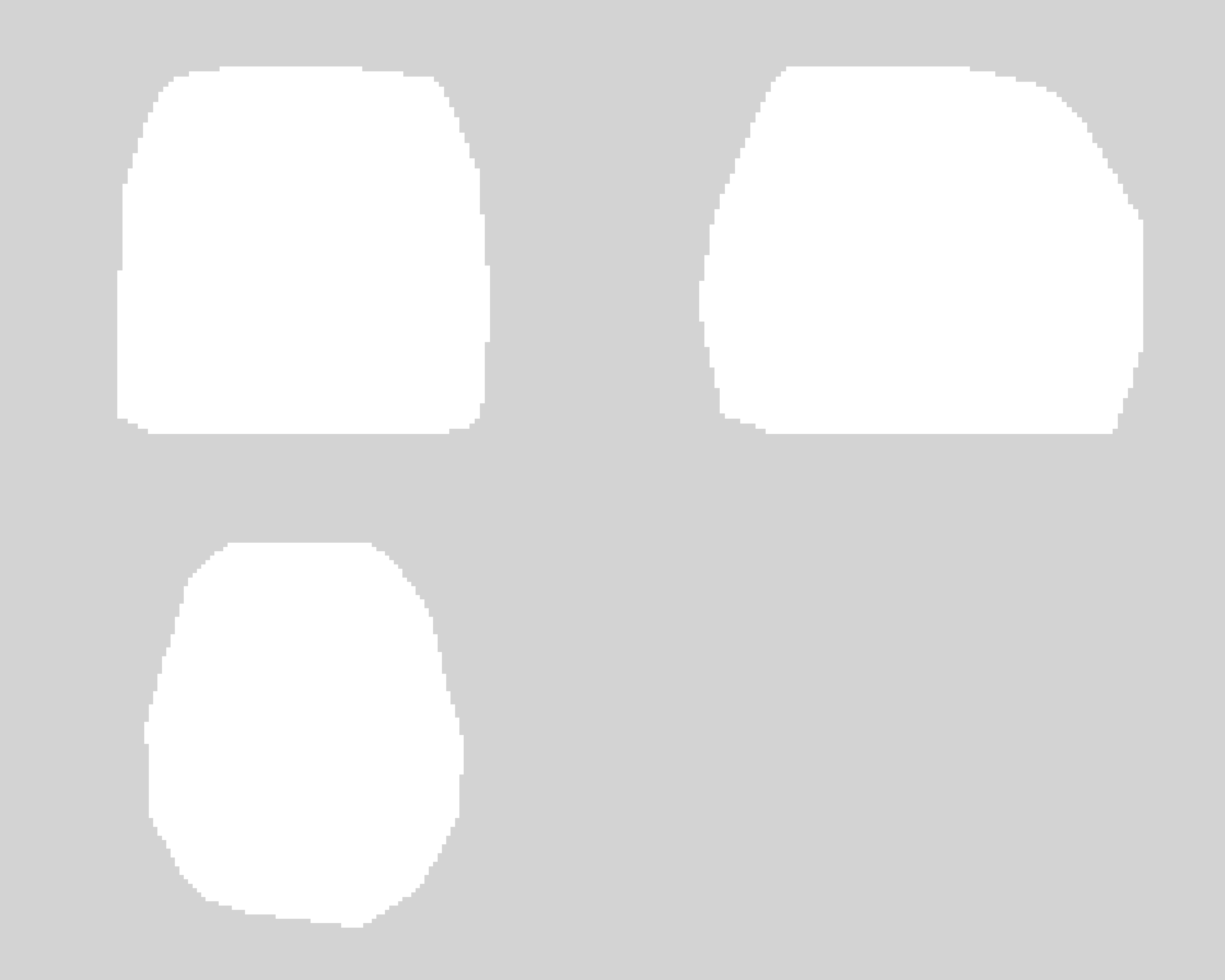} \\ [-1pt]
Difference & \includegraphics[width=.95in]{plots/app_e10/COVDIFF_INT_e10.png} &\includegraphics[width=.95in]{plots/app_e10/COVDIFF_SEX_e10.png} &\includegraphics[width=.95in]{plots/app_e10/COVDIFF_AGE_e10.png} &\includegraphics[width=.95in]{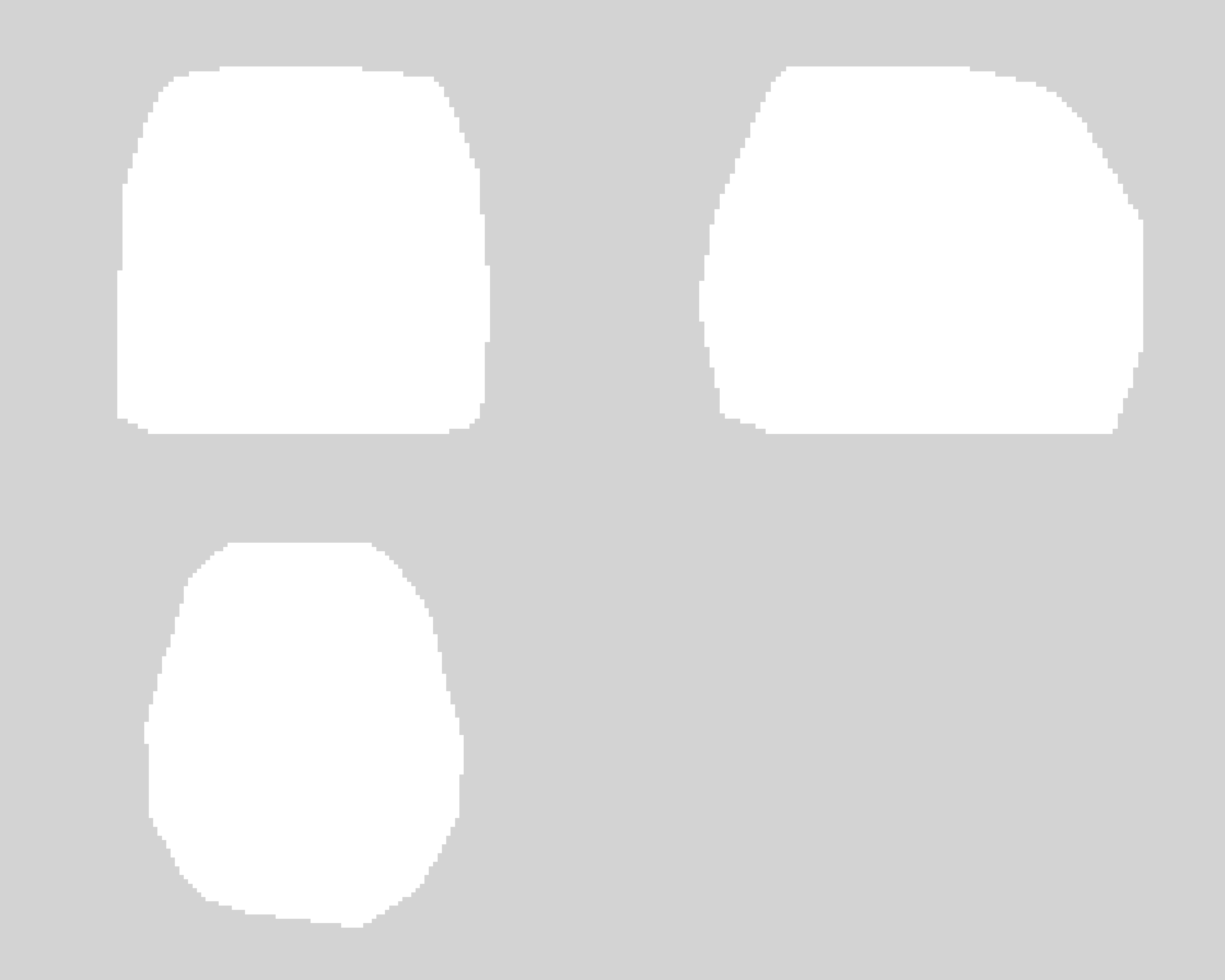} & \\[4pt]
& APOE$_1$ & APOE$_2$ & MMSE & MCI & AD \\[1pt]
ISR &\includegraphics[width=.95in]{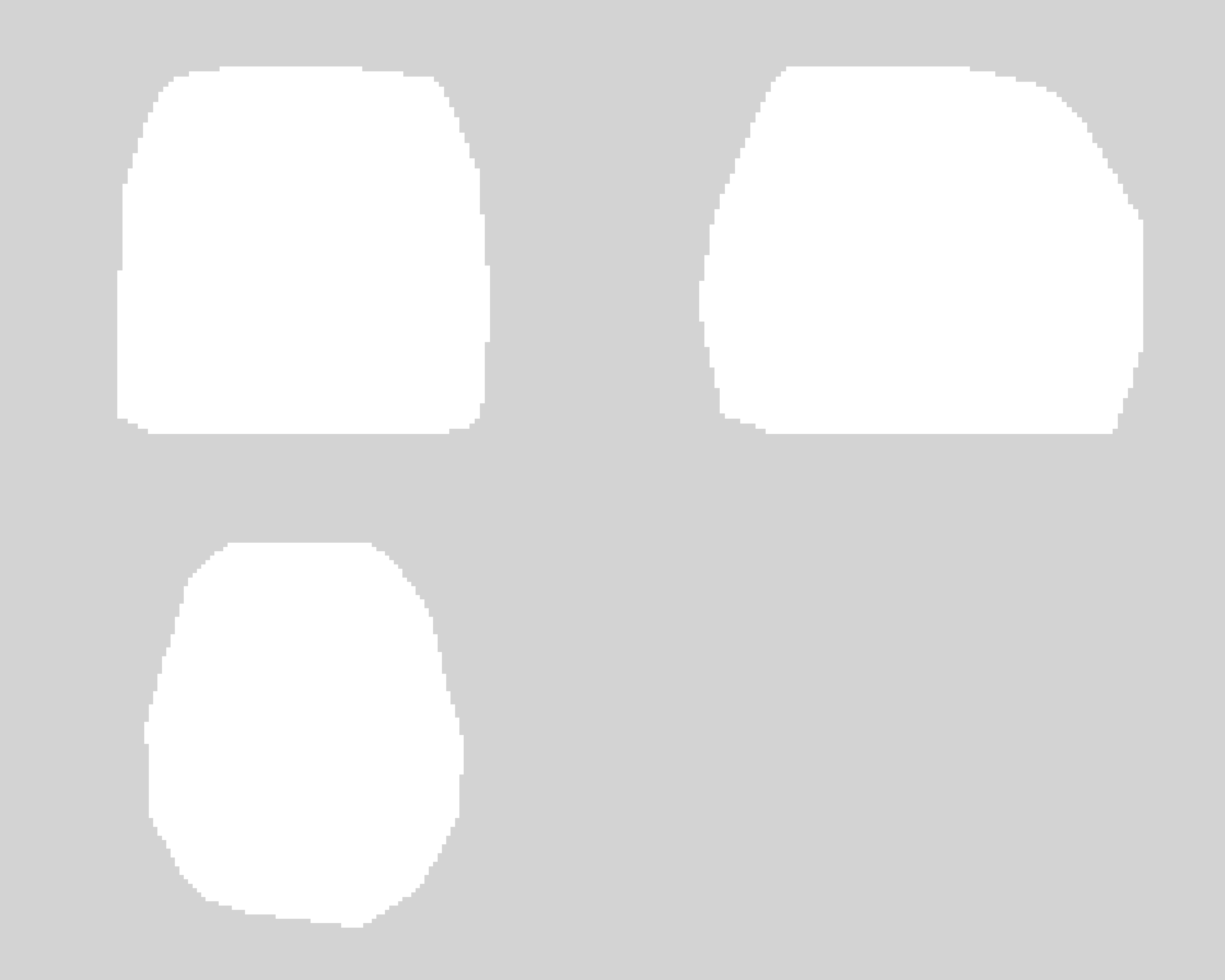} &\includegraphics[width=.95in]{plots/app_e10/SIG_APOE2_ISR_e10.png} &\includegraphics[width=.95in]{plots/app_e10/SIG_MMSE_ISR_e10.png} &\includegraphics[width=.95in]{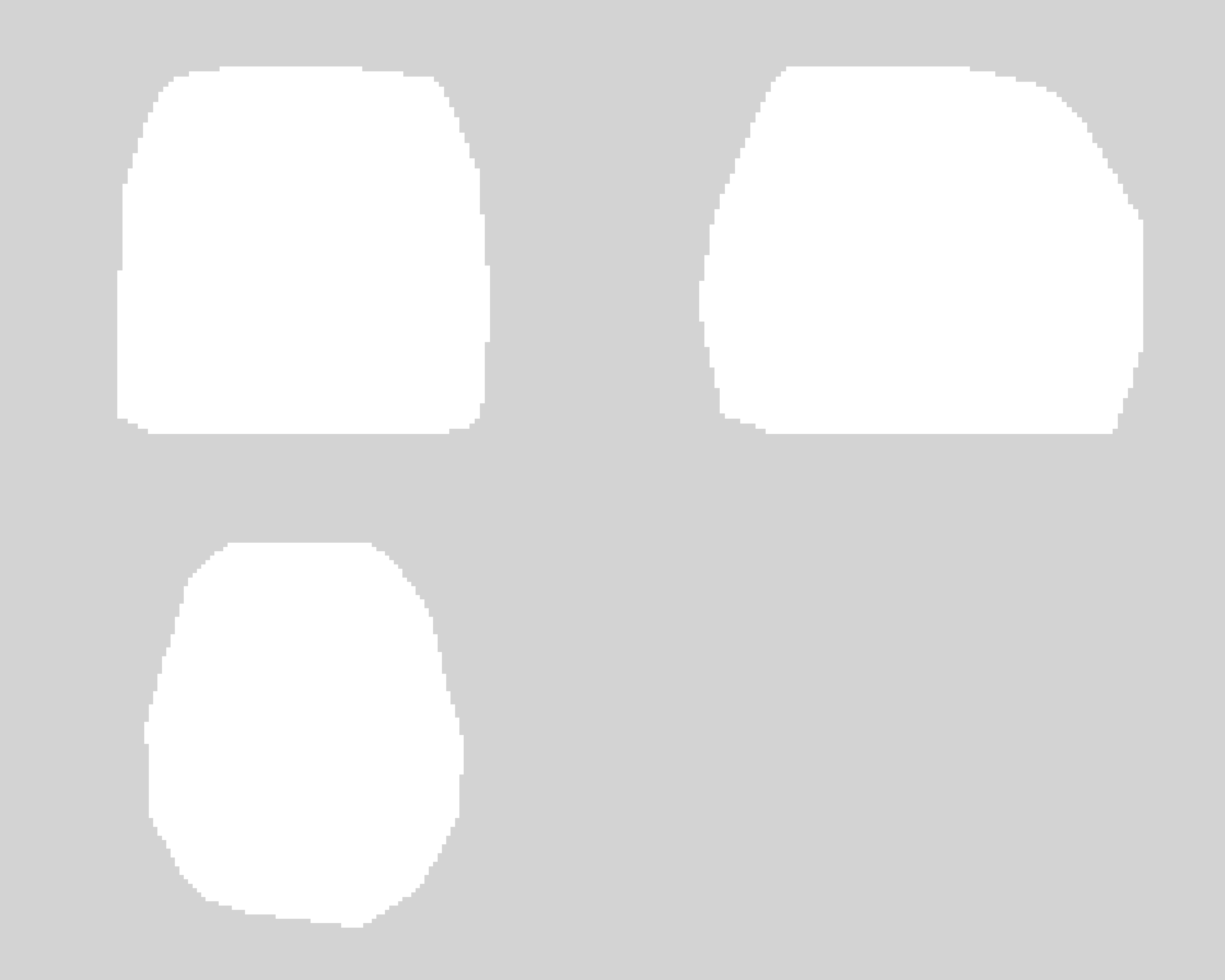} &\includegraphics[width=.95in]{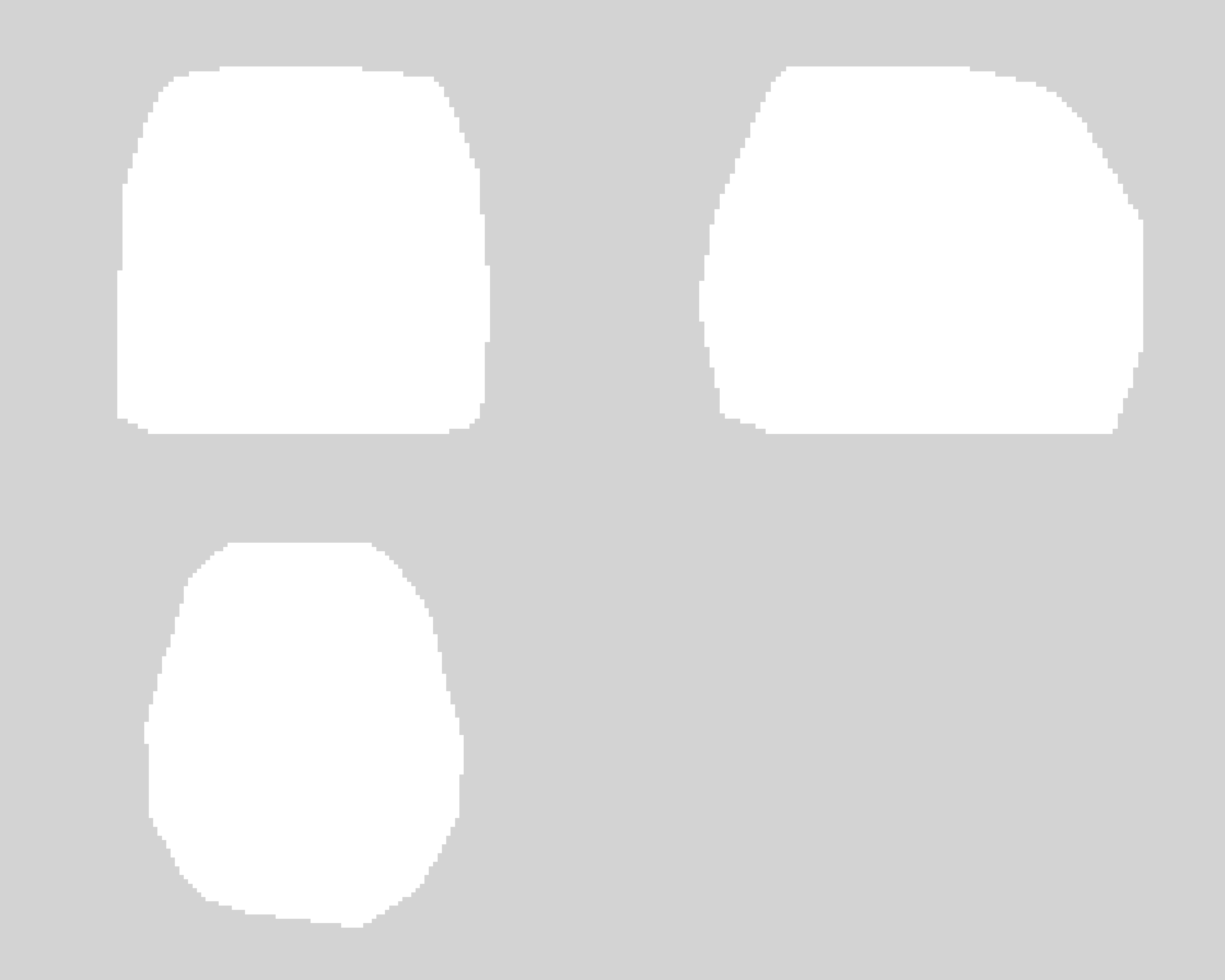} \\[-1pt]
SSFR &\includegraphics[width=.95in]{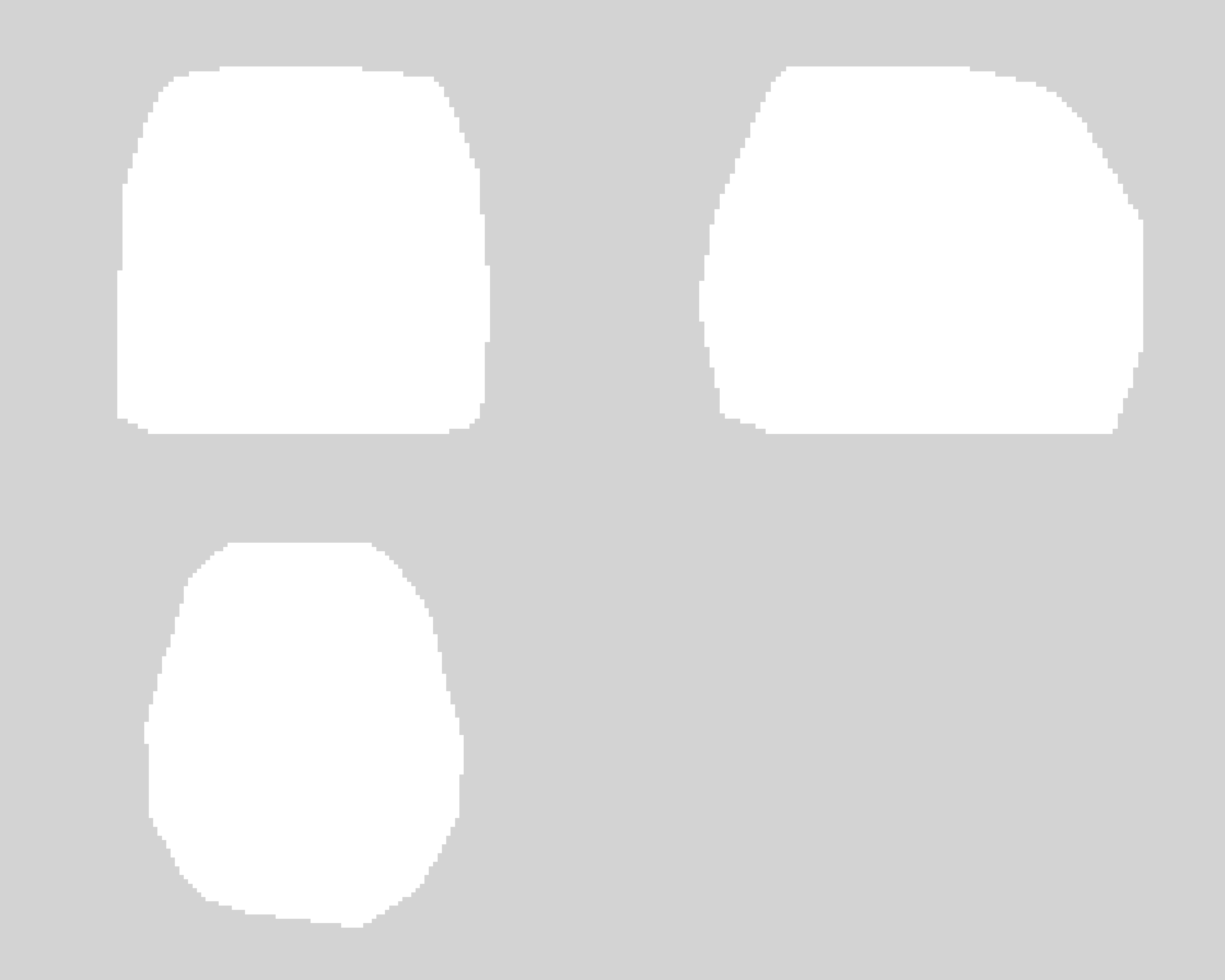} &\includegraphics[width=.95in]{plots/app_e10/SIG_APOE2_wSSFR_e10.png} &\includegraphics[width=.95in]{plots/app_e10/SIG_MMSE_wSSFR_e10.png} &\includegraphics[width=.95in]{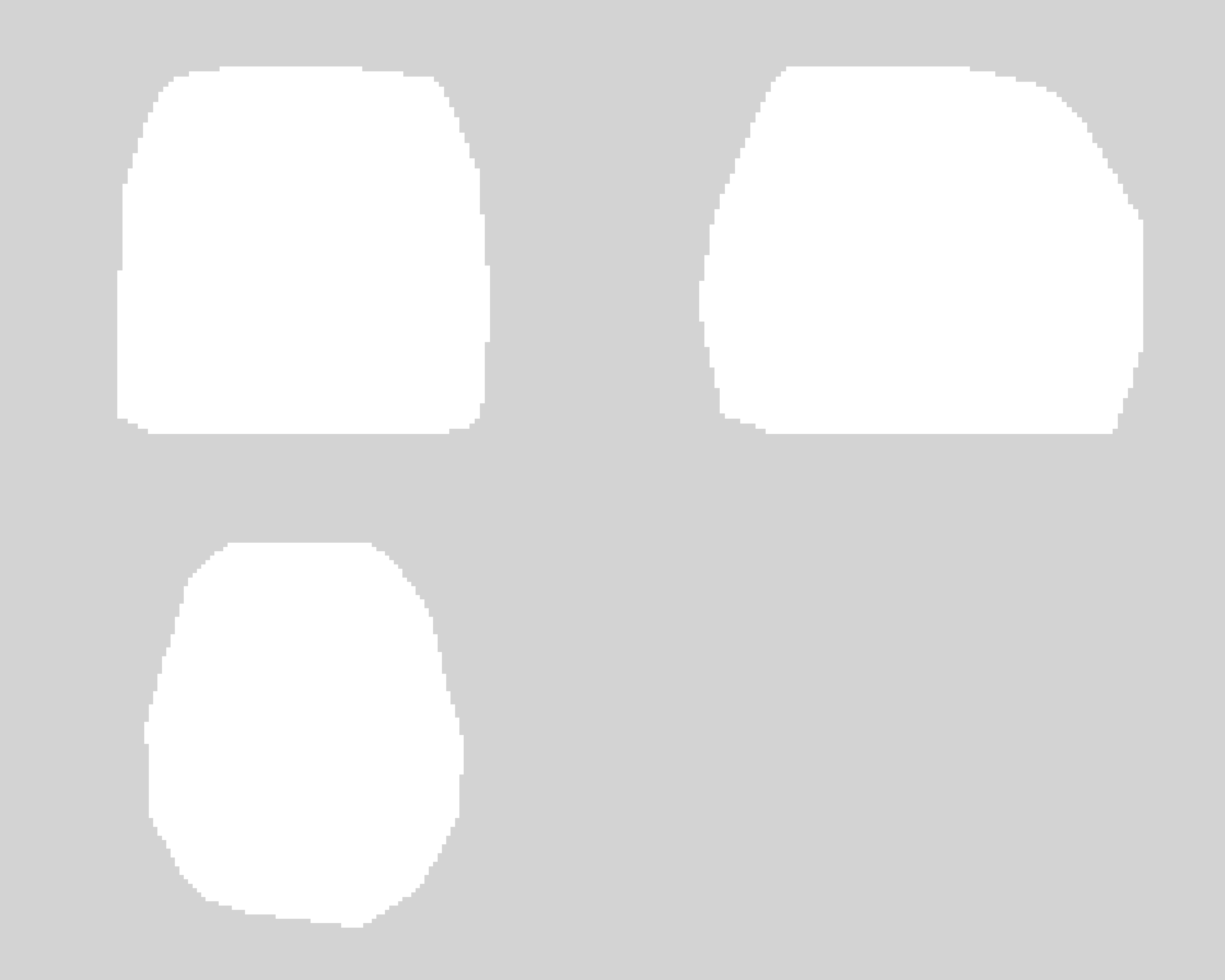} &\includegraphics[width=.95in]{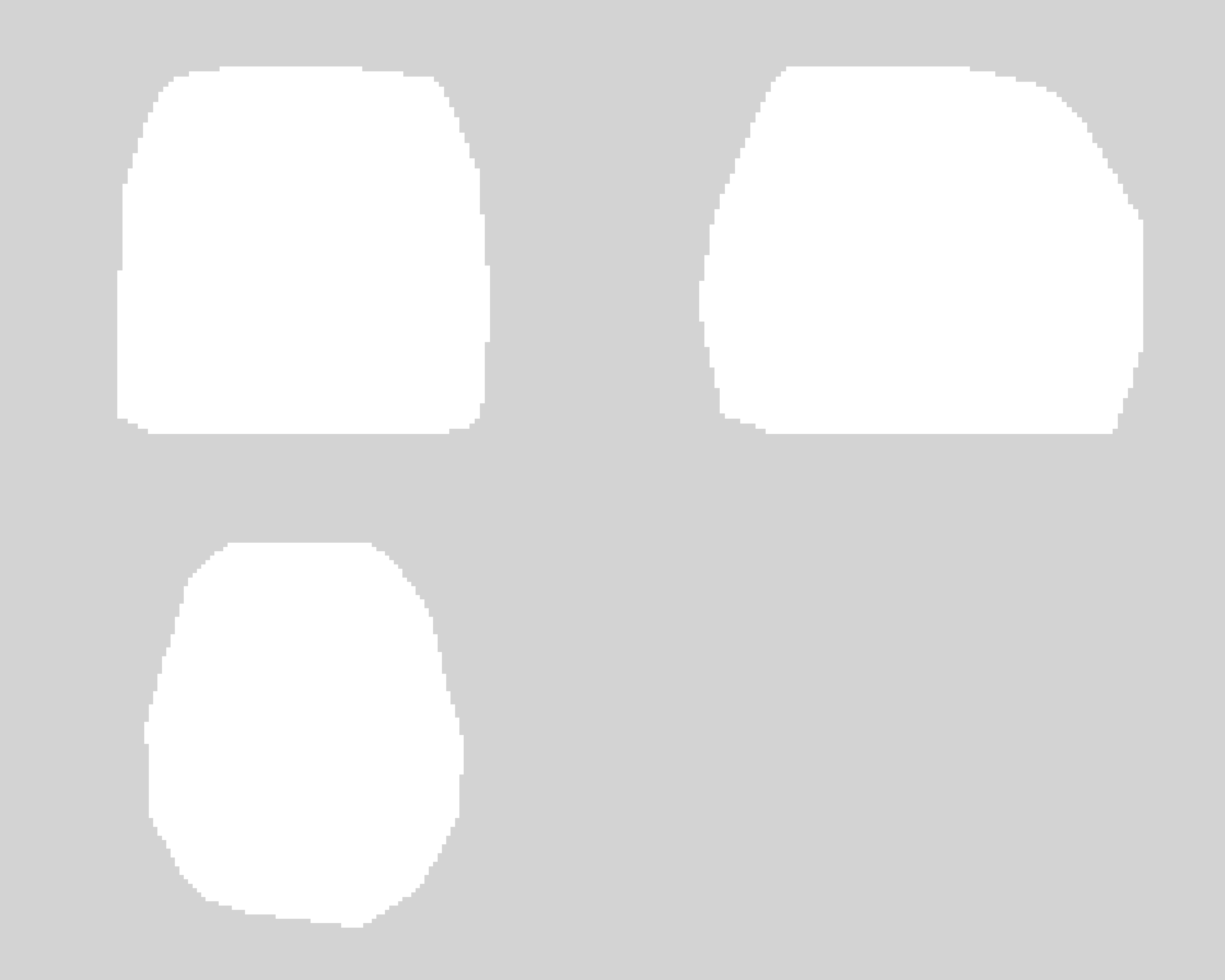} \\ [-1pt]
Difference &\includegraphics[width=.95in]{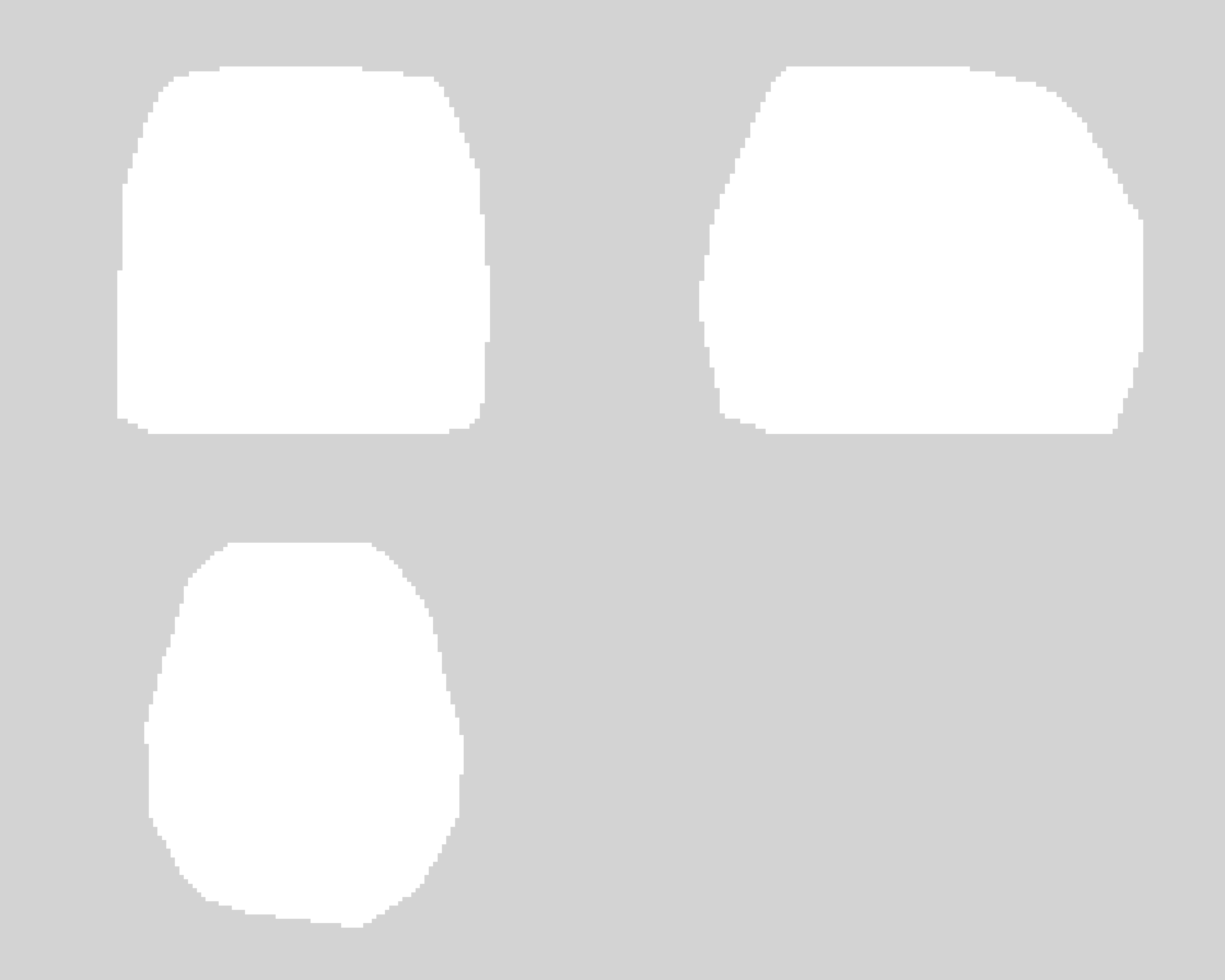} &\includegraphics[width=.95in]{plots/app_e10/COVDIFF_APOE2_e10.png} &\includegraphics[width=.95in]{plots/app_e10/COVDIFF_MMSE_e10.png} &\includegraphics[width=.95in]{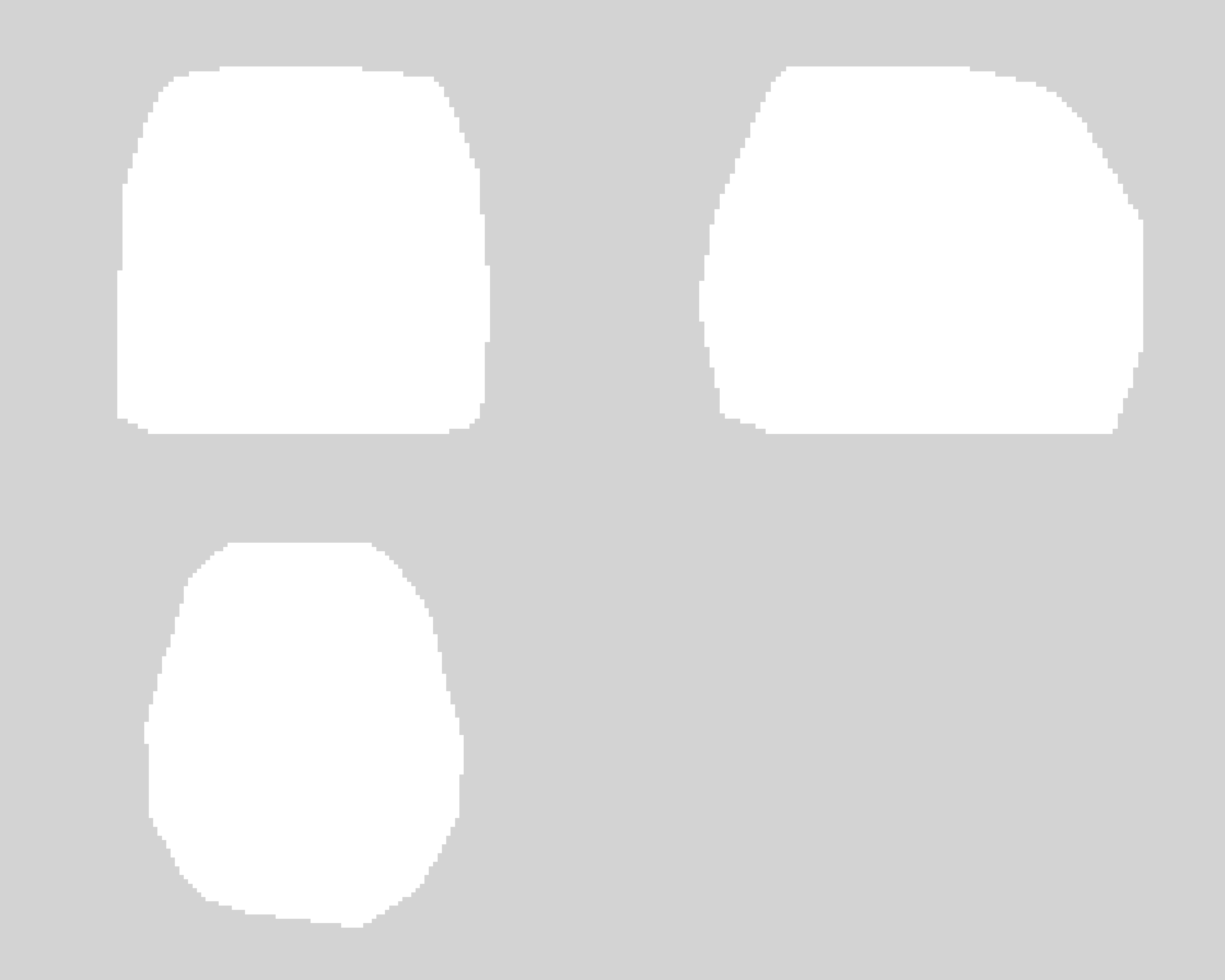} &\includegraphics[width=.95in]{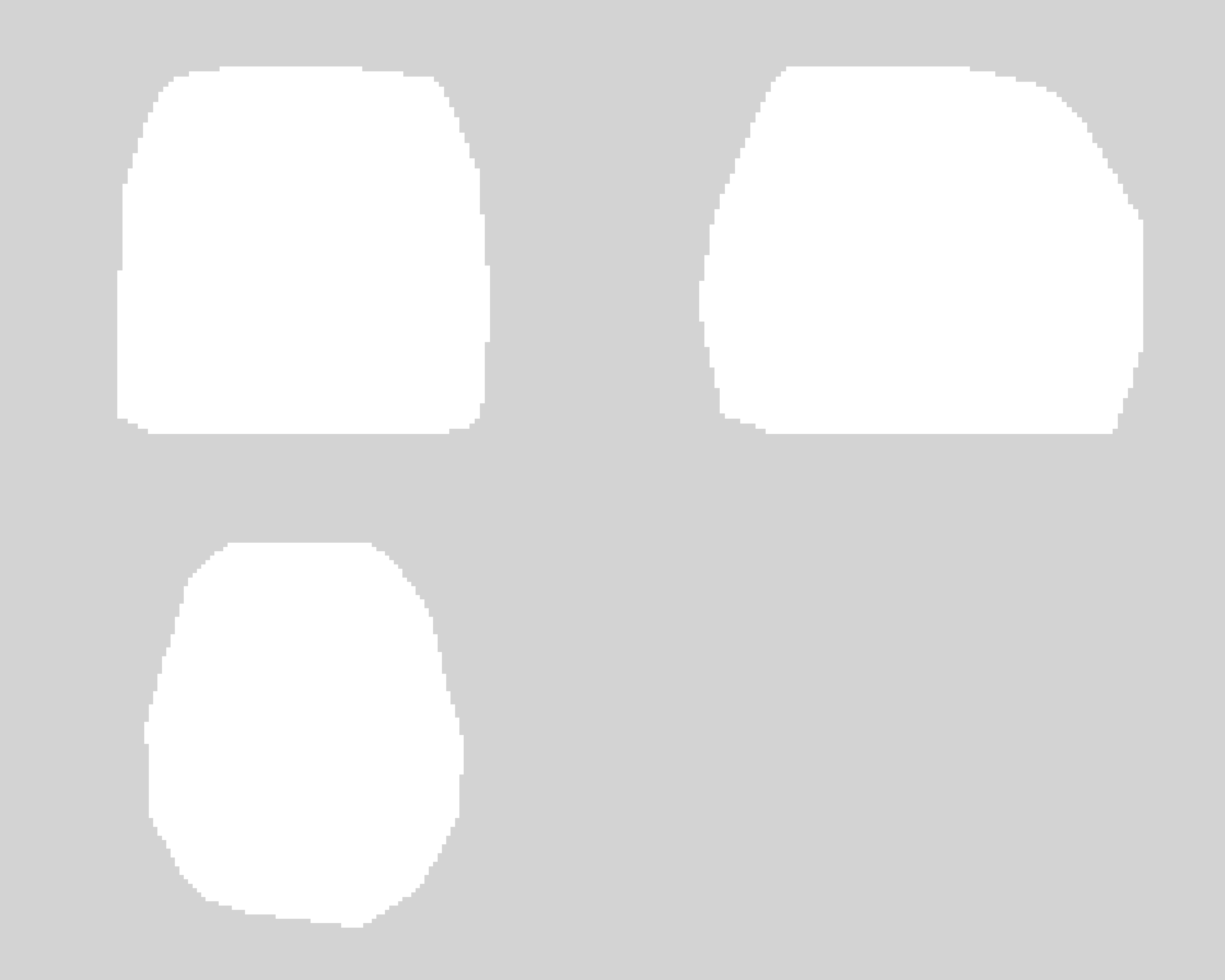} \\[-4pt]
\end{tabular}}
\caption{Significant regions identified by ISR and SSFR in sagittal, coronal, and axial views (left, right, and bottom), with darker blue/red indicating significantly negative/positive values for the coefficient functions. Difference row: dark grey, significant under both estimators; red, SSFR only; blue, ISR only.}
\label{FIG:supp-app-sigregions}
\end{figure}

Figure \ref{FIG:supp-app-sigregions} compares the simultaneously significant regions identified by ISR and SSFR for all nine coefficient functions and shows their voxelwise differences. The main paper displays the significance maps for five selected coefficients. At the 95\% level, neither estimator identifies any significant voxels for Education, APOE$_1$, or MCI. For AD, ISR identifies 94 significant voxels and SSFR identifies 164. For APOE$_2$, ISR identifies 351 significant voxels and SSFR identifies 15.

Overall, the additional figures are consistent with the conclusions reported in the main paper. Whenever surrogate information is informative, SSFR produces coefficient estimates that are comparable to those obtained by ISR while yielding more stable simultaneous inference. The remaining coefficient functions exhibit little evidence of association with the imaging outcome, regardless of the estimation procedure.

\renewcommand{\thesubsection}{C.\arabic{subsection}}
\renewcommand{\thesubsubsection}{C.\arabic{subsection}.\arabic{subsubsection}}
\renewcommand{\thetheorem}{{\sc C.\arabic{theorem}}}
\renewcommand{\thefigure}{C.\arabic{figure}}
\renewcommand{\thetable}{C.\arabic{table}}
\renewcommand{\theequation}{C.\arabic{equation}}

\renewcommand{\thealgorithm}{C.\arabic{algorithm}}
\setcounter{equation}{0}
\setcounter{theorem}{0}
\setcounter{figure}{0}
\setcounter{table}{0}
\setcounter{subsection}{0}
\setcounter{subsubsection}{0}

\section{Theoretical Results and Technical Proofs}
\label{SEC:proof}

This section collects the technical assumptions and the theoretical results for both stages of the proposed SSFR framework, along with their proofs. Section \ref{ssec:assumptions} states the assumptions. Section \ref{ssec:first-stage-theory} establishes the asymptotic properties of the first-stage estimator $\widehat{\beta}_{\ell}$ under the no-missing-data setting, and Sections \ref{sec:B-1}--\ref{sec:B-3} provide basic properties of trivariate splines together with additional results for the penalized trivariate spline estimators. Section \ref{ssec:ssfr-rates} establishes the convergence rate of the second-stage SSFR estimator (Theorem 4.2 of the main paper), and Section \ref{ssec:ssfr_efficiency} relates the resulting efficiency gain to residual variation explained by the surrogate.

In the following, we use $c$, $C$, $c_1$, $c_2$, $C_1$, $C_2$, etc., as generic constants, which may differ even in the same line. For any sequences $a_n$ and $b_n$, we write $a_n \asymp b_n$ if there exist two positive constants $c_1, c_2$ such that $c_1 |a_n| \le |b_n| \le c_2 |a_n|$, for all $n \ge 1$. For a real-valued vector $\bs{a}$, denote by $\|\bs{a}\|$ its Euclidean norm. For a matrix $\mathbf{A} = (a_{ij})$, denote $\|\mathbf{A}\|_{\infty} = \max_{i,j} |a_{ij}|$. For any real symmetric matrix $\mathbf{A}$, let $\lambda_{\min}(\mathbf{A})$ and $\lambda_{\max}(\mathbf{A})$ be the smallest and largest eigenvalues of $\mathbf{A}$. For the imaging response, it is assumed that the observed voxels are evenly distributed in the domain of interest $\Omega$, and the volume of each voxel is approximated by $1/N$.

For a vector-valued function $\bs{g}=(g_0, \ldots, g_p)^{\top}$ defined on domain $\Omega$, denote $\|\bs{g}\|_{L _2(\Omega)} = \left\{\sum_{\ell =0}^p \|g_{\ell}\|_{L _2(\Omega)}^2\right\}^{1/2}$ and $\|\bs{g}\|_{\infty} = \max_{0 \leq \ell \leq p} \|g_{\ell}\|_{\infty}$, where $\|g_{\ell}\|_{L_2(\Omega)}$ and $\|g_{\ell}\|_{\infty}$ are the norm $L_2$ and the supremum norm of $g_{\ell}$. Let $\bs{\alpha} = (\alpha_1, \alpha_2, \alpha_3)$ be a vector of nonnegative integers. Denote $D^{\bs{\alpha}} = D^{\alpha_1}_{z_1} D^{\alpha_2}_{z_2} D^{\alpha_3}_{z_3}$ and $|\bs{\alpha}| = \alpha_1 + \alpha_2 + \alpha_3$. In addition, denote $\|\bs{g}\|_{\upsilon, \infty} = \max_{0 \leq \ell \leq p}|g_{\ell}|_{\upsilon, \infty}$, where $|g_{\ell}|_{\upsilon, \infty} = \max_{|\bs{\alpha}| = \upsilon} \|D^{\bs{\alpha}}g_{\ell}(\bs{z})\|_{\infty}$. Let $\mathcal{W}^{d,\infty}(\Omega) = \left\{g:|g|_{k,\infty,\Omega} < \infty, 0 \le k \le d\right\}$ be the standard Sobolev space.

For $\bs{g}^{(1)}(\bs{z}) = (g_0^{(1)}(\bs{z}), \ldots, g_p^{(1)}(\bs{z}))^{\top}$, $\bs{g}^{(2)}(\bs{z})=(g_0^{(2)}(\bs{z}), \ldots, g_p^{(2)}(\bs{z}))^{\top}$, and any arbitrary domain $\Omega$, which may be the whole imaging domain or a subregion of it, define the empirical inner product as
\begin{align}
\label{DEF:DDC-empirical_product}
\langle\bs{g}^{(1)}, \bs{g}^{(2)}\rangle_{n} & = \frac{1}{nN} \sum_{\ell, \ell^{\prime} = 0}^p \sum_{i=1}^n \sum_{\bs{z}_j \in \Omega}X_{i\ell}X_{i\ell^{\prime}} g_{\ell}^{(1)}(\bs{z}_j) g_{\ell^{\prime}}^{(2)}(\bs{z}_j) \\
& = \frac{1}{n} \sum_{\ell, \ell^{\prime} = 0}^p \sum_{i=1}^n X_{i\ell}X_{i\ell^{\prime}} \times \frac{1}{N}\sum_{\bs{z}_j \in \Omega} g_{\ell}^{(1)}(\bs{z}_j) g_{\ell^{\prime}}^{(2)}(\bs{z}_j),  \notag
\end{align}
where $N^{-1}\sum_{\bs{z}_j \in \Omega} g_{\ell}^{(1)}(\bs{z}_j) g_{\ell^{\prime}}^{(2)}(\bs{z}_j)$ is an approximation of $\int_{\Omega} g_{\ell}^{(1)}(\bs{z}) g_{\ell^{\prime}}^{(2)}(\bs{z}) d\bs{z}$. The theoretical inner product is defined as
\begin{equation}
\langle\bs{g}^{(1)}, \bs{g}^{(2)}\rangle_{\Omega} = \sum_{\ell, \ell^{\prime} = 0}^p {\rm E}(X_{\ell}X_{\ell^{\prime}}) \int_{\Omega} g_{\ell}^{(1)}(\bs{z}) g_{\ell^{\prime}}^{(2)}(\bs{z}) d\bs{z}.
\label{DEF:DDC-theoretical_product}
\end{equation}
Furthermore, let $\|\cdot\|_{\mathcal{E}}$ be the norm introduced by the inner product $\langle \cdot, \cdot\rangle_{\mathcal{E}}$, where, for $\bs{g}^{(1)}(\bs{z})$ and $\bs{g}^{(2)}(\bs{z})$,
\begin{equation*}
\left\langle\bs{g}^{(1)}, \bs{g}^{(2)}\right\rangle_{\mathcal{E},\Omega} = \sum_{\ell = 0}^p \int_{\Omega} \left\{\sum_{i = 1}^3 \left(\nabla_{z_i}^2 g_{\ell}^{(1)}\right)^2\right\}^{1/2} \left\{\sum_{i = 1}^3 \left(\nabla_{z_i}^2 g_{\ell}^{(2)}\right)^2\right\}^{1/2}dz_1 dz_2 dz_3.
\label{DEF:DDC-energy_product}
\end{equation*}
For notational simplicity, we write $\langle\bs{g}^{(1)}, \bs{g}^{(2)}\rangle_{n} = \langle\bs{g}^{(1)}, \bs{g}^{(2)}\rangle_{n, \Omega}$, $\langle\bs{g}^{(1)}, \bs{g}^{(2)}\rangle = \langle\bs{g}^{(1)}, \bs{g}^{(2)}\rangle_{\Omega}$, and $\left\langle\bs{g}^{(1)}, \bs{g}^{(2)}\right\rangle_{\mathcal{E}} = \left\langle\bs{g}^{(1)}, \bs{g}^{(2)}\right\rangle_{\mathcal{E},\Omega}$. Note that the triangulation for different coefficient functions can be different from each other. In the following proof, we consider a common triangulation for all explanatory variables: $\mathbf{B}_0(\bs{z}) = \mathbf{B}_1(\bs{z}) = \cdots = \mathbf{B}_p(\bs{z}) = \mathbf{B}(\bs{z})$, and $\beta_{\ell}(\bs{z}) = \mathbf{B}^{\top}(\bs{z}) \bs{\gamma}_{\ell}$. Without loss of generality we assume $\lam = \lam_0 = \lam_1 = \ldots = \lam_p$ and $\lam_k = \lam_{0k} = \lam_{1k} = \ldots = \lam_{pk}$.

\subsection{Technical Assumptions}
\label{ssec:assumptions}

Throughout this section, $\triangle$ is a triangulation of $\Omega$ as defined in Section 3.2 of the main paper. For a nondegenerate tetrahedron $T\in\triangle$ with nonzero volume, let $|T|$ be its longest edge length and $\varrho_T$ the radius of its largest inscribed ball, and define the shape parameter $\pi_T=|T|/\varrho_T$, where a small $\pi_T$ indicates a well-shaped tetrahedron. Let $|\triangle|=\max_{T\in\triangle}|T|$ denote the mesh size of $\triangle$, that is, the length of its longest edge. Let $N$ be the number of observed voxel locations in $\Omega$, and let $\bs\Sigma_X=\E(\mathbf X\mathbf X^\top)$. Throughout, $d \ge 1$ denotes the spline degree and $r\ge 0$ the imposed smoothness across adjacent tetrahedra. We use $\mathcal W^{s,\infty}(\Omega)$ for the standard Sobolev space of functions on $\Omega$ with bounded weak derivatives through order $s$. The theory below is developed under the following eight conditions.

\begin{assumpA}[Triangulation and spatial resolution]
	\label{as:triangulation}
	The triangulation $\triangle$ is $\pi$-quasi-uniform: there exists a constant $\pi>0$, not depending on $\triangle$, such that $(\min_{T\in\triangle}\varrho_T)^{-1}|\triangle|\le \pi$, where $\varrho_T$ is the radius of the largest inscribed ball of $T$. As $n,N\to\infty$, $|\triangle|\to0$ and $N^{1/3}|\triangle|\to\infty$. In addition, for some $0<\kappa<1$, $N^{-1}n^{3/(2d+2)+\kappa}\to0$.
\end{assumpA}

\begin{assumpA}[Smoothness of the coefficient functions]
	\label{as:smoothness}
	For each $\ell=0,\ldots,p$, the first-stage surrogate coefficient function satisfies $\beta_{0,\ell}\in\mathcal W^{d+1,\infty}(\Omega)$, and the target coefficient function satisfies $\alpha_{0,\ell}\in\mathcal W^{d+1,\infty}(\Omega)$. The second-stage co-kriging kernel satisfies $U\in\mathcal W^{d+1,\infty}(\Omega\times\Omega)$ in the product sense, with bounded weak derivatives up to order $d+1$ in each spatial argument. Consequently, $\bs\nu_0(\bs z)=\bs\alpha_0(\bs z)-\int_\Omega U(\bs z,\bs z')\bs\beta_0(\bs z')d\bs z'$ belongs componentwise to $\mathcal W^{d+1,\infty}(\Omega)$.
\end{assumpA}

\begin{assumpA}[Error-process regularity]
	\label{as:errors}
	The error processes $\varepsilon_{O,i}$ and $\varepsilon_{S,i}$ are mean-zero and independent across subjects and of $\mathbf X_i$. Each process admits a Karhunen--Lo\`eve representation with continuously differentiable covariance/eigenfunction components; the corresponding scores have finite $(4+\delta)$th moments for some $\delta>0$. Any independent voxel-level error component is mean-zero, has finite $(4+\delta)$th moment, and has variance bounded above and away from zero uniformly over $\Omega$. The diagonal covariance functions of the smooth error components are bounded above and away from zero, and their covariance kernels and eigenfunctions are continuous on $\Omega\times\Omega$ and $\Omega$, respectively.
\end{assumpA}

\begin{assumpA}[Design regularity]
	\label{as:design}
	The covariate vectors $\mathbf X_i$, $i=1,\ldots,n$, are i.i.d., and for each $\ell=0,\ldots,p$ there exists $C_\ell<\infty$ such that $\E|X_{i\ell}|^8\le C_\ell$. The eigenvalues of $\bs\Sigma_X=\E(\mathbf X\mathbf X^\top)$ are bounded away from zero and infinity.
\end{assumpA}

\begin{assumpA}[Surrogate relevance]
	\label{as:relevance}
	Let $\varepsilon_{S,i}(\bs z)=\widetilde Y_i(\bs z)-\mathbf X_i^\top\bs\beta_0(\bs z)$ denote the surrogate deviation. Then $\varepsilon_{S,i}$ is independent of $\mathbf X_i$ and satisfies $\E\{\varepsilon_{S,i}(\bs z)\mid\mathbf X_i\}=0$ for all $\bs z\in\Omega$. The covariance operator $\Gamma_S$ of $\varepsilon_{S,i}$, restricted to the trivariate spline space $\mathcal S_d^r(\triangle)$, has eigenvalues bounded away from zero and infinity uniformly in $\triangle$. Equivalently, the surrogate-derived predictor is asymptotically non-collinear with the covariates on the spline space.
\end{assumpA}

\begin{assumpA}[Missingness and inverse-probability weights]
	\label{as:mar}
	The observation indicator $R_i=\mathbf 1(i\in\mathcal I_O)$ satisfies $R_i\perp Y_i(\cdot)\mid\mathbf X_i$ (MAR), with $\pi(\mathbf X_i)=\Pr(R_i=1\mid\mathbf X_i)\ge c_\pi>0$. The weights satisfy $w_i=\widehat\pi(\mathbf X_i)^{-1}$ and are uniformly consistent for $\pi(\mathbf X_i)^{-1}$, with $0<c_1\le w_i\le c_2<\infty$. Moreover, $n_O/n\to\bar\pi\in(0,1]$.
\end{assumpA}

\begin{assumpA}[Asymptotic growth and penalty conditions]
	\label{as:rates}
	As $n,N\to\infty$, assume that $n^{1/2}|\triangle|^{d+1}\to0$. The roughness penalties $\lambda_\ell$ satisfy $\lambda_\ell\to0$ and the following conditions
	\[
	\{\min_{0\le\ell\le p}\lambda_\ell\}^{-1/2}N^{-1}|\triangle|^{-3}\to0,
	\qquad
	\lambda_\ell n^{-1/2}N^{-1}|\triangle|^{-3}\to0,
	\]
	for every $\ell=0,\ldots,p$. In addition, the penalty contribution is asymptotically negligible at the approximation scale, so that $\lambda_{\max}/(nN|\triangle|^5)\to0$, where $\lambda_{\max}=\max_{0\le\ell\le p}\lambda_\ell$.
	Finally, the ridge parameter $\tau_U\ge0$ of the second-stage criterion (6) in the main paper satisfies
	\[
	r_\tau:=\frac{\tau_U}{N|\triangle|^{6}}=O\big(n^{-1/2}+|\triangle|^{d+1}\big).
	\]
\end{assumpA}

\begin{assumpA}[Second-stage residual regularity]
	\label{as:stage2}
	The co-kriging residual admits the decomposition $\zeta_i(\bs z)=\zeta_i^{\circ}(\bs z)+e_i(\bs z)$, where $\zeta_i^{\circ}$ is a mean-zero smooth process with covariance kernel in $\mathcal C^1(\Omega\times\Omega)$ and bounded diagonal, with Karhunen--Lo\`eve scores satisfying the moment condition in Assumption \ref{as:errors}. The voxel-level terms $e_i(\bs z_j)$ are independent across $i$ and $j$, mean-zero, have variances bounded above and away from zero, and possess finite $(4+\delta)$th moments for some $\delta>0$. The processes $\zeta_i$ are independent across subjects and independent of $(\mathbf X_i,\widetilde Y_i)$; in particular, $\E\{\zeta_i\mid\mathbf X_i,\widetilde Y_i\}=0$.
\end{assumpA}

The eight conditions may be read as follows. Assumption \ref{as:triangulation} asks that the triangulation be quasi-uniform and refine as $n$ and $N$ grow, with the voxel grid fine relative to the mesh, so that the spline space is rich enough to approximate the coefficient functions while remaining identifiable from the observed voxels. Assumption \ref{as:smoothness} gives the surrogate and target coefficient functions, and the co-kriging kernel $U$, bounded derivatives through order $d+1$, which is what delivers the $|\triangle|^{d+1}$ approximation term in Theorems 4.1 and 4.2. Assumptions \ref{as:errors} and \ref{as:design} are the usual moment and design conditions: mean-zero error processes with Karhunen--Lo\`eve scores having finite $(4+\delta)$th moments, and i.i.d.\ covariates with bounded eighth moments and a well-conditioned second-moment matrix $\bs\Sigma_X$. Assumption \ref{as:relevance} is the condition specific to the surrogate construction, and is discussed further below. Assumption \ref{as:mar} places the second stage in the standard MAR framework with inverse-probability weights bounded above and away from zero, so that $\sum_{i\in\mathcal I_O}w_i\asymp n_O$ and the rate in Theorem 4.2 may equivalently be stated in terms of either quantity. Assumption \ref{as:rates} makes the roughness penalties asymptotically negligible relative to the approximation error, so that the penalized and unpenalized estimators share a rate, and restricts the ridge parameter $\tau_U$ of (6) so that the shrinkage of the tensor-product coefficient vector $\bs b$ induces a bias of no larger order than the spline-approximation and stochastic errors (the quantity $r_\tau$ is the ratio of the ridge penalty to the data-fit term at the truth; see Remark \ref{rem:ridge-scale}). The ridge plays no role in identification, which rests on Assumption \ref{as:relevance}; it stabilizes the $|\mathcal M|^2$-dimensional $U$-block in finite samples, and Assumption \ref{as:stage2} asks of the second-stage co-kriging residual exactly the regularity that Assumption \ref{as:errors} asks of the first-stage errors.

Assumptions \ref{as:triangulation}--\ref{as:design} and \ref{as:rates} provide the regularity and rate conditions needed for the first-stage estimator, while Assumptions \ref{as:relevance}, \ref{as:mar}, and \ref{as:stage2} additionally govern the second-stage estimator under MAR. The surrogate exogeneity and relevance condition in Assumption \ref{as:relevance} is the key identification condition for the second-stage kernel $U$; it rules out the degenerate case in which the surrogate contributes no subject-specific stochastic information beyond the covariates.

\subsection{Asymptotic Properties of the First-Stage Estimator}
\label{ssec:first-stage-theory}

We investigate the asymptotic properties of the first-stage coefficient estimators $\widehat{\beta}_{\ell}(\cdot)$.

The following theorem provides the $L_2$ convergence rate of $\widehat{\beta}_{\ell}(\cdot)$, $\ell = 0, 1, \ldots, p$. 

\vspace{-8pt}
\begin{theorem}
\label{THM:beta-convergence}
Under Assumptions \ref{as:triangulation}--\ref{as:design} and \ref{as:rates}, if $N^{1/3}|\triangle| \rightarrow \infty$ as $N \rightarrow \infty$, then the first stage estimator $\widehat{\beta}_{\ell}(\cdot)$ is consistent. Moreover, for any $\ell = 0, 1, \ldots, p$, we have $\Vert \widehat{\beta}_{\ell} - \beta_{0,\ell}\Vert_{L^2(\Omega)} = O_P\{ \lam_{\max}(nN|\triangle|^3)^{-1}\|\bs{\beta}_0\|_{2,\infty} + \{1 + \lam_{\max}(nN|\triangle|^5)^{-1}\}|\triangle|^{d+1}\|\bs{\beta}_0\|_{d+1,\infty} + n^{-1/2}\}$, where $\lam_{\max} = \max_{\ell \in\{0, 1, \ldots, p\}} \lam_{\ell}$.
\end{theorem} 

\vspace{-16pt}
\begin{remark} \label{rem:beta-rate}
According to Theorem \ref{THM:beta-convergence}, three factors govern the convergence rate: spline approximation bias, estimation bias induced by the roughness penalty, and estimation variance. When $\lam_{\max} \rightarrow 0$, the convergence rate is primarily determined by $|\triangle|^{d+1} + n^{-1/2}$. Furthermore, the condition $N^{1/3}|\triangle| \to \infty$ imposes a lower bound on $|\triangle|$, indicating that a finer spatial resolution (i.e., more voxels) helps reduce the spline approximation bias.
\end{remark} 

\subsection{Properties of Trivariate Splines over Triangulations} 
\label{sec:B-1}

Let $T = \langle\bs{z}_1, \bs{z}_2, \bs{z}_3, \bs{z}_4\rangle$ be a nondegenerate tetrahedron with vertices $\bs{z}_1$, $\bs{z}_2$, $\bs{z}_3$, and $\bs{z}_4$. For any point $\bs{z} \in \mathbb{R}^3$, there exists a unique representation in the form: $\bs{z} = b_1\bs{z}_1 + b_2\bs{z}_2 + b_3\bs{z}_3 + b_4\bs{z}_4$, where $b_1 + b_2 + b_3 + b_4 = 1$. The coefficients $b_1$, $b_2$, $b_3$, and $b_4$ are called the \textit{barycentric coordinates} of the point $\bs{z}$ relative to the tetrahedron $T$. The Bernstein polynomials of degree $d$ relative to the tetrahedron $T$ are defined as: $B_{ijkl}^{T,d}(\bs{z}) = \frac{d!}{i!j!k!l!} b_1^i b_2^j b_3^k b_4^l$. For any $T \in \triangle$, we can express the polynomial piece of the spline $g$ restricted on $T \in \triangle$ as: $g|_T = \sum_{i+j+k+l=d} \gamma_{ijkl}^T B_{ijkl}^{T,d}$, where $\bs{\gamma}_T = \{\gamma_{ijkl}^T, i+j+k+l=d\}$ are called \textit{B-coefficients} of $g$.
 
For notational simplicity, let $\{B_m, m \in \mathcal{M}\}$ represent the set of degree-$d$ trivariate Bernstein basis polynomials for $\mathcal{S}_d^r(\triangle)$, with $\mathcal{M}$ denoting the index set of all Bernstein basis polynomials. Then, for any function $g \in \mathcal{S}_d^r(\triangle)$, we can express it using the following basis expansion: $g(\bs{z}) = \sum_{m \in \mathcal{M}} B_m(\bs{z})\gamma_m = \mathbf{B}(\bs{z})^{\top}\bs{\gamma}$, where $\bs{\gamma} = (\gamma_m, m \in \mathcal{M})^{\top}$ is the spline coefficient vector. 

The following lemmas summarize some key properties used in the proofs.

\begin{lemma}
\label{LEM:DDC-norm}
Let $\{ B_{m}\}_{m \in \mathcal{M}}$ be the set of degree $d$ trivariate Bernstein basis polynomials for the trivariate spline space $\mathcal S_d^r(\triangle_{\Omega})$, which is defined over a $\pi$-quasi-uniform triangulation $\Omega$ for each subregion $\Omega$. Then there exist positive constants $c$, $C$ depending on the smoothness $r, d, $ and the shape parameter $\pi$ such that $c|\triangle|^{3}\sum_{m \in \mathcal{M}}\gamma_{m}^{2} \leq \|\sum_{m \in \mathcal{M}}\gamma_m B_m \|^{2}_{L_2(\Omega)} \leq C|\triangle|^{3}\sum_{m \in \mathcal{M}}\gamma_m^{2}$.
\end{lemma}
\begin{proof}
Proof of Lemma \ref{LEM:DDC-norm} follows from Theorem 17.18 \citep{Lai:Schumaker:07}.
\end{proof} \vspace{-.2in}

\begin{lemma}
\textbf{(Theorem 3.5.2 in \cite{Lai:89} and Theorem 15.32 in \cite{Lai:Schumaker:07})} Under Assumption A1, for all $g(\cdot) \in \mathcal{W}^{d+1, \infty}(\Omega)$ with $1 \leq q \leq \infty$, $r \geq 0$ and $d \geq 6r + 3$, there exists a spline $s_g \in \mathcal{S}_d^r(\triangle)$ such that $\|D^{\bs{\alpha}}(g - s_g)\|_{L_q(\Omega)} \leq C |\triangle|^{d + 1 - |\bs{\alpha}|} |g|_{d+1, q, \Omega}$, for all $0 \leq |\bs{\alpha}| \leq d$, where $C$ is a positive constant independent of $g$ and $|\triangle|$ but depends on the geometry of $\triangle$.
\label{LEM:appord}
\end{lemma}

Lemma \ref{LEM:appord} shows that $\mathcal{S}_d^0(\triangle)$ and $\mathcal{S}_9^1(\triangle)$ can achieve the optimal convergence rate.

\begin{lemma}
\label{LEM:norm}
Let $\bs{g}(\bs{z}) = (g_0(\bs{z}), \ldots, g_p(\bs{z}))^{\top}$, where $g_{\ell}(\bs{z}) = \sum_{m \in \mathcal{M}} \gamma_{\ell m} B_m(\bs{z})$. Then, under Assumption A4, $\|\bs{g}\|_{\Omega} \asymp \sum_{\ell = 0}^p \|g_{\ell}\|_{L_2(\Omega)}$.
\end{lemma}
\begin{proof}
By the definition in (\ref{DEF:DDC-theoretical_product}), we have \vspace{-.1in}
\[
\|\bs{g}\|_{\Omega}^2 = \sum_{\ell, \ell^{\prime} = 0}^p {\rm E}(X_{\ell}X_{\ell^{\prime}}) \int_{\Omega} g_{\ell}(\bs{z}) g_{\ell^{\prime}}(\bs{z}) d\bs{z} = \int_{\Omega} \bs{g}^{\top}(\bs{z}) \bs{\Sigma}_{X} \bs{g}(\bs{z}) d\bs{z}.
\]
By Assumption A4, we have $\|\bs{g}\|_{\Omega}^2 \asymp \int_{\Omega} \bs{g}^{\top}(\bs{z}) \bs{g}(\bs{z}) d\bs{z} \asymp \sum_{\ell=0}^{p} \|g_{\ell}\|^2_{L_2(\Omega)}$.
\end{proof}

\begin{lemma}
\label{LEM:integration}
Under Assumptions \ref{as:triangulation} and \ref{as:errors}, for any Bernstein basis polynomials $B_{m}(\bs{z}), ~m \in \mathcal{M}$, of degree $d\geq0$ and the subregions $\Omega$ and $\Omega'$, we have
\begin{align}
& \bigg\vert
  \frac{1}{N}\sum_{\bs{z}_{j} \in \Omega} B_m^s(\bs{z}_{j})
  - \int_{\Omega} B_m^s(\bs{z}) d\bs{z}
  \bigg\vert = O_P\left(N^{-1/3}|\triangle|^2\right),
  ~\mathrm{for~any}~ s\geq1 \label{DDC-lemmaA18_1} \\
& \bigg\vert
  \frac{1}{N}\sum_{\bs{z}_{j} \in \Omega} B_{m}(\bs{z}_{j})B_{m'}(\bs{z}_{j})
  - \int_{\Omega} B_{m}(\bs{z})B_{m'}(\bs{z})d\bs{z}
  \bigg\vert = O_P\left(N^{-1/3}|\triangle|^2\right),
  \label{DDC-lemmaA18_2} \\
& \bigg\vert
  \frac{1}{N^2} \sum_{\bs{z}_{j} \in \Omega,\, \bs{z}_{j'} \in \Omega'}
  G_{\eta}(\bs{z}_{j}, \bs{z}_{j'}) B_{m}(\bs{z}_{j})B_{m'}(\bs{z}_{j'})
  \notag \\
&\qquad {}- \int_{\Omega \times \Omega'}
  G_{\eta}(\bs{z}, \bs{z}')B_{m}(\bs{z})B_{m'}(\bs{z}')d\bs{z} d\bs{z}'
  \bigg\vert = O\left(N^{-1/3}|\triangle|^{5}\right)
  \label{DDC-lemmaA18_3} \\
& \bigg\vert
  \frac{1}{N} \sum_{\bs{z}_{j} \in \Omega} \sigma^2(\bs{z}_{j})B_{m}^2(\bs{z}_{j})
  - \int_{\Omega}\sigma^2(\bs{z})B_{m}^2(\bs{z})d\bs{z}
  \bigg\vert = O\left(N^{-1/3}|\triangle|^{2}\right).
  \label{DDC-lemmaA18_4}
\end{align}
\end{lemma}

\begin{proof}
Note that there are $d^{\ast} = \binom{d+3}{3}$ Bernstein basis polynomials on each tetrahedron and $\int_{\Omega} B_m^s(\bs{z}) d\bs{z} = \int_{T_{\lceil m/d^{\ast} \rceil}} B_m^s(\bs{z}) d\bs{z}$, for any $s \geq 1$. Then, for piecewise constant basis functions, we have $B_m(\bs{z}) = I(\bs{z} \in T_m)$, then
\[
\bigg\vert \frac{1}{N} \sum_{\bs{z}_j \in \Omega} B_m^s(\bs{z}_j) - \int_{\Omega} B_m^s(\bs{z}) d\bs{z} \bigg\vert = \left|\frac{1}{N} \sum_{\bs{z}_j \in \Omega} I(\bs{z}_j \in T_m) - V(T_m)\right|.
\]
According to Assumption A1, we have
\[
\max_{m \in \mathcal{M}} \bigg\vert \frac{1}{N}\sum_{\bs{z}_{j} \in \Omega} B_m^s(\bs{z}_j) - \int_{\Omega} B_m^s(\bs{z}) d\bs{z} \bigg\vert \leq CN^{-1/3}|\triangle|^2.
\]	
For any $j = 1, \ldots, N$, let $\mathcal{V}_j$ be the $j$th voxel. It is clear that
\[
\bigg\vert \frac{1}{N}\sum_{\bs{z}_{j} \in \Omega} B_m^s(\bs{z}_j) - \int_{\Omega} B_m^s(\bs{z}) d\bs{z} \bigg\vert \leq \bigg\vert \sum_{\bs{z}_{j} \in \Omega} \int_{\mathcal{V}_j}\{B_{m}^{s}(\bs{z}_{j})-B_{m}^{s}(\bs{z})\}d\bs{z} \bigg\vert + \int_{\Omega\setminus\cup\mathcal{V}_j}B_{m}^{s}(\bs{z})d\bs{z}.
\]
If $d \geq 1$, by the properties of trivariate spline basis functions in \cite{Lai:Schumaker:07}, we have $\int_{\Omega \setminus \cup \mathcal{V}_j} B_m^s(\bs{z}) d\bs{z} = O(N^{-1/3}|\triangle|^2)$, and
\begin{eqnarray*}
\left|\sum_{\bs{z}_{j} \in \Omega} \int_{\mathcal{V}_j} \{B_m^s(\bs{z}_j) - B_m^s(\bs{z})\} d\bs{z}\right| &\leq& \sum_{\{j:\bs{z}_j \in {T_{\lceil m/d^{\ast}\rceil}}\}} \int_{\mathcal{V}_j} |B_m^s(\bs{z}_j) - B_m^s(\bs{z})| d\bs{z}\\
&\leq& C(N|\triangle|^3) \times N^{-1} \times (N^{-1/3}|\triangle|^{-1}) \leq CN^{-1/3}|\triangle|^2.
\end{eqnarray*}
Thus, (\ref{DDC-lemmaA18_1}) holds. The proof of (\ref{DDC-lemmaA18_2}) is similar to (\ref{DDC-lemmaA18_1}). 

\noindent Next, for any $m, m^{\prime} \in \mathcal{M}$, we obtain
\begin{align*}
& \frac{1}{N^2} \sum_{\bs{z}_j, \bs{z}_{j'} \in \Omega} G_{\eta}(\bs{z}_j, \bs{z}_{j^{\prime}}) B_m(\bs{z}_j) B_{m^{\prime}}(\bs{z}_{j^{\prime}}) - \int_{\Omega^2} G_{\eta}(\bs{z}, \bs{z}^{\prime}) B_m(\bs{z}) B_{m^{\prime}}(\bs{z}^{\prime}) d\bs{z} d\bs{z}^{\prime}\\
& = \sum_{\bs{z}_j, \bs{z}_{j'} \in \Omega} \int_{\mathcal{V}_j \times \mathcal{V}_{j^{\prime}}} \left\{G_{\eta}(\bs{z}_j, \bs{z}_{j^{\prime}}) B_m(\bs{z}_{j}) B_{m^{\prime}}(\bs{z}_{j^{\prime}}) - G_{\eta}(\bs{z}, \bs{z}^{\prime}) B_m(\bs{z}) B_{m^{\prime}}(\bs{z}^{\prime})\right\} d\bs{z} d\bs{z}^{\prime}\\
& \quad + \int_{\Omega^2 \setminus \cup_{j, j^{\prime}} \mathcal{V}_j \times \mathcal{V}_{j^{\prime}}} \left\{G_{\eta}(\bs{z}_j, \bs{z}_{j^{\prime}}) B_m(\bs{z}_j) B_{m^{\prime}}(\bs{z}_{j^{\prime}}) - G_{\eta}(\bs{z}, \bs{z}^{\prime}) B_m(\bs{z}) B_{m^{\prime}}(\bs{z}^{\prime})\right\} d\bs{z} d\bs{z}^{\prime}.
\end{align*}
As $N \rightarrow \infty$, we have
\[
\begin{aligned}
& \int_{\Omega^2 \setminus \cup_{j,j^{\prime}}
        \mathcal{V}_j \times \mathcal{V}_{j^{\prime}}}
  \Big\{
    G_{\eta}(\bs{z}_j,\bs{z}_{j^{\prime}})
    B_m(\bs{z}_j) B_{m^{\prime}}(\bs{z}_{j^{\prime}}) -
    G_{\eta}(\bs{z},\bs{z}^{\prime})
    B_m(\bs{z}) B_{m^{\prime}}(\bs{z}^{\prime})
  \Big\}\,d\bs{z}\,d\bs{z}^{\prime} \\
& \qquad = O(N^{-1/3}|\triangle|^5),
\end{aligned}
\]
\begin{align*}
& \bigg\vert \sum_{\bs{z}_j, \bs{z}_{j'} \in \Omega} \int_{\mathcal{V}_j\times \mathcal{V}_{j^{\prime}}} \left\{G_{\eta}(\bs{z}_{j},\bs{z}_{j^{\prime}})B_{m}(\bs{z}_{j})B_{m^{\prime}}(\bs{z}_{j^{\prime}}) - G_{\eta}(\bs{z},\bs{z}^{\prime})B_{m}(\bs{z})B_{m^{\prime}}(\bs{z}^{\prime})\right\} d\bs{z}d\bs{z}^{\prime} \bigg\vert\\
& \leq \sum_{\{(j,j^{\prime}): \bs{z}_{j} \in {T_{\lceil m/d^{\ast} \rceil}}, \bs{z}_{j^{\prime}} \in {T_{\lceil m/d^{\ast} \rceil}}\}} \int_{\mathcal{V}_j \times \mathcal{V}_{j^{\prime}}} \Omega_{jj^{\prime}}(G_{\eta}K_{mm'}, \sqrt{6}N^{-1/3}) d\bs{z}d\bs{z}^{\prime},
\end{align*}
where $K_{mm'}(\bs{z},\bs{z}^{\prime}) = B_{m}(\bs{z})B_{m'}(\bs{z}^{\prime})$ and $\Omega_{jj^{\prime}}(g,\varrho) = \sup_{\substack{(\bs{z}_{1},\bs{z}_{1}^{\prime}), (\bs{z}_{2},\bs{z}_{2}^{\prime}) \in \mathcal{V}_j \times \mathcal{V}_{j^{\prime}},\\ 
\|\bs{z}_{1} - \bs{z}_{2}\|^{2} + \|\bs{z}_{1}^{\prime} - \bs{z}_{2}^{\prime}\|^{2} = \varrho^{2}}} |g(\bs{z}_1,\bs{z}_{1}^{\prime}) - g(\bs{z}_2,\bs{z}_{2}^{\prime})|$ is the modulus of continuity of $g$ on $\mathcal{V}_j\times \mathcal{V}_{j^{\prime}}$. Therefore, by Assumption \ref{as:errors}, we have
\begin{align*}
& \bigg\vert \sum_{\bs{z}_j, \bs{z}_{j'} \in \Omega} \int_{\mathcal{V}_j \times \mathcal{V}_{j^{\prime}}}\left\{G_{\eta}(\bs{z}_{j},\bs{z}_{j^{\prime}})B_{m}(\bs{z}_{j})B_{m^{\prime}}(\bs{z}_{j^{\prime}}) - G_{\eta}(\bs{z},\bs{z}^{\prime})B_{m}(\bs{z})B_{m^{\prime}}(\bs{z}^{\prime})\right\}d\bs{z}d\bs{z}^{\prime} \bigg\vert\\
&\qquad \leq (N|\triangle|^3)^2 \times N^{-2} \times (N^{-1/3}|\triangle|^{-1}) = O(N^{-1/3}|\triangle|^{5}).
\end{align*}
Thus, (\ref{DDC-lemmaA18_3}) follows.

Finally, note that
\begin{align*}
& \bigg\vert
  \frac{1}{N}\sum_{\bs{z}_j\in\Omega}
  B_m^2(\bs{z}_j)\sigma^2(\bs{z}_j)
  - \int_\Omega B_m^2(\bs{z})\sigma^2(\bs{z})\,d\bs{z}
  \bigg\vert \\
&\qquad \leq
  \bigg\vert
  \sum_{\bs{z}_j\in\Omega}\int_{\mathcal{V}_j}
  \{B_m^2(\bs{z}_j)\sigma^2(\bs{z}_j)
  - B_m^2(\bs{z})\sigma^2(\bs{z})\}\,d\bs{z}
  \bigg\vert \\
&\qquad\qquad \phantom{\leq} {}+
  \int_{\Omega\setminus\cup\mathcal{V}_j}
  |B_m^2(\bs{z}_j)\sigma^2(\bs{z}_j)
  - B_m^2(\bs{z})\sigma^2(\bs{z})|\,d\bs{z}.
\end{align*}
It is easy to see that $\int_{\Omega\setminus\cup\mathcal{V}_j}|B_{m}^{2}(\bs{z}_{j})\sigma^{2}(\bs{z}_j) - B_{m}^{2}(\bs{z})\sigma^{2}(\bs{z})|d\bs{z} = O(N^{-1/3}|\triangle|^2)$. Denote $\Omega_j(g,\varrho) = \sup_{\bs{z},\bs{z}^{\prime}\in \mathcal{V}_j, \|\bs{z}-\bs{z}^{\prime}\|=\varrho}|g(\bs{z})-g(\bs{z}^{\prime})|$ is the modulus of continuity of $g$ on the $j$th voxel $\mathcal{V}_j$, then by Assumption \ref{as:errors}, we have
\begin{align*}
& \bigg\vert \sum_{\bs{z}_{j} \in \Omega} \int_{\mathcal{V}_j}\{B_{m}^{2}(\bs{z}_{j})\sigma^{2}(\bs{z}_j) - B_{m}^{2}(\bs{z})\sigma^{2}(\bs{z})\}d\bs{z} \bigg\vert \leq \sum_{\{j:\bs{z}_{j}\in {T_{\lceil m/d^{\ast} \rceil}}\}} \int_{\mathcal{V}_j}\Omega_j(B_{m}^{2}\sigma^2,\sqrt{3}N^{-1/3})d\bs{z}\\
& \qquad \leq C(N|\triangle|^3)\times N^{-1} \times (N^{-1/3}|\triangle|^{-1})\leq CN^{-1/3}|\triangle|^2.
\end{align*}
Therefore, we can obtain (\ref{DDC-lemmaA18_4}).
\end{proof}

\begin{lemma}
\label{LEM:inner product}
For any $m\in\mathcal{M}$ and $0\leq\ell,\ell^{\prime}\leq p$, define
\[
\Phi_{m,\ell,\ell^{\prime},\Omega}
  = {\rm E}(X_\ell X_{\ell^{\prime}})
    \int_\Omega B_m^2(\bs{z})\,d\bs{z}.
\]
Under Assumptions \ref{as:triangulation} and \ref{as:design},
if $N^{1/3}|\triangle|\rightarrow\infty$ as $N\rightarrow\infty$,
we have with probability $1$,
\begin{align*}
& \max_{m\in\mathcal{M}}
  \max_{0\leq\ell,\ell^{\prime}\leq p}
  \bigg\vert
  \frac{1}{nN}\sum_{i=1}^{n}\sum_{\bs{z}_j\in\Omega}
  B_m^2(\bs{z}_j)X_{i\ell}X_{i\ell^{\prime}}
  - \Phi_{m,\ell,\ell^{\prime},\Omega}
  \bigg\vert \\
&\qquad {}= O\left\{
  |\triangle|^3(\log n/n)^{1/2}
  + N^{-1/3}|\triangle|^2
  \right\}.
\end{align*}
\end{lemma}

\begin{proof}
Let $\varsigma_{i,m}\equiv\varsigma_{i,m,\ell,\ell^{\prime}, \Omega}=\frac{1}{N}\sum_{\bs{z}_j \in \Omega}B_{m}^{2}(\bs{z}_{j}) X_{i\ell}X_{i\ell^{\prime}}$. If $N^{1/3}|\triangle|\rightarrow \infty$ as $N\rightarrow \infty$, then by Lemma \ref{LEM:integration}, $E\varsigma_{i,m}=\frac{1}{N}\sum_{\bs{z}_j \in \Omega}B_{m}^{2}(\bs{z}_{j}){\rm E}(X_{\ell}X_{\ell^{\prime}}) \asymp  |\triangle|^3$, and ${\rm E}\left(\varsigma_{i,m}\right)^{2}=\left\{\frac{1}{N}\sum_{\bs{z}_j \in \Omega}B_{m}^{2}(\bs{z}_{j})\right\}^{2} {\rm E}\left(X_{\ell}X_{\ell^{\prime}}\right)^{2} \asymp |\triangle|^{6}$.

Next, define a sequence $D_{n}=n^{\alpha}$ with $\alpha\in(1/3,1/2)$. We make use of the following truncated and tail decomposition $X_{i\ell\ell^{\prime}}=X_{i\ell}X_{i\ell^{\prime}}=X_{i\ell\ell^{\prime},1}^{D_{n}}+X_{i\ell\ell^{\prime},2}^{D_{n}}$,
where $X_{i\ell\ell^{\prime},1}^{D_{n}}=X_{i\ell}X_{i\ell^{\prime}}I\left\{\left\vert
X_{i\ell}X_{i\ell^{\prime}}\right\vert >D_{n}\right\} $, $X_{i\ell\ell^{\prime
},2}^{D_{n}}=X_{i\ell}X_{i\ell^{\prime}}I\left\{\left\vert X_{i\ell}X_{i\ell^{\prime
}}\right\vert \leq D_{n}\right\}$. Correspondingly, the truncated and tail
parts of $\varsigma_{i,m}$ are $\varsigma
_{i,m,v}\equiv \varsigma
_{i,m,v,\ell,\ell^{\prime}}=\frac{1}{N}\sum_{\bs{z}_j \in \Omega}B_{m}^{2}(\bs{z}_{j})X_{i\ell\ell^{\prime},v}^{D_{n}}$, $v=1,2$. According to Assumption \ref{as:design}, for any $\ell,\ell^{\prime}=0,\ldots,p$,
\begin{equation*}
\sum_{n=1}^{\infty}P\left\{\left\vert X_{n\ell}X_{n\ell^{\prime}}\right\vert
>D_{n}\right\} \leq \sum_{n=1}^{\infty}\frac{{\rm E}
\left\vert X_{n\ell}X_{n\ell^{\prime}}\right\vert^{3}}{D_{n}^{3}}%
\leq C_{b}\sum_{n=1}^{\infty}D_{n}^{-3}<\infty .
\end{equation*}%
By the Borel-Cantelli Lemma, one has $N^{-1}\sum_{\bs{z}_j \in \Omega}B_{m}^{2}(\bs{z}_{j})X_{i\ell\ell^{\prime},1}^{D_{n}}=0$, almost surely.
So we obtain $\sup_{m,\ell,\ell^{\prime}}\left\vert n^{-1}\sum_{i=1}^{n}\varsigma
_{i,m,1}\right\vert =O_{a.s.}(n^{-k})$, $k\geq 1$. Since $N^{1/3}|\triangle|\rightarrow \infty$ as $N\rightarrow \infty$, for any $r\geq 2$,
\begin{align*}
|{\rm E}\varsigma_{i,m,1}| = &|{\rm E}(X_{i\ell\ell^{\prime},1}^{D_{n}})| \left\{\frac{1}{N}\sum_{\bs{z}_j \in \Omega}B_{m}^{2}(\bs{z}_{j})\right\} \leq D_{n}^{-2}{\rm E}\left\vert
X_{i\ell}X_{i\ell^{\prime}}\right\vert^{3}
\\
&\times \left\{\int_{\Omega}B_{m}^{2}(\bs{z}) d\bs{z}+O(N^{-1/3}|\triangle|^2)\right\} \leq c_bD_{n}^{-2}O(|\triangle|^3).
\end{align*}
Next, consider the truncated part $\varsigma_{i,m,2}$. Define $\varsigma_{i,m,2}^{\ast} = \varsigma_{i,m,2} - {\rm E}(\varsigma_{i,m,2})$, then ${\rm E}\varsigma_{i,m,2}^{\ast}=0$,
\[
{\rm E}(\varsigma_{i,m,2}^{\ast})^{2} = {\rm E}(\varsigma_{i,m,2})^{2} - ( {\rm E}\varsigma_{i,m,2} )^{2} = \left\{\frac{1}{N}\sum_{\bs{z}_j \in \Omega}B_{m}^{2}(\bs{z}_{j})\right\}^2 \left\{{\rm E}(X_{i\ell\ell^{\prime},2}^{D_{n}})^{2} - ({\rm E}X_{i\ell\ell^{\prime},2}^{D_{n}})^2\right\}.
\]
Note that
\[
{\rm E}(X_{i\ell\ell^{\prime},1}^{D_n})^2
  \leq D_n^{-1}{\rm E}\bigg\vert X_{i\ell}X_{i\ell^{\prime}}\bigg\vert^3
  \leq cD_n^{-1}.
\]
Thus,
\[
\begin{aligned}
{\rm E}(X_{i\ell\ell^{\prime},2}^{D_n})^2
  &= {\rm E}(X_{i\ell\ell^{\prime}})^2
     - {\rm E}(X_{i\ell\ell^{\prime},1}^{D_n})^2 = {\rm E}(X_{i\ell\ell^{\prime}})^2 - o(1).
\end{aligned}
\]
Therefore, there exists $c_{\varsigma}$ such that for large $n$,
\[
{\rm E}(\varsigma_{i,m,2}^{\ast})^2
  \geq c_{\varsigma}{\rm E}\left(X_{i\ell\ell^{\prime}}\right)^2
  \times \left\{N^{-1}\sum_{\bs{z}_j\in\Omega}
                B_m^2(\bs{z}_j)\right\}^2.
\]

Next, for any $k>2$,
\begin{align*}
{\rm E} \vert \varsigma_{i,m,2}^{\ast} \vert^{k} &= {\rm E} \vert \varsigma_{i,m,2} - {\rm E}\left( \varsigma_{i,m,2}\right) \vert^{k} \leq 2^{k-1}\left( {\rm E} \vert \varsigma_{i,m,2} \vert^{k} + \vert {\rm E}(\varsigma_{i,m,2}) \vert^{k}\right) \\
&= 2^{k-1}\left\{{\rm E} \bigg\vert X_{i\ell\ell^{\prime},2}^{D_{n}} \bigg\vert^{k} + O(1)\right\}\left\{\frac{1}{N} \sum_{\bs{z}_j \in \Omega}B_{m}^{2}(\bs{z}_{j})\right\}^{k},
\end{align*}
then there exists $C_{\varsigma}>0$ such that for any $k>2$ and large $n$,
\begin{align*}
{\rm E} \vert \varsigma_{i,m,2}^{\ast} \vert^{k} &\leq 2^{k-1}\left\{D_{n}^{k-2}{\rm E}\left(X_{i\ell\ell^{\prime}}\right)^{2} + O(1)\right\}\left\{\frac{1}{N}\sum_{\bs{z}_j \in \Omega}B_{m}^{2}(\bs{z}_{j})\right\}^{k}\\
&\leq 2^{k}D_{n}^{k-2} {\rm E}\left(\varsigma_{i,m,2}^{\ast}\right)^{2} \left\{\frac{1}{N}\sum_{\bs{z}_j \in \Omega}B_{m}^{2}(\bs{z}_{j})\right\}^{k-2} \leq \left(C_{\varsigma}D_{n}|\triangle|^{3}\right)^{k-2}k! {\rm E}\left(\varsigma_{i,m,2}^{\ast}\right)^{2},
\end{align*}
which implies that $\left\{\varsigma_{i,m,2}^{\ast}\right\}_{i=1}^{n}$ satisfies Cram\'{e}r's condition with constant $C_{\varsigma}D_{n}|\triangle|^{3}$. Applying Bernstein's inequality to $\sum_{i=1}^{n}\varsigma_{i,m,2}^{\ast}$, for $k>2$ and any large enough $\delta>0$,
\begin{align*}
& P\left\{
    \bigg\vert \frac{1}{n}\sum_{i=1}^{n}\varsigma_{i,m,2}^{\ast}
    \bigg\vert \geq \delta n^{-1/2}|\triangle|^{3}(\log n)^{1/2}
  \right\} \\
&\qquad \leq 2\exp\left[
    -\delta^{2}
    \{4+2C_{\varsigma}D_{n}\delta(\log n)^{1/2}n^{-1/2}\}^{-1}
    \log(n)
  \right].
\end{align*}
Assume that $|\triangle|^{-3} \asymp n^{\tau}$ for some $0<\tau<\infty$, we have
\begin{eqnarray*}
\sum_{n=1}^{\infty} &P\left\{\max_{m \in \mathcal{M}, 0 \leq \ell,\ell^{\prime} \leq p} \bigg\vert n^{-1}\sum_{i=1}^{n}\varsigma_{i,m,2}^{\ast} \bigg\vert \geq \delta n^{-1/2}|\triangle|^{3}(\log n)^{1/2}\right\}\\
&\leq 2\sum_{n=1}^{\infty}\sum_{m \in \mathcal{M}}\sum_{0 \leq \ell,\ell^{\prime} \leq p} n^{-2-\tau} < \infty.
\end{eqnarray*}
Thus, $\sup_{m,\ell,\ell^{\prime}} \left\vert n^{-1}\sum_{i=1}^{n}\varsigma_{i,m,2}^{\ast} \right\vert = O_{a.s.}\left\{n^{-1/2}|\triangle|^{3}(\log n)^{1/2}\right\}$ as $n \rightarrow \infty$ by the Borel-Cantelli Lemma. Furthermore, we have
\begin{align*}
\max_{m,\ell,\ell^{\prime}} &\left\vert n^{-1}\sum_{i=1}^{n}\varsigma_{i,m} - {\rm E}\varsigma_{i,m} \right\vert \leq \max_{m,\ell,\ell^{\prime}} \left\vert n^{-1}\sum_{i=1}^{n}\varsigma_{i,m,1} \right\vert + \max_{m,\ell,\ell^{\prime}} \left\vert n^{-1}\sum_{i=1}^{n}\varsigma_{i,m,2}^{\ast} \right\vert + \max_{m,\ell,\ell^{\prime}} \left\vert {\rm E}\varsigma_{i,m,1} \right\vert \\
&= O_{a.s.}(n^{-k}) + O_{a.s.}\left\{n^{-1/2}|\triangle|^{3}(\log n)^{1/2}\right\} + O\left(D_{n}^{-2}|\triangle|^{3}\right) \\
&= O_{a.s.}\left\{n^{-1/2}|\triangle|^{3}(\log n)^{1/2}\right\}.
\end{align*}
Finally, we notice that
\begin{align*}
&\max_{m\in\mathcal{M}}\max_{0 \leq \ell,\ell^{\prime}\leq p}\left| \frac{1}{nN}\sum_{i=1}^{n}\sum_{\bs{z}_j \in \Omega}B_{m}^{2}(\bs{z}_{j})X_{i\ell}X_{i\ell^{\prime}}-\Phi_{m,\ell,\ell^{\prime}}\right| \\
&=\max_{m\in\mathcal{M}}\max_{0 \leq \ell,\ell^{\prime} \leq p} \left\vert  n^{-1}\sum_{i=1}^{n}\varsigma_{i,m}-{\rm E}\varsigma_{i,m} \right\vert + \left\vert {\rm E}X_{i\ell}X_{i\ell^{\prime}} \right\vert \max_{m\in\mathcal{M}} \left\vert \frac{1}{N}\sum_{\bs{z}_j \in \Omega}B_{m}^{2}(\bs{z}_{j}) - \int_{\Omega}B_{m}^{2}(\bs{z})d\bs{z} \right\vert \\
&= O_{P}\left\{n^{-1/2}|\triangle|^{3}(\log n)^{1/2}\right\} + O(N^{-1/3}|\triangle|^2).
\end{align*}
Thus, we obtain the desired result.
\end{proof}

The following lemma provides the uniform convergence rate at which the empirical inner product in (\ref{DEF:DDC-empirical_product}) approximates the theoretical inner product in (\ref{DEF:DDC-theoretical_product}).
\begin{lemma}
\label{LEM:Rnorder-vec}
Let $g_{\ell}^{(1)}(\bs{z}) = \sum_{m\in \mathcal{M}}c_{\ell m}^{(1)}B_{m}(\bs{z})$, $g_{\ell}^{(2)}(\bs{z})=\sum_{m \in \mathcal{M}}c_{\ell m}^{(2)}B_{m}(\bs{z})$ be any spline functions in $\mathcal{S}_{d}^{r}(\triangle_{\Omega})$. Denote the vector of functions $\bs{g}(\bs{z})=(g_0(\bs{z}), \ldots, g_p(\bs{z}))^{\top}$ by $g_{\ell}\in \mathcal{S}_{d}^{r}(\triangle_{\Omega})$, for $\ell=0,\ldots, p$. Suppose Assumptions \ref{as:triangulation} and \ref{as:design} hold, and $N^{1/3}|\triangle|\rightarrow \infty$ as $N\rightarrow \infty$. For the subregion $\Omega$, we have
\begin{align*}
R_{n} &= \sup\limits_{\bs{g}^{(1)},\bs{g}^{(2)}\in \mathcal{S}_{d}^{r}(\triangle)}\left| \frac{\left\langle \bs{g}^{(1)},\bs{g}^{(2)}\right\rangle_{n} - \left\langle \bs{g}^{(1)},\bs{g}^{(2)}\right\rangle_{\Omega}}{\left\| \bs{g}^{(1)}\right\|_{L_2(\Omega)}\left\|\bs{g}^{(2)}\right\|_{L_2(\Omega)}} \right|\\
&= O_{P}\{n^{-1/2}(\log n)^{1/2} + N^{-1/3}|\triangle|^{-1}\}.
\end{align*}
\end{lemma}

\begin{proof}
It is easy to see
\begin{eqnarray*}
\langle \bs{g}^{(1)},\bs{g}^{(2)}\rangle_{n}\!\!\!&=&\!\!\!\frac{1}{nN}\sum_{i=1}^{n}\sum_{\bs{z}_{j} \in \Omega}\left\{\sum_{\ell =0}^{p}\sum_{m\in \mathcal{M}}c_{\ell m}^{(1)}X_{i\ell}B_{m}(\bs{z}_{j})\right\} \left\{\sum_{\ell ^\prime=0}^{p}\sum_{m^{\prime}\in \mathcal{M}}c_{\ell' m^\prime}^{(2)}X_{i\ell^\prime}B_{m^\prime}(\bs{z}_{j})\right\}\\
&=&\sum_{\ell,m}\sum_{\ell ^\prime,m^\prime}c_{\ell m}^{(1)}c_{\ell ^\prime m^\prime}^{(2)}\frac{1}{nN}\sum_{i=1}^{n}\sum_{\bs{z}_{j} \in \Omega} X_{i\ell}X_{i\ell^\prime}B_{m}(\bs{z}_{j})B_{m^\prime}(\bs{z}_{j}).
\end{eqnarray*}
Note that
\begin{align*}
	\|\bs{g}^{(1)}\|^{2}_{\Omega}&=\sum_{\ell,m}\sum_{\ell ^\prime,m^\prime}c_{\ell m}^{(1)}c_{\ell ^\prime m^\prime}^{(1)}{\rm E}(X_{\ell}X_{\ell^{\prime}})\int_{\Omega}B_{m}(\bs{z})B_{m^\prime}(\bs{z})d\bs{z},\\
	\|\bs{g}^{(2)}\|^{2}_{\Omega}&=\sum_{\ell,m}\sum_{\ell ^\prime,m^\prime}c_{\ell m}^{(2)}c_{\ell ^\prime m^\prime}^{(2)}{\rm E}(X_{\ell}X_{\ell^{\prime}})\int_{\Omega}B_{m}(\bs{z})B_{m^\prime}(\bs{z})d\bs{z}.
\end{align*}

It follows from Assumptions \ref{as:triangulation}, \ref{as:design}, Lemmas \ref{LEM:DDC-norm} and \ref{LEM:norm} that,
\[
	c_{v}|\triangle|^{3}\sum_{\ell,m}\{c_{\ell m}^{(v)}\}^2 \leq \Vert \bs{g}^{(v)}\Vert^{2}_{\Omega} \leq C_{v}|\triangle|^{3}\sum_{\ell,m}\{c_{\ell m}^{(v)}\}^2,
\]
\[
\begin{aligned}
C_{1}|\triangle|^{3}
\left[
  \sum_{\ell,m}\{c_{\ell m}^{(1)}\}^2
  \sum_{\ell',m'}\{c_{\ell' m'}^{(2)}\}^2
\right]^{1/2}
  &\leq \|\bs{g}^{(1)}\|_{\Omega}\,\|\bs{g}^{(2)}\|_{\Omega} \\
  &\leq C_{2}|\triangle|^{3}
    \left[
      \sum_{\ell,m}\{c_{\ell m}^{(1)}\}^2
      \sum_{\ell',m'}\{c_{\ell' m'}^{(2)}\}^2
    \right]^{1/2}.
\end{aligned}
\]
With the above preparation, we have
\begin{align}
\nonumber
&R_{n}\leq \frac{\sum_{\ell, \ell', |m-m'| \leq d^{\ast}} |c_{\ell m}^{(1)}c_{\ell ^\prime m^\prime}^{(2)}|}{C_{1}|\triangle|^{3}\left[\sum_{\ell,m}\{c_{\ell m}^{(1)}\}^2\sum_{\ell ^\prime,m^\prime}\{c_{\ell ^\prime m^\prime}^{(2)}\}^2\right]^{1/2}}\\
\nonumber
&\times\max_{
 \substack{m, m'\in\mathcal{M} \\ 0 \leq \ell,\ell^{\prime}\leq p}}\left| \frac{1}{nN}\sum_{i=1}^{n}\sum_{\bs{z}_j \in \Omega}B_{m}(\bs{z}_{j})B_{m'}(\bs{z}_{j})X_{i\ell}X_{i\ell^{\prime}}-{\rm E}(X_{\ell}X_{\ell^{\prime}})\int_{\Omega}B_{m}(\bs{z})B_{m'}(\bs{z})d\bs{z}\right|\\
&\leq C|\triangle|^{-3}\!\!\!\max_{
 \substack{m, m'\in\mathcal{M} \\ 0 \leq \ell,\ell^{\prime}\leq p}}\left| \frac{1}{nN}\sum_{i=1}^{n}\sum_{\bs{z}_j \in \Omega}B_{m}(\bs{z}_{j})B_{m'}(\bs{z}_{j})X_{i\ell}X_{i\ell^{\prime}}-{\rm E}(X_{\ell}X_{\ell^{\prime}})\int_{\Omega}B_{m}(\bs{z})B_{m'}(\bs{z})d\bs{z}\right|.
\label{EQ:R_n}
\end{align}
The desired result follows from (\ref{EQ:R_n}) and Lemma \ref{LEM:inner product}.	
\end{proof}

As a direct result of Lemma \ref{LEM:Rnorder-vec}, we can see that
\begin{eqnarray}
\sup_{\bs{g}\in \mathcal{S}_{d}^{r}(\triangle)}\left| \left. \left\| \bs{g}\right\|_{n}^{2}\right/ \Vert \bs{g}\Vert^{2}_\Omega-1\right| =O_{P}\{n^{-1/2}(\log n)^{1/2}+N^{-1/3}|\triangle|^{-1}\}.  \label{EQ:normratio-DDC}
\end{eqnarray}

\subsection{Uniform Convergence of the Unpenalized Spline Estimators} 
\label{sec:B-2}

In this section, we consider the unpenalized spline smoothing approach.
The unpenalized trivariate spline estimator of $\bs{\beta}^{o}=(\beta_{0}^{o},\ldots,\beta_{p}^{o})^{\top}$ within the subregion $\Omega$ is defined as
\begin{equation}
\widetilde{\bs{\beta}}=(\widetilde{\beta}_{0},\ldots,\widetilde{\beta}_{p})^{\top}
=\argmin_{\bs{\beta}\in \mathcal{G}^{(p+1)}} \sum_{i=1}^{n}\sum_{\bs{z}_j \in \Omega}\left\{Y_{i}(\bs{z}_{j})-\sum_{\ell =0}^{p}X_{i\ell}\beta_{\ell}(\bs{z}_{j})
\right\}^{2}.
\label{DEF:beta_tilde}
\end{equation}
Denote
\begin{align*}
\widetilde{\bs{\theta}}_{\mu}&=(\widetilde{\bs{\theta}}_{\mu,0}^{\top},\ldots,\widetilde{\bs{\theta}}_{\mu,p}^{\top})^{\top}
=\bs{\Gamma}_{n,0}^{-1}
\frac{1}{nN}\sum_{i=1}^{n}\sum_{\bs{z}_j \in \Omega}\left\{\widetilde{\mathbf{X}}_{i}\otimes \widetilde{\mathbf{B}}(\bs{z}_{j})\right\}\widetilde{\mathbf{X}}_{i}^{\top}\bs{\beta}^{o}(\bs{z}_j)\notag\\
\widetilde{\bs{\theta}}_{\eta}&=(\widetilde{\bs{\theta}}_{\eta,0}^{\top},\ldots,\widetilde{\bs{\theta}}_{\eta,p}^{\top})^{\top}
=\bs{\Gamma}_{n,0}^{-1}
\frac{1}{nN}\sum_{i=1}^{n}\sum_{\bs{z}_j \in \Omega} \left\{\widetilde{\mathbf{X}}_{i}\otimes \widetilde{\mathbf{B}}(\bs{z}_{j})\right\}\eta_{i}(\bs{z}_j),\notag\\
\widetilde{\bs{\theta}}_{\varepsilon}&=(\widetilde{\bs{\theta}}_{\varepsilon,0}^{\top},\ldots,
\widetilde{\bs{\theta}}_{\varepsilon,p}^{\top})^{\top}
=\bs{\Gamma}_{n,0}^{-1}
\frac{1}{nN}\sum_{i=1}^{n}\sum_{\bs{z}_j \in \Omega} \left\{\widetilde{\mathbf{X}}_{i}\otimes \widetilde{\mathbf{B}}(\bs{z}_{j})\right\}\sigma(\bs{z}_j)\varepsilon_{ij},
\end{align*}
where
\begin{equation}
\bs{\Gamma}_{n,0}=\frac{1}{nN}\sum_{i=1}^{n}\sum_{\bs{z}_j \in \Omega}(\widetilde{\mathbf{X}}_{i}\widetilde{\mathbf{X}}_{i}^{\top})\otimes
\{\widetilde{\mathbf{B}}(\bs{z}_{j})\widetilde{\mathbf{B}}(\bs{z}_{j})^{\top}\}.
\label{DEF:Gamma_0}
\end{equation}

\begin{lemma}
Under Assumptions \ref{as:triangulation} and \ref{as:design}, if $N^{1/3}|\triangle|\rightarrow \infty$ as $N\rightarrow \infty$, then there exist constants $0 < c_{\Gamma} < C_{\Gamma} < \infty$, such that with probability approaching 1 as $n\rightarrow \infty$,
$c_{\Gamma}|\triangle|^{3} \leq \lambda_{\min}(\bs{\Gamma}_{n,0}) \leq
\lambda_{\max}(\bs{\Gamma}_{n,0}) \leq  C_{\Gamma}|\triangle|^{3}$,
where $\bs{\Gamma}_{n,0}$ is given in (\ref{DEF:Gamma_0}).
\label{LEM:Gamma_0}
\end{lemma}
\begin{proof}
It is easy to see that for the vector $\bs{\theta}=(\bs{\theta}_{0}^{\top},\cdots,\bs{\theta}_{p}^{\top})^{\top}$ associated with the subregion $\Omega$,
\begin{align}
\bs{\theta}^{\top}\bs{\Gamma}_{n,0}\bs{\theta} &=\frac{1}{nN}\bs{\theta}^{\top}\sum_{i=1}^{n}\sum_{\bs{z}_j \in \Omega}(\widetilde{\mathbf{X}}_{i}
\widetilde{\mathbf{X}}_{i}^{\top})\otimes \{\widetilde{\mathbf{B}}_k^{\ast}(\bs{z}_{j})\widetilde{\mathbf{B}}^{\top}(\bs{z}_{j})\}\bs{\theta}\notag \\
&=\frac{1}{nN}\bs{\gamma}^{\top}\sum_{i=1}^{n}\sum_{\bs{z}_j \in \Omega}(\widetilde{\mathbf{X}}_{i}
\widetilde{\mathbf{X}}_{i}^{\top})\otimes \{\mathbf{B}(\bs{z}_{j})\mathbf{B}^{\top}(\bs{z}_{j})\}\bs{\gamma}
=\|\bs{g}_{\bs{\gamma}}\|_{n}^2, \label{EQ:g_gamma}
\end{align}
where $\bs{\gamma}=(\mathbf{I}_{p+1} \otimes \mathbf{Q}_2)\bs{\theta}$, and $\bs{g}_{\bs{\gamma}}=(g_{\bs{\gamma}_{0}}, \ldots, g_{\bs{\gamma}_{p}})^{\top}$ with $g_{\bs{\gamma}_{\ell}}=\sum \gamma_{\ell m}B_{m}$. By (\ref{EQ:normratio-DDC}), we have $
c (1-R_{n})|\triangle|^{3}\Vert \bs{\gamma}\Vert^{2}   \leq (1-R_{n}) \Vert \bs{g}_{\bs{\gamma}}\Vert_\Omega^{2}
\leq \Vert \bs{g}_{\bs{\gamma}}\Vert_{n}^{2}$, and
$\Vert \bs{g}_{\bs{\gamma}}\Vert_{n}^{2} \leq (1+R_{n}) \Vert \bs{g}_{\bs{\gamma}}\Vert_\Omega^{2}\leq C (1+R_{n})|\triangle|^{3}\Vert \bs{\gamma}\Vert^{2}$, in which we have used the stability conditions in Lemma \ref{LEM:DDC-norm}.
\end{proof}

Next, we consider the following decomposition $\widetilde{\bs{\beta}}(\bs{z})=\widetilde{\bs{\beta}}_{\mu}(\bs{z})
+\widetilde{\bs{\eta}}(\bs{z})+\widetilde{\bs{\varepsilon}}(\bs{z})$, where
\begin{align}
\widetilde{\bs{\beta}}_{\mu}(\bs{z})&=(\widetilde{\beta}_{\mu,0}(\bs{z}),\ldots,\widetilde{\beta}_{\mu,p}(\bs{z}))^{\top}
=\{\mathbf{I} \otimes \widetilde{\mathbf{B}}(\bs{z})\}^{\top}\widetilde{\bs{\theta}}_{\mu},\label{DEF:beta_tilde_mu}\\
\widetilde{\bs{\eta}}(\bs{z})&=(\widetilde{\eta}_{0}(\bs{z}), \ldots, \widetilde{\eta}_{p}(\bs{z}))^{\top}=\{\mathbf{I} \otimes \widetilde{\mathbf{B}}(\bs{z})\}^{\top}\widetilde{\bs{\theta}}_{\eta},
\label{DEF:eta_tilde}\\
\widetilde{\bs{\varepsilon}}(\bs{z})&=(\widetilde{\varepsilon}_{\Omega,0}(\bs{z}), \ldots, \widetilde{\varepsilon}_{p}(\bs{z}))^{\top}=\{\mathbf{I} \otimes \widetilde{\mathbf{B}}(\bs{z})\}^{\top}\widetilde{\bs{\theta}}_{\varepsilon}. \label{DEF:eps_tilde}
\end{align}
\begin{lemma}
\label{LEM:error-unif-order}
Under Assumptions \ref{as:triangulation}--\ref{as:rates}, if it holds that $N^{1/3}|\triangle|\rightarrow \infty$ as $N\rightarrow \infty$, $\|\sum_{t=1}^{\infty}\lambda_{t}^{1/2}\psi_{t}\|_{\infty} < \infty$, and $n^{1/(4+\delta)} \ll n^{1/2}N^{-1/2}|\triangle|^{-3/2}$ for some $\delta$, then for $\widetilde{\bs{\eta}}$ and $\widetilde{\bs{\epsilon}}$ in (\ref{DEF:eta_tilde}) and (\ref{DEF:eps_tilde}), we have $\|\widetilde{\bs{\eta}}\|_{\infty}=O_{P}\{n^{-1/2}(\log n)^{1/2}\}$ and $
\|\widetilde{\bs{\varepsilon}}\|_{\infty}=O_{P}\{(nN)^{-1/2}(\log n)^{1/2}|\triangle|^{-3/2}\}$.
\end{lemma}
\begin{proof}
Note that for any $\ell=0,1,\ldots,p$, $\widetilde{\eta}_{\ell}(\bs{z})=\sum_{m \in \mathcal{M}}\widetilde{\theta}_{\eta,\ell,m}\widetilde{B}_{m}(\bs{z})$ for some coefficients $\widetilde{\theta}_{\eta,\ell,m}$. Therefore, the order of $\widetilde{\eta}_{\ell}(\bs{z})$ is related to that of $\widetilde{\theta}_{\eta,\ell,m}$. In fact, 
\begin{align}
\|\widetilde{\bs{\eta}}\|_{\infty}&=\max_{0\leq\ell\leq p} \|\widetilde{\eta}_{\ell}\|_{\infty}\leq C_{\eta} \|\widetilde{\bs{\theta}}_{\eta,\ell}\|_{\infty}\\
&=\left\|(\mathbf{e}_{\ell}\otimes \mathbf{1})^{\top} \bs{\Gamma}_{n,0}^{-1}
\left[\frac{1}{nN} \sum_{i=1}^{n}\sum_{\bs{z}_j \in \Omega}\left\{\widetilde{\mathbf{X}}_{i}\otimes \widetilde{\mathbf{B}}_k(\bs{z}_{j})\right\}\eta_{i}(\bs{z}_j)\right] \right\|_{\infty}, \notag
\end{align}
where $\widetilde{\bs{\theta}}_{\eta}=(\widetilde{\theta}_{\eta,\ell,m})_{m \in \widetilde{\mathcal{M}}}$ with $\widetilde{\mathcal{M}}$ being an index set of the transformed Bernstein basis polynomials $\widetilde{B}_{m}(\bs{z})$ and $\bs{\Gamma}_{n,0}$ is the symmetric positive definite matrix defined in (\ref{DEF:Gamma_0}). Thus, by Lemma \ref{LEM:Gamma_0},
\[
\left\|\widetilde{\bs{\eta}}\right\|_{\infty} \leq C|\triangle|^{-3} \max_{0\leq\ell\leq p}\max_{m \in \widetilde{\mathcal{M}}}\left| \frac{1}{nN} \sum_{i=1}^{n}\sum_{\bs{z}_j \in \Omega} X_{i\ell}
\eta_{i}(\bs{z}_j)\widetilde{B}_{m}(\bs{z}_j)\right|,
\]
almost surely. Next, we show that with probability one
\begin{equation}
\max_{0\leq\ell\leq p}\max_{m \in \widetilde{\mathcal{M}}}\left| \frac{1}{nN} \sum_{i=1}^{n}\sum_{\bs{z}_j \in \Omega} X_{i\ell}\eta_{i}(\bs{z}_j)\widetilde{B}_{m}(\bs{z}_j)\right| =O_P\left\{n^{-1/2}|\triangle|^{3}(\log n)^{1/2}\right\}.
\label{EQ:BEsupnorm}
\end{equation}
To prove (\ref{EQ:BEsupnorm}), let
$\varpi_{i}=\varpi_{i,m}=N^{-1} \sum_{t=1}^{\infty}\lambda_t^{1/2}X_{i\ell}\xi_{it}\sum_{\bs{z}_j \in \Omega} \widetilde{B}_{m}(\bs{z}_j)\psi_{t}(\bs{z}_{j})$, where ${\rm E}(\varpi_{i})=0$ and ${\rm E}(\varpi_{i}^2)$ can be written as
\[
\frac{{\rm E}(X_{i\ell}^{2})}{N^2}\!\!\sum_{\bs{z}_j \in \Omega, \bs{z}_{j'} \in \Omega}\!\!\widetilde{B}_{m}(\bs{z}_j)\widetilde{B}_{m}(\bs{z}_{j^\prime})
G_{\eta}(\bs{z}_j,\bs{z}_{j^\prime}) \!\! \asymp \!\!
\int_{\Omega^2}G_{\eta}(\bs{z},\bs{z}^{\prime})B_m(\bs{z})B_{m}(\bs{z}^{\prime})
d\bs{z}d\bs{z}^{\prime} \asymp |\triangle|^6.
\]

We decompose the random variable $\varpi_{i}$ into a tail part and a truncated part,
\begin{align*}
\varpi_{i,1}^{D_{n}}&=\sum_{t=1}^{\infty}\lambda_t^{1/2}\left\{\frac{1}{N}\sum_{\bs{z}_j \in \Omega}
\widetilde{B}_{m}(\bs{z}_j)\psi_{t}(\bs{z}_{j})\right\} X_{i\ell}\xi_{it}I\left\{ \left|X_{i\ell}\xi_{it}\right|
>D_{n}\right\},\\
\varpi_{i,2}^{D_{n}}&=\sum_{t=1}^{\infty}\lambda_t^{1/2}\left\{\frac{1}{N}\sum_{\bs{z}_j \in \Omega}
\widetilde{B}_{m}(\bs{z}_j)\psi_{t}(\bs{z}_{j})\right\} X_{i\ell}\xi_{it}I\left\{ \left|X_{i\ell}\xi_{it}\right| \leq D_{n}\right\} -\mu_{i}^{D_{n}}, \\
\mu_{i}^{D_{n}}&=
\sum_{t=1}^{\infty}\lambda_t^{1/2}\left\{\frac{1}{N}\sum_{\bs{z}_j \in \Omega}\widetilde{B}_{m}(\bs{z}_j)
\psi_{t}(\bs{z}_{j})\right\} {\rm E}\left[X_{i\ell}\xi_{it}I\left\{ \left| X_{i\ell}\xi_{it}\right| \leq D_{n}\right\} \right],
\end{align*}
where $%
D_{n}=n^{\alpha }\left(1/(4+\delta_1) < \alpha < 1/2\right)$. At first, we show that the tail part vanishes almost surely. Note that, for any $k \geq 1$,
\begin{equation*}
\sum_{n=1}^{\infty }P\left\{ \left| X_{n\ell}\xi_{nk}\right| >D_{n}\right\} \leq
\sum_{n=1}^{\infty }\frac{{\rm E}\left|X_{n\ell}\xi_{nk} \right| ^{4+\delta_{1}}}{%
D_{n}^{4+\delta_1}}\leq \upsilon _{\delta_1 }\sum_{n=1}^{\infty }D_{n}^{-(4+\delta_{1})}<\infty .  \label{keyH}
\end{equation*}
By the Borel-Cantelli Lemma, we can show that $
\left|n^{-1}\sum_{i=1}^{n}\varpi_{i,1}^{D_{n}} \right| =O_{a.s.}\left( n^{-r}\right),\textrm{\ for\ any\ } r>0.$ As $E \varpi_{i} =0$, then it is straightforward to verify $\mu_{i}^{D_{n}}=-E\varpi_{i,1}^{D_{n}}=O\{D_n^{-2}|\triangle|^3\}$.

Next, notice that $E\varpi_{i,2}^{D_{n}}=0$. Then, we have
$\mathrm{Var}(\varpi_{i,2}^{D_{n}})={\rm E}(\varpi_{i}^{2}) - {\rm E}\left(\varpi_{i,1}^{D_{n}}\right)^2 -(\mu^{D_{n}})^{2}
\asymp |\triangle|^6$. Also, we have, for any $r\geq 3$,
\begin{align*}
E|\varpi_{i,2}^{D_{n}}|^r&={\rm E}\left|\sum_{t=1}^{\infty}\lambda_t^{1/2}\frac{1}{N}
\sum_{\bs{z}_j \in \Omega}\widetilde{B}_{m}(\bs{z}_j)\psi_{t}(\bs{z}_{j})
\left[X_{i\ell}\xi_{it}I\left\{ \left|X_{i\ell}\xi_{it}\right|
\leq D_{n}\right\} \right]
-\mu _{i}^{D_{n}}\right|^r\\
&\leq 2^{r-1}\Bigg[{\rm E}\left|\sum_{t=1}^{\infty}\lambda_r^{1/2}
\frac{1}{N}
\sum_{\bs{z}_j \in \Omega}\widetilde{B}_{m}(\bs{z}_j)\psi_{t}(\bs{z}_{j})
X_{i\ell}\xi_{it}I\left\{ \left|X_{i\ell}\xi_{it}\right| \leq D_{n}\right\} \right|^{r}
+\left( \mu _{i}^{D_{n}}\right)^{r}\Bigg]  \\
&\leq \left\{3D
_n
\frac{1}{N}
\sum_{\bs{z}_j \in \Omega}\widetilde{B}_{m}(\bs{z}_j)
\sum_{t=1}^{\infty}\lambda_{t}^{1/2}\psi_{t}(\bs{z}_{j})\right\}^{r-2}
E|\varpi_{i,2}^{D_{n}}|^2 \\
& \leq  (CD_n |\triangle|^3)^{r-2}E|\varpi_{i,2}^{D_{n}}|^2.
\end{align*}
Thus, ${\rm E}\left|\varpi_{i,2}/n\right|^{r}\leq \{Cn^{-1}D_{n}|\triangle|^{3}\}
^{r-2}r!{\rm E}(\varpi_{i,2}^{2}/n^2)<\infty $ with the Cramer constant $c^{*}=Cn^{-1}D_{n}|\triangle|^{3}$. By the Bernstein inequality, for any large enough $\delta >0$,
\[
\begin{aligned}
 P\left\{
    \left|\frac{1}{n}\sum_{i=1}^{n}\varpi_i^{D_n}\right|
    \geq \delta n^{-1/2}|\triangle|^3(\log n)^{1/2}
  \right\} &\leq 2\exp\left\{
    \frac{-\delta^2\log n}
         {4c+2\delta CD_n(\log n)^{1/2}n^{-1/2}}
  \right\} \\
  &\leq 2n^{-3}.
\end{aligned}
\]
Therefore, we obtain
\[
\sum_{n=1}^{\infty}P\left\{ \max_{0\leq\ell\leq p} \max_{m \in \mathcal{M}}\left| \frac{1}{n}\sum_{i=1}^{n}\varpi_{i} \right| \geq \delta n^{-1/2}|\triangle|^{3}(\log n)^{1/2}\right\} \leq C|\triangle|^{-3} \sum_{n=1}^{\infty}n^{-3}<\infty
\]
for such $\delta >0$. Thus, Borel-Cantelli's lemma implies that $\left\|\widetilde{\bs{\eta}} \right\|_{\infty}=O_{P}\{n^{-1/2}(\log n)^{1/2}\}$. $\left\|\widetilde{\bs{\epsilon}}_{\Omega}\right\|_{\infty}=O_{P}\{\{nN\}^{-1/2}(\log n)^{1/2}|\triangle|^{-3/2}\} = O_{P}\{(nN)^{-1/2}(\log n)^{1/2}|\triangle|^{-3/2}\}$ can be established similarly, and thus is omitted.
\end{proof}

For $\widetilde{\bs{\beta}}(\bs{z})$ in (\ref{DEF:beta_tilde}), Theorem \ref{THM:bias-rate} below provides its uniform convergence rate to $\bs{\beta}^{o}$.

\begin{theorem}
\label{THM:bias-rate}
Under Assumptions \ref{as:triangulation}--\ref{as:rates}, for $\widetilde{\bs{\beta}}(\bs{z})$ defined in (\ref{DEF:beta_tilde}), we have
\[
\|\widetilde{\bs{\beta}}-\bs{\beta}^{o}\|_{\infty}
=O_{P}\{|\triangle|^{d+1}\|\bs{\beta}^{o}\|_{d+1,\infty}+n^{-1/2}(\log n)^{1/2}\}.
\]
\end{theorem}
\begin{proof}
Note that $\|\widetilde{\bs{\beta}}-\bs{\beta}^{o}\|_{\infty}\leq
\|\widetilde{\bs{\beta}}_{\mu}-\bs{\beta}^{o}\|_{\infty}
+\|\widetilde{\bs{\eta}}\|_{\infty}
+\|\widetilde{\bs{\varepsilon}}\|_{\infty}$, where
\[
\widetilde{\bs{\beta}}_{\mu}=\argmin_{\bs{g}\in \mathcal{G}^{(p+1)}}\sum_{i=1}^{n}\sum_{\bs{z}_j \in \Omega}\left\{\sum_{\ell =0}^{p}X_{i\ell}(\beta_{0,\ell}-g_{\ell})(\bs{z}_{j})\right\}^{2}.
\]
Let $\bs{\beta}^{\ast}=(\beta_{0}^{\ast},\ldots, \beta_{p}^{\ast})^{\ast}\in \mathcal{G}^{(p+1)}$, where $\beta_{\ell}^{\ast}$'s are the best approximation to $\beta_{0,\ell}$'s with the approximation rate $\|\beta_{\ell}^{\ast}-\beta_{0,\ell}\|_{\infty}\leq C
|\triangle|^{d+1}\|\bs{\beta}^{o}\|_{d+1,\infty}$ for any $\ell=0,\ldots,p$. According to \cite{Lai:Wang:13}, we have
\begin{equation}
\|\widetilde{\bs{\beta}}_{\mu}-\bs{\beta}^{o}\|_{\infty}\leq
\|\widetilde{\bs{\beta}}_{\mu}-\bs{\beta}^{\ast}\|_{\infty} + \|\bs{\beta}^{\ast}-\bs{\beta}^{o}\|_{\infty}
\leq C
|\triangle|^{d+1}\|\bs{\beta}^{o}\|_{d+1,\infty}.
\label{EQ:beta_tilde_bias}
\end{equation}
The desired result follows from Lemma \ref{LEM:error-unif-order}.
\end{proof}

\subsection{Asymptotic Properties of Penalized Spline Estimators} 
\label{sec:B-3}

Denote by
\begin{equation}
\bs{\Gamma}_{n,\lam}=\frac{1}{nN}\sum_{i=1}^{n}\sum_{\bs{z}_j \in \Omega}(\widetilde{\mathbf{X}}_{i}\widetilde{\mathbf{X}}_{i}^{\top})\otimes
\{\widetilde{\mathbf{B}}_k^{\ast}(\bs{z}_{j})\widetilde{\mathbf{B}}^{\top}(\bs{z}_{j})\}
+\frac{\lam}{nN} \mathbf{I}_{p} \otimes \mathbf{Q}_2^{\top} \mathbf{P} \mathbf{Q}_2.
\label{DEF:Gamma_rho_DDC}
\end{equation}
It is a symmetric positive definite matrix.
Next, we define
\begin{align}
\widehat{\bs{\theta}}_{\mu}&=(\widehat{\bs{\theta}}_{\mu,0}^{\top},\ldots,\widehat{\bs{\theta}}_{\mu,p}^{\top})^{\top}
=\bs{\Gamma}_{n,\lam}^{-1}\frac{1}{nN}\sum_{i=1}^{n}\sum_{\bs{z}_j \in \Omega} \left\{\widetilde{\mathbf{X}}_{i}\otimes \widetilde{\mathbf{B}}_k(\bs{z}_{j})\right\} \widetilde{\mathbf{X}}_{i}^{\top}\bs{\beta}^{o}(\bs{z}_j), \notag\\
\widehat{\bs{\theta}}_{\eta}&=(\widehat{\bs{\theta}}_{\eta,0}^{\top},\ldots,\widehat{\bs{\theta}}_{\eta,p}^{\top})^{\top}
=\bs{\Gamma}_{n,\lam}^{-1}
\frac{1}{nN}\sum_{i=1}^{n}\sum_{\bs{z}_j \in \Omega} \left\{\widetilde{\mathbf{X}}_{i}\otimes \widetilde{\mathbf{B}}_k(\bs{z}_{j})\right\}\sum_{t=1}^{\infty}\sqrt{\lambda_t}\xi_{it}\psi_{t}(\bs{z}_j),\notag\\
\widehat{\bs{\theta}}_{\varepsilon}&=(\widehat{\bs{\theta}}_{\varepsilon,0}^{\top},\ldots,
\widehat{\bs{\theta}}_{\varepsilon,p}^{\top})^{\top}
=\bs{\Gamma}_{n,\lam}^{-1}
\frac{1}{nN}\sum_{i=1}^{n}\sum_{\bs{z}_j \in \Omega} \left\{\widetilde{\mathbf{X}}_{i}\otimes \widetilde{\mathbf{B}}_k(\bs{z}_{j})\right\}\sigma(\bs{z}_j)\varepsilon_{ij}.
\label{DEF:theta-eta-eps}
\end{align}

Note that, for any $\ell=0,\ldots,p$, the penalized trivariate spline estimator $\widehat{\beta}_{\ell}$ within the subregion $\Omega$ can be written as the following:
\begin{equation}
\widehat{\beta}_{\ell}(\bs{z})=\widehat{\beta}_{\mu,\ell}(\bs{z})+\widehat{\eta}_{\ell}(\bs{z})
+\widehat{\varepsilon}_{\ell}(\bs{z}),
\label{EQ:decompose2}
\end{equation}
where
\begin{equation}
\widehat{\beta}_{\mu,\ell}(\bs{z})=\widetilde{\mathbf{B}}(\bs{z})^{\top}\widehat{\bs{\theta}}_{\mu,\ell},~~
\widehat{\eta}_{\ell}(\bs{z})=\widetilde{\mathbf{B}}(\bs{z})^{\top}\widehat{\bs{\theta}}_{\eta,\ell},~~
\widehat{\varepsilon}_{\ell}(\bs{z})=\widetilde{\mathbf{B}}(\bs{z})^{\top}\widehat{\bs{\theta}}_{\varepsilon,\ell}.
\label{DEF:pls_estimator}
\end{equation}
Therefore,
\begin{equation}
\widehat{\beta}_{\ell}(\bs{z})-\beta^{o}_{\ell}(\bs{z})=\widehat{\beta}_{\mu,\ell}(\bs{z})-\beta^{o}_{\ell}(\bs{z})
+\widehat{\eta}_{\ell}(\bs{z})+\widehat{\varepsilon}_{\ell}(\bs{z}).
\label{EQ:decompose3}
\end{equation}

\begin{lemma}
\label{LEM:DDC-Gamma_rho}
Under Assumptions \ref{as:triangulation}, \ref{as:design}, and \ref{as:rates}, if $N^{1/3}|\triangle|\rightarrow \infty$ as $N\rightarrow \infty$, then there exist constants $0 < c_{\Gamma} < C_{\Gamma} < \infty$, such that with probability approaching 1 as $n\rightarrow \infty$,
\[
c_{\Gamma}|\triangle|^{3} \leq \lambda_{\min}(\bs{\Gamma}_{n,\lam}) \leq
\lambda_{\max}(\bs{\Gamma}_{n,\lam}) \leq  C_{\Gamma}\left\{|\triangle|^{3}+(nN|\triangle|)^{-1}\lam\right\}.
\]
\end{lemma}
\begin{proof}
By (\ref{EQ:g_gamma}), it is easy to see that, for any vector $\bs{\theta}=(\bs{\theta}_0^{\top},\cdots,\bs{\theta}_p^{\top})^{\top}$,
$\bs{\theta}^{\top}\bs{\Gamma}_{n,\lam}\bs{\theta} =\|\bs{g}_{\bs{\gamma}}\|_{n}^2
+(nN)^{-1}\lam\sum_{\ell =0}^{p} \bs{\gamma}_{\ell}^{\top}\mathbf{P}_{\Omega} \bs{\gamma}_{\ell}$, where $\bs{\gamma}=(\bs{\gamma}_0,\ldots, \bs{\gamma}_p)^{\top}=\mathbf{Q}_2\bs{\theta} $ with $\bs{\gamma}_{\ell}=\{\gamma_{\ell m},m\in\mathcal{M}(\Omega)\}^{\top}$.
Using Markov's inequality, we have
\[
\sum_{\ell =0}^{p}\left\|\sum_{m\in \mathcal{M}}\gamma_{\ell m}B_{m}\right\|_{\mathcal{E}}^2\leq \frac{C}{|\triangle|^{4}}\sum_{\ell =0}^{p}\left\Vert \sum_{m\in \mathcal{M}}\gamma_{\ell m}B_{m}\right\Vert_{L _2} ^{2}\leq \frac{C}{|\triangle|}\Vert\bs{\gamma}\Vert^{2}.
\]
Thus, the largest eigenvalue of $\bs{\Gamma}_{n,\lam}$ in (\ref{DEF:Gamma_rho_DDC}) satisfies that $\lambda_{\max}(\bs{\Gamma}_{n,\lam}) \leq C\{(1+R_{n})
|\triangle|^{3}+(nN|\triangle|)^{-1}\lam)\}$. Thus, with probability approaching 1, we have
$\lambda_{\max}(\bs{\Gamma}_{n,\lam})\leq C_{\Gamma} (|\triangle|^{3}+(nN|\triangle|)^{-1}\lam)$
for some positive constant $C_{\Gamma}$.

On the other hand, we use Lemma
\ref{LEM:DDC-norm} and equation (\ref{EQ:normratio-DDC})  to have $\Vert \bs{g}_{\bs{\gamma}}\Vert_{n}^{2} =(1-R_{n})
\Vert\bs{g}_{\bs{\gamma}}\Vert^{2}_{\Omega} \geq c (1-R_{n})
|\triangle|^{3}\Vert \bs{\gamma}\Vert
^{2}$.

Therefore, the smallest eigenvalue of $\bs{\Gamma}_{n,\lam}$ is greater than $c(1-R_{n})|\triangle|^{3}=c_{\Gamma}|\triangle|^{3}$. Thus, we can obtain Lemma \ref{LEM:DDC-Gamma_rho}.
\end{proof}

In the following, Lemma \ref{LEM:DDC-uniformbiasrate} establishes the uniform convergence rate of $\widehat{\beta}_{\mu,\ell}(\bs{z})$ to $\beta^{o}_{\ell}(\bs{z})$. Additionally, Lemmas \ref{LEM:DDC-thetatilde-eta} and \ref{LEM:DDC-thetatilde-eps} provide the asymptotic convergence rates for the spline coefficients.

\begin{lemma}
\label{LEM:DDC-uniformbiasrate}
Under Assumptions \ref{as:triangulation}, \ref{as:smoothness}, and \ref{as:design}, if $N^{1/3}|\triangle| \to \infty$, we have
\[
\|\widehat{\bs{\beta}}_{\mu} - \bs{\beta}^{o}\|_{\infty} = O_{P}\left\{ \frac{\lam}{nN|\triangle|^{3}}\|\bs{\beta}^{o}\|_{2,\infty} + \left(1+\frac{\lam}{nN|\triangle|^{5}}\right) |\triangle|^{d+1}\|\bs{\beta}^{o}\|_{d+1,\infty} \right\}.
\]
\end{lemma}

\vspace{-6pt}
\begin{lemma}
\label{LEM:DDC-thetatilde-eta}
Suppose Assumptions \ref{as:triangulation}--\ref{as:errors} hold and $N^{1/3}|\triangle| \rightarrow \infty$ as $N \rightarrow \infty$, then $\|\widehat{\bs{\theta}}_{\eta}\|^2 = O_P\{n^{-1}|\triangle|^{-3}\}$.
\end{lemma}

\vspace{-6pt}
\begin{lemma}
\label{LEM:DDC-thetatilde-eps}
Suppose Assumptions \ref{as:triangulation}--\ref{as:errors} hold and $N^{1/3}|\triangle| \rightarrow \infty$ as $N \rightarrow \infty$, then $\|\widehat{\bs{\theta}}_{\varepsilon}\|^2 = O_P\{(nN)^{-1}|\triangle|^{-6}\}$.
\end{lemma}

\begin{proof}[Proof of Lemma \ref{LEM:DDC-uniformbiasrate}]
Define
\begin{equation}
A_{n} = \sup_{\bs{g} \in \mathcal{G}^{(p+1)}} \left\{ \frac{\|\bs{g}\|_{\infty}}{\|\bs{g}\|_{n}},\|\bs{g}\|_{n} \neq 0 \right\}, ~~\overline{A}_{n} = \sup_{\bs{g} \in \mathcal{G}^{(p+1)}} \left\{ \frac{\|\bs{g}\|_{\mathcal{E}}}{\|\bs{g}\|_{n}},\|\bs{g}\|_{n} \neq 0 \right\},
\label{DEF:DDC-An}
\end{equation}
where random variables $A_{n}$ and $\overline{A}_{n}$ depend on the collection of $X_{i\ell}$'s, $i=1,\ldots,n$, $\ell=0,\ldots,p$. It is clear that $\|\bs{\beta}^{o} - \widehat{\bs{\beta}}_{\mu}\|_{\infty} \leq \|\bs{\beta}^{o}-\widetilde{\bs{\beta}}_{\mu}\|_{\infty} + \|\widetilde{\bs{\beta}}_{\mu} - \widehat{\bs{\beta}}_{\mu}\|_{\infty}$, where $\widetilde{\bs{\beta}}_{\mu}$ is given in (\ref{DEF:beta_tilde_mu}), and $\|\widetilde{\bs{\beta}}_{\mu} - \bs{\beta}^{o}\|_{\infty} \leq C|\triangle|^{d+1}\|\bs{\beta}^{o}\|_{d+1,\infty}$ according to (\ref{LEM:appord}).

By the definition of $A_{n}$ in (\ref{DEF:DDC-An}), we have
\begin{equation}
\|\widetilde{\bs{\beta}}_{\mu} - \widehat{\bs{\beta}}_{\mu}\|_{\infty} \leq A_{n}\|\widetilde{\bs{\beta}}_{\mu} - \widehat{\bs{\beta}}_{\mu}\|_{n}.
\label{EQ:DDC-pls-ps-sup}
\end{equation}
Note that the penalized spline $\widehat{\bs{\beta}}_{\mu}$ of $\bs{\beta}^{o}$ is characterized by the orthogonality relations
\begin{eqnarray}
nN\langle\bs{\beta}^{o} - \widehat{\bs{\beta}}_{\mu},\bs{g}\rangle_{n} = \lam \langle \widehat{\bs{\beta}}_{\mu},\bs{g} \rangle_{\mathcal{E}}, ~\textrm{for all}~ \bs{g} \in \mathcal{G}^{(p+1)},
\label{EQ:DDC-pls-bias}
\end{eqnarray}
while $\widetilde{\bs{\beta}}_{\mu}$ is characterized by
\begin{eqnarray}
\langle\bs{\beta}^{o} - \widetilde{\bs{\beta}}_{\mu},\bs{g}\rangle_{n} = 0, ~\textrm{for all}~ \bs{g} \in \mathcal{G}^{(p+1)}.
\label{EQ:DDC-ls-bias}
\end{eqnarray}
By (\ref{EQ:DDC-pls-bias}) and (\ref{EQ:DDC-ls-bias}), we have $nN\langle \widetilde{\bs{\beta}}_{\mu} - \widehat{\bs{\beta}}_{\mu},\bs{g}\rangle_{n} = \lam \langle \widehat{\bs{\beta}}_{\mu},\bs{g} \rangle_{\mathcal{E}}$, for all $\bs{g}\in \mathcal{G}^{(p+1)}$. Inserting $\bs{g} = \widetilde{\bs{\beta}}_{\mu} - \widehat{\bs{\beta}}_{\mu}$ yields that
\begin{eqnarray}
nN\|\widetilde{\bs{\beta}}_{\mu} - \widehat{\bs{\beta}}_{\mu}\|_{n}^{2} = \lam \langle \widehat{\bs{\beta}}_{\mu},\widetilde{\bs{\beta}}_{\mu} - \widehat{\bs{\beta}}_{\mu} \rangle_{\mathcal{E}}.  \label{EQ:DDC-pls-ls}
\end{eqnarray}
Thus, by Cauchy-Schwarz inequality and the definition of $\overline{A}_{n}$, we have
\[
nN\|\widetilde{\bs{\beta}}_{\mu} - \widehat{\bs{\beta}}_{\mu}\|_{n}^{2} \leq \lam \|\widehat{\bs{\beta}}_{\mu}\|_{\mathcal{E}}\|\widetilde{\bs{\beta}}_{\mu} - \widehat{\bs{\beta}}_{\mu}\|_{\mathcal{E}} \leq \lam \overline{A}_{n} \|\widehat{\bs{\beta}}_{\mu}\|_{\mathcal{E}}\| \widetilde{\bs{\beta}}_{\mu} - \widehat{\bs{\beta}}_{\mu}\|_{n}.
\]
Similarly, using (\ref{EQ:DDC-pls-ls}), we have $nN\|\widetilde{\bs{\beta}}_{\mu} - \widehat{\bs{\beta}}_{\mu}\|_{n}^{2} = \lam \{\langle \widehat{\bs{\beta}}_{\mu},\widetilde{\bs{\beta}}_{\mu} \rangle_{\mathcal{E}} - \langle \widehat{\bs{\beta}}_{\mu},\widehat{\bs{\beta}}_{\mu} \rangle_{\mathcal{E}}\} \geq 0$. Thus, by Cauchy-Schwarz inequality, $\|\widehat{\bs{\beta}}_{\mu}\|_{\mathcal{E}}^{2} \leq \langle \widehat{\bs{\beta}}_{\mu},\widetilde{\bs{\beta}}_{\mu} \rangle_{\mathcal{E}} \leq \|\widehat{\bs{\beta}}_{\mu}\|_{\mathcal{E}}\|\widetilde{\bs{\beta}}_{\mu}\|_{\mathcal{E}}$, which implies that $\|\widehat{\bs{\beta}}_{\mu}\|_{\mathcal{E}} \leq \|\widetilde{\bs{\beta}}_{\mu}\|_{\mathcal{E}}$. Therefore,
\begin{eqnarray}
\|\widetilde{\bs{\beta}}_{\mu} - \widehat{\bs{\beta}}_{\mu}\|_{n} \leq \lam (nN)^{-1}\overline{A}_{n} \|\widetilde{\bs{\beta}}_{\mu}\|_{\mathcal{E}}. \label{EQ:DDC-pls-ps2}
\end{eqnarray}
Combining (\ref{EQ:DDC-pls-ps-sup}) and (\ref{EQ:DDC-pls-ps2}) yields that
$$
\|\widetilde{\bs{\beta}}_{\mu} - \widehat{\bs{\beta}}_{\mu}\|_{\infty,\Omega} \leq A_{n} \|\widetilde{\bs{\beta}}_{\mu} - \widehat{\bs{\beta}}_{\mu}\|_{n} \leq \lam (nN)^{-1}A_{n} \overline{A}_{n} \|\widetilde{\bs{\beta}}_{\mu}\|_{\mathcal{E}}.
$$
By Lemma \ref{LEM:appord}, we have
\[
\|\widetilde{\bs{\beta}}_{\mu}\|_{\mathcal{E}} = C_{1}
\{\|\bs{\beta}^{o}\|_{2,\infty} + \sum_{i=1}^3\|\nabla_{z_{i}}^2(\bs{\beta}^{o} - \widetilde{\bs{\beta}}_{\mu})\|_{\infty}\} \leq C_{2}(\|\bs{\beta}^{o}\|_{2,\infty} + |\triangle|^{d-1}\|\bs{\beta}^{o}\|_{d +1,\infty}).
\]
It follows
\begin{eqnarray}
\label{EQ:DDC-pls-ps-sup-2}
\|\widetilde{\bs{\beta}}_{\mu}-\widehat{\bs{\beta}}_{\mu}\|_{\infty,\Omega}=\lam (nN)^{-1}A_{n}\overline{A}_{n}
C_{2}(\left\|\bs{\beta}^{o}\right\|_{2,\infty}+|\triangle|^{d -1}\|\bs{\beta}^{o}\|
_{d+1,\infty}) .
\end{eqnarray}
Next, we derive the order of $A_{n}$ and $\overline{A}_{n}$. By Markov's inequality, for any $\bs{g}\in \mathcal{G}^{(p+1)}$, $\|\bs{g}\|_{\infty}\leq C|\triangle| ^{-1}\|\bs{g}\|_{L_2(\Omega)}$, $\|\bs{g}\|_{\mathcal{E}}\leq C|\triangle|^{-2}\|\bs{g}\|_{L_2(\Omega)}$. Equation (\ref{EQ:normratio-DDC}) implies that
\[
\sup_{\bs{g}\in\mathcal{G}(\triangle)}\left\{ \left. \|\bs{g}\|_{n}\right/
\|\bs{g}\|_{L_2(\Omega)}\right\} \geq \left[1-O_{P}\left\{(\log n)^{1/2}{n}^{-1/2}+ N^{-1/3}|\triangle|^{-1}\right\}\right]^{1/2}.
\]
Thus, we have
\begin{align*}
A_{n} &\leq C|\triangle| ^{-1}\left[1-O_{P}\left\{(\log n)^{1/2}{n}^{-1/2}+ N^{-1/3}|\triangle|^{-1}\right\} \right]
^{-1/2}=O_{P}\left( |\triangle| ^{-1}\right) , \\
\overline{A}_{n} &\leq C|\triangle|
^{-2}\left[1-O_{P}\left\{(\log n)^{1/2}{n}^{-1/2}+ N^{-1/3}|\triangle|^{-1}\right\}
\right]^{-1/2}=O_{P}\left( |\triangle| ^{-2}\right).
\end{align*}
Plugging the order of $A_{n}$ and $\overline{A}_{n}$ into (\ref{EQ:DDC-pls-ps-sup-2}) yields that
\[
\|\widetilde{\bs{\beta}}_{\mu}-\widehat{\bs{\beta}}_{\mu}\|_{\infty,\Omega}
=O_{P}\left\{ \frac{\lam}{nN|\triangle| ^{3}}(\|\bs{\beta}^{o}\|_{2,\infty}
+|\triangle|^{d -1}\|\bs{\beta}^{o}\|_{d +1,\infty}) \right\}.
\]
Therefore, we can obtain
\[
\|\widehat{\bs{\beta}}_{\mu}-\bs{\beta}^{o}\|_{\infty,\Omega}\leq C_{1}|\triangle|^{d +1}\|\bs{\beta}^{o}\|_{d+1,\infty}+O_{P}\left\{ \frac{\lam}{nN|\triangle|^{3}}
\left(\|\bs{\beta}^{o}\|_{2,\infty}+|\triangle|^{d -1}\|
\bs{\beta}^{o}\|_{d +1,\infty}\right) \right\} .
\]
Therefore, Lemma \ref{LEM:DDC-uniformbiasrate} is established.
\end{proof}

\begin{proof}[Proof of Lemma \ref{LEM:DDC-thetatilde-eta}]
Note that 
\[
\widehat{\bs{\theta}}_{\eta}=(\bs{\Gamma}_{n,\lam})^{-1}
\frac{1}{nN}\sum_{i=1}^{n}\sum_{\bs{z}_j \in \Omega} \left\{\widetilde{\mathbf{X}}_{i}\otimes \widetilde{\mathbf{B}}_k(\bs{z}_{j})\right\}\sum_{t=1}^{\infty}\sqrt{\lambda_t}\xi_{it}\psi_{t}(\bs{z}_j).
\]
By Lemma \ref{LEM:DDC-Gamma_rho}, we have
\[
\begin{aligned}
& \|\widehat{\bs{\theta}}_{\eta}\|^2
  \asymp \frac{V^2(\Omega)}{n^2N^2|\triangle|^6}
  \sum_{\substack{i,i^{\prime}=1,\ldots,n \\
                   \bs{z}_j, \bs{z}_{j'} \in \Omega}}
  \left\{\widetilde{\mathbf{X}}_{i}
  \otimes\widetilde{\mathbf{B}}(\bs{z}_j)\right\}^{\top}
  \sum_{t=1}^{\infty}\sqrt{\lambda_t}\xi_{it}\psi_t(\bs{z}_j) \\
&\qquad {}\times
  \left\{\widetilde{\mathbf{X}}_{i^{\prime}}
  \otimes\widetilde{\mathbf{B}}(\bs{z}_{j^{\prime}})\right\}
  \sum_{t=1}^{\infty}\sqrt{\lambda_t}\xi_{i^\prime t}
  \psi_t(\bs{z}_{j^\prime}).
\end{aligned}
\]
Note that
\[
\begin{aligned}
 \widetilde{\mathbf{X}}_{i}\otimes\widetilde{\mathbf{B}}(\bs{z}_j)
  \sum_{t=1}^{\infty}\sqrt{\lambda_t}\xi_{it}\psi_t(\bs{z}_j)
  &= \Bigg(
    X_{i0}\widetilde{\mathbf{B}}(\bs{z}_j)^\top
    \sum_{t=1}^{\infty}\sqrt{\lambda_t}\xi_{it}\psi_t(\bs{z}_j),
    \ldots, \\
&\qquad
    X_{ip}\widetilde{\mathbf{B}}(\bs{z}_j)^\top
    \sum_{t=1}^{\infty}\sqrt{\lambda_t}\xi_{it}\psi_t(\bs{z}_j)
  \Bigg)^\top.
\end{aligned}
\]
Therefore, one has $\|\widehat{\bs{\theta}}_{\eta}\|^2$ follows that
\[
\begin{aligned}
 \|\widehat{\bs{\theta}}_{\eta}\|^2
  &\asymp \frac{1}{n^2N^2|\triangle|^6}
  \sum_{\ell=0}^{p}\sum_{i,i^{\prime}=1}^{n}
  \sum_{\bs{z}_j,\bs{z}_{j'}\in\Omega}
  X_{i\ell}X_{i^{\prime}\ell}
  \widetilde{\mathbf{B}}(\bs{z}_j)^\top
  \widetilde{\mathbf{B}}(\bs{z}_{j^{\prime}}) \\
&\qquad \times
  \sum_{t,t^{\prime}=1}^{\infty}
  \sqrt{\lambda_t\lambda_{t'}}
  \xi_{it}\psi_t(\bs{z}_j)
  \xi_{i^{\prime}t^{\prime}}\psi_{t^{\prime}}(\bs{z}_{j^{\prime}}).
\end{aligned}
\]
Notice that the eigenvalues of $\mathbf{Q}_2\mathbf{Q}_2^{\top}$ are either 0 or 1. By Assumptions \ref{as:errors} and \ref{as:design},
for any $\ell,i$, we have
\begin{align*}
\frac{1}{N^2}&\sum_{\bs{z}_j, \bs{z}_{j'} \in \Omega}
{\rm E}\left\{X_{i\ell}^{2}\widetilde{\mathbf{B}}(\bs{z}_j)^{\top}\widetilde{\mathbf{B}}(\bs{z}_{j^{\prime}})\sum_{t, t^{\prime}=1}^{\infty}\sqrt{\lambda_t\lambda_{t'}}\xi_{it}\psi_{t}(\bs{z}_j)
\xi_{it^{\prime}}\psi_{t^{\prime}}(\bs{z}_{j^{\prime}})\right\}\\
& \leq \frac{1}{N^2}\sum_{\bs{z}_j, \bs{z}_{j'} \in \Omega} {\rm E}\left\{X_{i\ell}^{2}\sum_{m\in \mathcal{M}_{\Omega_k}}B_{m}(\bs{z}_j)B_{m}(\bs{z}_{j^{\prime}})\sum_{t=1}^{\infty}\lambda_t\xi_{it}^{2}
\psi_{t}(\bs{z}_j)\psi_{t}(\bs{z}_{j^{\prime}})\right\}\\
&\leq C \sum_{m\in \mathcal{M}}
\frac{1}{N^2}\sum_{\bs{z}_j, \bs{z}_{j'} \in \Omega} B_{m}(\bs{z}_j)B_{m}(\bs{z}_{j^{\prime}})
G_{\eta}(\bs{z}_j,\bs{z}_{j^{\prime}}).
\end{align*}
Assumption \ref{as:triangulation} and (\ref{DDC-lemmaA18_3}) imply that
\begin{align*}
\frac{1}{N^2}&\sum_{j\neq j^{\prime}} B_{m}(\bs{z}_j)B_{m}(\bs{z}_{j^{\prime}})
G_{\eta}(\bs{z}_j,\bs{z}_{j^{\prime}})=\int_{T_{m}\times T_{m}}G_{\eta}(\bs{z},\bs{z}^{\prime})B_m(\bs{z})B_m(\bs{z}^{\prime})
d\bs{z}d\bs{z}^{\prime}\\
& \times \{1+O(N^{-1/3}|\triangle|^{5})\}=O(|\triangle|^{6}).
\end{align*}
Thus, we have
\[
\frac{1}{N^2}\sum_{\bs{z}_j, \bs{z}_{j'} \in \Omega} {\rm E}\left\{X_{i\ell}^{2}\mathbf{B}(\bs{z}_j)^{\top}\mathbf{B}(\bs{z}_{j^{\prime}})\sum_{t, t^{\prime}=1}^{\infty}\sqrt{\lambda_t\lambda_{t'}}\xi_{it}\psi_{t}(\bs{z}_j)
\xi_{it^{\prime}}\psi_{t^{\prime}}(\bs{z}_{j^{\prime}})\right\} \leq C|\triangle|^{3}.
\]
Next, for any $\ell$, $i\neq i^{\prime}$, $j$, $j^{\prime}$, we have
\begin{align*}
{\rm E}X_{i\ell}X_{i^{\prime}\ell}& \mathbf{B}(\bs{z}_j)^{\top}
\mathbf{B}(\bs{z}_{j^{\prime}}) \sum_{t, t^{\prime}=1}^{\infty}\sqrt{\lambda_t\lambda_{t'}}\xi_{it}\psi_{t}(\bs{z}_j)
\xi_{it^{\prime}}\psi_{t^{\prime}}(\bs{z}_{j^{\prime}})\\
&={\rm E}X_{i\ell}X_{i^{\prime}\ell}\sum_{m\in \mathcal{M}}B_{m}^{2}(\bs{z}_j)B_{m}^{2}(\bs{z}_{j^{\prime}}) \sum_{t,t^{\prime}} {\rm E}\sqrt{\lambda_{t}\lambda_{t'}}\xi_{it}\xi_{i^{\prime}t^{\prime}}\psi_{t}(\bs{z}_{j})\psi_{t^{\prime}}(\bs{z}_{j^{\prime}})=0.
\end{align*}
Therefore, ${\rm E}\|\widehat{\bs{\theta}}_{\eta}\|^2
\leq C p \{n^{-1}|\triangle|^{-3}\}$. The conclusion of the lemma follows.
\end{proof}

\begin{proof}[Proof of Lemma \ref{LEM:DDC-thetatilde-eps}]
By the definition of $\widehat{\bs{\theta}}_{\varepsilon}$ in (\ref{DEF:theta-eta-eps}), we have
\begin{align*}
	\|\widehat{\bs{\theta}}_{\varepsilon}\|^2&=\frac{1}{n^2N^2|\triangle|^6}\left\{|\triangle|^{-3}\bs{\Gamma}_{n,\lam}\right\}^{-1}
\sum_{i=1}^{n}\sum_{\bs{z}_{j} \in \Omega} \left\{\widetilde{\mathbf{X}}_{i}\otimes \widetilde{\mathbf{B}}_k(\bs{z}_{j})\right\}^{\top}\sigma(\bs{z}_j)\varepsilon_{ij}\\
	&\quad \times\left\{|\triangle|^{-3}\bs{\Gamma}_{n,\lam}\right\}^{-1}
	\sum_{i=1}^{n} \sum_{\bs{z}_{j'} \in \Omega} \left\{\widetilde{\mathbf{X}}_{i}\otimes \widetilde{\mathbf{B}}(\bs{z}_{j^{\prime}})\right\}\sigma(\bs{z}_{j^{\prime}})\varepsilon_{ij^{\prime}}.
\end{align*}
	By Lemma \ref{LEM:DDC-Gamma_rho}, we can have
	\[
\|\widehat{\bs{\theta}}_{\varepsilon}\|^2 \asymp \frac{1}{n^2N^2|\triangle|^6} \sum_{i,i^{\prime}=1}^{n}\sum_{\bs{z}_j, \bs{z}_{j'} \in \Omega}\left\{\widetilde{\mathbf{X}}_{i}
\otimes\widetilde{\mathbf{B}}(\bs{z}_j)\right\}^{\top}\sigma(\bs{z}_j)\varepsilon_{ij} \left\{\mathbf{X}_{i^{\prime}}\otimes\widetilde{\mathbf{B}}(\bs{z}_{j^{\prime}})\right\}
\sigma(\bs{z}_{j^{\prime}})\varepsilon_{i^{\prime}j^{\prime}}.
\]
 Note that
\[
\begin{aligned}
 \widetilde{\mathbf{X}}_i\otimes
  \widetilde{\mathbf{B}}(\bs{z}_j)\sigma(\bs{z}_j)\varepsilon_{ij}
  &= \Big(
    X_{i0}\widetilde{\mathbf{B}}(\bs{z}_j)^\top
    \sigma(\bs{z}_j)\varepsilon_{ij},
    X_{i1}\widetilde{\mathbf{B}}(\bs{z}_j)^\top
    \sigma(\bs{z}_j)\varepsilon_{ij},\ldots, \\
  &\qquad
    X_{ip}\widetilde{\mathbf{B}}(\bs{z}_j)^\top
    \sigma(\bs{z}_j)\varepsilon_{ij}
  \Big)^\top.
\end{aligned}
\]
Then, we have
\[
\|\widehat{\bs{\theta}}_{\varepsilon}\|^2  \asymp \frac{1}{n^2N^2|\triangle|^6} \sum_{\ell =0}^{p} \sum_{i,i^{\prime}=1}^{n}\sum_{\bs{z}_j, \bs{z}_{j'} \in \Omega} X_{i\ell}X_{i^{\prime}\ell}\widetilde{\mathbf{B}}(\bs{z}_j)^{\top}
\widetilde{\mathbf{B}}(\bs{z}_{j^{\prime}}) \sigma(\bs{z}_j) \sigma(\bs{z}_{j^{\prime}})\varepsilon_{ij} \varepsilon_{i^{\prime}j}.
 \]
Because the eigenvalues of $\mathbf{Q}_2\mathbf{Q}_2^{\top}$ are either 0 or 1. Under Assumptions \ref{as:triangulation} and \ref{as:errors}, for any $\ell,i$, we have
\begin{align*}
& \frac{1}{N}\sum_{\bs{z}_j\in\Omega}
  \widetilde{\mathbf{B}}(\bs{z}_j)^\top
  \widetilde{\mathbf{B}}(\bs{z}_j)\sigma^2(\bs{z}_j)
  = \frac{1}{N}\sum_{\bs{z}_j\in\Omega}
    \mathbf{B}(\bs{z}_j)^\top\mathbf{Q}_2\mathbf{Q}_2^\top
    \mathbf{B}(\bs{z}_j)\sigma^2(\bs{z}_j) \\
& \leq C\!\!\sum_{m\in\mathcal{M}}\frac{1}{N}
    \sum_{\bs{z}_j\in\Omega}B_m^2(\bs{z}_j)\sigma^2(\bs{z}_j) \\
& \leq C\!\!\sum_{m\in\mathcal{M}}
    \int_{T_{\lceil m/d^{\ast}\rceil}}\!\!
    \sigma^2(\bs{z})B_m^2(\bs{z})\,d\bs{z}
    \{1+O(N^{-1/3}|\triangle|^2)\}
    \leq C.
\end{align*}
Next, note that for any $\ell,i,j\neq j^{\prime}$,
${\rm E}\{X_{i\ell}^{2}\widetilde{\mathbf{B}}(\bs{z}_j)^{\top}
    \widetilde{\mathbf{B}}(\bs{z}_{j^{\prime}})\varepsilon_{ij} \varepsilon_{ij^{\prime}}\}=0$,
and for any $\ell$, $i\neq i^{\prime}$, $j$, $j^{\prime}$, ${\rm E}\{X_{i\ell}X_{i^{\prime}\ell}\widetilde{\mathbf{B}}(\bs{z}_j)^{\top} \widetilde{\mathbf{B}}(\bs{z}_{j^{\prime}})\sigma(\bs{z}_j)\sigma(\bs{z}_{j^{\prime}})
\varepsilon_{ij}\varepsilon_{i^{\prime}j^{\prime}}\}=0$.
Therefore, we have
\[
{\rm E}\|\widehat{\bs{\theta}}_{\varepsilon}\|^2 \asymp \frac{1}{nN^2|\triangle|^6} \sum_{\ell =0}^{p} {\rm E}(X_{i\ell}^{2})\sum_{\bs{z}_j \in \Omega} \widetilde{\mathbf{B}}(\bs{z}_j)^{\top}
\widetilde{\mathbf{B}}(\bs{z}_{j}) \sigma^{2}(\bs{z}_j)
\leq Cp (nN)^{-1}  |\triangle|^{-6}.
\]
The conclusion of the lemma follows.
\end{proof}

\begin{proof}[Proof of Theorem \ref{THM:beta-convergence}]
Let $\Omega = \Omega$. Then, Theorem \ref{THM:beta-convergence} follows by combining Lemmas \ref{LEM:DDC-uniformbiasrate}, \ref{LEM:DDC-thetatilde-eta}, and \ref{LEM:DDC-thetatilde-eps}.
\end{proof}

\subsection{Convergence Rates and Consistency of the SSFR estimator}
\label{ssec:ssfr-rates}

Write $\Ssp=\mathcal S_d^r(\tri)$ for the space of degree-$d$, $\mathcal C^r$ trivariate
splines over the triangulation $\tri$, let $\mathbf B(\bs z)=\{B_m(\bs z)\}_{m\in\mathcal M}$
be its Bernstein--B\'ezier basis with index set $\mathcal M$ and cardinality
$|\mathcal M|\asymp|\tri|^{-3}$, and let $\Stp$ denote the tensor-product space
$\Ssp\otimes\Ssp$ with basis $\Bzz=\Bz\otimes\mathbf B(\bs z')$.  Let
\[
\delta_i=\mathbf 1(i\in\mathcal I_O),\qquad
n_O=\sum_{i=1}^n \delta_i,\qquad
W_O=\sum_{i=1}^n \delta_i w_i .
\]

The surrogate-adjusted (second-stage) model is, for the observed subjects
$i\in\IO$,
\begin{equation}
\label{eq:ssfr-adj}
Y_i(\bs z)=\mathbf{X}_i^\top\bs\nu_0(\bs z)+C(\bs z,\widetilde Y_i)+\zeta_i(\bs z),
\qquad
C(\bs z,\widetilde Y_i)=\int_\Omega U(\bs z,\bs z')\,\widetilde Y_i(\bs z')\,d\bs z',
\end{equation}
where $\bs\nu_0(\bs z)=\bs\alpha_0(\bs z)-\int_\Omega U(\bs z,\bs z')\bs\beta_0(\bs z')\,d\bs z'$.
 We approximate $\bs\nu(\bs z)\approx\mathbf I_{p+1}\otimes\Bz^\top\bs a$
and $U(\bs z,\bs z')\approx\Bzz^\top\bs b$.
Stacking over $(i,j)$, $i\in\IO$, $j=1,\dots,N$, define the two design blocks
\begin{align}
 \label{eq:blocks} 
 & \mathbb{Z}:\ \text{rows }\ \widetilde{\mathbf X}_{ij}^\top=\mathbf X_i^\top\otimes\mathbf B(\bs z_j)^\top,  \\
 \notag
& \mathbb{X}:\ \text{rows }\ \mathbf C_{ij}^\top,~
\mathbf C_{ij}=\Big\{B_m(\bs z_j)\!\int_\Omega\! B_{m'}(\bs z')\widetilde Y_i(\bs z')d\bs z'\Big\}_{m,m'\in\mathcal M},
\end{align}
so that $\widetilde{\mathbf X}_{ij}^\top\bs a=\mathbf X_i^\top\bs\nu(\bs z_j)$ and
$\mathbf C_{ij}^\top\bs b=\int_\Omega \mathbf{B}( \bs z_j, \bs z)^\top\bs b\,\widetilde Y_i(\bs z')\,d\bs z'$.
Let $\mathbf W=\diag(w_i)$ (constant within subject) collect the IPW weights,
and let
\[
\mathbf H_x=\mathbb{X}(\mathbb{X}^\top\mathbf W\mathbb{X})^{-1}\mathbb{X}^\top\mathbf W,
\qquad
\mathbf H_z=\mathbb{Z}(\mathbb{Z}^\top\mathbf W\mathbb{Z})^{-1}\mathbb{Z}^\top\mathbf W
\]
be the $\mathbf W$-weighted projections onto the column spaces of $\mathbb X$
and $\mathbb Z$.
In the notation of this section, the stage-two criterion (6) of the main paper reads
\begin{equation}
\label{eq:L-stage2}
L(\bs a,\bs b)=\frac{1}{W_O}\sum_{i\in\IO} w_i\sum_{j=1}^N \big(Y_{ij}-\widetilde{\mathbf X}_{ij}^\top\bs a-\mathbf C_{ij}^\top\bs b\big)^2
+\sum_{\ell=0}^p\lambda_\ell\,\bs a_\ell^\top\mathbf P_\ell\bs a_\ell
+\tau_U\,\bs b^\top\bs b ,
\end{equation}
with the roughness penalties acting on the $\nu$-block and the ridge penalty $\tau_U\bs b^\top\bs b$, $\tau_U\ge0$, acting on the tensor-product coefficient vector $\bs b$. The two penalties play different roles in the proofs. The roughness penalties add a lower-order term to the $\nu$-block Gram matrix, exactly as in Lemma \ref{LEM:DDC-Gamma_rho}, and are handled verbatim as in Section \ref{sec:B-3}; we therefore suppress them below. The ridge penalty cannot be suppressed in the same way, for two reasons: it alters the Frisch--Waugh algebra, because the ridge smoother that replaces $\mathbf H_x$ no longer annihilates $\mathbb X$, and it introduces a shrinkage bias in the oracle (noiseless) fits. We therefore track it explicitly. Write
\begin{equation}
\label{eq:tau-def}
\tau=W_O\,\tau_U ,
\qquad
r_\tau=\frac{\tau_U}{N|\tri|^{6}} ,
\end{equation}
for the ridge parameter on the scale of the unnormalized normal equations and for the ridge-to-fit ratio of Assumption \ref{as:rates}; the role of $r_\tau$ is explained in Remark \ref{rem:ridge-scale}.
Following the $1/\sum_{i\in\IO}w_i$ normalization of the data-fit term in \eqref{eq:L-stage2}, the
(unpenalized) partialled Gram matrices that drive the second-stage estimators
are
\begin{equation}
\label{eq:AzAx}
\begin{aligned}
\mathbf A_z
  &= \frac{1}{N\sum_{i\in\IO}w_i}\,
     \mathbb{Z}^\top\mathbf W(\mathbf I-\mathbf H_x)\mathbb{Z}
     \quad\text{(for }\widehat{\bs\nu}\text{)}, \\
\mathbf A_x
  &= \frac{1}{N\sum_{i\in\IO}w_i}\,
     \mathbb{X}^\top\mathbf W(\mathbf I-\mathbf H_z)\mathbb{X}
     \quad\text{(for }\widehat U\text{)},
\end{aligned}
\end{equation}
and their ridge counterparts are
\begin{equation}
\label{eq:AzAx-tau}
\mathbf A_{z,\tau}=\frac{1}{NW_O}\,\mathbb Z^\top\mathbf W(\mathbf I-\mathbf H_{x,\tau})\mathbb Z,
\qquad
\mathbf A_{x,\tau}=\frac{1}{NW_O}\,\big\{\mathbb X^\top\mathbf W(\mathbf I-\mathbf H_z)\mathbb X+\tau\mathbf I\big\}
=\mathbf A_x+\frac{\tau_U}{N}\,\mathbf I ,
\end{equation}
where
\[
\mathbf H_{x,\tau}=\mathbb X\big(\mathbb X^\top\mathbf W\mathbb X+\tau\mathbf I\big)^{-1}\mathbb X^\top\mathbf W
\]
is the $\mathbf W$-weighted ridge smoother onto $\mathrm{col}(\mathbb X)$; $\mathbf H_{x,0}=\mathbf H_x$, $\mathbf A_{z,0}=\mathbf A_z$ and $\mathbf A_{x,0}=\mathbf A_x$. Indeed, setting the gradient of $W_O L(\bs a,\bs b)$ (roughness penalties suppressed) to zero gives the normal equations
$\mathbb Z^\top\mathbf W(\mathbf Y-\mathbb Z\bs a-\mathbb X\bs b)=\bm0$ and
$\mathbb X^\top\mathbf W(\mathbf Y-\mathbb Z\bs a-\mathbb X\bs b)=\tau\bs b$, with $\mathbf Y=(Y_{ij})$ stacked as in \eqref{eq:blocks}; eliminating $\bs b$ from the first equation and $\bs a$ from the second yields the ridge Frisch--Waugh representation
\begin{equation}
\label{eq:FWL-tau}
\begin{aligned}
\widehat{\bs a}
  &= \big\{\mathbb Z^\top\mathbf W(\mathbf I-\mathbf H_{x,\tau})\mathbb Z\big\}^{-1}
     \mathbb Z^\top\mathbf W(\mathbf I-\mathbf H_{x,\tau})\mathbf Y, \\
\widehat{\bs b}
  &= \big\{\mathbb X^\top\mathbf W(\mathbf I-\mathbf H_z)\mathbb X+\tau\mathbf I\big\}^{-1}
     \mathbb X^\top\mathbf W(\mathbf I-\mathbf H_z)\mathbf Y.
\end{aligned}
\end{equation}
For $\tau_U=0$ this is the usual Frisch--Waugh--Lovell form; for $\tau_U>0$ the residual-maker $\mathbf I-\mathbf H_{x,\tau}$ in the $\nu$-block is a contraction rather than a projection, and the $U$-block Gram is shifted by $\tau\mathbf I$. The algebraic properties of $\mathbf H_{x,\tau}$, $\mathbf A_{z,\tau}$ and $\mathbf A_{x,\tau}$ used in the proofs are collected in Lemma \ref{LEM:ridge} below.

\medskip
\noindent\textit{IPW de-biasing.}
Under MAR with $w_i=1/\pi_i$, $\pi_i=\Pr(R_{Y_i}=1\mid\mathbf X_i)$, the
inverse-probability identity gives, for any integrable
$g=g(\mathbf X,\widetilde Y)$ that depends only on the \emph{fully observed}
covariates and surrogate,
\begin{equation}
\label{eq:debias}
\frac{1}{\sum_{i\in\IO}w_i}\sum_{i\in\IO}w_i\,g(\mathbf X_i,\widetilde Y_i)
\xrightarrow{P}\E\{g(\mathbf X,\widetilde Y)\},
\qquad
\frac1n\sum_{i\in\IO}w_i\xrightarrow{P}\E \Big(\frac{R_Y}{\pi}\Big)=1 .
\end{equation}
Because both design blocks \eqref{eq:blocks} are built from $\mathbf X_i$ and the surrogate $\widetilde Y_i$ alone -- which carry \emph{no} missingness -- the entries of $\mathbf A_z$ and $\mathbf A_x$ are exactly of this form.  Hence the weights \emph{de-bias} the design Gram to its full-population, weight-free limit: the matrix $\mathbf W$ enters the estimator but \emph{disappears} from the probability limit of $\mathbf A_z,\mathbf A_x$.  This is the mechanism by which $\mathbf W$ (and any per-subject $\Omega_i$) drops out of the second-stage conditioning theory; see Lemma \ref{LEM:ULLN}.

\medskip
\noindent\textit{Key structural identity.}
Writing $\widetilde Y_i(\bs z)=\mathbf X_i^\top\bs\beta_0(\bs z)+\varepsilon_{S,i}(\bs z)$
for the surrogate deviation $\varepsilon_{S,i}$, the $U$-block field decomposes as
\begin{equation}
\label{eq:reparam}
\int_\Omega V(\bs z,\bs z')\widetilde Y_i(\bs z')\,d\bs z' = \mathbf X_i^\top\underbrace{\int_\Omega V(\bs z,\bs z')\bs\beta_0(\bs z')\,d\bs z'}_{=:\ \bs\mu_V(\bs z)} \; + \;\int_\Omega V(\bs z,\bs z')\varepsilon_{S,i}(\bs z')\,d\bs z'.
\end{equation}
The first term is collinear with the $\nu$-block (it is $\mathbf X_i^\top$ times a coefficient function), and if $V\in\Stp$ then $\bs\mu_V\in\Ssp^{\,p+1}$ exactly. Thus the surrogate's \emph{mean} structure carries no information about $U$ once $\mathbf X$ is in the model; identification of $U$ rests entirely on the surrogate's stochastic deviation $\varepsilon_S$.  This is the mechanism behind Lemmas~\ref{LEM:Az} and~\ref{LEM:Ax}.

We first record the limiting behaviour of the weighted empirical design Gram over the spline class; it is the joint-design, IPW-weighted extension of Lemma \ref{LEM:Rnorder-vec}, and it is what the conditioning lemmas consume.

\begin{lemma}[Limiting design functional]
\label{LEM:ULLN}
Suppose Assumptions \ref{as:triangulation}--\ref{as:stage2} hold, and let
$D_{\bs\eta,V,i}(\bs z)=\mathbf X_i^\top\bs\eta(\bs z)+\int_\Omega V(\bs z,\bs z')\varepsilon_{S,i}(\bs z')d\bs z'$.
Define the quadratic functional
$\mathcal Q(\bs\eta,V)=\E\{\|D_{\bs\eta,V}\|_{L_2(\Omega)}^2\}$ on
$\Ssp^{\,p+1}\times\Stp$.  Then, uniformly over $(\bs\eta,V)\neq(\bm0,\bm0)$,
\[
\frac{1}{N\sum_{i\in\IO}w_i}\sum_{i\in\IO}\sum_{j=1}^N w_i\,D_{\bs\eta,V,i}^2(\bs z_j)
=\mathcal Q(\bs\eta,V)\,\{1+o_P(1)\}.
\]
Moreover $\mathcal Q$ separates as $\mathcal Q(\bs\eta,V)=\mathcal Q_\nu(\bs\eta)+\mathcal Q_U(V)$
with
$\mathcal Q_\nu(\bs\eta)=\int_\Omega\bs\eta^\top\bs\Sigma_X\bs\eta$ and
$\mathcal Q_U(V)=\int_\Omega\langle V(\bs z,\cdot),\Gamma_S V(\bs z,\cdot)\rangle d\bs z$,
and there are constants $0<c_\dagger\le C_\dagger<\infty$ with
\[
c_\dagger\|\bs\eta\|_{L_2(\Omega)}^2\le\mathcal Q_\nu(\bs\eta)\le C_\dagger\|\bs\eta\|_{L_2(\Omega)}^2,
\qquad
c_\dagger\|V\|_{L_2(\Omega^2)}^2\le\mathcal Q_U(V)\le C_\dagger\|V\|_{L_2(\Omega^2)}^2 .
\]
\end{lemma}

\begin{proof}
\emph{De-biasing.}  The summand $D_{\bs\eta,V,i}^2$ is a function of
$(\mathbf X_i,\widetilde Y_i)$ only---both fully observed, since the surrogate has
no missingness and $\varepsilon_{S,i}=\widetilde Y_i-\mathbf X_i^\top\bs\beta_0$.  Hence the
inverse-probability identity \eqref{eq:debias} applies to the per-subject
average $h_i=N^{-1}\sum_{j}D_{\bs\eta,V,i}^2(\bs z_j)$, giving
$\big(\sum_{\IO}w_i\big)^{-1}\sum_{\IO}w_i h_i\xrightarrow{P}\E(h)$, the
full-population, \emph{weight-free} expectation.  This is why $\mathbf W$ does not
appear in $\mathcal Q$.

\emph{Voxel-to-integral and uniformity.}  Expanding
$D^2=(\mathbf X^\top\bs\eta)^2+2(\mathbf X^\top\bs\eta)\!\int V\varepsilon_S+(\int V\varepsilon_S)^2$
and writing each piece in the Bernstein basis, Lemma \ref{LEM:integration} replaces $N^{-1}\sum_j(\cdot)(\bs z_j)$ by $\int_\Omega(\cdot)$ with relative error $O_P(N^{-1/3}|\tri|^2)$ on the covariate piece and $O_P(N^{-1/3}|\tri|^{5})$ on the surrogate piece, both $o_P(1)$ since $N^{1/3}|\tri|\to\infty$. Uniformity over the (scale-invariant) normalized class follows from Lemma \ref{LEM:Rnorder-vec} for the covariate piece and a maximal inequality over the finite-dimensional space $\Stp$
for the surrogate piece, using the FPC moment bounds in Assumption \ref{as:errors}, exactly as in Lemmas A.3--A.4 of \cite{cao2020estimation}. Thus the displayed limit holds with $\mathcal Q(\bs\eta,V)=\E\{\|D_{\bs\eta,V}\|^2\}$.

\emph{Separation and equivalence.} For $\bs\eta\in\Ssp^{\,p+1}$ and $V\in\Stp$, write the design field
$D_{\bs\eta,V}=\mathbf X^\top\bs\eta+\int_\Omega V(\cdot,\bs z')\varepsilon_S(\bs z')d\bs z'$. By $\E(\varepsilon_S\mid\mathbf X)=0$ the cross
piece has mean zero, so
\[
Q(\bs\eta,V)
=\mathcal Q_\nu+\mathcal Q_U = \int_\Omega\bs\eta(\bs z)^\top\bs\Sigma_X\bs\eta(\bs z)\,d\bs z
+\int_\Omega\big\langle V(\bs z,\cdot),\,\Gamma_S V(\bs z,\cdot)\big\rangle\,d\bs z.
\]
Then $\mathcal Q_\nu(\bs\eta)=\int_\Omega\bs\eta^\top\bs\Sigma_X\bs\eta\asymp\|\bs\eta\|^2$ by Assumption \ref{as:design}, and $\mathcal Q_U(V)=\int_\Omega\langle V,\Gamma_S V\rangle\asymp\|V\|_{L_2(\Omega^2)}^2$ by Assumption \ref{as:relevance}; the constants $c_\dagger,C_\dagger$ are the corresponding eigenvalue
bounds.
\end{proof}

\begin{remark}
\label{rem:debias-cond}
Only positivity ($w_i$ bounded) is needed for the two-sided conclusion: even
without exact de-biasing, $1\le w_i\le C_w$ sandwiches the weighted average
between fixed multiples of the unweighted one, whose limit is
$\asymp\|\bs\eta\|^2+\|V\|^2$ by the same argument.  Consequently the conditioning
Lemmas~\ref{LEM:Az}--\ref{LEM:Ax} are robust to using an estimated propensity
$\widehat\pi_i$ in $w_i$, provided $\widehat\pi_i$ stays bounded below; the de-biased
form merely identifies the cleaner weight-free limit.
\end{remark}

\begin{lemma}[$\nu$-block conditioning; analogue of Lemma A.6 of \cite{cao2020estimation}]
\label{LEM:Az}
Under Assumptions \ref{as:triangulation}--\ref{as:stage2}, if $N^{1/3}|\tri|\to\infty$ as $N\to\infty$, then
there exist constants $0<c_3\le C_3<\infty$ such that, except on an event
whose probability tends to zero, all eigenvalues of $|\tri|^{-3}\mathbf A_z$
lie in $[c_3,C_3]$.  In particular $\mathbf A_z$ is invertible and
$\|\mathbf A_z^{-1}\|_2=O_P(|\tri|^{-3})$.
\end{lemma}

\begin{proof}
Fix $\bs a\neq\bm0$ and set $\bs\eta_{\bs a}(\bs z)=\mathbf I_{p+1}\otimes\Bz^\top\bs a\in\Ssp^{\,p+1}$,
so $\mathbf X_i^\top\bs\eta_{\bs a}(\bs z_j)=(\mathbb Z\bs a)_{ij}$.  Because
$\mathbf W^{1/2}\mathbf H_x\mathbf W^{-1/2}$ is the orthogonal projection onto
$\mathrm{col}(\mathbf W^{1/2}\mathbb X)$, the partialled quadratic form admits the
variational (residual-sum-of-squares) representation
\begin{align}
\notag
\bs a^\top\mathbf A_z\bs a
& =\min_{\bs b}\frac{1}{N\sum_{i\in\IO}w_i}\big\|\mathbf W^{1/2}(\mathbb Z\bs a-\mathbb X\bs b)\big\|_2^2 \\
 \label{eq:varAz}
& =\min_{V\in\Stp}\frac{1}{N\sum_{i\in\IO}w_i}\sum_{i\in\IO}\sum_{j=1}^N w_i
\Big\{\mathbf X_i^\top\bs\eta_{\bs a}(\bs z_j)-\!\int_\Omega\! V(\bs z_j,\bs z')\widetilde Y_i(\bs z')d\bs z'\Big\}^2 .
\end{align}
Insert \eqref{eq:reparam}: the bracket equals
$\mathbf X_i^\top\{\bs\eta_{\bs a}-\bs\mu_V\}(\bs z_j)-\int V(\bs z_j,\cdot)\varepsilon_{S,i}
=D_{\bs\eta_{\bs a}-\bs\mu_V,\,-V,\,i}(\bs z_j)$, with $\bs\mu_V\in\Ssp^{\,p+1}$.  By
Lemma~\ref{LEM:ULLN} and its separation, uniformly in $\bs a$,
\begin{align*}
    \bs a^\top\mathbf A_z\bs a
& =\{1+o_P(1)\}\min_{V\in\Stp}\big\{\mathcal Q_\nu(\bs\eta_{\bs a}-\bs\mu_V)+\mathcal Q_U(V)\big\}
\\
& \ge\ \{1+o_P(1)\}\,c_\dagger\min_{V\in\Stp}\big(\|\bs\eta_{\bs a}-\bs\mu_V\|^2+\|V\|_{L_2(\Omega^2)}^2\big).
\end{align*}
The map $L:V\mapsto\bs\mu_V$ is linear with $\|L\|\le\|\bs\beta_0\|$ (Cauchy--Schwarz),
so the Tikhonov identity
$\min_V\{\|\bs\eta-LV\|^2+\|V\|^2\}=\langle\bs\eta,(LL^*+\mathbf I)^{-1}\bs\eta\rangle\ge(\|L\|^2+1)^{-1}\|\bs\eta\|^2$
gives
$\min_{V}(\|\bs\eta_{\bs a}-\bs\mu_V\|^2+\|V\|_{L_2(\Omega^2)}^2)\ge(1+\|\bs\beta_0\|^2)^{-1}\|\bs\eta_{\bs a}\|^2$.
With the norm equivalence (Lemma~\ref{LEM:DDC-norm})
$\|\bs\eta_{\bs a}\|^2\asymp|\tri|^3\|\bs a\|_2^2$,
\[
\bs a^\top\mathbf A_z\bs a\ \ge\ \{1+o_P(1)\}\,\frac{c_\dagger}{1+\|\bs\beta_0\|^2}\,\|\bs\eta_{\bs a}\|^2
\ \ge\ c_3\,|\tri|^3\,\|\bs a\|_2^2 \quad\text{w.p.}\to1 .
\]
For the upper bound, take $V=0$ in \eqref{eq:varAz} (drop the projection):
$$\bs a^\top\mathbf A_z\bs a\le\frac{1}{N\sum_{i\in\IO}w_i}\bs a^\top\mathbb Z^\top\mathbf W\mathbb Z\bs a
=\{1+o_P(1)\}\,\mathcal Q_\nu(\bs\eta_{\bs a})\le C_\dagger\|\bs\eta_{\bs a}\|^2\le C_3|\tri|^3\|\bs a\|_2^2. $$
As $\bs a$ was arbitrary, the eigenvalues of $|\tri|^{-3}\mathbf A_z$ lie in
$[c_3,C_3]$ w.p.\ $\to1$.
\end{proof}

\begin{lemma}[$U$-block conditioning; analogue of Lemma A.5 of \cite{cao2020estimation}]
\label{LEM:Ax}
Under Assumptions \ref{as:triangulation}--\ref{as:stage2}, if $N^{1/3}|\tri|\to\infty$ as $N\to\infty$, then
there exist constants $0<c_4\le C_4<\infty$ such that, except on an event
whose probability tends to zero, all eigenvalues of $|\tri|^{-6}\mathbf A_x$
lie in $[c_4,C_4]$.  In particular $\mathbf A_x$ is invertible and
$\|\mathbf A_x^{-1}\|_2=O_P(|\tri|^{-6})$.
\end{lemma}

\begin{proof}
Fix $\bs b\neq\bm0$ and set $V_{\bs b}=\Bzz^\top\bs b\in\Stp$.  As in
\eqref{eq:varAz},
\[
\bs b^\top\mathbf A_x\bs b
=\min_{\bs\eta\in\Ssp^{\,p+1}}\frac{1}{N\sum_{i\in\IO}w_i}\sum_{i\in\IO}\sum_{j=1}^N w_i
\Big\{\!\int_\Omega\! V_{\bs b}(\bs z_j,\bs z')\widetilde Y_i(\bs z')d\bs z'-\mathbf X_i^\top\bs\eta(\bs z_j)\Big\}^2 .
\]
By \eqref{eq:reparam} the bracket is
$\mathbf X_i^\top\{\bs\mu_{V_{\bs b}}-\bs\eta\}(\bs z_j)+\int V_{\bs b}(\bs z_j,\cdot)\varepsilon_{S,i}
=D_{\bs\mu_{V_{\bs b}}-\bs\eta,\,V_{\bs b},\,i}(\bs z_j)$, with $\bs\mu_{V_{\bs b}}\in\Ssp^{\,p+1}$.
By Lemma~\ref{LEM:ULLN} and its separation,
\[
\bs b^\top\mathbf A_x\bs b
=\{1+o_P(1)\}\min_{\bs\eta\in\Ssp^{\,p+1}}\big\{\mathcal Q_\nu(\bs\mu_{V_{\bs b}}-\bs\eta)+\mathcal Q_U(V_{\bs b})\big\}.
\]
Since $\bs\mu_{V_{\bs b}}\in\Ssp^{\,p+1}$, choosing $\bs\eta=\bs\mu_{V_{\bs b}}$ annihilates the
first term ($\mathcal Q_\nu(\bm0)=0$); that is, partialling out the $\nu$-block removes
the entire surrogate-mean contribution, leaving
\[
\bs b^\top\mathbf A_x\bs b=\{1+o_P(1)\}\,\mathcal Q_U(V_{\bs b}).
\]
By Lemma~\ref{LEM:ULLN}, $c_\dagger\|V_{\bs b}\|_{L_2(\Omega^2)}^2\le\mathcal Q_U(V_{\bs b})\le C_\dagger\|V_{\bs b}\|_{L_2(\Omega^2)}^2$;
and the tensor-product norm equivalence (Lemma~\ref{LEM:DDC-norm}, applied
to $\Bz\otimes\mathbf B(\bs z')$ via Fubini) gives
$\|V_{\bs b}\|_{L_2(\Omega^2)}^2\asymp|\tri|^3\cdot|\tri|^3\|\bs b\|_2^2=|\tri|^6\|\bs b\|_2^2$.
Hence the eigenvalues of $|\tri|^{-6}\mathbf A_x$ lie in $[c_4,C_4]$ w.p.\ $\to1$.
Finally, taking instead $\bs\eta=\bm0$ in the display above (i.e.\ dropping the
$\nu$-projection) and using $\|\bs\mu_{V_{\bs b}}\|\le\|\bs\beta_0\|\|V_{\bs b}\|_{L_2(\Omega^2)}$
gives the companion upper bound on the \emph{un-partialled} $U$-Gram,
$\frac{1}{N\sum_{i\in\IO}w_i}\bs b^\top\mathbb X^\top\mathbf W\mathbb X\bs b
=\{1+o_P(1)\}\{\mathcal Q_\nu(\bs\mu_{V_{\bs b}})+\mathcal Q_U(V_{\bs b})\}\le C|\tri|^6\|\bs b\|_2^2$,
that is $\lambda_{\max}(\mathbb X^\top\mathbf W\mathbb X)\le C\sum_{i\in\mathcal{I}_O} w_iN|\tri|^6$ (used in
Lemma~\ref{LEM:stoch-U}); the analogous $V=0$ step above gives
$\lambda_{\max}(\mathbb Z^\top\mathbf W\mathbb Z)\le C\sum_{i\in\mathcal{I}_O} w_iN|\tri|^3$.
\end{proof}

The scalings $\mathbf A_z\asymp|\tri|^3$ and $\mathbf A_x\asymp|\tri|^6$ are the
trivariate-triangulation images of the results of \cite{cao2020estimation}
``the eigenvalues of $J_n A_z$ (resp.\ $J_n^2 A_x$) are bounded,'' under the
correspondence $J_n\leftrightarrow|\tri|^{-3}$ (one-dimensional knot count
$\leftrightarrow$ trivariate basis count).  No sampling weight survives in the
limit: because the design depends only on the fully observed $(\mathbf X,\widetilde Y)$,
the inverse-probability weights \emph{de-bias} the Gram to its weight-free
population moment \eqref{eq:debias}, so $\mathbf W$ (and any per-subject $\Omega_i$)
disappears from $\mathcal Q$ and hence from the eigenvalue scalings; cf.\
Remark~\ref{rem:debias-cond} for the weaker bounded-weight argument.

We now record how the ridge penalty in \eqref{eq:L-stage2} modifies the two Gram matrices. Statements (a)--(d) below are algebraic and hold for every realization of the design (for $\tau=0$, whenever $\mathbb X^\top\mathbf W\mathbb X$ is invertible, which holds with probability tending to one by Lemma \ref{LEM:Ax}); the last statement is probabilistic.

\begin{lemma}[Ridge-partialled Gram matrices]
\label{LEM:ridge}
Let $\tau=W_O\tau_U\ge0$, and let $\mathbf H_{x,\tau}$, $\mathbf A_{z,\tau}$ and $\mathbf A_{x,\tau}$ be as in \eqref{eq:AzAx-tau}.
\begin{itemize}
\item[(a)] (Self-adjointness and contraction.) The matrix $\mathbf W\mathbf H_{x,\tau}$ is symmetric. Moreover, $\widetilde{\mathbf H}_{x,\tau}:=\mathbf W^{1/2}\mathbf H_{x,\tau}\mathbf W^{-1/2}$ is symmetric with eigenvalues in $[0,1)$. Consequently $\mathbf W(\mathbf I-\mathbf H_{x,\tau})$ is symmetric positive semidefinite, $\{(\mathbf I-\mathbf H_{x,\tau})\mathbb Z\}^\top\mathbf W=\mathbb Z^\top\mathbf W(\mathbf I-\mathbf H_{x,\tau})$, and
\[
\{(\mathbf I-\mathbf H_{x,\tau})\mathbb Z\}^\top\mathbf W\{(\mathbf I-\mathbf H_{x,\tau})\mathbb Z\}
\ \preceq\ \mathbb Z^\top\mathbf W(\mathbf I-\mathbf H_{x,\tau})\mathbb Z
\ =\ NW_O\,\mathbf A_{z,\tau},
\]
with equality when $\tau=0$.
\item[(b)] (Variational form and sandwich.) For every $\bs a$,
\[
\bs a^\top\mathbf A_{z,\tau}\bs a
=\min_{\bs b}\frac{1}{NW_O}\Big\{\big\|\mathbf W^{1/2}(\mathbb Z\bs a-\mathbb X\bs b)\big\|_2^2+\tau\|\bs b\|_2^2\Big\},
\]
so that $\mathbf A_z\preceq\mathbf A_{z,\tau}\preceq(NW_O)^{-1}\mathbb Z^\top\mathbf W\mathbb Z$.
\item[(c)] (Leakage.) $(\mathbf I-\mathbf H_{x,\tau})\mathbb X=\tau\,\mathbb X\big(\mathbb X^\top\mathbf W\mathbb X+\tau\mathbf I\big)^{-1}$; in particular, the ridge residual-maker annihilates $\mathbb X$ only when $\tau=0$.
\item[(d)] (Cross-block coupling.) $\kappa_{z,\tau}:=\big\|\mathbb Z^\top\mathbf W\mathbb X(\mathbb X^\top\mathbf W\mathbb X+\tau\mathbf I)^{-1}\big\|_2$ satisfies
\[
\kappa_{z,\tau}^2\le\frac{\lambda_{\max}(\mathbb Z^\top\mathbf W\mathbb Z)}{\lambda_{\min}(\mathbb X^\top\mathbf W\mathbb X)+\tau}.
\]
\end{itemize}
Moreover, under Assumptions \ref{as:triangulation}--\ref{as:stage2}, if $N^{1/3}|\tri|\to\infty$ as $N\to\infty$, then, except on an event whose probability tends to zero and uniformly in $\tau_U\ge0$: all eigenvalues of $|\tri|^{-3}\mathbf A_{z,\tau}$ lie in $[c_3,C_3]$; all eigenvalues of $|\tri|^{-6}\mathbf A_{x,\tau}$ lie in $[c_4+r_\tau,\,C_4+r_\tau]$; and $\kappa_{z,\tau}^2\le C|\tri|^{-3}$, where $c_3,C_3,c_4,C_4$ are the constants of Lemmas \ref{LEM:Az}--\ref{LEM:Ax}. In particular, $\|\mathbf A_{z,\tau}^{-1}\|_2=O_P(|\tri|^{-3})$ and $\|\mathbf A_{x,\tau}^{-1}\|_2\le\|\mathbf A_x^{-1}\|_2=O_P(|\tri|^{-6})$.
\end{lemma}

\begin{proof}
Write $\mathbf S=\mathbf W^{1/2}\mathbb X$, $\mathbf T=\mathbf W^{1/2}\mathbb Z$ and $\mathbf G=\mathbf S^\top\mathbf S=\mathbb X^\top\mathbf W\mathbb X$, and let $\mathbf S=\mathbf U\mathbf D\mathbf V^\top$ be a singular value decomposition with $\mathbf V$ orthogonal, $\mathbf D=\diag(\sigma_k)$, $\sigma_k\ge0$, and $\mathbf U^\top\mathbf U=\mathbf I$, so that $\mathbf G+\tau\mathbf I=\mathbf V(\mathbf D^2+\tau\mathbf I)\mathbf V^\top$.

(a) $\mathbf W\mathbf H_{x,\tau}=\mathbf W\mathbb X(\mathbf G+\tau\mathbf I)^{-1}\mathbb X^\top\mathbf W$ is symmetric, which gives $\{(\mathbf I-\mathbf H_{x,\tau})\mathbb Z\}^\top\mathbf W=\mathbb Z^\top(\mathbf I-\mathbf H_{x,\tau}^\top)\mathbf W=\mathbb Z^\top\mathbf W(\mathbf I-\mathbf H_{x,\tau})$. Next,
$\widetilde{\mathbf H}_{x,\tau}=\mathbf S(\mathbf G+\tau\mathbf I)^{-1}\mathbf S^\top=\mathbf U\,\diag\{\sigma_k^2/(\sigma_k^2+\tau)\}\,\mathbf U^\top$ is symmetric with eigenvalues in $[0,1)$ when $\tau>0$; for $\tau=0$ it is the orthogonal projection $\widetilde{\mathbf H}_x=\mathbf U\mathbf U^\top$ onto $\mathrm{col}(\mathbf S)$. Hence $\mathbf I-\widetilde{\mathbf H}_{x,\tau}$ is symmetric with eigenvalues in $(0,1]$, so $(\mathbf I-\widetilde{\mathbf H}_{x,\tau})^2\preceq\mathbf I-\widetilde{\mathbf H}_{x,\tau}$, with equality in the idempotent case $\tau=0$. Since $\mathbf W(\mathbf I-\mathbf H_{x,\tau})=\mathbf W^{1/2}(\mathbf I-\widetilde{\mathbf H}_{x,\tau})\mathbf W^{1/2}$ and
\[
\begin{aligned}
\{(\mathbf I-\mathbf H_{x,\tau})\mathbb Z\}^{\top}
  \mathbf W\{(\mathbf I-\mathbf H_{x,\tau})\mathbb Z\}
  &= \mathbf T^\top(\mathbf I-\widetilde{\mathbf H}_{x,\tau})^2\mathbf T
     \preceq \mathbf T^\top(\mathbf I-\widetilde{\mathbf H}_{x,\tau})\mathbf T \\
  &= \mathbb Z^\top\mathbf W(\mathbf I-\mathbf H_{x,\tau})\mathbb Z,
\end{aligned}
\]
(a) follows.

(b) For fixed $\bs a$, the strictly convex (for $\tau>0$) function $\bs b\mapsto\|\mathbf W^{1/2}(\mathbb Z\bs a-\mathbb X\bs b)\|_2^2+\tau\|\bs b\|_2^2$ is minimized at $\bs b_{\bs a}=(\mathbf G+\tau\mathbf I)^{-1}\mathbb X^\top\mathbf W\mathbb Z\bs a$, and its minimum value is
$\bs a^\top\mathbb Z^\top\mathbf W\mathbb Z\bs a-\bs a^\top\mathbb Z^\top\mathbf W\mathbb X(\mathbf G+\tau\mathbf I)^{-1}\mathbb X^\top\mathbf W\mathbb Z\bs a
=\bs a^\top\mathbb Z^\top\mathbf W(\mathbf I-\mathbf H_{x,\tau})\mathbb Z\bs a$; dividing by $NW_O$ gives the variational form (for $\tau=0$ it reduces to \eqref{eq:varAz}). The minimum value is nondecreasing in $\tau$, whence $\mathbf A_z=\mathbf A_{z,0}\preceq\mathbf A_{z,\tau}$, and taking $\bs b=\bm0$ gives $\mathbf A_{z,\tau}\preceq(NW_O)^{-1}\mathbb Z^\top\mathbf W\mathbb Z$.

(c) $(\mathbf I-\mathbf H_{x,\tau})\mathbb X=\mathbb X-\mathbb X(\mathbf G+\tau\mathbf I)^{-1}\mathbf G=\mathbb X(\mathbf G+\tau\mathbf I)^{-1}\{(\mathbf G+\tau\mathbf I)-\mathbf G\}=\tau\,\mathbb X(\mathbf G+\tau\mathbf I)^{-1}$.

(d) $\kappa_{z,\tau}^2=\|\mathbf T^\top\mathbf S(\mathbf G+\tau\mathbf I)^{-2}\mathbf S^\top\mathbf T\|_2$ and
$\mathbf S(\mathbf G+\tau\mathbf I)^{-2}\mathbf S^\top=\mathbf U\,\diag\{\sigma_k^2/(\sigma_k^2+\tau)^2\}\,\mathbf U^\top\preceq\{\min_k\sigma_k^2+\tau\}^{-1}\mathbf U\mathbf U^\top\preceq\{\lambda_{\min}(\mathbf G)+\tau\}^{-1}\mathbf I$, because $\sigma_k^2/(\sigma_k^2+\tau)^2\le(\sigma_k^2+\tau)^{-1}$ and $\min_k\sigma_k^2=\lambda_{\min}(\mathbf G)$. Hence $\kappa_{z,\tau}^2\le\{\lambda_{\min}(\mathbf G)+\tau\}^{-1}\lambda_{\max}(\mathbf T^\top\mathbf T)$, which is (d).

For the probabilistic statements, recall from Lemma \ref{LEM:Az} and the $V=0$ step of its proof that, w.p.\ $\to1$,
\[
c_3|\tri|^3\le\lambda_{\min}(\mathbf A_z)
\qquad\text{and}\qquad
\lambda_{\max}\big\{(NW_O)^{-1}\mathbb Z^\top\mathbf W\mathbb Z\big\}\le C_3|\tri|^3 ;
\]
by the sandwich in (b), the eigenvalues of $|\tri|^{-3}\mathbf A_{z,\tau}$ therefore lie in $[c_3,C_3]$ w.p.\ $\to1$, for every $\tau_U\ge0$. The statement for $\mathbf A_{x,\tau}=\mathbf A_x+(\tau_U/N)\mathbf I$ is immediate from Lemma \ref{LEM:Ax}, since $(\tau_U/N)|\tri|^{-6}=r_\tau$. Finally, the proof of Lemma \ref{LEM:Ax} gives
\[
\lambda_{\max}(\mathbb Z^\top\mathbf W\mathbb Z)\le C\,W_ON|\tri|^3,~
\lambda_{\min}(\mathbb X^\top\mathbf W\mathbb X)\ge\lambda_{\min}\{\mathbb X^\top\mathbf W(\mathbf I-\mathbf H_z)\mathbb X\}\ge c\,W_ON|\tri|^6 ,
\]
w.p.\ $\to1$, so that (d) yields $\kappa_{z,\tau}^2\le C|\tri|^{-3}$, uniformly in $\tau\ge0$.
\end{proof}

\begin{remark}[Scale of the ridge penalty]
\label{rem:ridge-scale}
In \eqref{eq:L-stage2} the data-fit term is normalized by $W_O=\sum_{i\in\IO}w_i$ but still sums over the $N$ voxels, so at the truth it is of order $N$, whereas the ridge term evaluated at the spline-approximant coefficients $\bs b^{*}$ of $U$ (defined in Step 1 of the proof of Theorem \ref{THM:nuU-rate}) is $\tau_U\|\bs b^{*}\|_2^2\asymp\tau_U|\tri|^{-6}$, by the tensor-product norm equivalence (Lemma \ref{LEM:DDC-norm}) and $\|U^{*}\|_{L_2(\Omega\times\Omega)}=O(1)$. Their ratio is $r_\tau=\tau_U/(N|\tri|^6)$, and this is the quantity through which the ridge enters every bound below: it shifts the $U$-block Gram eigenvalues by $r_\tau|\tri|^6$ (Lemma \ref{LEM:ridge}) and it produces a shrinkage bias of order $r_\tau$ in root-mean-square units in both second-stage estimators (Theorem \ref{THM:nuU-rate}). The condition $r_\tau=O(n^{-1/2}+|\tri|^{d+1})$ of Assumption \ref{as:rates} is therefore the exact analogue, for the ridge penalty, of the requirement that the roughness penalties be negligible at the approximation scale; it holds, for instance, whenever $\tau_U=O(N|\tri|^{d+7})$. No condition on $\tau_U$ is needed for the conditioning of the Gram matrices (Lemma \ref{LEM:ridge}) or for the stochastic-error bounds (Lemmas \ref{LEM:stoch-nu}--\ref{LEM:stoch-U}): the ridge can only improve the conditioning of the $U$-block and only shrinks the noise. The condition is needed solely to control the shrinkage bias in the oracle fits.
\end{remark}

Let $\mathbf Y^{*}=(Y^{*}_{ij},\,1\le j\le N,\,i\in\IO)^{\top}$ collect the
noiseless responses,
\begin{equation}
\label{eq:Ystar}
Y^{*}_{ij}=\mathbf X_i^\top\bs\nu_0(\bs z_j)
+\int_\Omega U(\bs z_j,\bs z')\,\widetilde Y_i(\bs z')\,d\bs z',
\end{equation}
so that, by \eqref{eq:ssfr-adj}, $Y_i(\bs z_j)=Y^{*}_{ij}+\zeta_i(\bs z_j)$.
Define the oracle (noiseless) coefficient fits by applying the ridge Frisch--Waugh maps \eqref{eq:FWL-tau} to $\mathbf Y^{*}$,
\begin{equation}
\label{eq:oracle}
\begin{aligned}
\widetilde{\bs a}
  &= \big\{\mathbb Z^\top\mathbf W(\mathbf I-\mathbf H_{x,\tau})\mathbb Z\big\}^{-1}
     \mathbb Z^\top\mathbf W(\mathbf I-\mathbf H_{x,\tau})\mathbf Y^{*}, \\
\widetilde{\bs b}
  &= \big\{\mathbb X^\top\mathbf W(\mathbf I-\mathbf H_z)\mathbb X+\tau\mathbf I\big\}^{-1}
     \mathbb X^\top\mathbf W(\mathbf I-\mathbf H_z)\mathbf Y^{*},
\end{aligned}
\end{equation}
and set $\widetilde{\bs\nu}(\bs z)=(\mathbf I_{p+1}\otimes\Bz^\top)\widetilde{\bs a}$ and
$\widetilde U(\bs z,\bs z')=\Bzz^\top\widetilde{\bs b}$.  Because the ridge Frisch--Waugh
maps in \eqref{eq:oracle} are linear in the response, the stage-two spline
coefficients decompose as
$\widehat{\bs a}=\widetilde{\bs a}+\widehat{\bs a}^{\rm st}$ and
$\widehat{\bs b}=\widetilde{\bs b}+\widehat{\bs b}^{\rm st}$, where the
\emph{stochastic} parts are the partialled (ridge-)projections of the noise
$\bs\zeta=(\zeta_i(\bs z_j))$,
\begin{equation}
\label{eq:stoch}
\begin{aligned}
\widehat{\bs a}^{\rm st}
  &= \big\{\mathbb Z^\top\mathbf W(\mathbf I-\mathbf H_{x,\tau})\mathbb Z\big\}^{-1}
     \mathbb Z^\top\mathbf W(\mathbf I-\mathbf H_{x,\tau})\bs\zeta, \\
\widehat{\bs b}^{\rm st}
  &= \big\{\mathbb X^\top\mathbf W(\mathbf I-\mathbf H_z)\mathbb X+\tau\mathbf I\big\}^{-1}
     \mathbb X^\top\mathbf W(\mathbf I-\mathbf H_z)\bs\zeta,
\end{aligned}
\end{equation}
and the oracle parts carry the spline-approximation remainder and the ridge shrinkage bias (Steps 3--5 of the proof of Theorem \ref{THM:nuU-rate}).  Set
$\widehat{\bs\nu}^{\rm st}=\widehat{\bs\nu}-\widetilde{\bs\nu}
=(\mathbf I_{p+1}\otimes\Bz^\top)\widehat{\bs a}^{\rm st}$
and $\widehat U^{\rm st}=\widehat U-\widetilde U=\Bzz^\top\widehat{\bs b}^{\rm st}$.
The normalization
$1/(N\sum_{i\in\IO}w_i)$ cancels between the inverse Gram and the score, so \eqref{eq:stoch} is scale-free (the ridge enters only through $\tau/(NW_O)=\tau_U/N$, cf.\ \eqref{eq:AzAx-tau}); and by Assumption \ref{as:stage2}, the noise is independent of the design with $\E(\bs\zeta\mid\text{design})=\bm0$, so the variance computations
below condition on the design (in particular treating the random $U$-block
$\mathbb X$ as fixed).

\begin{lemma}[$\nu$-block stochastic error; analogue of Lemma A.7 of \cite{cao2020estimation}]
\label{LEM:stoch-nu}
Under Assumptions \ref{as:triangulation}--\ref{as:stage2}, if $N^{1/3}|\tri|\to\infty$ as $N\to\infty$, then
$\|\widehat{\bs\nu}^{\rm st}\|_{L_2(\Omega)}^2=O_P\{(\sum_{i\in\mathcal{I}_O} w_i)^{-1}\}$.
\end{lemma}

\begin{proof}
Condition throughout on the design $\{(\mathbf X_i,\widetilde Y_i)\}$; by Assumption \ref{as:stage2} the
noise $\bs\zeta$ is independent of it with mean zero, and $\bs\zeta=\bs\zeta^\circ+\mathbf e$
(smooth $+$ white).  Recall $\widehat{\bs\nu}^{\rm st}=(\mathbf I_{p+1}\otimes\Bz^\top)\widehat{\bs a}^{\rm st}$
with $\widehat{\bs a}^{\rm st}=M_{z,\tau}^{-1}\mathbb Z^\top\mathbf W(\mathbf I-\mathbf H_{x,\tau})\bs\zeta$ and
$M_{z,\tau}=\mathbb Z^\top\mathbf W(\mathbf I-\mathbf H_{x,\tau})\mathbb Z=(N\textstyle\sum_{i\in\IO}w_i)\mathbf A_{z,\tau}$; we also write $M_z=M_{z,0}=(N\sum_{i\in\IO}w_i)\mathbf A_z$ and $M_x=\mathbb X^\top\mathbf W(\mathbf I-\mathbf H_z)\mathbb X=(N\sum_{i\in\IO}w_i)\mathbf A_x$.

\emph{Step 1 (reduction to the un-partialled scores).}  By Lemma~\ref{LEM:ridge} (which reduces to Lemma~\ref{LEM:Az} when $\tau_U=0$) the
eigenvalues of $\mathbf A_{z,\tau}$ lie in $[c_3|\tri|^3,C_3|\tri|^3]$ for every $\tau_U\ge0$, so $M_{z,\tau}$ has bounded
condition number with $\lambda_{\min}(M_{z,\tau})\ge\lambda_{\min}(M_z)$ and $\lambda_{\min}(M_{z,\tau})\asymp \sum_{i\in\mathcal{I}_O} w_iN|\tri|^3$, and
\begin{equation}\label{eq:stnu1}
\|\widehat{\bs a}^{\rm st}\|_2^2\le\lambda_{\min}(M_{z,\tau})^{-2}\,\|\mathbb Z^\top\mathbf W(\mathbf I-\mathbf H_{x,\tau})\bs\zeta\|_2^2 .
\end{equation}
Writing $\mathbf H_{x,\tau}\bs\zeta=\mathbb X(\mathbb X^\top\mathbf W\mathbb X+\tau\mathbf I)^{-1}\mathbb X^\top\mathbf W\bs\zeta$,
\begin{equation}
\label{eq:stnu2}
\begin{aligned}
\|\mathbb Z^\top\mathbf W(\mathbf I-\mathbf H_{x,\tau})\bs\zeta\|_2^2
  &\le 2\|\mathbb Z^\top\mathbf W\bs\zeta\|_2^2
      + 2\kappa_{z,\tau}^2\,\|\mathbb X^\top\mathbf W\bs\zeta\|_2^2, \\
\kappa_{z,\tau}
  &:= \|\mathbb Z^\top\mathbf W\mathbb X
        (\mathbb X^\top\mathbf W\mathbb X+\tau\mathbf I)^{-1}\|_2.
\end{aligned}
\end{equation}
By Lemma \ref{LEM:ridge}(d),
$$\kappa_{z,\tau}^2\le\lambda_{\max}(\mathbb Z^\top\mathbf W\mathbb Z)/\{\lambda_{\min}(\mathbb X^\top\mathbf W\mathbb X)+\tau\}\le\lambda_{\max}(\mathbb Z^\top\mathbf W\mathbb Z)/\lambda_{\min}(\mathbb X^\top\mathbf W\mathbb X);$$
for $\tau_U=0$ this is the familiar bound for a $\mathbf W$-orthogonal projection, $\mathbb Z^\top\mathbf W\mathbf H_x\mathbb Z\preceq\mathbb Z^\top\mathbf W\mathbb Z$.
The upper bounds established en route to Lemmas~\ref{LEM:Az}--\ref{LEM:Ax} (via
Lemma~\ref{LEM:ULLN}) give $\lambda_{\max}(\mathbb Z^\top\mathbf W\mathbb Z)\le C\sum_{i\in\mathcal{I}_O} w_iN|\tri|^3$, while
$\mathbb X^\top\mathbf W\mathbb X\succeq M_x$ gives $\lambda_{\min}(\mathbb X^\top\mathbf W\mathbb X)\ge c\,\sum_{i\in\mathcal{I}_O} w_iN|\tri|^6$;
hence $\kappa_{z,\tau}^2\le C|\tri|^{-3}$, uniformly in $\tau_U\ge0$.

\emph{Step 2 (second moments of the un-partialled scores).}
The $(\ell,m)$ entry of $\mathbb Z^\top\mathbf W\bs\zeta$ is
\[
\sum_{i\in\IO} w_i X_{i\ell} \sum_j B_m(\bs z_j)\zeta_{ij}.
\]
As the $\zeta_i$ are independent across $i$, mean zero, and independent of
$(\mathbf X_i,\widetilde Y_i)$, the cross-subject terms vanish and
\[
\E\big(\mathbb Z^\top\mathbf W\bs\zeta\big)_{\ell m}^2
=\sum_{i\in\IO}w_i^2\,\E(X_{i\ell}^2)\sum_{j,j'}B_m(\bs z_j)B_m(\bs z_{j'})\big\{G_\zeta(\bs z_j,\bs z_{j'})+\sigma_\zeta^2(\bs z_j)\mathbb 1_{j=j'}\big\}.
\]
By Lemma \ref{LEM:integration},
\begin{align*}
\sum_{j,j'} B_mB_m G_\zeta
  &= \{1+o(1)\}N^2\!\int_{T_m^2} B_mB_m G_\zeta
   = O(N^2|\tri|^6), \\
\sum_j B_m^2\sigma_\zeta^2
  &= \{1+o(1)\}N\!\int B_m^2\sigma_\zeta^2
   = O(N|\tri|^3).
\end{align*} 
With Assumption \ref{as:design},
bounded weights, $|\IO|\asymp n$, summing the $(p+1)|\mathcal M|\asymp|\tri|^{-3}$ entries,
\begin{equation}\label{eq:Zscore}
\E\|\mathbb Z^\top\mathbf W\bs\zeta\|_2^2=O\big(\sum_{i\in\mathcal{I}_O} w_iN^2|\tri|^3+\sum_{i\in\mathcal{I}_O} w_iN\big).
\end{equation}
For the $U$-block, $\big(\mathbb X^\top\mathbf W\bs\zeta\big)_{m m'}=\sum_{i\in\IO} w_i\phi_{im'}\sum_j B_m(\bs z_j)\zeta_{ij}$
with $\phi_{im'}=\int_\Omega B_{m'}\widetilde Y_i$.  Since $\bs\zeta\perp\widetilde Y$, the same
vanishing holds and $\E[\phi_{im'}^2(\cdot)]=\E[\phi_{im'}^2]\E[(\cdot)]$; using
$\E\phi_{im'}^2=\int\!\int B_{m'}B_{m'}\E\{\widetilde Y\widetilde Y\}=O(|\tri|^6)$ and summing the
$|\mathcal M|^2\asymp|\tri|^{-6}$ entries,
\begin{equation}\label{eq:Xscore}
\E\|\mathbb X^\top\mathbf W\bs\zeta\|_2^2=O\big(\sum_{i\in\mathcal{I}_O} w_iN^2|\tri|^6+\sum_{i\in\mathcal{I}_O} w_iN|\tri|^3\big).
\end{equation}

\emph{Step 3 (combine).}  As $\kappa_{z,\tau}^2\,\E\|\mathbb X^\top\mathbf W\bs\zeta\|_2^2=O(\sum_{i\in\mathcal{I}_O} w_iN^2|\tri|^3+\sum_{i\in\mathcal{I}_O} w_iN)$
matches \eqref{eq:Zscore}, \eqref{eq:stnu1}--\eqref{eq:Xscore} give
\[
\begin{aligned}
\E\|\widehat{\bs a}^{\rm st}\|_2^2
  &\le \lambda_{\min}(M_{z,\tau})^{-2}\,
    O\Big(\sum_{i\in\mathcal{I}_O} w_iN^2|\tri|^3
      + \sum_{i\in\mathcal{I}_O} w_iN\Big) \\
  &= O\!\Big(
    \tfrac{1}{\sum_{i\in\mathcal{I}_O} w_i|\tri|^3}
    + \tfrac{1}{\sum_{i\in\mathcal{I}_O} w_iN|\tri|^6}
    \Big).
\end{aligned}
\]
By the norm equivalence (Lemma~\ref{LEM:DDC-norm})
$\|\widehat{\bs\nu}^{\rm st}\|^2\asymp|\tri|^3\|\widehat{\bs a}^{\rm st}\|_2^2$, so
\[
\E\|\widehat{\bs\nu}^{\rm st}\|_{L_2(\Omega)}^2=O\!\Big(\tfrac1{\sum_{i\in\mathcal{I}_O} w_i}+\underbrace{\tfrac{|\tri|^{-3}}{\sum_{i\in\mathcal{I}_O} w_iN}}_{=\,o(n^{-1})}\Big)=O\left\{(\sum_{i\in\mathcal{I}_O} w_i)^{-1}\right\},
\]
the white term $|\tri|^{-3}/(\sum_{i\in\mathcal{I}_O} w_iN)=\{\sum_{i\in\mathcal{I}_O} w_i\}^{-1}(N|\tri|^3)^{-1}=o\left\{(\sum_{i\in\mathcal{I}_O} w_i)^{-1}\right\}$ since
$N|\tri|^3\to\infty$.  Markov's inequality gives the claim.
\end{proof}

\begin{lemma}[$U$-block stochastic error; analogue of Lemma A.8 of \cite{cao2020estimation}]
\label{LEM:stoch-U}
Under Assumptions \ref{as:triangulation}--\ref{as:stage2}, if $N^{1/3}|\tri|\to\infty$ as $N\to\infty$, then
$\|\widehat U^{\rm st}\|_{L_2(\Omega\times\Omega)}^2=O_P(n^{-1})$.
\end{lemma}

\begin{proof}
The argument mirrors Lemma~\ref{LEM:stoch-nu}, with the roles of the blocks
exchanged.  Here $\widehat U^{\rm st}=\Bzz^\top\widehat{\bs b}^{\rm st}$,
$\widehat{\bs b}^{\rm st}=M_{x,\tau}^{-1}\mathbb X^\top\mathbf W(\mathbf I-\mathbf H_z)\bs\zeta$,
$M_{x,\tau}=M_x+\tau\mathbf I=(N\sum_{i\in\IO}w_i)\mathbf A_{x,\tau}$ with $M_x=(N\sum_{i\in\IO}w_i)\mathbf A_x$, and Lemma~\ref{LEM:Ax} gives
$\lambda_{\min}(M_x)\asymp \sum_{i\in\mathcal{I}_O} w_iN|\tri|^6$ with bounded condition number; since the ridge can only increase the smallest eigenvalue, $\lambda_{\min}(M_{x,\tau})\ge\lambda_{\min}(M_x)$ for every $\tau_U\ge0$ (Lemma \ref{LEM:ridge}), so
$$\|\widehat{\bs b}^{\rm st}\|_2^2\le\lambda_{\min}(M_{x,\tau})^{-2}\|\mathbb X^\top\mathbf W(\mathbf I-\mathbf H_z)\bs\zeta\|_2^2\le\lambda_{\min}(M_x)^{-2}\|\mathbb X^\top\mathbf W(\mathbf I-\mathbf H_z)\bs\zeta\|_2^2.$$
As in \eqref{eq:stnu2},
$$\|\mathbb X^\top\mathbf W(\mathbf I-\mathbf H_z)\bs\zeta\|_2^2\le 2\|\mathbb X^\top\mathbf W\bs\zeta\|_2^2+2\kappa_x^2\|\mathbb Z^\top\mathbf W\bs\zeta\|_2^2$$
with
\[
\begin{aligned}
\kappa_x^2
  &= \|\mathbb X^\top\mathbf W\mathbb Z
       (\mathbb Z^\top\mathbf W\mathbb Z)^{-1}\|_2^2 \le \lambda_{\max}(\mathbb X^\top\mathbf W\mathbb X)/
       \lambda_{\min}(\mathbb Z^\top\mathbf W\mathbb Z)
       \le C|\tri|^{3},
\end{aligned}
\]
now using the bounds
\[
\begin{aligned}
\lambda_{\max}(\mathbb X^\top\mathbf W\mathbb X)
  &\le C\sum_{i\in\mathcal{I}_O} w_iN|\tri|^6, \\
\lambda_{\min}(\mathbb Z^\top\mathbf W\mathbb Z)
  &\ge \lambda_{\min}(M_z)
   \ge c\sum_{i\in\IO}w_iN|\tri|^3.
\end{aligned}
\]
Since
$$\kappa_x^2\,\E\|\mathbb Z^\top\mathbf W\bs\zeta\|_2^2=O(\sum_{i\in\mathcal{I}_O} w_iN^2|\tri|^6+\sum_{i\in\mathcal{I}_O} w_iN|\tri|^3)$$
matches \eqref{eq:Xscore}, the moment bounds \eqref{eq:Zscore}--\eqref{eq:Xscore} give
\[
\E\|\widehat{\bs b}^{\rm st}\|_2^2\le\lambda_{\min}(M_x)^{-2}\,O\big(\sum_{i\in\mathcal{I}_O} w_iN^2|\tri|^6+\sum_{i\in\mathcal{I}_O} w_iN|\tri|^3\big)
\]
By the tensor-product norm equivalence (Lemma~\ref{LEM:DDC-norm} applied to
$\Bz\otimes\mathbf B(\bs z')$) $\|\widehat U^{\rm st}\|_{L_2(\Omega^2)}^2\asymp|\tri|^6\|\widehat{\bs b}^{\rm st}\|_2^2$,
whence
\[
\E\|\widehat U^{\rm st}\|_{L_2(\Omega\times\Omega)}^2=O\!\Big(\tfrac1n+\tfrac{|\tri|^{-3}}{nN}\Big)=O(n^{-1}),
\]
the white term again $o(n^{-1})$ under $N|\tri|^3\to\infty$; Markov's inequality
concludes.
\end{proof}

The dominant (smooth) part of each stochastic error is the parametric rate
$O_P(n^{-1})$, governed by the $n$ subjects rather than the $N$ voxels: the smooth
residual is perfectly correlated within a subject, so voxel averaging does not
reduce its variance.  The white-measurement part is $O_P\{(nN|\tri|^{3})^{-1}\}$, negligible under the standing resolution condition $N^{1/3}|\tri|\to\infty$. Neither bound depends on the ridge parameter $\tau_U$: the ridge only enlarges the smallest eigenvalue of the $U$-block Gram and leaves the $\nu$-block conditioning unchanged (Lemma \ref{LEM:ridge}). The two lemmas furnish the variance term of Theorem~\ref{THM:nuU-rate}; the remaining spline-approximation (bias) term, of order $|\tri|^{2d+2}$, and the ridge-shrinkage bias, of order $r_\tau^2$, are supplied in Section C.4.

We are now in a position to derive the $L_2$ convergence rates of the second-stage spline estimators and, as a consequence, the asymptotic consistency of the two-stage least-squares estimator of $\bs\alpha_0$. The proof follows the oracle decomposition in the proof of Theorem 1 of \cite{cao2020estimation}: the estimation error is split into a stochastic term, bounded by Lemmas~\ref{LEM:stoch-nu} and~\ref{LEM:stoch-U}; a noiseless estimation term driven by the spline-approximation remainder and by the ridge shrinkage of $\bs b$, controlled through the conditioning Lemmas~\ref{LEM:Az}, \ref{LEM:Ax} and \ref{LEM:ridge}; and a pure spline-approximation term, bounded via Lemma~\ref{LEM:appord} under Assumption \ref{as:smoothness}.

\begin{theorem}[Convergence of the second-stage estimators]
\label{THM:nuU-rate}
Suppose Assumptions \ref{as:triangulation}--\ref{as:stage2} hold, and the roughness-penalty parameters and the ridge parameter $\tau_U$ in \eqref{eq:L-stage2} satisfy Assumption \ref{as:rates}. If $N^{1/3}|\tri|\to\infty$ as $N\to\infty$, then the following hold, where $r_\tau=\tau_U/(N|\tri|^6)$ as in \eqref{eq:tau-def}.
\begin{itemize}
\item[(i)] The spline estimator $\widehat{\bs\nu}$ is $L_2(\Omega)$ consistent;
that is,
\[
\big\|\widehat{\bs\nu}-\bs\nu_0\big\|_{L_2(\Omega)}^2
=O_P\!\big\{(\sum_{i\in\mathcal{I}_O} w_i)^{-1}+|\tri|^{2d+2}+r_\tau^2\big\}
=O_P\!\big\{(\sum_{i\in\mathcal{I}_O} w_i)^{-1}+|\tri|^{2d+2}\big\}.
\]
\item[(ii)] The spline estimator $\widehat U$ is $L_2(\Omega\times\Omega)$
consistent; that is,
\[
\big\|\widehat U-U\big\|_{L_2(\Omega\times\Omega)}^2
=O_P\!\big\{(\sum_{i\in\mathcal{I}_O} w_i)^{-1}+|\tri|^{2d+2}+r_\tau^2\big\}
=O_P\!\big\{(\sum_{i\in\mathcal{I}_O} w_i)^{-1}+|\tri|^{2d+2}\big\}.
\]
\end{itemize}
In each display the first expression holds for every $\tau_U\ge0$ and isolates the ridge-shrinkage bias $r_\tau^2$; the second equality uses the ridge condition $r_\tau=O(n^{-1/2}+|\tri|^{d+1})$ of Assumption \ref{as:rates} together with $\sum_{i\in\IO}w_i\asymp n$ (Assumption \ref{as:mar}).
\end{theorem}

\begin{proof}
We first prove (i).

\smallskip
\emph{Step 1 (spline approximation of the targets).} Let $\mathcal Q:\mathcal W^{d+1,\infty}(\Omega)\to\Ssp$ denote the linear quasi-interpolation operator underlying Lemma \ref{LEM:appord}; see Theorem 15.32 of \cite{Lai:Schumaker:07}), which satisfies, for any $g\in\mathcal W^{d+1,\infty}(\Omega)$,
\begin{equation}
\label{eq:QI}
\|g-\mathcal Qg\|_{\infty,\Omega}\le C\,|\tri|^{d+1}\,|g|_{d+1,\infty,\Omega},
\qquad
\|\mathcal Qg\|_{\infty,\Omega}\le C\,\|g\|_{\infty,\Omega}.
\end{equation}
For $\ell=0,\dots,p$, set $\nu^{*}_{\ell}=\mathcal Q\,\nu_{0,\ell}\in\Ssp$, and
write $\bs\nu^{*}(\bs z)=(\mathbf I_{p+1}\otimes\Bz^\top)\,\bs a^{*}$ for the stacked spline coefficient vector $\bs a^{*}$. By Assumption \ref{as:smoothness}, the first inequality in \eqref{eq:QI},
\begin{align}
\|\bs\nu^{*}-\bs\nu_0\|_{\infty}
&=\max_{0\le\ell\le p}\|\nu^{*}_{\ell}-\nu_{0,\ell}\|_{\infty,\Omega}
=O\big(|\tri|^{d+1}\big),
\notag\\
\|\bs\nu^{*}-\bs\nu_0\|_{L_2(\Omega)}^2
&\le(p+1)\,\|\bs\nu^{*}-\bs\nu_0\|_{\infty}^2
=O\big(|\tri|^{2d+2}\big),
\label{eq:nu-approx}
\end{align}
where $\|\bs g\|_{\infty}=\max_{0\le\ell\le p}\|g_{\ell}\|_{\infty,\Omega}$.  For
the kernel, apply $\mathcal Q$ in each argument separately and set
$U^{*}=\mathcal Q_{\bs z}\mathcal Q_{\bs z'}U=\Bzz^\top\bs b^{*}\in\Stp$.  Writing
$U-U^{*}=(U-\mathcal Q_{\bs z}U)+\mathcal Q_{\bs z}(U-\mathcal Q_{\bs z'}U)$ and using the product smoothness in Assumption \ref{as:smoothness} slice-wise, together with the sup-norm boundedness of $\mathcal Q$ in \eqref{eq:QI},
\begin{align}
\|U-U^{*}\|_{\infty,\Omega\times\Omega}
&\le\big\|U-\mathcal Q_{\bs z}U\big\|_{\infty,\Omega\times\Omega}
+C\,\big\|U-\mathcal Q_{\bs z'}U\big\|_{\infty,\Omega\times\Omega}
=O\big(|\tri|^{d+1}\big),
\notag\\
\|U-U^{*}\|_{L_2(\Omega\times\Omega)}^2
&\le\|U-U^{*}\|_{\infty,\Omega\times\Omega}^2=O\big(|\tri|^{2d+2}\big).
\label{eq:U-approx}
\end{align}
We shall also need the size of the approximant coefficient vector $\bs b^{*}$ itself, because the ridge penalty shrinks it. By the sup-norm boundedness in \eqref{eq:QI} applied in each argument, $\|U^{*}\|_{\infty,\Omega\times\Omega}\le C\|U\|_{\infty,\Omega\times\Omega}<\infty$ under Assumption \ref{as:smoothness}, and the tensor-product norm equivalence (Lemma \ref{LEM:DDC-norm} applied to $\Bz\otimes\mathbf B(\bs z')$ via Fubini, as in the proof of Lemma \ref{LEM:Ax}) gives
\begin{equation}
\label{eq:bstar}
\|\bs b^{*}\|_2^2\asymp|\tri|^{-6}\,\|U^{*}\|_{L_2(\Omega\times\Omega)}^2\le C\,|\tri|^{-6},
\qquad\text{i.e.}\qquad
\|\bs b^{*}\|_2=O\big(|\tri|^{-3}\big).
\end{equation}
This is the generic size of a Bernstein--B\'ezier coefficient vector of a bounded function on $\Omega\times\Omega$: $|\mathcal M|^2\asymp|\tri|^{-6}$ coefficients, each $O(1)$.

\smallskip
\emph{Step 2 (error decomposition).}
Recall the oracle fit $\widetilde{\bs\nu}=(\mathbf I_{p+1}\otimes\Bz^\top)\widetilde{\bs a}$
from \eqref{eq:oracle}.  By the triangle inequality,
\begin{equation}
\label{eq:tri-nu}
\|\widehat{\bs\nu}-\bs\nu_0\|_{L_2(\Omega)}
\le\|\widehat{\bs\nu}-\widetilde{\bs\nu}\|_{L_2(\Omega)}
+\|\widetilde{\bs\nu}-\bs\nu^{*}\|_{L_2(\Omega)}
+\|\bs\nu^{*}-\bs\nu_0\|_{L_2(\Omega)}.
\end{equation}
The third term is $O(|\tri|^{d+1})$ by \eqref{eq:nu-approx}, and the first term
satisfies
$\|\widehat{\bs\nu}-\widetilde{\bs\nu}\|_{L_2(\Omega)}^2
=\|\widehat{\bs\nu}^{\rm st}\|_{L_2(\Omega)}^2=O_P\{(\sum_{i\in\mathcal{I}_O} w_i)^{-1}\}$
by Lemma~\ref{LEM:stoch-nu}.  It remains to bound the middle, noiseless
estimation term.

\smallskip
\emph{Step 3 (representation of the noiseless estimation error).}
By \eqref{eq:Ystar}, \eqref{eq:blocks}, and the definitions of
$\bs a^{*},\bs b^{*}$ in Step~1, the noiseless response decomposes as
$\mathbf Y^{*}=\mathbb Z\bs a^{*}+\mathbb X\bs b^{*}+\mathbf R$, where
$\mathbf R=(R_{ij})$ collects the spline-approximation remainder
\begin{equation}
\label{eq:remainder}
R_{ij}
=\mathbf X_i^\top\big\{\bs\nu_0-\bs\nu^{*}\big\}(\bs z_j)
+\int_\Omega\big\{U-U^{*}\big\}(\bs z_j,\bs z')\,\widetilde Y_i(\bs z')\,d\bs z' .
\end{equation}
Plugging this into \eqref{eq:oracle} and noting that
$\{\mathbb Z^\top\mathbf W(\mathbf I-\mathbf H_{x,\tau})\mathbb Z\}^{-1}
\mathbb Z^\top\mathbf W(\mathbf I-\mathbf H_{x,\tau})\mathbb Z\,\bs a^{*}=\bs a^{*}$, while by Lemma \ref{LEM:ridge}(c)
$(\mathbf I-\mathbf H_{x,\tau})\mathbb X\bs b^{*}=\tau\,\mathbb X(\mathbb X^\top\mathbf W\mathbb X+\tau\mathbf I)^{-1}\bs b^{*}$, we obtain
\begin{align}
\bs u:&=\widetilde{\bs a}-\bs a^{*}=\bs u_R+\bs u_\tau,
\notag\\
\bs u_R&=\mathbf A_{z,\tau}^{-1}\,\frac{1}{N\sum_{i\in\IO}w_i}\,
\mathbb Z^\top\mathbf W(\mathbf I-\mathbf H_{x,\tau})\,\mathbf R ,
\quad
\bs u_\tau=\tau\,M_{z,\tau}^{-1}\,\mathbb Z^\top\mathbf W\mathbb X\big(\mathbb X^\top\mathbf W\mathbb X+\tau\mathbf I\big)^{-1}\bs b^{*},
\label{eq:utilde}
\end{align}
with $M_{z,\tau}=(N\sum_{i\in\IO}w_i)\mathbf A_{z,\tau}$ as in the proof of Lemma \ref{LEM:stoch-nu}. When $\tau_U=0$, $\bs u_\tau=\bm0$ and \eqref{eq:utilde} reduces to the exact Frisch--Waugh identity, $(\mathbf I-\mathbf H_x)\mathbb X=\mathbf 0$. For $\tau_U>0$ the ridge lets part of the $U$-block signal $\mathbb X\bs b^{*}$ leak into the $\nu$-block equation, and $\bs u_\tau$ is this leakage term.

\emph{The remainder term $\bs u_R$.} Write $\widetilde{\mathbb Z}_\tau=(\mathbf I-\mathbf H_{x,\tau})\mathbb Z$ and
$\|\bs v\|_{\mathbf W}^2=\bs v^\top\mathbf W\bs v$.  By Lemma \ref{LEM:ridge}(a),
\[
\widetilde{\mathbb Z}_\tau^\top\mathbf W=\mathbb Z^\top\mathbf W(\mathbf I-\mathbf H_{x,\tau})
\qquad\text{and}\qquad
\widetilde{\mathbb Z}_\tau^\top\mathbf W\widetilde{\mathbb Z}_\tau
\preceq\mathbb Z^\top\mathbf W(\mathbf I-\mathbf H_{x,\tau})\mathbb Z
=\Big(N\sum_{i\in\IO}w_i\Big)\mathbf A_{z,\tau},
\]
with equality when $\tau=0$ ($\mathbf H_x$ is then idempotent). Hence \eqref{eq:utilde} reads
\[
\mathbf A_{z,\tau}\bs u_R
  = \Big(N\sum_{i\in\IO}w_i\Big)^{-1}
    \widetilde{\mathbb Z}_\tau^\top\mathbf W\mathbf R,
\]
and the Cauchy--Schwarz inequality in the $\mathbf W$-inner product gives
\begin{equation*}
\bs u_R^\top\mathbf A_{z,\tau}\bs u_R
=\frac{(\widetilde{\mathbb Z}_\tau\bs u_R)^\top\mathbf W\mathbf R}{N\sum_{i\in\IO}w_i}
\le\frac{\|\widetilde{\mathbb Z}_\tau\bs u_R\|_{\mathbf W}\,\|\mathbf R\|_{\mathbf W}}
{N\sum_{i\in\IO}w_i}
\le\bigg(\frac{\bs u_R^\top\mathbf A_{z,\tau}\bs u_R}{N\sum_{i\in\IO}w_i}\bigg)^{1/2}
\|\mathbf R\|_{\mathbf W},
\end{equation*}
where the last inequality uses
$\|\widetilde{\mathbb Z}_\tau\bs u_R\|_{\mathbf W}^2
=\bs u_R^\top\widetilde{\mathbb Z}_\tau^\top\mathbf W\widetilde{\mathbb Z}_\tau\bs u_R
\le(N\sum_{i\in\IO}w_i)\,\bs u_R^\top\mathbf A_{z,\tau}\bs u_R$.  Squaring and cancelling
(the bound being trivial if the left-hand side vanishes), and then applying
the eigenvalue bound $\|\bs u_R\|_2^2\le\lambda_{\min}^{-1}(\mathbf A_{z,\tau})\,
\bs u_R^\top\mathbf A_{z,\tau}\bs u_R$, we arrive at the energy bound
\begin{equation}
\label{eq:energy}
\bs u_R^\top\mathbf A_{z,\tau}\bs u_R\le\frac{\|\mathbf R\|_{\mathbf W}^2}{N\sum_{i\in\IO}w_i},
\qquad
\|\bs u_R\|_2^2\le\big\|\mathbf A_{z,\tau}^{-1}\big\|_2\,
\frac{\|\mathbf R\|_{\mathbf W}^2}{N\sum_{i\in\IO}w_i}
\le\big\|\mathbf A_{z}^{-1}\big\|_2\,
\frac{\|\mathbf R\|_{\mathbf W}^2}{N\sum_{i\in\IO}w_i},
\end{equation}
the last step by the sandwich $\mathbf A_{z,\tau}\succeq\mathbf A_z$ of Lemma \ref{LEM:ridge}(b). Thus the ridge does not enlarge the remainder-driven part of the noiseless error.

\emph{The leakage term $\bs u_\tau$.} By the definition of $\kappa_{z,\tau}$ in Lemma \ref{LEM:ridge}(d),
\begin{align}
\|\bs u_\tau\|_2
&\le\tau\,\lambda_{\min}(M_{z,\tau})^{-1}\,\kappa_{z,\tau}\,\|\bs b^{*}\|_2
\le\frac{\sum_{i\in\IO}w_i\,\tau_U}{c\,\sum_{i\in\IO}w_i\,N|\tri|^{3}}\cdot C|\tri|^{-3/2}\cdot C|\tri|^{-3}
\notag\\
&= C\,\frac{\tau_U}{N|\tri|^{15/2}}
= C\,r_\tau\,|\tri|^{-3/2}
\label{eq:leak}
\end{align}
with probability tending to one, where we used $\tau=\sum_{i\in\IO}w_i\,\tau_U$ from \eqref{eq:tau-def}, $\lambda_{\min}(M_{z,\tau})\ge\lambda_{\min}(M_z)\ge c\sum_{i\in\IO}w_iN|\tri|^3$ (Lemmas \ref{LEM:Az} and \ref{LEM:ridge}), $\kappa_{z,\tau}^2\le C|\tri|^{-3}$ (Lemma \ref{LEM:ridge}), and $\|\bs b^{*}\|_2\le C|\tri|^{-3}$ from \eqref{eq:bstar}.

\smallskip
\emph{Step 4 (order of the remainder).}
By the elementary inequality $(a+b)^2\le 2a^2+2b^2$ together with the
Cauchy--Schwarz bounds
$\{\mathbf X_i^\top\bs g(\bs z_j)\}^2\le(p+1)\|\mathbf X_i\|^2\,\|\bs g\|_{\infty}^2$
and
$\{\int_\Omega(U-U^{*})(\bs z_j,\bs z')\widetilde Y_i(\bs z')d\bs z'\}^2
\le\|U-U^{*}\|_{\infty,\Omega\times\Omega}^2\,\|\widetilde Y_i\|_{L_2(\Omega)}^2$,
we have
\begin{align}
\label{eq:Rsplit}
\|\mathbf R\|_{\mathbf W}^2
& =\sum_{i\in\IO}w_i\sum_{j=1}^N R_{ij}^2
\le 2N(p+1)\,\|\bs\nu_0-\bs\nu^{*}\|_{\infty}^2\sum_{i\in\IO}w_i\|\mathbf X_i\|^2 \notag \\
& \quad +2N\,\|U-U^{*}\|_{\infty,\Omega\times\Omega}^2
\sum_{i\in\IO}w_i\big\|\widetilde Y_i\big\|_{L_2(\Omega)}^2 .
\end{align}
The functions $\|\mathbf X\|^2$ and $\|\widetilde Y\|_{L_2(\Omega)}^2$ depend only on
the fully observed covariates and surrogate, and they have finite means: the former by Assumption \ref{as:design}, and the latter by Assumptions\ref{as:smoothness}, \ref{as:design} and \ref{as:relevance}, since $\widetilde Y=\mathbf X^\top\bs\beta_0+\varepsilon_S$ with $\sup_{\bs z\in\Omega}\E\{\varepsilon_S^2(\bs z)\}<\infty$.  Hence the
inverse-probability identity \eqref{eq:debias} yields
$\sum_{i\in\IO}w_i\|\mathbf X_i\|^2=O_P\big(\sum_{i\in\IO}w_i\big)$ and
$\sum_{i\in\IO}w_i\|\widetilde Y_i\|_{L_2(\Omega)}^2
=O_P\big(\sum_{i\in\IO}w_i\big)$, so that, combining \eqref{eq:nu-approx},
\eqref{eq:U-approx}, and \eqref{eq:Rsplit},
\begin{equation}
\label{eq:Rbound}
\|\mathbf R\|_{\mathbf W}^2
=O_P\Big\{\Big(N\sum_{i\in\IO}w_i\Big)\,|\tri|^{2d+2}\Big\}.
\end{equation}

\smallskip
\emph{Step 5 (combining).}
By Lemma~\ref{LEM:Az}, $\|\mathbf A_z^{-1}\|_2=O_P(|\tri|^{-3})$.  Together with
\eqref{eq:energy} and \eqref{eq:Rbound},
$\|\bs u_R\|_2^2
=O_P\big(|\tri|^{-3}\cdot|\tri|^{2d+2}\big)=O_P\big(|\tri|^{2d-1}\big)$, while \eqref{eq:leak} gives $\|\bs u_\tau\|_2^2=O_P(r_\tau^2|\tri|^{-3})$, and the
trivariate-spline norm equivalence (Lemma~\ref{LEM:DDC-norm}) gives
\begin{equation}
\label{eq:bias-nu}
\|\widetilde{\bs\nu}-\bs\nu^{*}\|_{L_2(\Omega)}^2
\asymp|\tri|^{3}\,\|\widetilde{\bs a}-\bs a^{*}\|_2^2
\le2|\tri|^{3}\big(\|\bs u_R\|_2^2+\|\bs u_\tau\|_2^2\big)
=O_P\big(|\tri|^{2d+2}+r_\tau^2\big).
\end{equation}
Combining \eqref{eq:tri-nu}, \eqref{eq:nu-approx}, \eqref{eq:bias-nu}, and
Lemma~\ref{LEM:stoch-nu}, we obtain
\[
\|\widehat{\bs\nu}-\bs\nu_0\|_{L_2(\Omega)}^2=O_P\Big\{\Big(\sum_{i\in\IO}w_i\Big)^{-1}+|\tri|^{2d+2}+r_\tau^2\Big\},
\]
which is the first expression in (i); the ridge condition $r_\tau=O(n^{-1/2}+|\tri|^{d+1})$ of Assumption \ref{as:rates}, with $\sum_{i\in\IO}w_i\asymp n$, makes $r_\tau^2=O\{(\sum_{i\in\IO}w_i)^{-1}+|\tri|^{2d+2}\}$, which
proves (i).

\smallskip
Next, we prove (ii).  By the triangle inequality,
\begin{equation}
\label{eq:tri-U}
\|\widehat U-U\|_{L_2(\Omega\times\Omega)}
\le\|\widehat U-\widetilde U\|_{L_2(\Omega\times\Omega)}
+\|\widetilde U-U^{*}\|_{L_2(\Omega\times\Omega)}
+\|U^{*}-U\|_{L_2(\Omega\times\Omega)},
\end{equation}
where the first term is $O_P(n^{-1/2})$ by Lemma~\ref{LEM:stoch-U} and the
third is $O(|\tri|^{d+1})$ by \eqref{eq:U-approx}.  For the middle term, note
that $(\mathbf I-\mathbf H_z)\mathbb Z=\mathbf 0$ (the $\nu$-block carries no ridge, so the Frisch--Waugh annihilation is exact in this direction), so that applying the algebra of
Step~3 to $\mathbf Y^{*}=\mathbb Z\bs a^{*}+\mathbb X\bs b^{*}+\mathbf R$, with the
\emph{same} remainder $\mathbf R$ as in \eqref{eq:remainder}, yields
\begin{align}
\bs v:=\widetilde{\bs b}-\bs b^{*}
&=\big(M_x+\tau\mathbf I\big)^{-1}\big\{M_x\bs b^{*}+\mathbb X^\top\mathbf W(\mathbf I-\mathbf H_z)\mathbf R\big\}-\bs b^{*}
=\bs v_R+\bs v_\tau ,
\notag\\
\bs v_R&=\big(M_x+\tau\mathbf I\big)^{-1}\mathbb X^\top\mathbf W(\mathbf I-\mathbf H_z)\,\mathbf R ,
\qquad
\bs v_\tau=-\tau\big(M_x+\tau\mathbf I\big)^{-1}\bs b^{*},
\label{eq:vtilde}
\end{align}
where $M_x=\mathbb X^\top\mathbf W(\mathbf I-\mathbf H_z)\mathbb X=(N\sum_{i\in\IO}w_i)\mathbf A_x$. The term $\bs v_\tau$ is the ridge shrinkage bias of the tensor-product coefficients; it vanishes when $\tau_U=0$, in which case \eqref{eq:vtilde} reduces to $\bs v=\mathbf A_x^{-1}(N\sum_{i\in\IO}w_i)^{-1}\mathbb X^\top\mathbf W(\mathbf I-\mathbf H_z)\mathbf R$.

\emph{The remainder term $\bs v_R$.} Writing $\widetilde{\mathbb X}=(\mathbf I-\mathbf H_z)\mathbb X$, the
$\mathbf W$-self-adjointness and idempotence of $\mathbf H_z$ give
$\widetilde{\mathbb X}^\top\mathbf W=\mathbb X^\top\mathbf W(\mathbf I-\mathbf H_z)$ and $\widetilde{\mathbb X}^\top\mathbf W\widetilde{\mathbb X}=M_x$, so that $(M_x+\tau\mathbf I)\bs v_R=\widetilde{\mathbb X}^\top\mathbf W\mathbf R$ and the argument leading to
\eqref{eq:energy} applies:
\begin{equation*}
\bs v_R^\top(M_x+\tau\mathbf I)\bs v_R
=(\widetilde{\mathbb X}\bs v_R)^\top\mathbf W\mathbf R
\le\big(\bs v_R^\top M_x\bs v_R\big)^{1/2}\|\mathbf R\|_{\mathbf W}
\le\big\{\bs v_R^\top(M_x+\tau\mathbf I)\bs v_R\big\}^{1/2}\|\mathbf R\|_{\mathbf W},
\end{equation*}
whence $\bs v_R^\top(M_x+\tau\mathbf I)\bs v_R\le\|\mathbf R\|_{\mathbf W}^2$ and
\begin{equation*}
\|\bs v_R\|_2^2
\le\frac{\|\mathbf R\|_{\mathbf W}^2}{\lambda_{\min}(M_x)+\tau}
\le\big\|\mathbf A_x^{-1}\big\|_2\,\frac{\|\mathbf R\|_{\mathbf W}^2}{N\sum_{i\in\IO}w_i}
=O_P\big(|\tri|^{-6}\cdot|\tri|^{2d+2}\big)
=O_P\big(|\tri|^{2d-4}\big),
\end{equation*}
using $\|\mathbf A_x^{-1}\|_2=O_P(|\tri|^{-6})$ from Lemma~\ref{LEM:Ax} and the
remainder bound \eqref{eq:Rbound}. As for $\bs u_R$, the ridge does not enlarge this term.

\emph{The shrinkage term $\bs v_\tau$.} Since $\lambda_{\min}(M_x+\tau\mathbf I)=\lambda_{\min}(M_x)+\tau$,
\begin{equation}
\label{eq:shrink}
\begin{aligned}
\|\bs v_\tau\|_2
  &\le \frac{\tau}{\lambda_{\min}(M_x)+\tau}\,\|\bs b^{*}\|_2
  \le \frac{\tau}{\lambda_{\min}(M_x)}\,\|\bs b^{*}\|_2 \le \frac{\sum_{i\in\IO}w_i\,\tau_U}
             {c\,\sum_{i\in\IO}w_i\,N|\tri|^{6}}
    \cdot C|\tri|^{-3} \\
  & = C\,r_\tau\,|\tri|^{-3},
\end{aligned}
\end{equation}
with probability tending to one, by $\lambda_{\min}(M_x)\ge c\sum_{i\in\IO}w_iN|\tri|^6$ (Lemma \ref{LEM:Ax}), $\tau=\sum_{i\in\IO}w_i\,\tau_U$, and \eqref{eq:bstar}. The first inequality in \eqref{eq:shrink} also shows that the shrinkage bias is never larger than $\|\bs b^{*}\|_2$ itself, whatever the value of $\tau_U$.

The tensor-product norm equivalence
(Lemma~\ref{LEM:DDC-norm}, applied in each argument via Fubini's
theorem, as in the proof of Lemma~\ref{LEM:Ax}) then gives
\begin{equation*}
\|\widetilde U-U^{*}\|_{L_2(\Omega\times\Omega)}^2
\asymp|\tri|^{6}\,\|\widetilde{\bs b}-\bs b^{*}\|_2^2
\le2|\tri|^{6}\big(\|\bs v_R\|_2^2+\|\bs v_\tau\|_2^2\big)
=O_P\big(|\tri|^{2d+2}+r_\tau^2\big).
\end{equation*}
Combining the three bounds in \eqref{eq:tri-U} gives $\|\widehat U-U\|_{L_2(\Omega\times\Omega)}^2=O_P\{(\sum_{i\in\IO}w_i)^{-1}+|\tri|^{2d+2}+r_\tau^2\}$, and the ridge condition of Assumption \ref{as:rates} again absorbs $r_\tau^2$ into the leading terms. This establishes (ii), and the proof
of Theorem~\ref{THM:nuU-rate} is complete.
\end{proof}

\begin{remark}
\label{rem:typo}
The squared $L_2$ error bounds in Theorem~\ref{THM:nuU-rate} have approximation term $|\tri|^{2d+2}$, matching the first-stage rate in Theorem~\ref{THM:beta-convergence}. Taking square roots gives
\[
\|\widehat{\bs\nu}-\bs\nu_0\|_{L_2(\Omega)},\quad
\|\widehat U-U\|_{L_2(\Omega\times\Omega)}
=O_P\big\{(\sum_{i\in\IO}w_i)^{-1/2}+|\tri|^{d+1}\big\}.
\]
\end{remark}

\medskip
The two-stage least-squares (2SLS) estimator of $\bs\alpha_0$ is
\begin{equation}
\label{eq:alpha-2sls}
\widehat{\bs\alpha}(\bs z)
=\widehat{\bs\nu}(\bs z)
+\int_\Omega\widehat U(\bs z,\bs z')\,\widehat{\bs\beta}(\bs z')\,d\bs z',
\end{equation}
where $\widehat{\bs\beta}$ is the first-stage estimator. The next theorem shows that $\widehat{\bs\alpha}$ is consistent whenever its three
ingredients are, and quantifies how the plug-in errors combine. The statement is deterministic given the rates $r_\nu$, $r_\beta$, $r_U$, so it applies verbatim to any plug-in estimators attaining these rates.

\begin{theorem}[Consistency of the 2SLS estimator $\widehat{\bs\alpha}$]
\label{THM:alpha}
Suppose that $U\in L_2(\Omega\times\Omega)$ and
$\|\bs\beta_0\|_{L_2(\Omega)}<\infty$, and that
\[
\big\|\widehat{\bs\nu}-\bs\nu_0\big\|_{L_2(\Omega)}=O_P(r_\nu),
\qquad
\big\|\widehat{\bs\beta}-\bs\beta_0\big\|_{L_2(\Omega)}=O_P(r_\beta),
\qquad
\big\|\widehat U-U\big\|_{L_2(\Omega\times\Omega)}=O_P(r_U).
\]
Then
\[
\big\|\widehat{\bs\alpha}-\bs\alpha_0\big\|_{L_2(\Omega)}
=O_P\big(r_\nu+r_\beta+r_U+r_U r_\beta\big).
\]
In particular, if $r_\nu\to0$, $r_\beta\to0$, and $r_U\to0$, then
$\|\widehat{\bs\alpha}-\bs\alpha_0\|_{L_2(\Omega)}=o_P(1)$; that is,
$\widehat{\bs\alpha}$ is $L_2(\Omega)$ consistent.
\end{theorem}

\begin{proof}
By the definition of $\bs\nu_0$ following \eqref{eq:ssfr-adj}, we have
$$\bs\alpha_0(\bs z)
=\bs\nu_0(\bs z)+\int_\Omega U(\bs z,\bs z')\,\bs\beta_0(\bs z')\,d\bs z'.$$
Therefore,
\begin{equation*}
\widehat{\bs\alpha}(\bs z)-\bs\alpha_0(\bs z)
=\big\{\widehat{\bs\nu}(\bs z)-\bs\nu_0(\bs z)\big\}
+\int_\Omega\big\{\widehat U(\bs z,\bs z')\,\widehat{\bs\beta}(\bs z')
-U(\bs z,\bs z')\,\bs\beta_0(\bs z')\big\}\,d\bs z' .
\end{equation*}
The second term can be decomposed as
$\bs T_1(\bs z)+\bs T_2(\bs z)+\bs T_3(\bs z)$, where
\begin{align*}
\bs T_1(\bs z)
&=\int_\Omega\big\{\widehat U(\bs z,\bs z')-U(\bs z,\bs z')\big\}\,
\bs\beta_0(\bs z')\,d\bs z',\\
\bs T_2(\bs z)
&=\int_\Omega U(\bs z,\bs z')\,
\big\{\widehat{\bs\beta}(\bs z')-\bs\beta_0(\bs z')\big\}\,d\bs z',\\
\bs T_3(\bs z)
&=\int_\Omega\big\{\widehat U(\bs z,\bs z')-U(\bs z,\bs z')\big\}
\big\{\widehat{\bs\beta}(\bs z')-\bs\beta_0(\bs z')\big\}\,d\bs z' .
\end{align*}
For a vector-valued function the $L_2(\Omega)$ norm is taken componentwise,
$\|\bs T\|_{L_2(\Omega)}^2=\sum_{\ell=0}^p\|T_{\ell}\|_{L_2(\Omega)}^2$.  For
$\bs T_1$, the Cauchy--Schwarz inequality applied to the inner integral gives,
for each component $\ell=0,\dots,p$ and each $\bs z\in\Omega$,
\begin{equation*}
|T_{1,\ell}(\bs z)|
=\Big|\int_\Omega\big\{\widehat U(\bs z,\bs z')-U(\bs z,\bs z')\big\}\,
\beta_{0,\ell}(\bs z')\,d\bs z'\Big|
\le\big\|\widehat U(\bs z,\cdot)-U(\bs z,\cdot)\big\|_{L_2(\Omega)}\,
\|\beta_{0,\ell}\|_{L_2(\Omega)} .
\end{equation*}
Squaring, integrating over $\bs z\in\Omega$, and summing over $\ell$,
\begin{equation}
\label{eq:T1}
\begin{aligned}
\|\bs T_1\|_{L_2(\Omega)}^2
  &\le \Big(\int_\Omega
    \big\|\widehat U(\bs z,\cdot)-U(\bs z,\cdot)\big\|_{L_2(\Omega)}^2
    \,d\bs z\Big)\,\|\bs\beta_0\|_{L_2(\Omega)}^2 \\
  &= \big\|\widehat U-U\big\|_{L_2(\Omega\times\Omega)}^2
    \,\|\bs\beta_0\|_{L_2(\Omega)}^2
    = O_P\big(r_U^2\big),
\end{aligned}
\end{equation}
since $\|\bs\beta_0\|_{L_2(\Omega)}<\infty$.  The same argument, with the roles
of the fixed and estimated factors exchanged, gives
\begin{align}
\|\bs T_2\|_{L_2(\Omega)}
&\le\|U\|_{L_2(\Omega\times\Omega)}\,
\big\|\widehat{\bs\beta}-\bs\beta_0\big\|_{L_2(\Omega)}
=O_P(r_\beta),
\notag\\
\|\bs T_3\|_{L_2(\Omega)}
&\le\big\|\widehat U-U\big\|_{L_2(\Omega\times\Omega)}\,
\big\|\widehat{\bs\beta}-\bs\beta_0\big\|_{L_2(\Omega)}
=O_P(r_U r_\beta),
\label{eq:T23}
\end{align}
the first using $U\in L_2(\Omega\times\Omega)$.  Combining the above bounds via
the triangle inequality,
\[
\begin{aligned}
\|\widehat{\bs\alpha}-\bs\alpha_0\|_{L_2(\Omega)}
  &\le \|\widehat{\bs\nu}-\bs\nu_0\|_{L_2(\Omega)}
  + \|\bs T_1\|_{L_2(\Omega)}
  + \|\bs T_2\|_{L_2(\Omega)}
  + \|\bs T_3\|_{L_2(\Omega)} \\
  &= O_P\big(r_\nu+r_U+r_\beta+r_U r_\beta\big),
\end{aligned}
\]
which proves the stated bound; the consistency claim is immediate.
\end{proof}

\begin{corollary}[Convergence rate of the 2SLS estimator]
\label{COR:alpha-rate}
Suppose the assumptions of Theorem \ref{THM:beta-convergence} and Theorem~\ref{THM:nuU-rate} hold, with the roughness parameters chosen so that
the penalty-induced bias terms in Theorem \ref{THM:beta-convergence} are dominated (cf.\ Remark \ref{rem:beta-rate}), and with the ridge parameter $\tau_U$ of \eqref{eq:L-stage2} satisfying the condition $r_\tau=O(n^{-1/2}+|\tri|^{d+1})$ of Assumption \ref{as:rates}.
Then
\[
\big\|\widehat{\bs\alpha}-\bs\alpha_0\big\|_{L_2(\Omega)}
=O_P\big\{(\sum_{i\in\mathcal{I}_O} w_i)^{-1/2}+|\tri|^{d+1}\big\}.
\]
\end{corollary}

\begin{proof}
Under Assumption \ref{as:smoothness}, the hypotheses $\|\bs\beta_0\|_{L_2(\Omega)}<\infty$ and
$U\in L_2(\Omega\times\Omega)$ of Theorem~\ref{THM:alpha} hold automatically.
Theorem \ref{THM:beta-convergence} gives $r_\beta=n^{-1/2}+|\tri|^{d+1}$, and Theorem \ref{THM:nuU-rate} (in root-mean-square units; Remark~\ref{rem:typo}) gives $r_\nu=r_U=(\sum_{i\in\mathcal{I}_O} w_i)^{-1/2}+|\tri|^{d+1}$, the ridge-shrinkage bias $r_\tau$ being of no larger order by Assumption \ref{as:rates}.  The cross term $r_U r_\beta=O\{(n^{-1/2}+|\tri|^{d+1})\{(\sum_{i\in\mathcal{I}_O} w_i)^{-1/2}+|\tri|^{d+1}\}\}$ is of smaller order, and Theorem \ref{THM:alpha} yields the conclusion.
\end{proof}

Corollary \ref{COR:alpha-rate} shows that the 2SLS estimator
$\widehat{\bs\alpha}$ attains the same $L_2$ convergence rate as each of its
ingredients: the rate is governed by the spline-approximation bias
$|\tri|^{d+1}$ and the parametric variance order $(\sum_{i\in\mathcal{I}_O} w_i)^{-1/2}$, exactly as in the
first stage. The ridge penalty $\tau_U\bs b^\top\bs b$ in \eqref{eq:L-stage2} does not alter this conclusion. Its only effects on the theory are (i) a shrinkage bias of root-mean-square order $r_\tau=\tau_U/(N|\tri|^6)$ in both second-stage estimators, which is of no larger order than the spline-approximation and stochastic errors under Assumption \ref{as:rates} and enters Theorem \ref{THM:alpha} only through $r_\nu$ and $r_U$, and (ii) an improvement in the conditioning of the $U$-block Gram matrix (Lemma \ref{LEM:ridge}), which is what motivates its use in practice: the tensor-product block has $|\mathcal M|^2$ coefficients, and the ridge regularizes it in finite samples without affecting the asymptotic rate. Without the ridge condition, the first expressions in Theorem \ref{THM:nuU-rate} still hold and the rate of $\widehat{\bs\alpha}$ becomes $O_P\{(\sum_{i\in\IO}w_i)^{-1/2}+|\tri|^{d+1}+r_\tau\}$.

Theorem~\ref{THM:alpha} uses the integrated rate
$\|\widehat U-U\|_{L_2(\Omega\times\Omega)}=O_P(r_U)$
established in Theorem~\ref{THM:nuU-rate}.
The bound in \eqref{eq:T1} follows by applying Cauchy--Schwarz for each $\bs z$ and then integrating over $\bs z$.

\subsection{Efficiency Gains from Synthetic Surrogate Information}
\label{ssec:ssfr_efficiency}

The synthetic surrogate data can improve the efficiency of SSFR by providing additional information about the component of the target imaging phenotype that is not explained by the observed covariates. Importantly, the surrogate image is not treated as a substitute for the target image. Instead, it enters SSFR as an auxiliary functional predictor, while the second-stage regression continues to calibrate the surrogate information using the observed target images.

Recall the joint model
\[
Y_i(\bs z) = \mathbf X_i^\top\bs\alpha_0(\bs z) + \varepsilon_{O,i}(\bs z), \qquad i \in \mathcal I_O,
\]
and
\[
\widetilde Y_i(\bs z) = \mathbf X_i^\top\bs\beta_0(\bs z) + \varepsilon_{S,i}(\bs z), \qquad i=1,\ldots,n.
\]
Under the joint Gaussian co-kriging model,
\[
E\{\varepsilon_{O,i}(\bs z) \mid \varepsilon_{S,i}\} = \int_\Omega U(\bs z,\bs z') \varepsilon_{S,i}(\bs z')\,d\bs z',
\]
where $U = \Sigma_{OS}\Sigma_{SS}^{-1}$.
Consequently, the target residual admits the decomposition
\[
\varepsilon_{O,i}(\bs z) = \int_\Omega U(\bs z,\bs z') \varepsilon_{S,i}(\bs z')\,d\bs z' + \zeta_i(\bs z),
\]
where $\zeta_i$ is the residual component that remains after conditioning on the surrogate residual process. Its covariance operator is
\begin{equation}
\Sigma_\zeta = \Sigma_{OO} - \Sigma_{OS}\Sigma_{SS}^{-1}\Sigma_{SO}.
\label{eq:conditional-covariance}
\end{equation}
Since $\Sigma_{OS}\Sigma_{SS}^{-1}\Sigma_{SO}$ is positive semidefinite,
\begin{equation}
\Sigma_\zeta \preceq \Sigma_{OO}.
\label{eq:variance-reduction}
\end{equation}
Thus, at the population level, conditioning on informative surrogate information cannot increase the residual covariance relative to an observed-target-only analysis.

This covariance reduction provides an intuitive explanation for the efficiency advantage of SSFR. Let $h$ be any admissible direction in the target residual space and define the direction-specific residual
explanation ratio
\begin{equation}
R_h^2 = \frac{ \langle h,\Sigma_{OS}\Sigma_{SS}^{-1}\Sigma_{SO}h \rangle}{ \langle h,\Sigma_{OO}h \rangle }.
\end{equation}
Then
\begin{equation}
\operatorname{Var}\{\langle h,\zeta_i\rangle\} = \operatorname{Var}\{\langle h,\varepsilon_{O,i} \rangle\} (1-R_h^2).
\end{equation}
Hence, surrogate information is most useful in directions for which the surrogate residual process explains a substantial fraction of the target residual variation. In the scalar analogue, if $\operatorname{Corr}(\varepsilon_O,\varepsilon_S)=\rho$, then
\[
\operatorname{Var}(\zeta) = \sigma_O^2(1-\rho^2),
\]
which makes the variance-reduction mechanism transparent.

A second source of potential efficiency gain is sample size. The surrogate images are available for all $n$ subjects, whereas the target images are observed only for $n_O$ subjects. Thus, the first-stage estimator $\widehat{\bs\beta}$ is based on the full surrogate cohort, while the calibration of $\widehat{\bs\nu}$ and $\widehat U$ uses the observed target cohort through the second-stage regression. When $n_O/n\to\pi_O\in(0,1]$, the first-stage estimation error is of order $n^{-1/2}$, whereas the second-stage stochastic error is of order $n_O^{-1/2}$. Consequently, when the surrogate cohort is substantially larger than the observed target cohort, the contribution of first-stage estimation uncertainty can be relatively small.

Importantly, efficiency improvement does not require $\bs\beta_0(\bs z)=\bs\alpha_0(\bs z)$. SSFR permits
\begin{equation}
\bs\beta_0(\bs z) \neq \bs\alpha_0(\bs z)
\end{equation}
and estimates the discrepancy-adjusted component
\begin{equation}
\bs\nu_0(\bs z) = \bs\alpha_0(\bs z) - \int_\Omega U(\bs z,\bs z') \bs\beta_0(\bs z')\,d\bs z'.
\end{equation}
Thus, the surrogate information is used to reduce estimation variability rather than to impose the potentially incorrect restriction that the surrogate and target coefficient functions coincide.

The theoretical results established in Section 4 of the main paper and Supplement Section~C characterize the consistency and convergence rates of the SSFR estimator. We do not pursue a separate closed-form asymptotic variance expansion or a pointwise asymptotic normality result here. Equation (10) in the main text is used as an approximate relationship for summarizing the potential efficiency effects of surrogate information. A formal variance derivation for the full estimator would require additional analysis to account for the increasing-dimensional spline spaces and the propagation of first-stage estimation error. Instead, we use the weighted bootstrap procedure described in Section 5 to quantify uncertainty and construct simultaneous confidence corridors for the spatially varying coefficient functions.

Overall, the efficiency interpretation of SSFR is therefore qualitative but clear: synthetic surrogate information can reduce the residual variation that must be learned from the observed target images, while the second-stage calibration preserves consistency even when the surrogate coefficient functions are misspecified. The magnitude of the efficiency gain is expected to increase with the surrogate explanatory power and with the size of the surrogate cohort.

\bibliographystyle{plain}
\bibliography{reference}

\end{document}